\documentclass[a4paper,12pt]{article}
\usepackage{ulem}
\usepackage{subcaption}

\usepackage[margin=1in]{geometry}    
\usepackage{graphicx}               
\usepackage{sectsty}                
\usepackage{fancyhdr}               
\usepackage{hyperref}               
\usepackage{xcolor}                 
\usepackage{listings}               
\usepackage{enumitem}               
\usepackage{caption}                
\usepackage{tikz}                   
\usepackage{float}
\usepackage{array}                  
\usepackage{xcolor}                 
\usepackage{caption}                
\usepackage{booktabs}               
\usepackage{tabularx}               
\usepackage{makecell} 
\usepackage{listings}
\usepackage{graphicx} 
\usepackage{rotating} 
\usepackage[export]{adjustbox} 
\usepackage{atbegshi}
\usepackage{listings}
\hypersetup{
    colorlinks=true,
    linkcolor=blue,
    filecolor=magenta,
    urlcolor=cyan,
}

\sectionfont{\color{blue}\bfseries\Large}
\subsectionfont{\color{blue}\bfseries\normalsize}
\subsubsectionfont{\color{blue}\bfseries\small}
\newcolumntype{C}[1]{>{\centering\textbackslash}m{#1}}  

\AtBeginShipout{%
  \AtBeginShipoutAddToBoxForeground{%
    \begin{tikzpicture}[remember picture,overlay]
      \node[opacity=0.15, rotate=45] at (current page.center) {
           \scalebox{8}{\textbf{\textcolor{black}{}}} 
      };
    \end{tikzpicture}%
  }%
}

\lstdefinelanguage{yaml}{
  keywords={true, false, null, yes, no, description, title, logsource, detection, condition, fields, tags, falsepositives, mitre, author, date, id, status},
  keywordstyle=\color{blue}\bfseries,
  basicstyle=\ttfamily\small,        
  comment=[l]{\#},                   
  commentstyle=\color{gray}\itshape, 
  morestring=[b]',                   
  morestring=[b]",                   
  stringstyle=\color{red},           
}

\begin{document}

\begin{titlepage}
        \centering
    \vspace*{1cm}
    {\Large\textcolor{black}{\textbf{Lazarus Group Targets Crypto-Wallets and Financial Data while employing new Tradecrafts}}} \\[1cm]
    Alessio Di Santo (alessio.disanto@graduate.univaq.it)\\
    Università degli Studi dell’Aquila, L’Aquila, Abruzzo, Italy  \\
    \textbf{Date:} November 26,2024 \\
    \vfill
    {\Large\textcolor{gray!60}{\it "Non videmus ea quae mox futura sunt"}} \\[0.5cm]
    {\small\textcolor{gray!60}{(We do not see the things that will soon be) — Marcus Tullius Cicero}} \\
    \vfill
\end{titlepage}

\tableofcontents
\newpage

\section{Executive Summary}

\section{Introduction}
\subsection{Objective}
The objective of this \textit{Malware Analysis Report} is to provide an in-depth understanding of the behavior, architecture, and intent of a malicious software instance. At its core, this report serves as a crucial tool for identifying the characteristics and operations of the \textit{threat}, offering detailed insights that can be used to map the broader attack landscape. By dissecting the capabilities and infrastructure of the malware, analysts are able to build a clear picture of its functionality, origin, and potential impact.

Mapping a \textit{threat} accurately is of paramount importance for defenders. A well-crafted malware analysis report helps connect individual malicious artifacts with broader attack campaigns and identifies common \textit{Techniques, Tactics, and Procedures} (\textit{TTPs}) employed by adversaries. This intelligence feeds into a larger knowledge base that allows cybersecurity teams to understand how threats evolve, recognize new campaigns with similar signatures, and anticipate potential next steps of attackers. The report is not merely an exercise in detailing technical specifics but also a way of enriching the collective understanding of a \textit{Threat Actor}'s capabilities, motivations, and behaviors.

Actionable \textit{Threat Intelligence} derived from malware analysis is particularly valuable because it enables proactive defenses. With a structured understanding of the malware’s \textit{Indicators of Compromise} (\textit{IOCs}), behavioral patterns, and infrastructure, \textit{Threat Hunting} and \textit{Monitoring} teams are equipped with the context needed to seek out malicious activity before it fully manifests. \textit{Threat Hunters} can leverage this intelligence to identify adversarial presence across their environments more effectively, while \textit{Monitoring} teams can enhance detection logic and fine-tune alerts to identify these threats more accurately in real time. This coordinated approach bolsters an organization’s defense posture, making it possible to detect and respond to even well-structured, sophisticated threats that are designed to evade traditional security mechanisms.

Ultimately, a comprehensive malware analysis report provides not only a retrospective view of what a threat has done but also equips defenders with the tools and knowledge to better \textit{predict}, \textit{detect}, and \textit{prevent} future attacks. This knowledge empowers security teams to make informed decisions, prioritize vulnerabilities, and improve their capabilities against \textit{Advanced Persistent Threats} (\textit{APTs}).

\subsection{Infection Chain}

\begin{figure}[H]
    \centering
    \includegraphics[width=1\linewidth,frame]{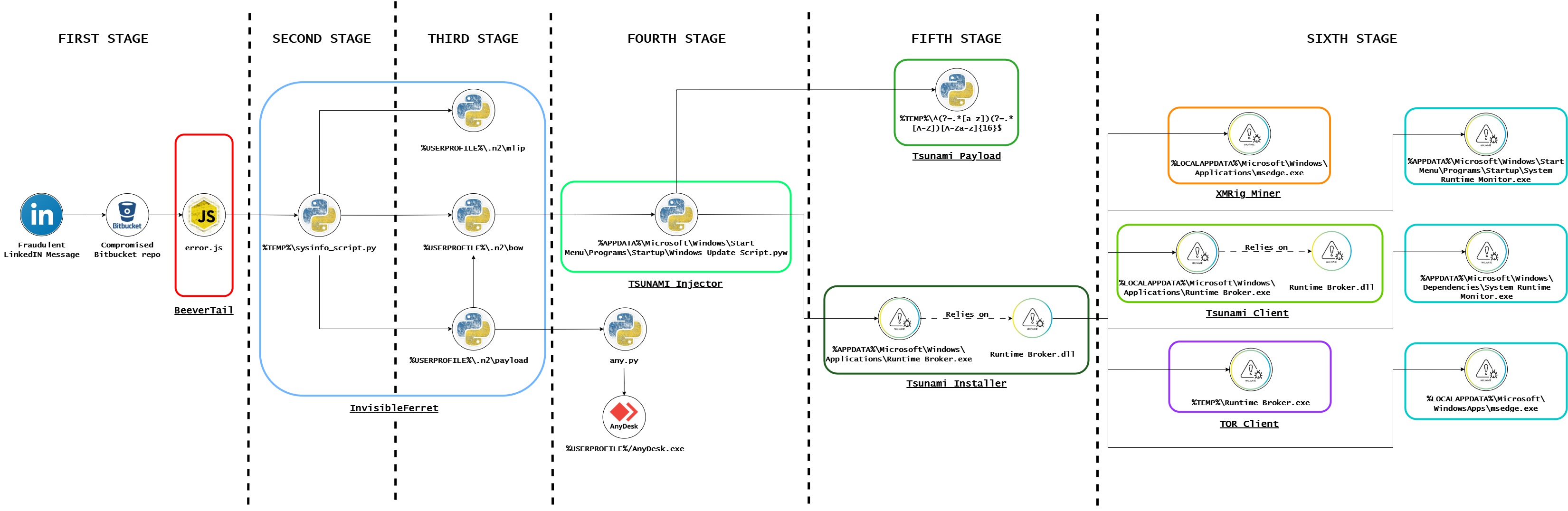}
    \caption{Infection Chain Diagram}
    \label{fig:00}
\end{figure}

\newpage

\section{Methodology}
Analyzing the malware involved a comprehensive approach utilizing both static and dynamic analysis techniques to thoroughly understand its structure, behavior, and potential impact. By combining these two approaches, it is possible to gain a comprehensive understanding of the malware's capabilities and objectives. Static analysis provided insights into its structure and obfuscation methods, while dynamic analysis revealed its real-time behavior and interactions with the system. This dual approach was essential in developing effective detection and mitigation strategies against this sophisticated threat.
\subsection{Static Analysis}
Static analysis is a fundamental technique in malware analysis that involves examining the code of malicious software without executing it. This approach focuses on understanding the structure, logic, and intent of the malware through methods such as \textit{disassembling}, \textit{decompiling}, and reviewing its binary or script content. By analyzing the static properties of malware, such as strings, embedded resources, file headers, and imported functions, researchers can gather valuable insights into its capabilities, communication patterns, and potential targets.

The main goal of static analysis is to dissect the malware's inner workings, identify hardcoded \textit{Indicators of Compromise} (\textit{IoCs}) like IP addresses, URLs, or file paths, and infer its behavior without the risk of executing harmful code. This method is particularly useful for uncovering obfuscation techniques, encrypted payloads, and multi-stage architectures, which are often employed by modern malware to hinder direct analysis.

However, static analysis comes with its challenges. Advanced malware frequently uses obfuscation, packing, or encryption to conceal its code and deter examination. Analysts must rely on specialized tools and techniques, such as deobfuscation scripts, unpackers, and cryptographic analysis, to overcome these barriers. Moreover, analyzing assembly-level or machine code demands a high level of expertise, as the complexity of the malware's logic can obscure its true intent.

Despite its limitations, static analysis is invaluable as it allows analysts to preemptively assess a malware sample’s potential threats, providing critical intelligence without the inherent risks of execution. Combined with dynamic analysis, it forms a comprehensive approach to malware investigation, equipping defenders with the necessary understanding to develop effective detection and mitigation strategies.

\subsection{Dynamic Analysis}
Dynamic analysis is a cornerstone of malware analysis, enabling researchers to observe the behavior of malicious software in real-time by executing it within a controlled, isolated environment. This approach is particularly valuable for analyzing modern malware that employs sophisticated \textit{obfuscation techniques}, rendering static analysis alone insufficient. By simulating realistic conditions, analysts can examine how malware interacts with the file system, registry, processes, network, and system \textit{API}s, providing direct insights into its functionality and intent.

The objective of dynamic analysis is to uncover the behavioral profile of the malware, revealing actions such as \textit{data exfiltration}, \textit{Command-and-Control} communication, \textit{credential theft}, and \textit{persistence mechanisms}. It also aids in identifying \textit{Indicators of Compromise} (\textit{IoCs}), such as IP addresses, domains, and modified system configurations, which are crucial for detection and response efforts. This method is not without challenges, as modern malware often incorporates \textit{anti-analysis techniques} designed to detect and evade \textit{Sandboxed Environments}, \textit{Virtual Machines}, or \textit{Debugging Tools}. These measures include delaying execution, checking for artifacts indicative of analysis environments, and employing runtime obfuscation to conceal its activities.

Despite these difficulties, dynamic analysis remains a critical tool in the fight against advanced threats. Its ability to reveal runtime behavior complements static analysis, providing a comprehensive understanding of the malware’s objectives and capabilities. While the process can be resource-intensive and time-consuming, its contributions to cybersecurity are indispensable, offering valuable intelligence to counteract and mitigate malicious campaigns effectively.
\newpage
\section{Analysis Results}
\subsection{Malware Distribution}
On November 13, 2024, an attempted social engineering attack was detected involving \textit{LinkedIn}, a widely trusted professional networking platform. The target, a Web3 and blockchain developer, was approached by an individual posing as a representative of a reputable company in the NFT and blockchain space. The attacker initially framed their approach as a business opportunity, inviting the target to participate in an NFT gaming project, as extensively reported by Luca Di Domenico on his \href{https://lucadidomenico.notion.site/Someone-tried-to-rug-me-on-Linkedin-13d8b714b46c80968c32c77737287e54}{Notion website}.

 \begin{figure}[H]
     \centering
     \includegraphics[width=0.5\linewidth,frame]{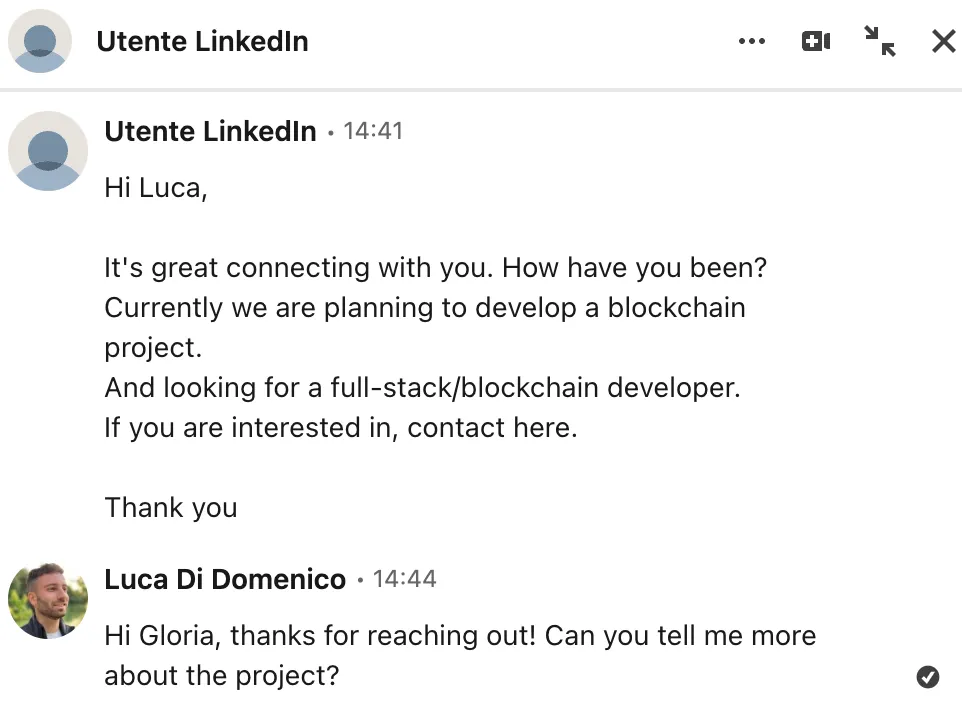}
     \caption{Attacker trying to engage its victim.}
     \label{fig:1}
 \end{figure}
 
The interaction began with what appeared to be a standard recruitment message, containing project details that aligned with the target’s professional expertise and current industry trends. The attacker followed up by requesting that the target download and run a codebase hosted on \textit{Bitbucket}, presented as part of a skill assessment process. However, as communication progressed, subtle signs raised suspicion, prompting the target to further investigate the provided code.

 \begin{figure}[H]
     \centering
     \includegraphics[width=0.9\linewidth,frame]{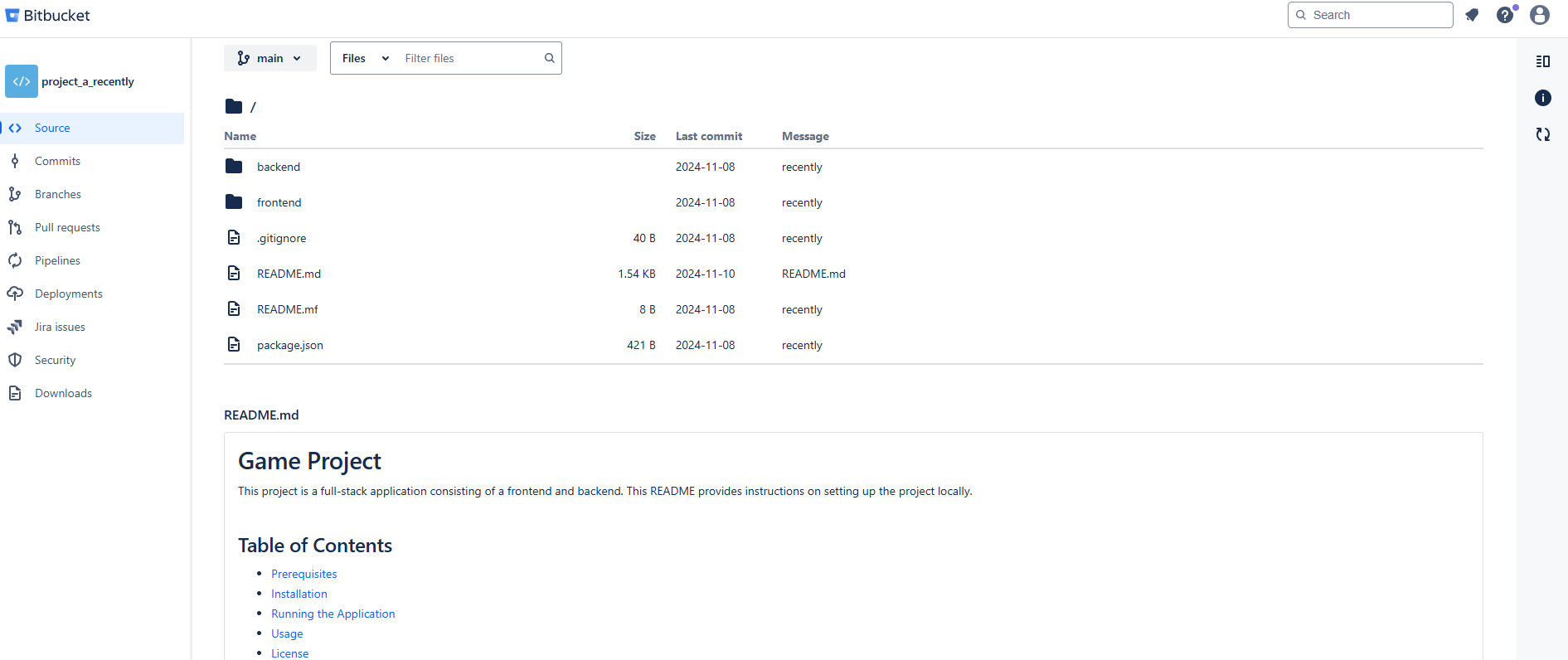}
     \caption{\textit{BitBucket} malicious repository.}
     \label{fig:2}
 \end{figure}
  
Upon examination, the codebase was found to contain obfuscated scripts designed to perform unauthorized actions on the target’s system. This discovery revealed the true nature of the message: a well-crafted attempt to execute malicious code under the guise of a professional opportunity. The following report outlines the timeline of events, initial detection, and subsequent findings, detailing the approach used by the attacker and the potential risks identified.

\begin{figure}[H]
    \centering
    \includegraphics[width=1\linewidth,frame]{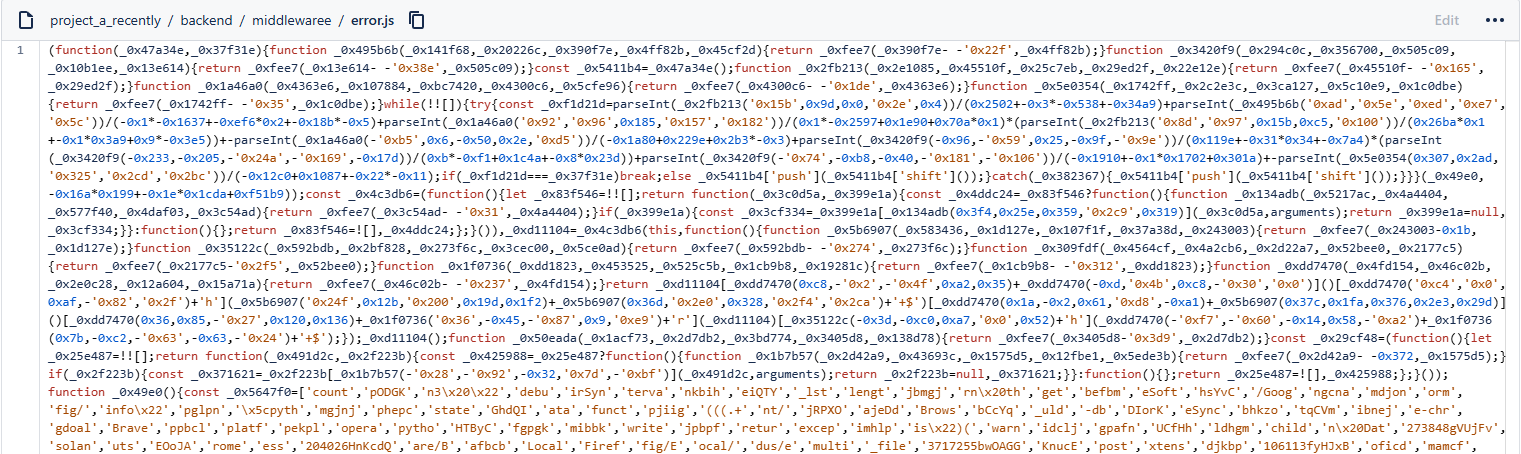}
\textit{}    \caption{Obfuscated malicious code posed inside the \textbf{\textit{error.js}} Middleware module.}
    \label{fig:3}
\end{figure} 
\subsection{First-Stage}
\begin{figure} [H]
    \centering
    \includegraphics[width=0.65\linewidth,frame]{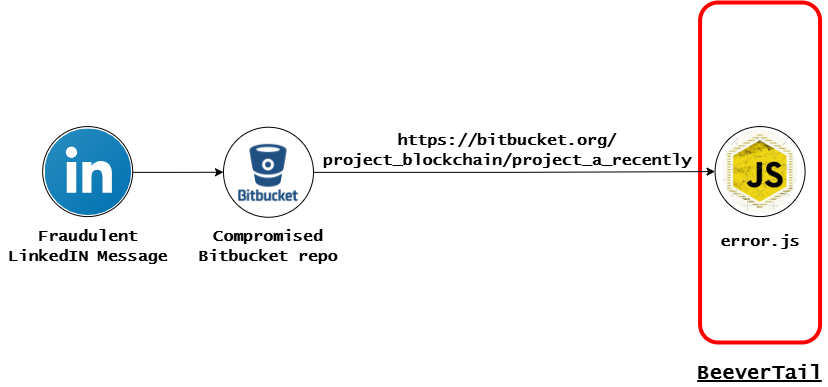}
    \caption{\textit{First Stage}}
    \label{fig:FS}
\end{figure}

The initial \textit{JavaScript} code is a highly obfuscated script crafted to execute malicious operations, including the \textit{deployment of additional payloads}, \textit{collection of sensitive data} and its subsequent \textit{exfiltration} to a remote server under the attacker's control. The obfuscation layers serve to conceal its true intent, complicating analysis and detection efforts. By targeting critical data such as \textit{credentials} and \textit{cryptocurrency wallets}, the script demonstrates a deliberate focus on \textit{financial and personal information theft}, aligning with its malicious objectives.
\subsubsection{Code Obfuscation}
In this section, there will be explored the various obfuscation techniques and decoy mechanisms utilized in the code to hinder reverse engineering and analysis efforts.
One of the primary methodologies used is the adoption of meaningless and non-descriptive variable and function names. Variables such as \textit{\_0x5647f0}, \textit{\_0x49e0}, and functions like \textbf{\textit{\_0xb038d0}} are prevalent throughout the script. This practice obscures the code's intent, making it challenging for a human reader to discern the purpose of different variables and functions.

\begin{figure}[H]
    \centering
    \includegraphics[width=0.9\linewidth,frame]{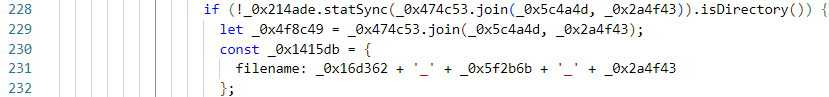}
    \caption{Variables are renamed to avoid leaking any useful insight.}
    \label{fig:4}
\end{figure}

In addition to meaningless naming, the code employs string encoding and lookup tables. Functions like \textbf{\textit{\_0xfee7}} and \textbf{\textit{\_0x49e0}} map obfuscated strings to their actual values using a lookup table, which is an array of strings that are themselves difficult to interpret. This method effectively hides string literals and function names, complicating static analysis.

\begin{figure}[H]
    \centering
    \includegraphics[width=0.6\linewidth,frame]{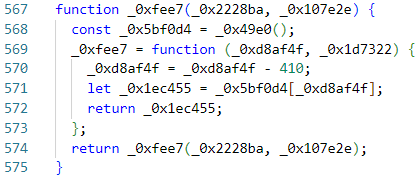}
    \caption{Code employs lookup-tables for strings to reduce code understandability.}
    \label{fig:5}
\end{figure}

\begin{figure}[H]
    \centering
    \includegraphics[width=1\linewidth,frame]{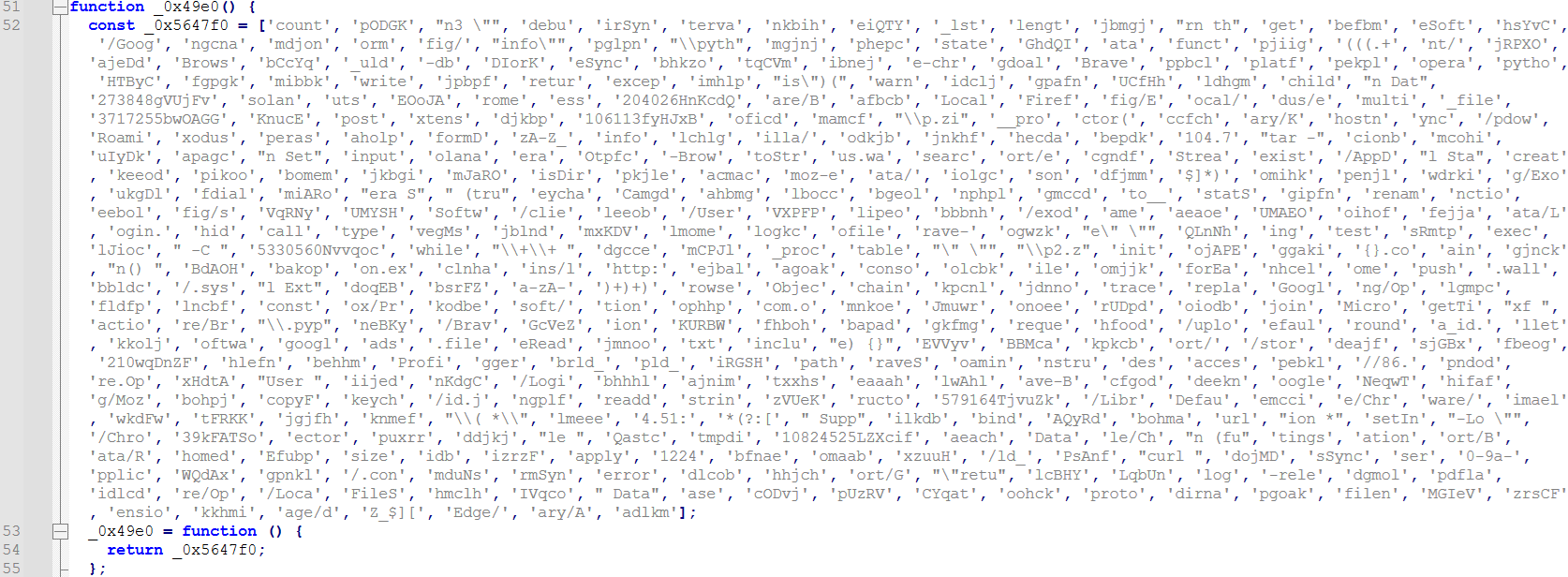}
    \caption{Lookup-table content}
    \label{fig:6}
\end{figure}

The script makes extensive use of \textit{self-invoking} functions and \textit{nested} function wrappers. These patterns complicate the control flow and make it harder to follow the sequence of execution. By encapsulating code within multiple layers of functions that immediately invoke themselves, the script hides the true entry points and interconnections between different parts of the code.

\begin{figure}[H]
    \centering
    \includegraphics[width=0.8\linewidth,frame]{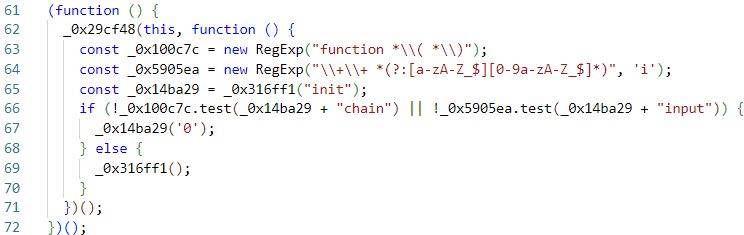}
    \caption{An example of \textit{self-invoking} and \textit{wrapped} functions.}
    \label{fig:7}
\end{figure}

Another obfuscation technique introduced is the use of the \textit{function constructor} for dynamic code execution. By constructing new functions at runtime, the script can generate and execute code that is not visible in its static form, thereby concealing the actual operations being performed. This method hinders static analysis tools, which rely on examining the code as it appears without executing it.

\begin{figure}[H]
    \centering
    \includegraphics[width=0.8\linewidth,frame]{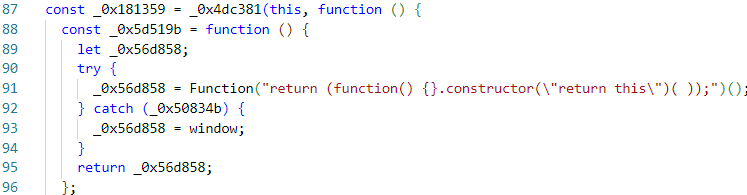}
    \caption{Functions are instantiated at runtime to make it harder analyze source code.}
    \label{fig:8}
\end{figure}

\textit{Anti-debugging} and \textit{Anti-Tampering} techniques are also employed. The script includes functions designed to detect if it is being debugged and alter its behavior, accordingly, potentially interfering with debugging efforts, by even invoking the \textit{debugger} statement dynamically, which can cause debuggers to pause execution unexpectedly or enter infinite loops.

\begin{figure}[H]
    \centering
    \includegraphics[width=0.7\linewidth,frame]{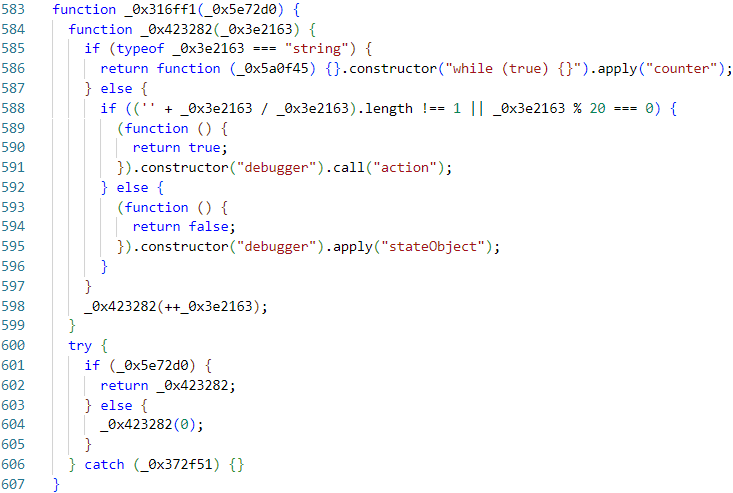}
    \caption{\textit{Anti-Debugging} functionalities}
    \label{fig:9}
\end{figure}

It also utilizes \textit{Opaque Predicates} and \textit{Dead Code}. These are conditions and code blocks that do not affect the overall program logic but are intended to confuse the analyst. \textit{Opaque predicates} are conditions that always evaluate to true or false, making it difficult to determine the actual execution path, while \textit{Dead Code} is never invoked. 

\begin{figure}[H]
    \centering
    \includegraphics[width=0.8\linewidth,frame]{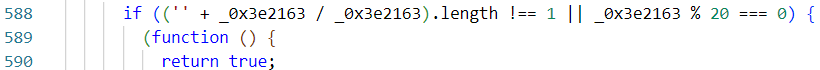}
    \caption{Example of \textit{Opaque Predicate}.}
    \label{fig:10}
\end{figure}

\textit{Control flow flattening} is another technique used to obfuscate the code. By rearranging the normal execution flow and breaking it into smaller blocks with indirect jumps and calls, the script makes it challenging to follow the logical sequence of operations. This method obscures the natural structure of the code, hindering attempts to map out its functionality.
Numeric literals are often encoded in hexadecimal or expressed as computations, making it harder to interpret constants directly. This adds an additional layer of complexity, as analysts must compute the actual numeric values to understand the code's behavior.

\begin{figure}[H]
    \centering
    \includegraphics[width=0.8\linewidth,frame]{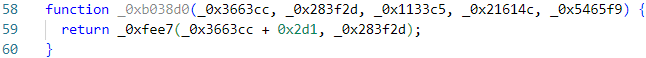}
    \caption{Numbers are hex-encoded to add complexity to code analysis.}
    \label{fig:11} 
\end{figure} 

\textit{Confusing naming conventions} are also used as a decoy strategy. The use of similar or repeating variable names with slight variations, such as \textbf{\textit{\_0x214ade}} and \textbf{\textit{\_0x2f409e}}, can cause confusion. This practice makes it difficult to track variables and understand their roles in the code.

\begin{figure}[H]
    \centering
    \includegraphics[width=1\linewidth,frame]{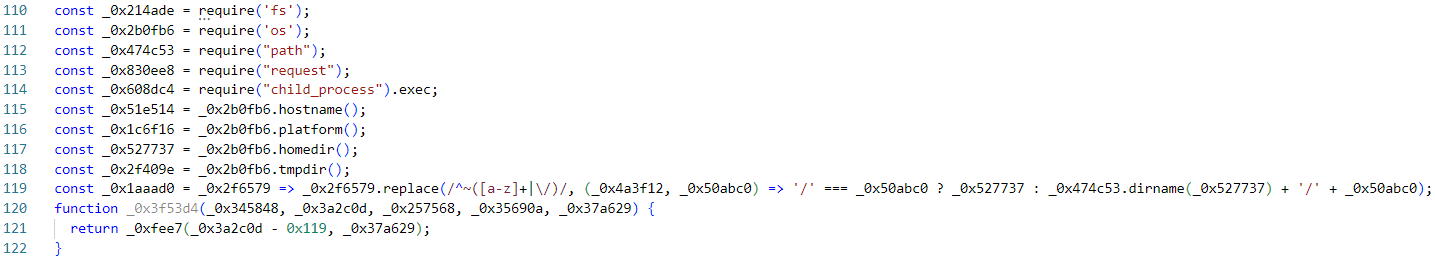}
    \caption{Usage of confusing naming conventions for script imports.}
    \label{fig:12}
\end{figure} 

Additionally, the script introduces unnecessary complex mathematical operations, including mathematical computations or expressions that serve no purpose can obfuscate the actual logic and mislead analysts into thinking they are significant when they are not.

By \textit{nesting functions} and using \textit{self-invoking patterns}, the script creates multiple layers of execution that hide the entry point and make it harder to trace the execution \textit{path}. Analysts may need to unravel several layers before reaching the core functionality, increasing the effort required for analysis. The use of dynamic code generation with the \textit{function constructor} serves as a decoy by obscuring the actual code being executed until runtime. This makes static analysis less effective, as the code's behavior cannot be fully understood without executing it.

The primary goal of these obfuscation techniques and decoy mechanisms is to prevent easy reading and understanding of the code. By making it difficult to interpret, the attacker aims to prevent quick detection of the malicious activities. The obfuscated code can evade detection by static analysis tools that rely on pattern matching or signature-based detection. Furthermore, by increasing its complexity, the attacker delays reverse engineering efforts. This added difficulties and pitfalls increases the time and effort required for analysts to de-obfuscate the code, which may allow the attacker more time to exploit the compromised system. The inclusion of decoy code and unnecessary complexity helps hide the malicious intent within layers of confusing code, potentially leading analysts down incorrect paths and causing them to misinterpret the code's purpose or miss critical malicious components.

\subsubsection{Code Analysis - error.js}

By investigating a refactored version of this code, it is possible to gather how the execution begins with the invocation of the main function, which serves as the orchestrator of the script's activities.

\begin{figure}[H]
    \centering
    \includegraphics[width=0.7\linewidth,frame]{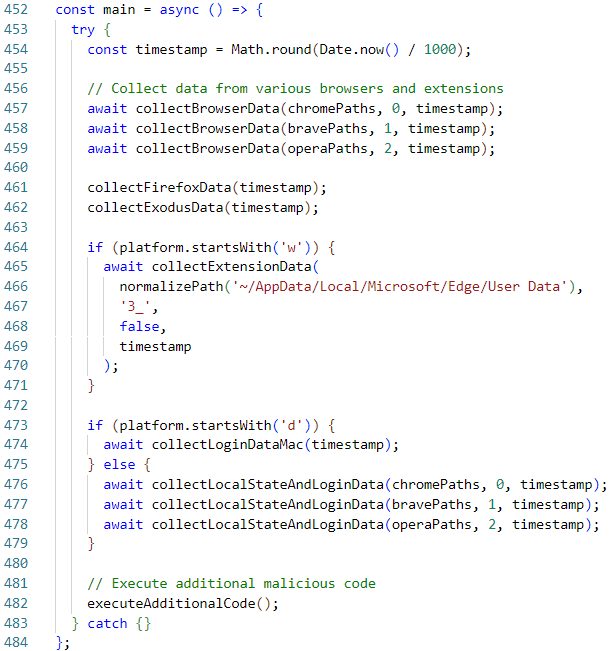}
    \caption{Refactored main routine of the malicious \textit{JS} file.}
    \label{fig:13}
\end{figure}

Inside this section the script first generates a \textit{UNIX timestamp} to tag the exfiltrated data uniquely. It then proceeds to collect information from various browsers by invoking \textbf{\textit{collectBrowserData}} for \textit{Chrome}, \textit{Brave}, and \textit{Opera} browsers. The \textbf{\textit{collectBrowserData}} function determines the appropriate base directory for each browser based on the operating system and then calls \textbf{\textit{collectExtensionData}} to harvest data from targeted extensions.

\begin{figure}[H]
    \centering
    \includegraphics[width=0.8\linewidth,frame]{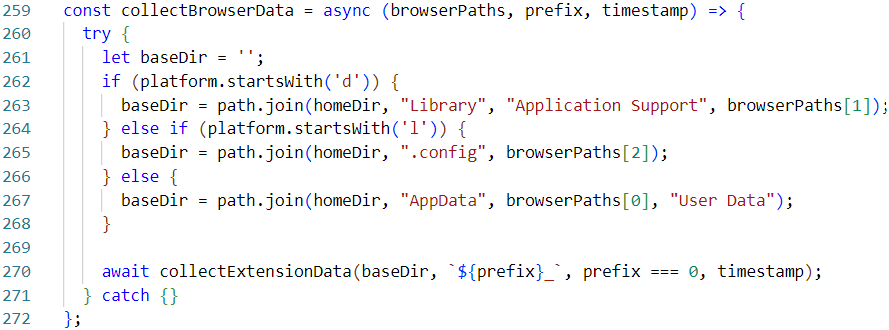}
    \caption{Snippet of the refactored capabilities of \textbf{\textit{collectBrowserData}}.}
    \label{fig:14}
\end{figure}

\begin{figure}[H]
    \centering
    \includegraphics[width=0.95\linewidth,frame]{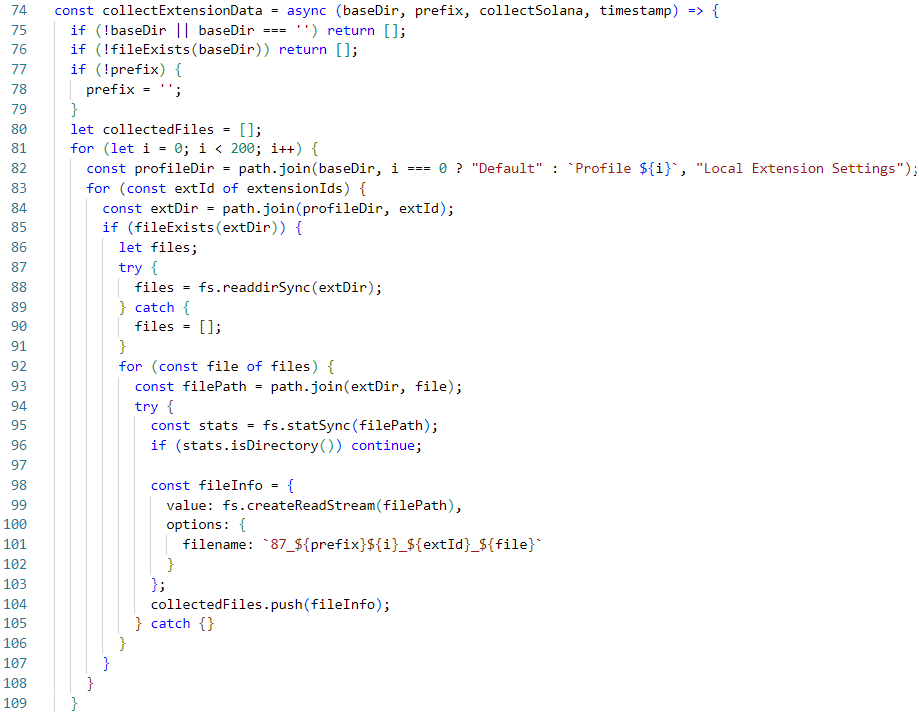}
    \caption{Snippet of the refactored capabilities of \textbf{\textit{collectExtensionData}}.}
    \label{fig:15}
\end{figure}

\textbf{\textit{collectExtensionData}} scans through multiple browser profiles, attempting to find and collect data from extensions specified in the \textit{extensionIds} array, which includes popular cryptocurrency wallets like \textit{MetaMask}. For each profile and extension the identified \textit{threat} constructs the path to the extension's data directory and, if it exists, reads the files within. Each file is read and stored in an array along with its \textit{metadata}, such as the \textit{filename} constructed from the \textit{browser prefix}, \textit{profile number}, \textit{extension ID}, and \textit{original filename}. A complete list of all the extensions tracked is provided below:
\begin{itemize}
\item \textit{nkbihfbeogaeaoehlefnkodbefgpgknn} - MetaMask (A widely used cryptocurrency wallet for Ethereum and ERC-20 tokens);
\item \textit{ejbalbakoplchlghecdalmeeeajnimhm} - TronLink (The official wallet for the TRON blockchain);
\item \textit{fhbohimaelbohpjbbldcngcnapndodjp} - LastPass: Free Password Manager (Helps users store and manage passwords securely);
\item \textit{ibnejdfjmmkpcnlpebklmnkoeoihofec} - Binance Chain Wallet (Official wallet for Binance Chain, Binance Smart Chain, and Ethereum);
\item \textit{bfnaelmomeimhlpmgjnjophhpkkoljpa} - Coinbase Wallet Extension (Allows users to interact with decentralized applications (dApps) on the browser);
\item \textit{aeachknmefphepccionboohckonoeemg} - Jaxx Liberty Wallet (A multi-currency, multi-platform cryptocurrency wallet);
\item \textit{hifafgmccdpekplomjjkcfgodnhcellj} - Exodus Wallet (Provides a user-friendly interface for managing multiple cryptocurrencies);
\item \textit{jblndlipeogpafnldhgmapagcccfchpi} - BitPay Wallet (Allows users to manage Bitcoin and other cryptocurrencies);
\item \textit{acmacodkjbdgmoleebolmdjonilkdbch} - Nifty Wallet (Designed for interacting with Ethereum and related dApps);
\item \textit{dlcobpjiigpikoobohmabehhmhfoodbb} - Authy (A two-factor authentication (2FA) app to secure online accounts);
\item \textit{mcohilncbfahbmgdjkbpemcciiolgcge} - Guarda Wallet (A non-custodial wallet supporting multiple cryptocurrencies);
\item \textit{agoakfejjabomempkjlepdflaleeobhb} - Ledger Wallet (A hardware wallet extension for managing cryptocurrencies securely);
\item \textit{omaabbefbmiijedngplfjmnooppbclkk} - OneKey Wallet (A hardware wallet extension providing secure cryptocurrency storage);
\item \textit{aholpfdialjgjfhomihkjbmgjidlcdno} - Math Wallet (Supports numerous blockchains and provides dApp support);
\item \textit{nphplpgoakhhjchkkhmiggakijnkhfnd} - SafePal Wallet (Offers secure cryptocurrency management with hardware and software solutions);
\item \textit{penjlddjkjgpnkllboccdgccekpkcbin} - Yoroi Wallet (A light wallet for Cardano (ADA) cryptocurrency);
\item \textit{lgmpcpglpngdoalbgeoldeajfclnhafa} - Phantom Wallet (A friendly Solana wallet built for DeFi and NFTs);
\item \textit{fldfpgipfncgndfolcbkdeeknbbbnhcc} - Brave Wallet (The built-in crypto wallet of the Brave browser);
\item \textit{bhhhlbepdkbapadjdnnojkbgioiodbic} - Ronin Wallet (Used for the Axie Infinity game and manages NFTs and tokens on the Ronin network);
\item \textit{gjnckgkfmgmibbkoficdidcljeaaaheg} - XDEFI Wallet (A cross-chain wallet extension supporting multiple blockchains);
\item \textit{afbcbjpbpfadlkmhmclhkeeodmamcflc} - MEW CX (MyEtherWallet Extension) (Provides access to Ethereum accounts directly in the browser).
\end{itemize}

\begin{figure} [H]
    \centering
    \includegraphics[width=0.35\linewidth,frame]{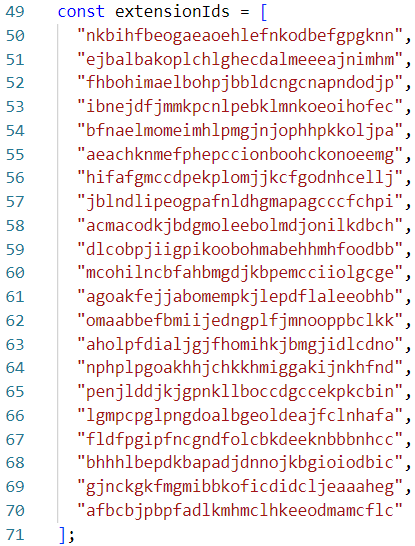}
    \caption{Crypto-related browser extensions list.}
    \label{fig:16}
\end{figure}

If the \textit{collectSolana} flag is true, the script also attempts to collect the \textit{Solana id.json} file from the user's home directory. This file often contains sensitive wallet information.
After this information gathering activity is completed, the script calls \textbf{\textit{sendData}} to exfiltrate the collected files to the attacker's server.

\begin{figure}[H]
    \centering
    \includegraphics[width=0.8\linewidth,frame]{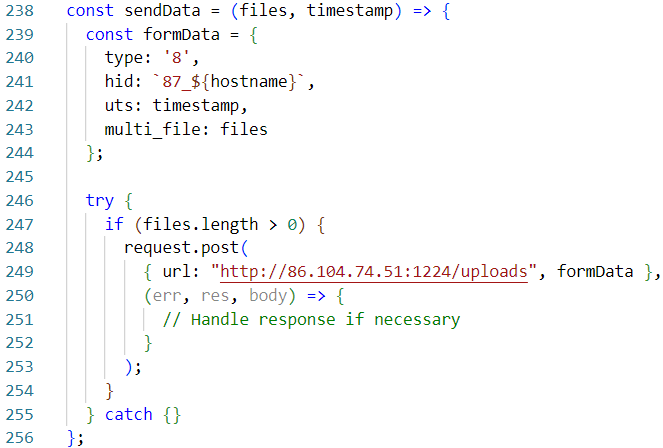}
    \caption{Malicious function designed to exfiltrate data to remote \textit{C2 server}.}
    \label{fig:17}
\end{figure}
\break
The \textbf{\textit{sendData}} function constructs a form data object containing the \textit{type}, a \textit{unique host identifier}, the \textit{timestamp}, and the \textit{array of collected files}. It then uses the request module to perform an \textit{HTTP POST request} to the attacker's server, effectively transmitting the stolen data.

Returning to the main function (\figurename~\ref{fig:13}), the script also calls \textbf{\textit{collectFirefoxData}} to target \textit{Mozilla Firefox} profiles. This function navigates through Firefox's profile directories, specifically those containing \textit{-release} in their names, and searches for extension data within the storage/default directory. It targets \textit{moz-extension} directories and collects \textit{IndexedDB} files used by extensions, which may contain sensitive information.

\begin{figure}[H]
    \centering
    \includegraphics[width=1\linewidth,frame]{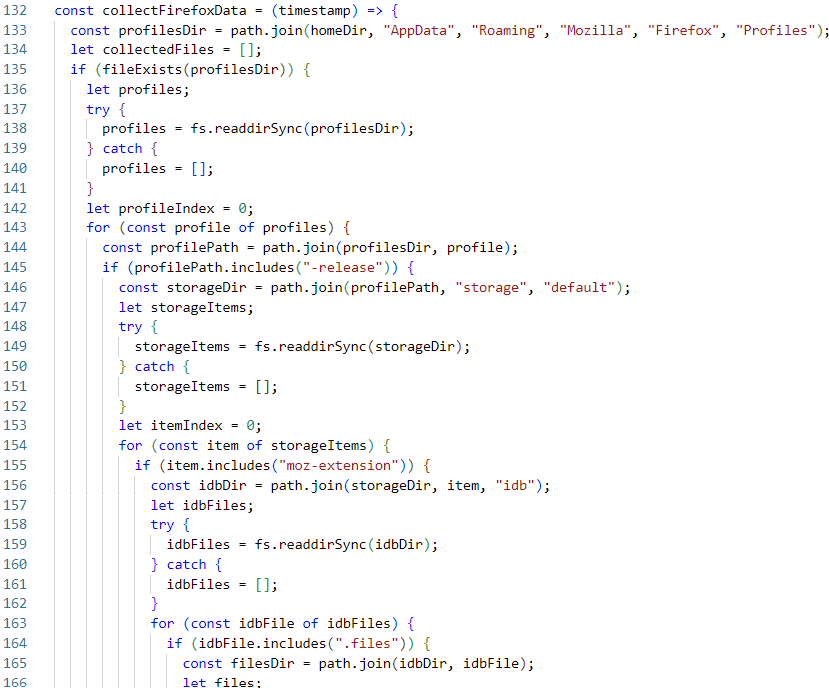}
    \caption{Function designed to collect Firefox profiles and extensions.}
    \label{fig:100}
\end{figure}

The script further attempts to collect data from the \textit{Exodus cryptocurrency wallet} by invoking \textbf{\textit{collectExodusData}}. Depending on the operating system, it constructs the path to the \textit{exodus.wallet} directory and collects any files found within it. These may contain \textit{wallet data}, \textit{private keys}, or \textit{transaction histories}.

\begin{figure}[H]
    \centering
    \includegraphics[width=1\linewidth,frame]{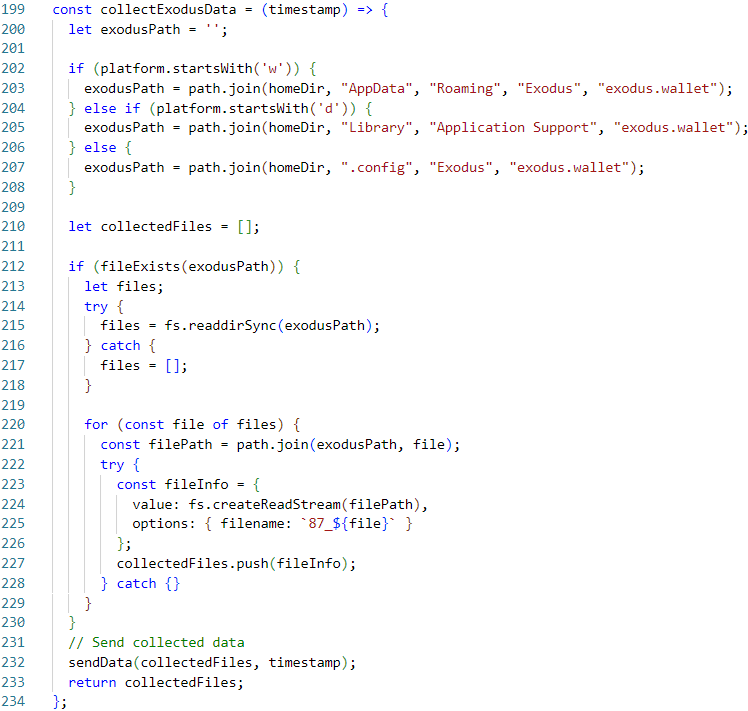}
    \caption{Function designed to collect the \textit{Exodus Cryptowallet} information.}
    \label{fig:18}
\end{figure}

For Windows systems, the script additionally targets \textit{Microsoft Edge} by calling \textbf{\textit{collectExtensionData}} with the appropriate path. This increases the scope of data collection to include users who primarily use \textit{Edge}.
The script then performs a platform check to determine whether to collect login data. On \textit{macOS systems} (platform starting with 'd'), it calls \textbf{\textit{collectLoginDataMac}} to collect the \textit{macOS keychain} file (\textit{login.keychain} or \textit{login.keychain-db}) and the \textit{Login Data} files from \textit{Chrome} and \textit{Brave} browsers. The \textit{keychain} may contain \textit{passwords}, \textit{certificates}, and \textit{secure notes}, while the \textit{Login Data} files store \textit{saved login credentials}.

\begin{figure}[H]
    \centering
    \includegraphics[width=1\linewidth,frame]{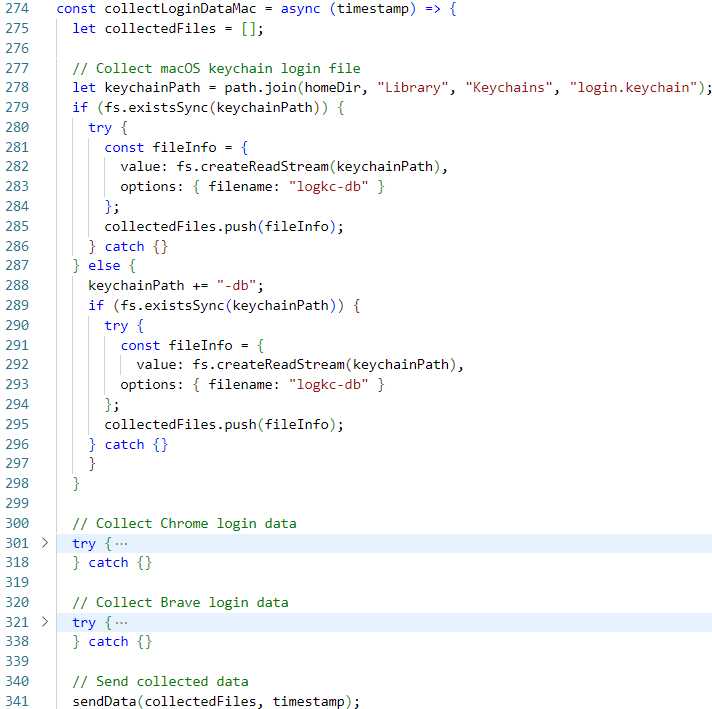}
    \caption{Function designed to collect the \textit{macOS Keychain} and browser's login data.}
    \label{fig:19}
\end{figure}

For other platforms, the script calls \textbf{\textit{collectLocalStateAndLoginData}} for \textit{Chrome}, \textit{Brave}, and \textit{Opera} browsers. This function collects the \textit{Local State file}, which contains browser settings and encryption keys, and the \textit{Login Data files} from each browser profile. By collecting these files, the attacker aims to access encrypted passwords and other sensitive data stored by the browsers.

\begin{figure}[H]
    \centering
    \includegraphics[width=0.77\linewidth,frame]{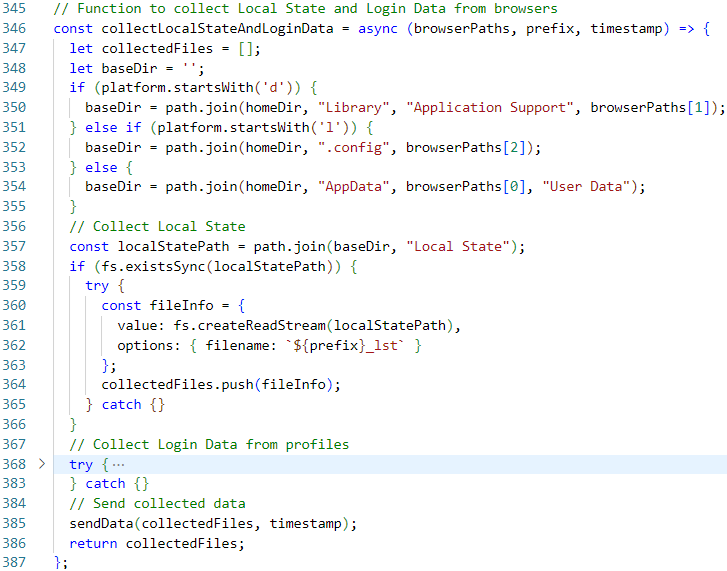}
    \caption{Function designed to collect credentials from different Browser.}
    \label{fig:20}
\end{figure}

After completing the data collection, the script calls \textbf{\textit{executeAdditionalCode}} to download and execute further malicious code.

\begin{figure}[H]
    \centering
    \includegraphics[width=0.77\linewidth,frame]{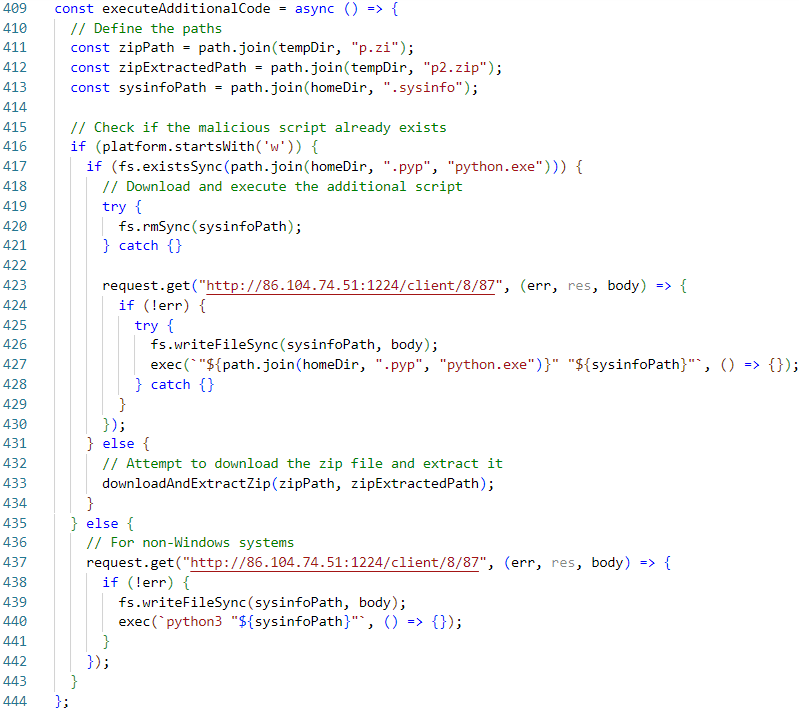}
    \caption{Function related to download and execution of the subsequent infection stages.}
    \label{fig:21}
\end{figure}

In \textbf{\textit{executeAdditionalCode}}, the script checks if it is running on a Windows system and whether a Python interpreter exists at \textit{~/.pyp/Python.exe}. If it does, the script downloads a Python script from the attacker's server and executes it using the available interpreter. If the latter is not present, the script calls \textbf{\textit{downloadAndExtractZip}} to download and extract a legit \textit{Python3.11} archive named \textbf{\textit{p.zip}}.
For non-Windows systems, the script directly downloads a \textit{.py} script and executes it using \textit{Python3}. This allows the attacker to execute additional code on the victim's machine.

\begin{figure}[H]
    \centering
    \includegraphics[width=0.85\linewidth,frame]{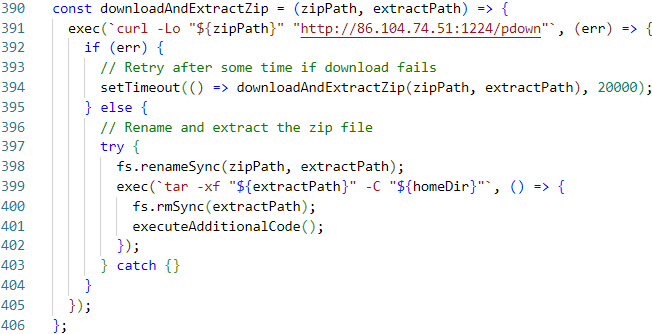}
    \caption{Function designed to download and extract a compressed \textit{Python 3.11 interpreter} if not available on target machine.}
    \label{fig:22}
\end{figure}

This function uses the \textit{curl} command to download an archive from the attacker's server. If the download fails, it retries after 20 seconds. Upon successful download, it renames and extracts the archive into the user's home directory,  then proceeds to execute the additional code and remove the stored archive.
Throughout the script, helper functions such as \textbf{\textit{normalizePath}} and \textbf{\textit{fileExists}} are used to handle file paths and check for the existence of files or directories.

\begin{figure}[H]
    \centering
    \includegraphics[width=0.85\linewidth,frame]{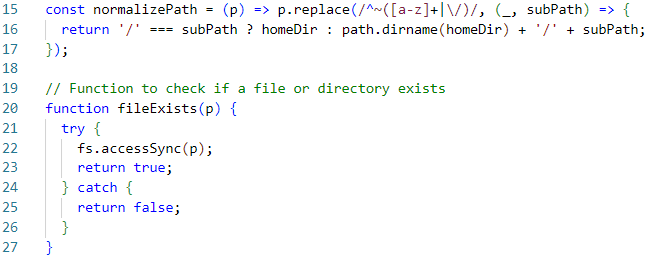}
    \caption{\textbf{\textit{normalizePath}} and \textbf{\textit{fileExists}} functions snippet.}
    \label{fig:23}
\end{figure}

These functions ensure that the script can correctly navigate the file system across different operating systems, enhancing its effectiveness and portability.
At the end of the script, an interval is set to repeat the main function every five minutes, up to a total of three executions. By repeatedly executing the \textit{main} function, the script ensures that it can capture any new data that may have been added since the last execution, such as newly saved passwords or wallet transactions. This repetition increases the chances of collecting valuable information over time.

\begin{figure}[H]
    \centering
    \includegraphics[width=0.7\linewidth,frame]{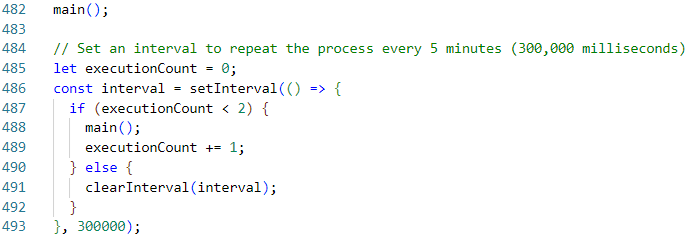}
    \caption{Main function is executed every 5 minutes on the compromised host.}
    \label{fig:24}
\end{figure}

Based on the observed behaviors and \href{https://www.esentire.com/blog/bored-beavertail-invisibleferret-yacht-club-a-lazarus-lure-pt-2}{technical characteristics} of the analyzed \textit{JavaScript} code, it is plausible to associate the subjected threat with the \textbf{\textit{BeeverTail}} malware family. The latter is recognized for its \textit{advanced data-stealing capabilities}, particularly targeting \textit{browser extensions} and \textit{cryptocurrency wallets}. The code operates by \textit{infiltrating systems} and \textit{scanning browser profiles} across multiple web browsers, including \textit{Google Chrome}, \textit{Brave}, \textit{Opera}, \textit{Mozilla Firefox}, and \textit{Microsoft Edge}. It specifically targets extensions associated with popular cryptocurrency wallets such as \textit{MetaMask}, \textit{TronLink}, and \textit{Exodus Wallet}. By accessing data stored by these extensions, the malware aims to \textit{extract sensitive information} like \textit{private keys}, \textit{seed phrases}, and \textit{wallet files}, potentially compromising users' cryptocurrency assets.
Additionally, the malware harvests \textit{login credentials} and \textit{browser data} by accessing files like \textit{Login Data} and \textit{Local State} from browser profiles. These files may contain \textit{encrypted usernames}, \textit{passwords}, and \textit{session cookies}. The exfiltration of collected data to remote servers controlled by the attackers, typically using \textit{HTTP POST requests}, aligns with the data exfiltration methods employed by \textbf{\textit{BeeverTail}}. Also the usage of \textit{port 1224}, \textit{known URL path} as \textit{/pdown/} and a Python-based second-stage payload.
To evade detection and hinder analysis, malware employs advanced obfuscation techniques. It uses meaningless variable and function names, making the code difficult to read and understand. Strings are encoded and utilized through lookup tables to conceal actual values and function calls. \textit{Control flow flattening} is used to alter the logical flow of the program, complicating efforts to follow the execution path. Moreover, dynamic code execution is implemented using the \textit{function constructor} and \textit{self-invoking functions}, allowing the malware to execute code dynamically at runtime.
The analyzed code also demonstrates the capability to download and execute additional malicious code from remote servers. By installing legitimate-looking software, such as a Python interpreter, it can run further scripts without raising suspicion. This modular approach allows the malware to enhance its capabilities, maintain persistence, and adapt to different environments, which is consistent with \textbf{\textit{BeeverTail}}'s behavior.

The malware known as \textbf{\textit{BeeverTail}} has often been utilized as a delivery mechanism for subsequent stages, notably deploying the malware family \href{https://malpedia.caad.fkie.fraunhofer.de/details/py.invisibleferret}{\textbf{\textit{InvisibleFerret}}}. The behavior exhibited by the \textbf{\textit{error.js}} file, which was analyzed in this report, aligns closely with this pattern. The specific set of \textit{Tactics, Techniques, and Procedures (TTPs)} and \textit{Indicators of Compromise} (\textit{IoCs}) associated with this file have been extensively \href{https://unit42.paloaltonetworks.com/north-korean-threat-actors-lure-tech-job-seekers-as-fake-recruiters/}{documented} as characteristic of the \textit{DPRK Threat Actor} \textbf{\textit{Lazarus Group}}.

\subsection{Second-Stage}

\begin{figure}[H]
    \centering
    \includegraphics[width=0.8\linewidth,frame]{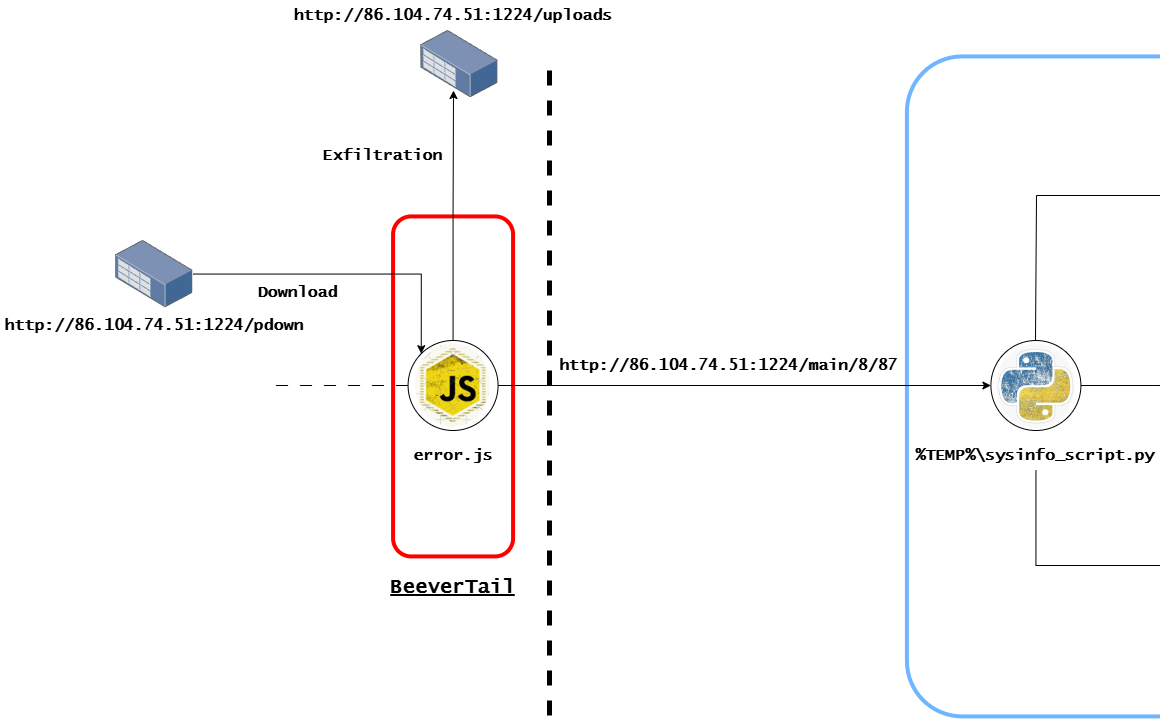}
    \caption{Moving from \textit{First} to \textit{Second Stage}.}
    \label{fig:ss}
\end{figure}

\subsubsection{Code Obfuscation} \label{Sec:Obf}
The identified Second-Stage payload is located inside a Python script, stored as \textbf{\textit{\%TEMP\%\\\textbackslash sysinfo\_script.py}} and downloaded from \textit{hxxp[:]//86.104.74[.]51:1224/client/8/87}, carrying the initial stage of the \textbf{\textit{InvisibleFerret}} malware family.

\begin{figure}[H]
    \centering
    \includegraphics[width=0.9\linewidth,frame]{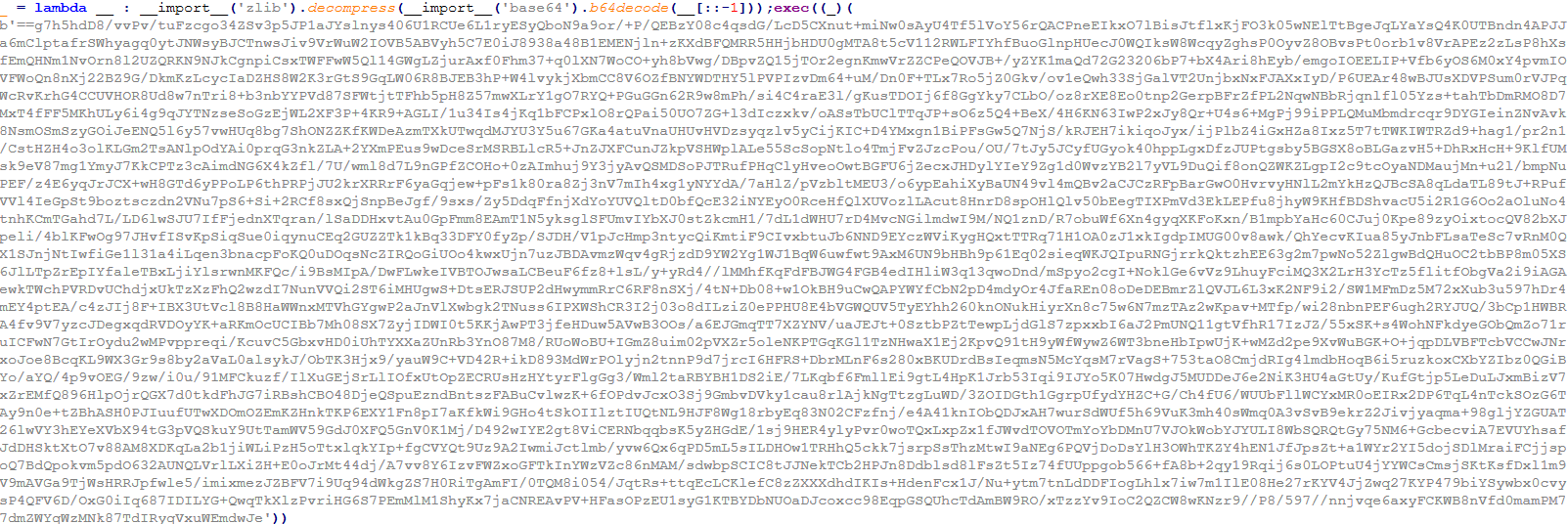}
    \caption{Second-Stage payload content.}
    \label{fig:25}
\end{figure}

As observed in the preceding image, the malware employs a sophisticated obfuscation strategy designed to hinder analysis. To reveal the underlying payload, analysts must reverse the provided string, decode it using \textit{base64}, and decompress the resulting output. This sequence of operations must be repeated fifty times before the actual malicious payload becomes accessible.

This obfuscation technique is consistently applied across nearly all subsequent Python scripts identified in the malware’s progression. Even scripts initially stored in clear text at earlier stages are later written to disk using the same obfuscation mechanism. This deliberate and systematic use of layered obfuscation underscores the attacker’s intent to evade static detection and impede reverse engineering attempts.

\subsubsection{Code Analysis - sys\_info.py}

\begin{figure}[H]
    \centering
    \includegraphics[width=0.8\linewidth,frame]{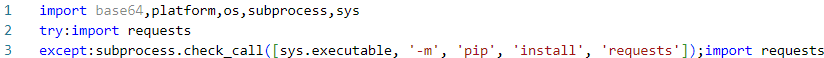}
    \caption{Second-Stage imported modules}
    \label{fig:26}
\end{figure}

Identified script defines several variables and sets up the environment. It uses the \textit{platform} module to determine the operating system type, which is stored in the variable \textit{ot}. This information is subsequently used to decide how the payloads will be handled. The user's home directory is determined and stored in the variable home, and a hidden folder named \textbf{\textit{.n2}} is created within this directory to store the downloaded payloads. By storing the payloads in a hidden folder, the script aims to avoid detection by the user.

\begin{figure}[H]
    \centering
    \includegraphics[width=0.35\linewidth,frame]{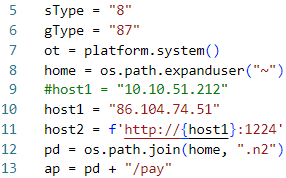}
    \caption{Remote connection configurations}
    \label{fig:27}
\end{figure}

The first payload is handled by the function \textbf{\textit{download\_payload()}}. It checks if the payload file, named \textbf{\textit{\%USERPROFILE\%\textbackslash.n2\textbackslash pay}}, already exists in the hidden directory. If it does, it attempts to remove it. Then, the script ensures that the directory \textbf{\textit{.n2}} is created if it does not already exist. The payload is downloaded from \textbf{\textit{86.104.74[.]51:1224}}, with additional parameters (\textit{sType} and \textit{gType}, campaign identifiers) passed in the URL. The downloaded content is saved in the hidden directory, and once the download is successful, the script proceeds to execute the payload. If the system is Windows, the payload is executed using the \textit{subprocess.Popen()} method with specific flags to suppress the console window and create a new process group, making the execution less noticeable. Otherwise, for \textit{macOS} systems, the payload is executed without these flags.

\begin{figure}[H]
    \centering
    \includegraphics[width=0.9\linewidth,frame]{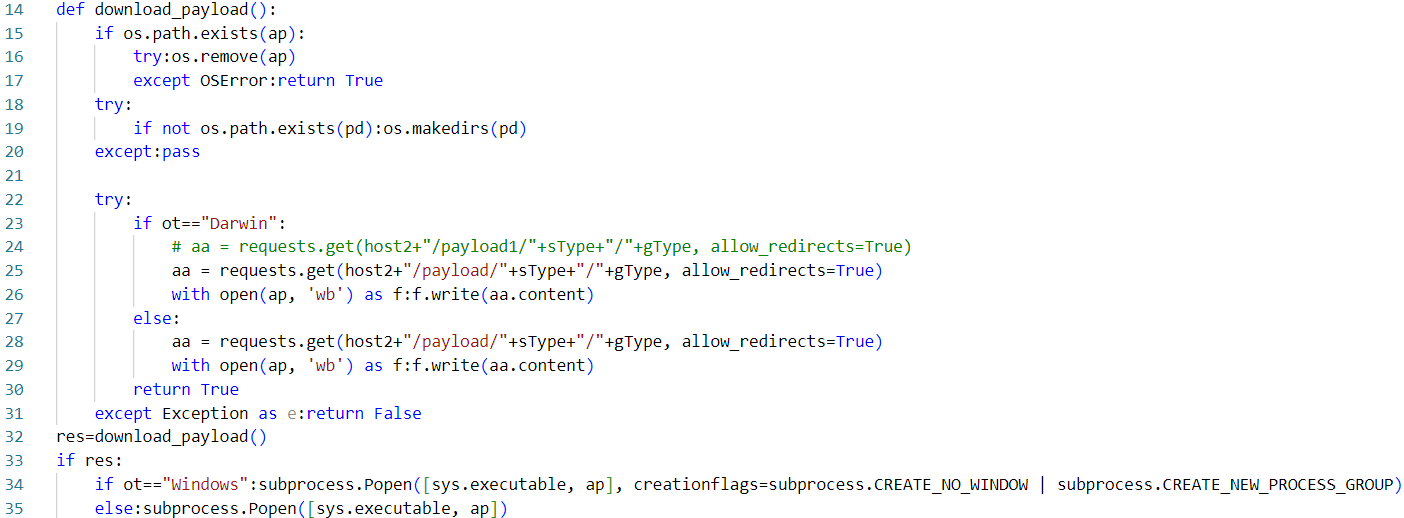}
    \caption{Malicious function designed to retrieve and run \textit{pay} Python script.}
    \label{fig:28}
\end{figure}

A specific condition is implemented for \textit{macOS} systems, identified by platform as \textit{Darwin}. After the first payload is downloaded and executed, the script terminates if it is running on \textit{macOS}, implying that subsequent parts of the script are not meant to be executed on this platform.

The script then continues to download and execute two additional payloads through the functions \textbf{\textit{download\_browse()}} and \textbf{\textit{download\_mclip()}}. Like the process described for the first payload, each of these functions first checks whether the corresponding file already exists, removing it if necessary. It also ensures that the hidden directory \textbf{\textit{.n2}} is present. The second payload, named \textbf{\textit{\%USERPROFILE\%\textbackslash.n2\textbackslash bow}}, still a Python script, is downloaded from a different endpoint on the same server, and the content is saved and executed in the same way as before.

\begin{figure}[H]
    \centering
    \includegraphics[width=1\linewidth,frame]{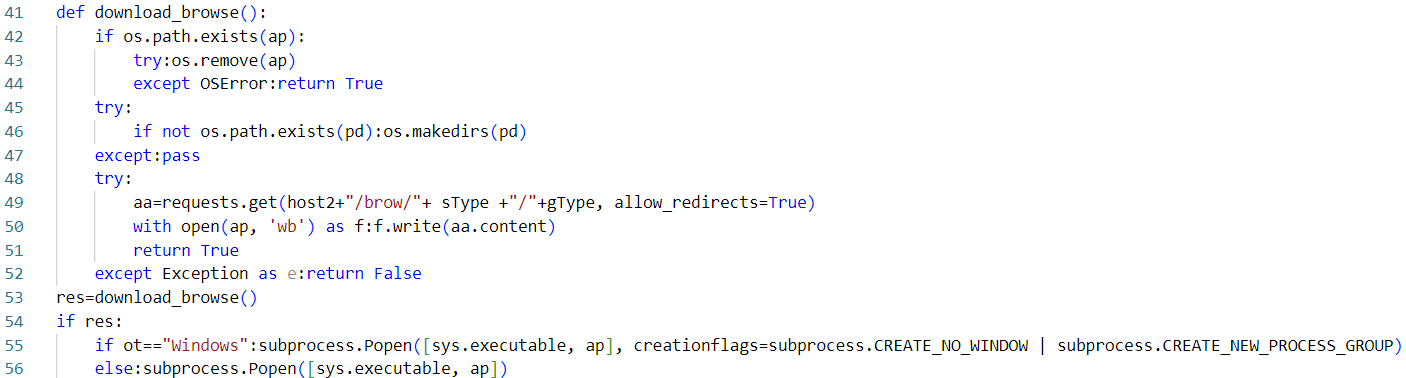}
    \caption{An additional Python payload is downloaded from the same \textit{C2 server}.}
    \label{fig:30}
\end{figure}

The third payload, named \textbf{\textit{\%USERPROFILE\%\textbackslash.n2\textbackslash mlip}}, follows the same download, save, and execute procedure, using yet another endpoint on the server and still employing a Python script.

\begin{figure}[H]
    \centering
    \includegraphics[width=1\linewidth,frame]{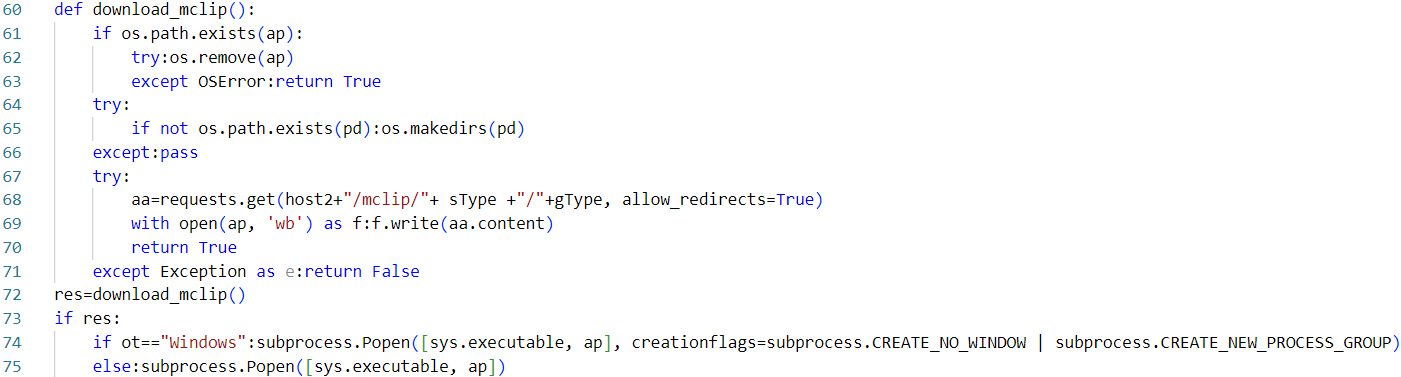}
    \caption{A third Python script is then downloaded and executed.}
    \label{fig:31}
\end{figure}

Additionally, as illustrated in \figurename~\ref{fig:27} and \figurename~\ref{fig:28}, the \textit{Threat Actor} appears to have left behind comments within the code that point to potential debugging targets. The inclusion of a \textit{private IP address} and an alternative \textit{URL} for retrieving the \textit{pay} script suggests that the attacker might have been testing the functionality of this \textit{threat}. Alternatively, this could indicate a rushed deployment, where programmers neglected to remove these debugging artifacts prior to release. Regardless of the reason, these elements provide valuable intelligence, offering insight into the attacker's development process and potentially aiding in attribution or threat profiling.

\subsection{Third-Stage}

\begin{figure} [H]
    \centering
    \includegraphics[width=0.7\linewidth,frame]{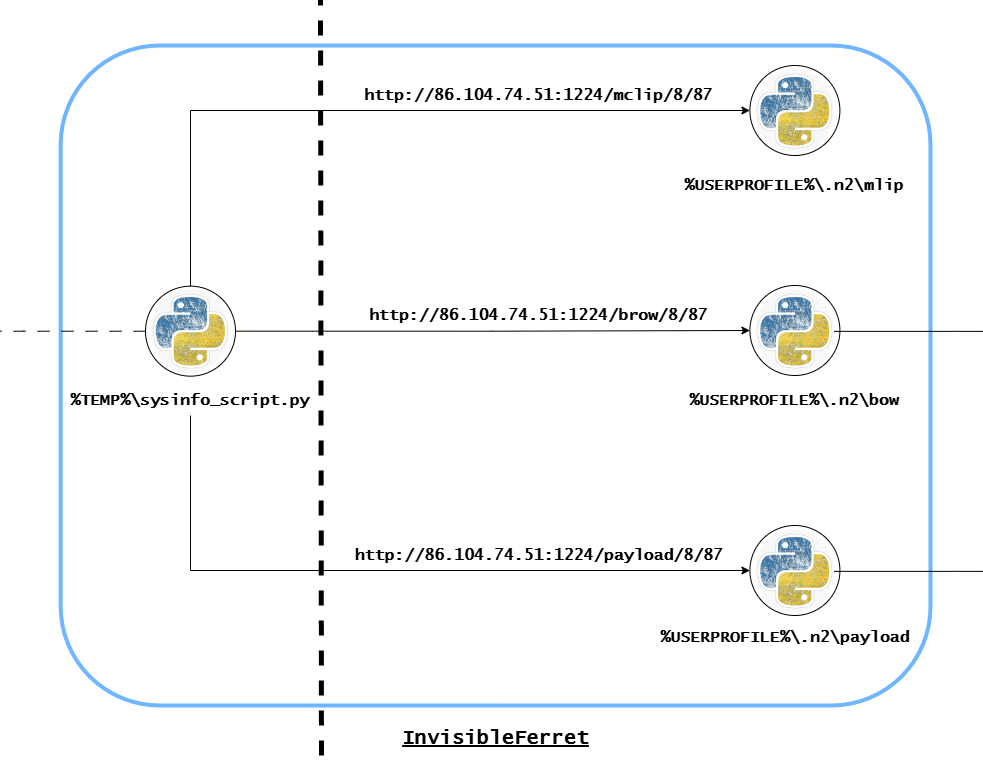}
    \caption{Moving from Second to Third Stage.}
    \label{fig:TS}
\end{figure}

As previously noted, this specific \textit{infection-stage} provides a clear indication of the new \textit{tradecrafts} being employed by the \textbf{\textit{Lazarus Group}} in this campaign. Notably, the introduction of a new Python script, \textbf{\textit{mlip}}, first \href{https://www.esentire.com/blog/bored-beavertail-invisibleferret-yacht-club-a-lazarus-lure-pt-2}{identified} only a few weeks prior to the discovery of this campaign, signifies a deliberate evolution in their operational approach. Additionally, an unprecedented payload embedded within the \textbf{\textit{bow}} script was identified during this investigation, further underscoring the group’s intent to expand their arsenal of malicious tools.

These developments suggest that the \textit{Threat Actor} is actively seeking to extend their capabilities, aligning with their shift in focus over recent years. While \textit{Lazarus} historically targeted industry leaders, such as \textit{Sony} and \textit{Blockbuster}, their operations have increasingly pivoted toward exploiting individuals and organizations within the cryptocurrency and technology sectors. This strategic redirection leverages a combination of social engineering, sophisticated malware, and multi-stage attack chains, marking a significant departure from their earlier campaigns focused on traditional industrial targets.

\subsubsection{Code Obfuscation}
All of the Python scripts involved in this stage are obfuscated with the same technique described in Section \ref{Sec:Obf}.
\subsubsection{Code Analysis - mlip} \label{sec:mlip}
\textbf{\textit{mlip}} defines a malicious script designed to \textit{capture sensitive information} from a user's system, specifically targeting \textit{cryptocurrency data} such as \textit{private keys} and \textit{mnemonic phrases}. It functions as a \textit{keylogger} and \textit{clipboard monitor}, intercepting \textit{keystrokes} and \textit{clipboard contents} when the user interacts with certain web browsers, and then transmitting this data to a remote server.

At the beginning of the script, the main section attempts to import several modules necessary for its operation. If any of these modules are not present, the script automatically installs them using \textit{pip}. This ensures that all dependencies are met without user intervention.

\begin{figure} [H]
    \centering
    \includegraphics[width=1\linewidth,frame]{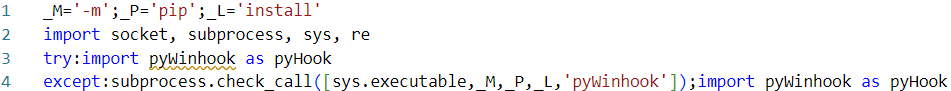}
    \caption{\textbf{\textit{mlip}} imports and missing libraries installation.}
    \label{fig:32}
\end{figure}

This pattern repeats for modules like \textit{psutil}, \textit{win32process}, \textit{win32gui}, \textit{win32api}, \textit{win32con}, \textit{win32clipboard}, \textit{requests}, and \textit{wx}. The script uses these modules to interact with \textit{Windows system APIs}, handle \textit{HTTP requests}, and interact with \textit{GUI applications}.

As first it initializes several global variables, including the server's \textit{IP address} and port to which the stolen data will be sent, and a list of targeted web browsers. Thus, indicating that the script specifically monitors these processes. Developers also left a commented-out \textit{HOST}, highlighting how \textit{localhost} was probably used for testing purposes. 

\begin{figure}[H]
    \centering
    \includegraphics[width=0.3\linewidth,frame]{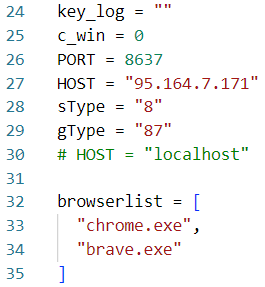}
    \caption{Hard-coded very useful information}
    \label{fig:33}
\end{figure}

The \textbf{\textit{act\_win\_pn()}} function retrieves information about the active window, such as the \textit{process ID}, \textit{process name}, and \textit{window caption}. These information is used to determine if the user is interacting with one of the targeted browsers.

\begin{figure} [H]
    \centering
    \includegraphics[width=0.8\linewidth,frame]{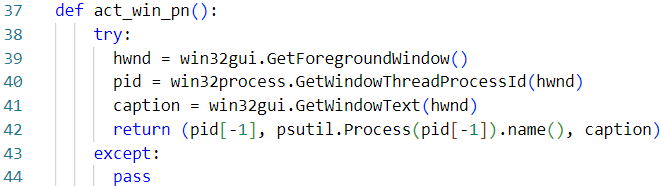}
    \caption{\textbf{\textit{act\_win\_pn()}} function code snippet}
    \label{fig:34}
\end{figure}

The script then defines several utility functions to check the state of control keys and to save logs. Indeed, \textbf{\textit{save\_log()}} function is particularly important as it sends the captured data to the remote server using an \textit{HTTP POST request}.

\begin{figure} [H]
    \centering
    \includegraphics[width=0.45\linewidth,frame]{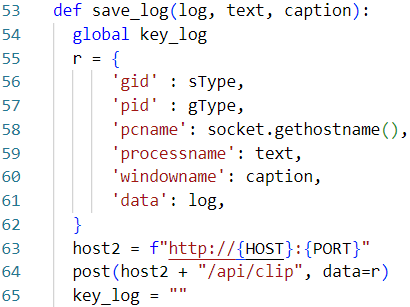}
    \caption{\textit{C\&C Server} URL and exfiltration parameters.}
    \label{fig:35}
\end{figure}

The \textbf{\textit{OnKeyboardEvent()}} function is a callback that is triggered on every keyboard event. It checks if the active process is one of the targeted browsers and captures the keystrokes. This function also intercepts clipboard data when the user pastes content using \textit{Ctrl+V}, invoking \textbf{\textit{GetTextFromClipboard()}} to process the clipboard contents. Additionally, the script sets up a keyboard hook using \textbf{\textit{pyHook}} to monitor all keyboard events.

\begin{figure} [H]
    \centering
    \includegraphics[width=0.55\linewidth,frame]{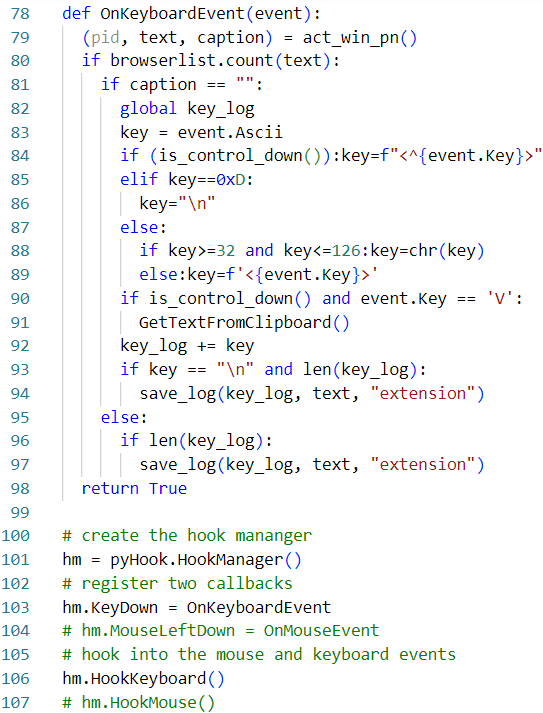}
    \caption{Callback function to trigger \textit{Keylogging} activity.}
    \label{fig:36}
\end{figure}

In addition to keystroke logging, the script defines the \textit{TestFrame} class, which inherits from \textit{wx.Frame}. This class sets up a clipboard viewer that monitors changes to the clipboard.

\begin{figure} [H]
    \centering
    \includegraphics[width=0.8\linewidth,frame]{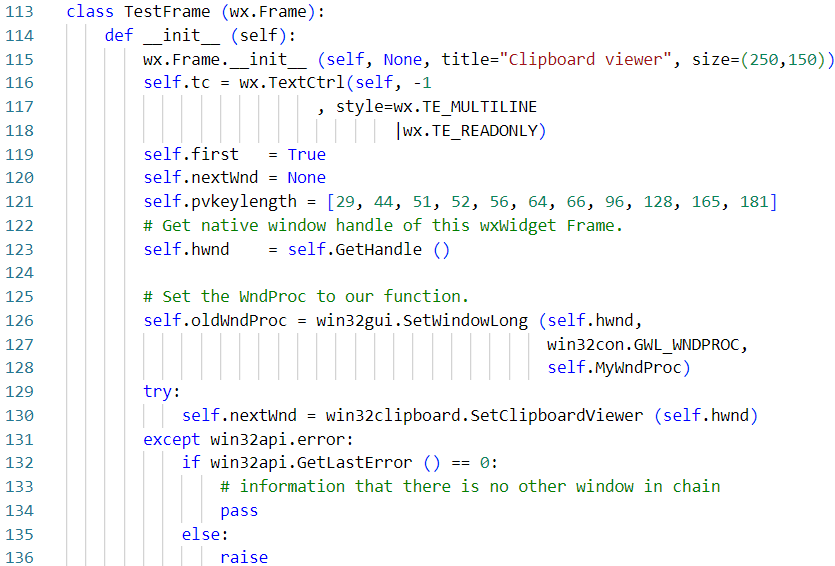}
    \caption{\textbf{\textit{TestFrame}} class initialization}
    \label{fig:37}
\end{figure}

Within this class, the \textbf{\textit{OnDrawClipboard()}} method is called whenever the clipboard content changes. It processes the new clipboard data to detect potential \textit{private keys} or \textit{mnemonic} phrases.

\begin{figure} [H]
    \centering
    \includegraphics[width=0.7\linewidth,frame]{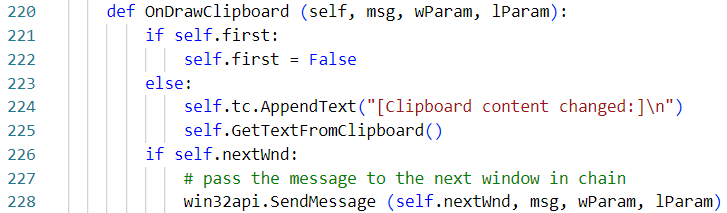}
    \caption{\textbf{\textit{OnDrawClipboard()}} code snippet}
    \label{fig:38}
\end{figure}

The \textbf{\textit{GetTextFromClipboard()}} method retrieves the clipboard text and checks if it contains sensitive information.

\begin{figure}[H]
    \centering
    \includegraphics[width=0.7\linewidth,frame]{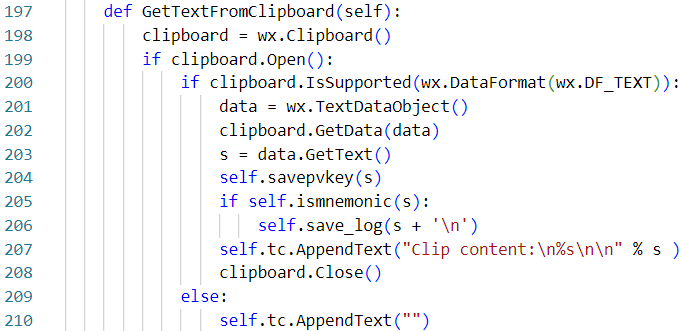}
    \caption{Function designed to capture and retrieve sensitive data.}
    \label{fig:39}
\end{figure} 
The \textbf{\textit{savepvkey()}} method searches for hexadecimal strings of specific lengths that may represent private keys. Similarly, the \textbf{\textit{ismnemonic()}} method checks if the clipboard content consists of \textit{12}, \textit{16}, or \textit{24 words}, which are common lengths for mnemonic seed phrases in \textit{cryptocurrency wallets}.

\begin{figure} [H]
    \centering 
    \includegraphics[width=0.75\linewidth,frame]{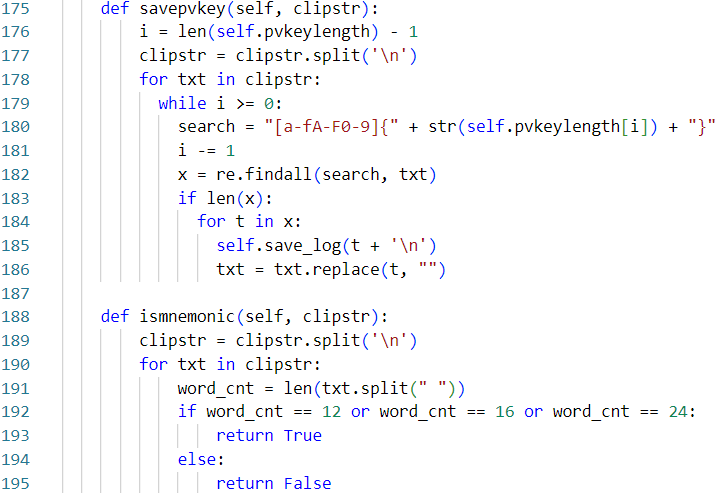}
    \caption{\textbf{\textit{savepvkey()}} and \textbf{\textit{ismnemonic()}} implementations}
    \label{fig:40}
\end{figure}

Finally, the \textit{main loop} of the script creates an instance of the \textit{TestFrame} class and starts the application. This ensures that the clipboard monitoring continues to run as long as the application is active.

\begin{figure} [H]
    \centering
    \includegraphics[width=0.35\linewidth,frame]{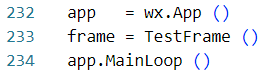}
    \caption{Main loop}
    \label{fig:41}
\end{figure}

In conclusion, the script operates by covertly \textit{logging keystrokes} and \textit{clipboard contents} when the user interacts with specific web browsers. It specifically targets data that resembles \textit{cryptocurrency private keys} or \textit{mnemonic phrases}. The captured data is then transmitted to a \textit{remote server} without the user's consent, representing a significant security and privacy threat.

Unused code in the script appears minimal, as most functions and classes are integral to its malicious functionality. However, certain error handling or exception cases might not be fully fleshed out, potentially causing the script to fail silently under unexpected conditions.

Additionally, further \textit{OSINT} investigations revealed how this code was built by incorporating code available on some Online-Forums (\href{https://code.activestate.com/recipes/355593-windows-clipboard-viewer/}{\textit{ActiveState}} and \href{https://www.douban.com/note/179506326/?_i=1869827RAEkQG2,2001849RAEkQG2}{Douban}). In both the provided websites, there is available the exact same code the attacker embedded in its \textit{threat} to interact with the compromised system's clipboard.

\begin{figure} [H]
    \centering
    \includegraphics[width=0.9\linewidth,frame]{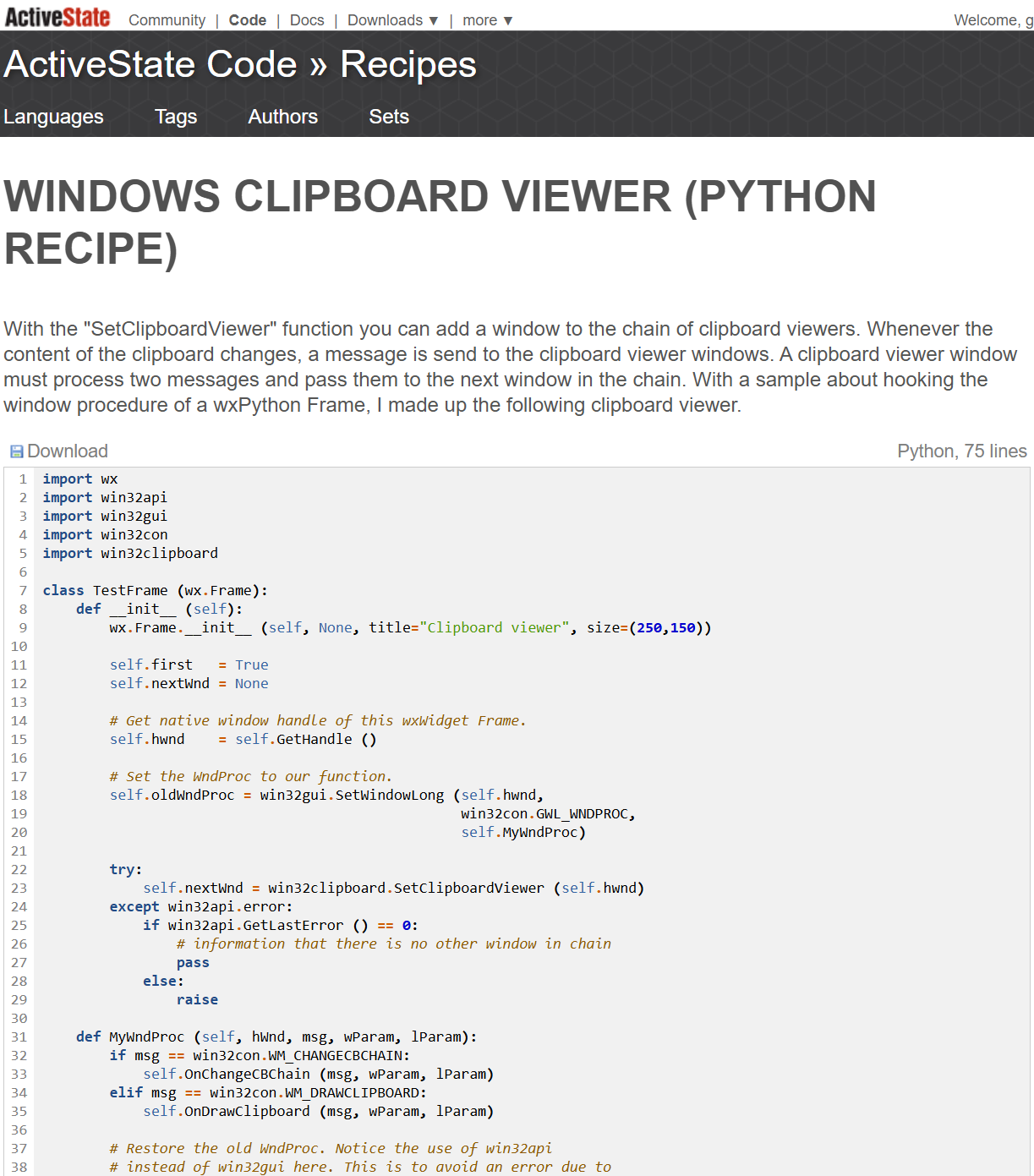}
    \caption{Code shown in \figurename~\ref{fig:37} was found on Online Python Forums.}
    \label{fig:XX}
\end{figure}

\subsubsection{Code Analysis - pay}
Proposed script is a malicious program designed to infiltrate a victim's computer, \textit{gather sensitive information}, and \textit{establish persistent remote control}. It combines several malicious functionalities, including \textit{system reconnaissance}, \textit{data exfiltration}, \textit{remote command execution}, \textit{keylogging}, and \textit{clipboard monitoring}. The malware is crafted to operate on both Windows and non-Windows systems, adapting its behavior while also being able to download and execute the aforementioned \textbf{\textit{bow}} Python script. This indeed highlight the enhanced resilience the \textit{Threat Actor} employed in its \textit{tradecrafts}.

Starting from the main execution point, the script initiates its malicious activities by importing essential modules and defining global variables that will be used throughout its operation. It begins by importing modules such as \textit{base64}, \textit{socket}, \textit{uuid}, \textit{hashlib}, \textit{getpass}, \textit{platform}, and \textit{time}. These imports are crucial for network communication, system information retrieval, and cryptographic functions.

\begin{figure} [H]
    \centering
    \includegraphics[width=0.55\linewidth,frame]{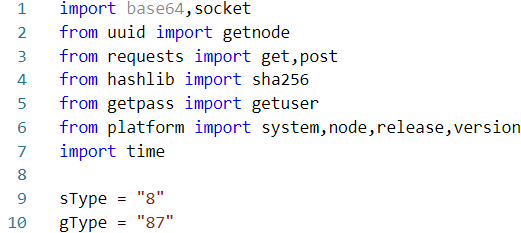}
    \caption{\textbf{\textit{pay}} script's imports}
    \label{fig:42}
\end{figure}

The script defines \textit{sType} and \textit{gType}, constants in this campaign and used to uniquely define it within their various compromising activities.

The \textit{main function} of the script is encapsulated within the \textbf{\textit{run\_comm()}} function, which initiates the transmission of collected system and network information to the attacker's server. It does so by creating an instance of the \textit{Trans} class and calling its \textbf{\textit{contact\_server()}} method.

\begin{figure} [H]
    \centering
    \includegraphics[width=0.7\linewidth,frame]{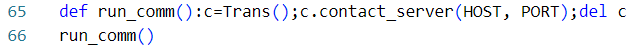}
    \caption{Snippet of \textbf{\textit{run\_comm()}} function}
    \label{fig:43}
\end{figure}

\begin{figure} [H]
    \centering
    \includegraphics[width=1\linewidth,frame]{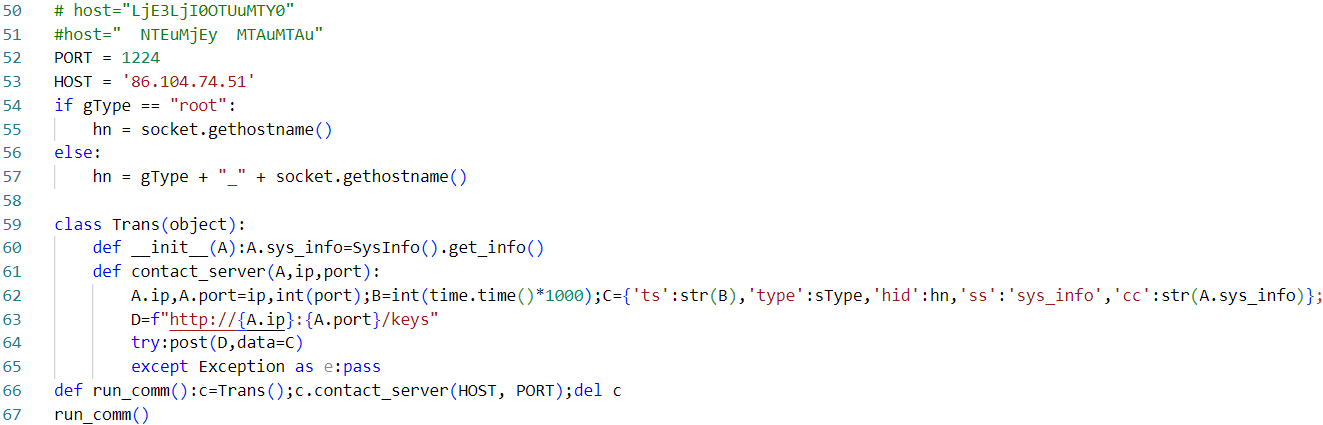}
    \caption{\textbf{\textit{Trans}} class initialization}
    \label{fig:44}
\end{figure}

\begin{figure} [H]
    \centering
    \includegraphics[width=0.8\linewidth,frame]{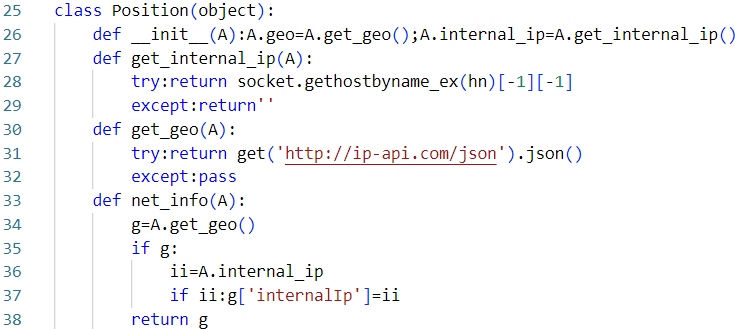}
    \caption{\textit{Position} class is designed to gather geographical information from victim's \textit{IP}.}
    \label{fig:46}
\end{figure}

Within the \textit{Trans} class, the \textit{\_\_init\_\_} method collects system and network information by instantiating the \textit{SysInfo} class and calling its \textbf{\textit{get\_info()}} method. This method aggregates system information such as the \textit{operating system}, \textit{hostname}, \textit{release version}, and \textit{user details}, as well as network information like \textit{IP address} and \textit{geolocation data}. Additionally, by comparing information provided in \figurename~\ref{fig:44} and \figurename~\ref{fig:46} it is possible to gather how the attacker set up two different ports to achieve two different malicious purposes. Port \textit{1224} is used to extract geographical victim's information, while Port \textit{2247} will be used as a remote \textit{C2 Endpoint} to bind an interactive shell between the \textit{Threat Actor} and the victim's system. It is also interesting to highlight how \figurename~\ref{fig:44} shows two commented \textit{host} variable containing seemingly \textit{base64} encoded information. As it will be discussed in Sec. \ref{Sec:any}, this same string is manipulated to retrieve the remote \textit{C2 Server}. Thus, denoting a possible on-the-fly change applied to the inner workings of their scripts, either due to changing their habits or experimenting obfuscation boundaries for \textit{AV detection}.

The \textbf{SysInfo} class leverages the \textit{HostInfo} and \textit{Position} classes to gather this information. The \textit{HostInfo} class collects system-related data, while the \textit{Position} class retrieves network-related information.

\begin{figure} [H]
    \centering
    \includegraphics[width=1\linewidth,frame]{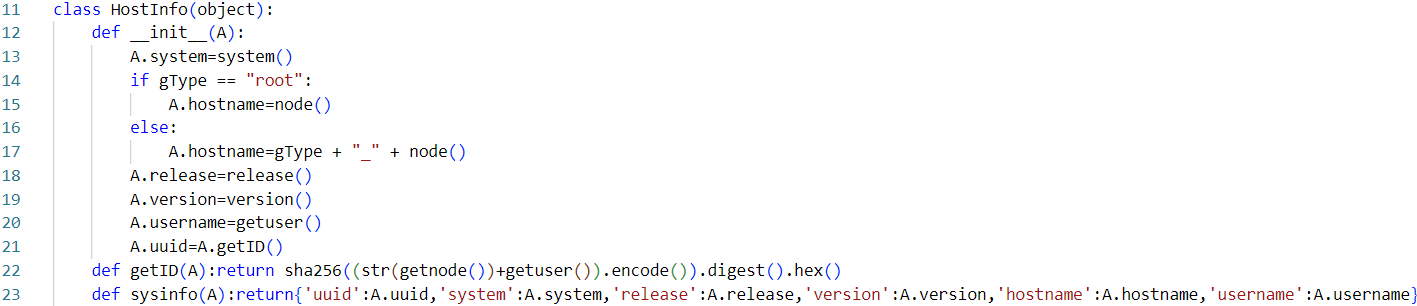}
    \caption{\textit{HostInfo} class maps host information into a dictionary to be exfiltrated}
    \label{fig:45}
\end{figure}

In the \textit{HostInfo} class, the \textbf{\textit{getID()}} method generates a unique identifier for the victim's machine by hashing the \textit{MAC address} and \textit{username}. This \textit{UUID} is used to uniquely identify the infected system.

The Position class retrieves the internal \textit{IP address} and \textit{geolocation} data by making a request to \textit{hxxp[:]//ip-api[.]com/json}, which returns the public \textit{IP} and associated \textit{geolocation} information.

After collecting all the necessary information, the \textit{Trans} class's \textbf{\textit{contact\_server()}} method sends this data to the attacker's server using an \textit{HTTP POST request} \figurename~\ref{fig:44}.

Furthermore, developers introduced a dictionary, \textit{C}, which contains a \textit{timestamp}, the \textit{type identifier}, \textit{host identifier}, a label \textit{sys\_info}, and the collected \textit{system} and \textit{network information}. This data is then sent to the attacker's server at the specified \textit{HOST} and \textit{PORT}.

Following the initial data exfiltration, the script attempts to establish a persistent connection to the attacker's \textit{Command and Control} (\textit{C2}) \textit{server} to receive further instructions. It defines the \textit{Client} class, which handles the connection setup and maintains the communication loop.

\begin{figure} [H]
    \centering
    \includegraphics[width=1\linewidth,frame]{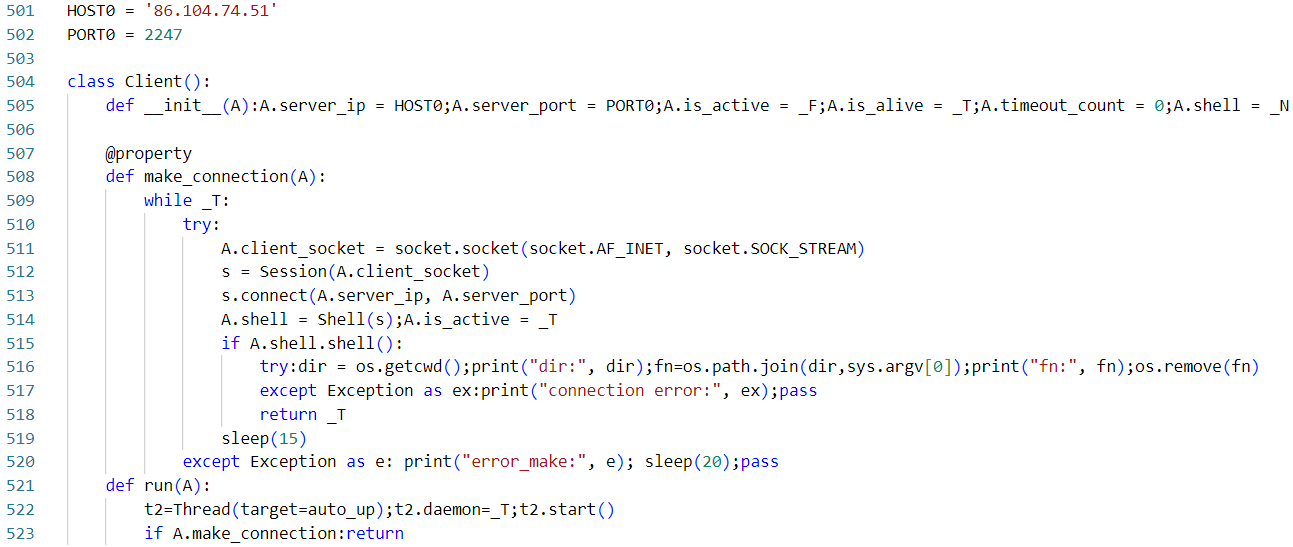}
    \caption{\textit{Client} class setups an interpretative connection to attacker's servers.}
    \label{fig:47}
\end{figure}

The \textbf{\textit{make\_connection()}} method attempts to establish a socket connection to the attacker's server. If successful, it creates a \textit{Session} object for low-level communication and a \textit{Shell} object to handle commands. The \textit{Shell} class contains methods for executing various commands received from the attacker, such as running shell commands, uploading files, and manipulating processes.

The \textbf{\textit{Shell}} class is responsible for interpreting and executing various commands sent by the attacker, effectively acting as a remote shell. It maintains the session state, handles incoming commands, and dispatches them to the appropriate methods.

\begin{figure} [H]
    \centering
    \includegraphics[width=1\linewidth,frame]{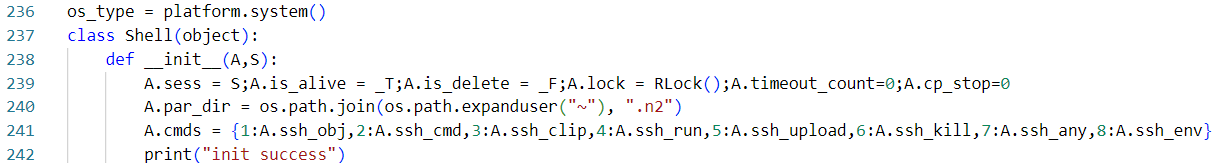}
    \caption{\textit{Shell} class provides the attacker with \textit{RAT} capabilities.}
    \label{fig:48}
\end{figure}

In the \textit{Shell} class's constructor, it initializes various attributes and defines a dictionary \textit{self.cmds} that maps command codes to their corresponding methods. These methods handle different functionalities such as executing shell commands, terminating processes, uploading files, and more.

The \textbf{\textit{listen\_recv()}} method continuously listens for incoming commands from the attacker.

\begin{figure} [H]
    \centering
    \includegraphics[width=1\linewidth,frame]{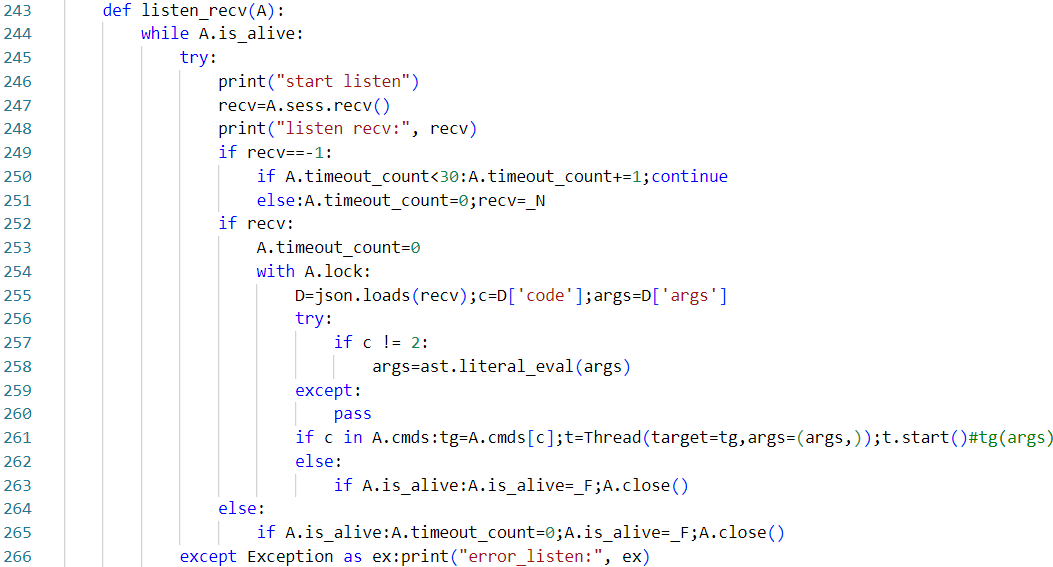}
    \caption{Function \textbf{\textit{listen\_recv()}} code snippet}
    \label{fig:49}
\end{figure}

The method receives data from the \textit{session}, parses it, and dispatches it to the appropriate \textit{handler} method based on the command code. It uses threading to handle commands concurrently.

The \textbf{\textit{shell()}} method starts the listener thread and keeps the \textit{shell} active until it's terminated.

\begin{figure} [H]
    \centering
    \includegraphics[width=0.7\linewidth,frame]{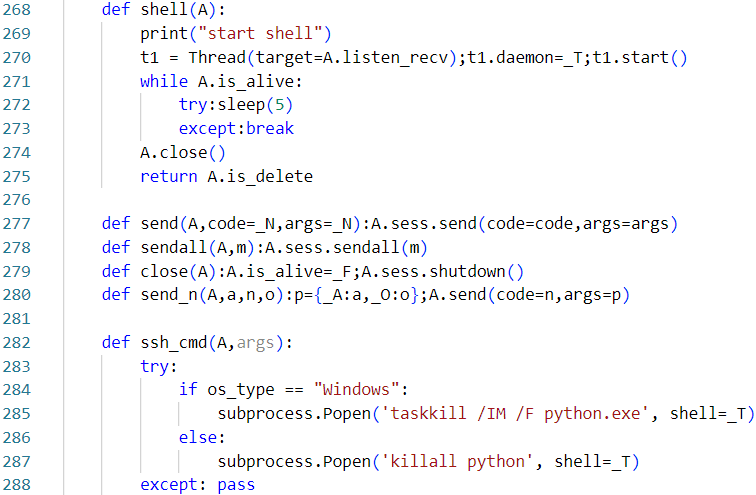}
    \caption{\textbf{\textit{Shell()}} translates attacker's command into ones to be executed on target.}
    \label{fig:50}
\end{figure}
Below are some of the handler methods in the Shell class:

\begin{itemize}
    \item \textbf{\textit{ssh\_obj(self, args)}}: This method allows the attacker to execute arbitrary shell commands on the victim's machine and returns the output;
    \item \textbf{\textit{ssh\_cmd(self, args)}}: Terminates Python processes running on the victim's machine;
    \item \textbf{\textit{ssh\_clip(self, args)}}: Sends the contents of the clipboard to the attacker;
    \item \textbf{\textit{ssh\_upload(self, args)}}: This method provides the attacker with the ability to search for and exfiltrate files from the victim's system;
    \item \textbf{\textit{ssh\_kill(self, args)}}: Terminates specific processes, such as web browsers;
    \item \textbf{\textit{ssh\_any(self, args)}}: These methods collectively enable the attacker to perform a wide range of malicious activities on the victim's machine, from executing commands and terminating processes to uploading and downloading files.
\end{itemize}

\begin{figure} [H]
    \centering
    \includegraphics[width=1\linewidth,frame]{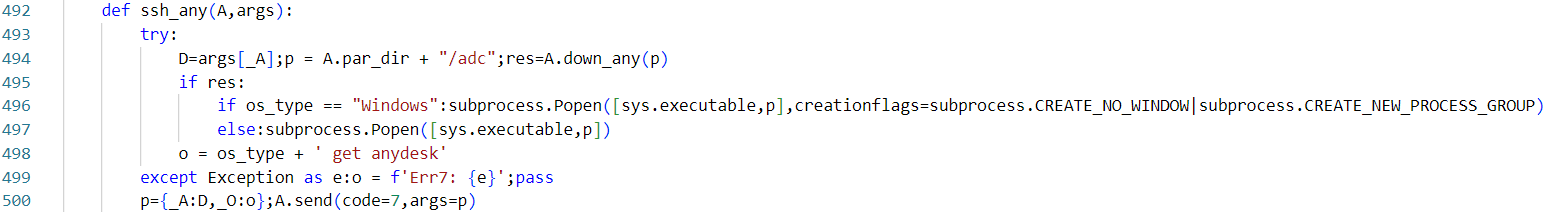}
    \caption{Function used to download \textbf{\textit{any.py}}, which gets and runs AnyDesk.}
    \label{fig:51}
\end{figure}

An additional essential part of the malware's operation is its capability to search for and exfiltrate sensitive files from the victim's system. It defines patterns and exclusion lists to target specific files while avoiding others. The \textbf{\textit{ld()}} function recursively lists files in directories, excluding those that match the specified patterns. It collects file paths that are then used by the \textbf{\textit{ups()}} function to upload the files to the attacker's server.

\begin{figure} [H]
    \centering
    \includegraphics[width=1\linewidth,frame]{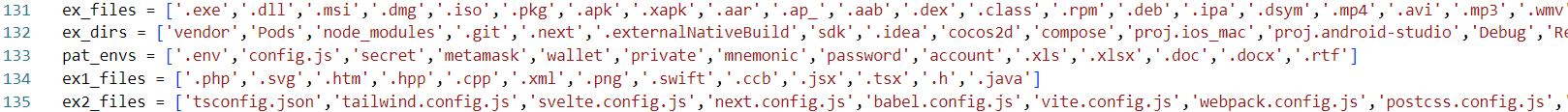}
    \caption{Arrays embedding file's extensions to be serached on target system.}
    \label{fig:52}
\end{figure}

The \textbf{\textit{ups()}} function handles the file upload process, sending the collected files to the attacker's server via \textit{HTTP POST requests}.

\begin{figure} [H]
    \centering
    \includegraphics[width=1\linewidth,frame]{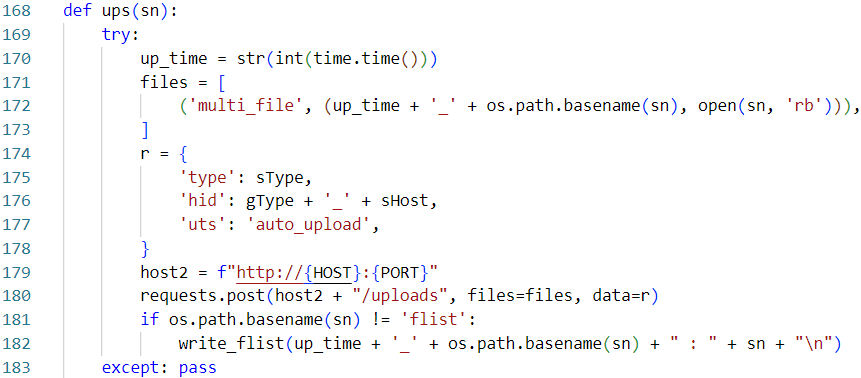}
    \caption{How files with known extensions are exfiltrated.}
    \label{fig:53}
\end{figure}

The malware also incorporates \textit{keylogging} and \textit{clipboard monitoring} capabilities. As first it ensures that these modules are present, then malware can interact with the \textit{Windows API} to \textit{capture keystrokes} and \textit{clipboard content}. The \textit{keylogging} functionality is initiated in the \textbf{\textit{run\_client()}} function, which starts a thread to hook keyboard and mouse events.

\begin{figure} [H]
    \centering
    \includegraphics[width=0.6\linewidth,frame]{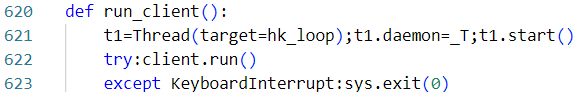}
    \caption{\textbf{\textit{run\_client()}} deploys keyboard hooking functionality.}
    \label{fig:54}
\end{figure}

The \textbf{\textit{hk\_loop()}} function sets up the hooks for keyboard and mouse events using \textit{pyHook}. Within the \textit{event handlers}, the script captures keystrokes and writes them to a buffer. It also captures clipboard content when the user performs copy or paste actions. In the \textbf{\textit{hkb()}} function, the script checks if control keys are pressed and handles special keys accordingly. It also sets up timers to capture clipboard content shortly after copy or paste actions are detected.

\begin{figure} [H]
    \centering
    \includegraphics[width=1\linewidth,frame]{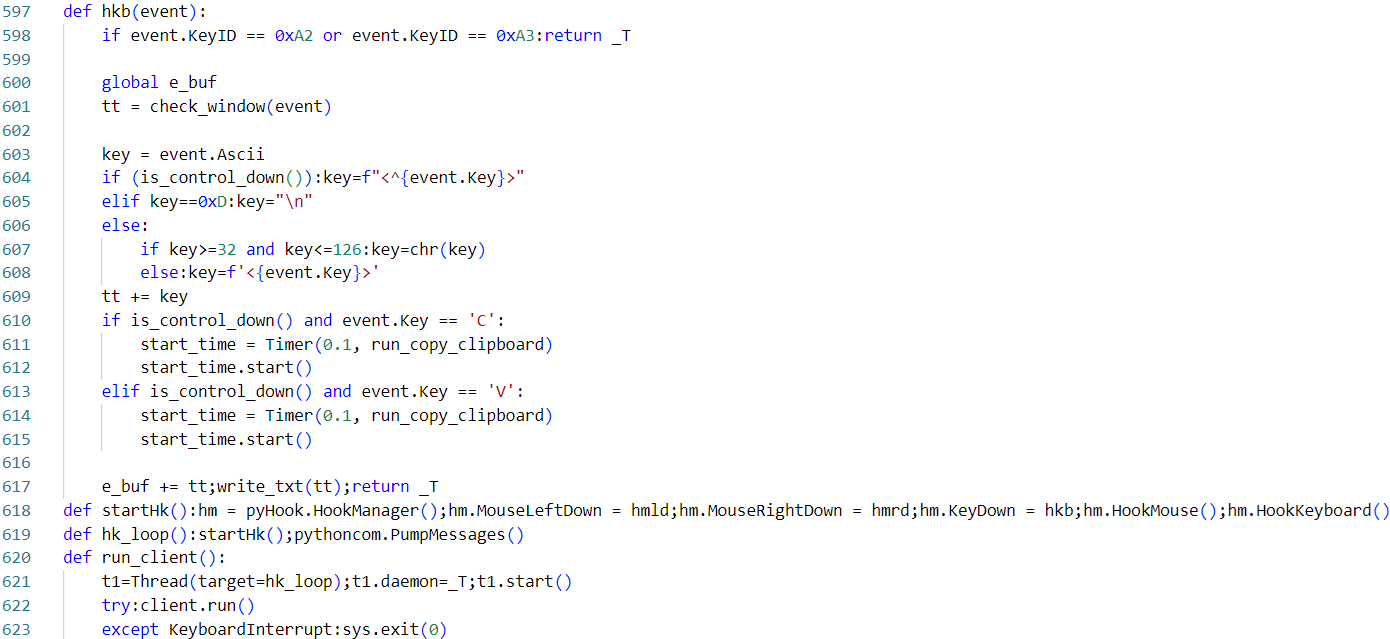}
    \caption{Main \textit{Hooking} routine}
    \label{fig:55}
\end{figure}

At the end of the script, the \textbf{\textit{run\_client()}} function is called within the \textit{\_\_main\_\_} block to start the malware's execution.

Regarding unused code, there are several sections where function calls are commented out, such as in the \textbf{\textit{auto\_up()}} function. This function is intended to search for files with patterns related to cryptocurrency wallets and configuration files, but the calls are commented out, possibly to avoid immediate detection or to be activated under certain conditions.

\begin{figure} [H]
    \centering
    \includegraphics[width=0.45\linewidth,frame]{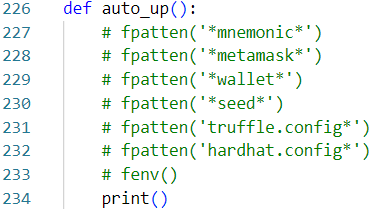}
    \caption{\textit{Crypto-Wallet} related patten which have been commented out.}
    \label{fig:56}
\end{figure}

Additionally, the \textbf{\textit{write\_txt()}} function is defined but does not perform any operation. It may have been intended to log captured keystrokes or clipboard content to a file but remains unused.

\begin{figure} [H]
    \centering
    \includegraphics[width=0.4\linewidth,frame]{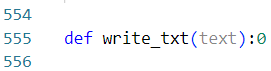}
    \caption{Unused function \textbf{\textit{write\_txt()}}}
    \label{fig:57}
\end{figure}

In conclusion, the script is a complex piece of malware that performs multiple malicious activities, including \textit{system information gathering},\textit{data exfiltration}, \textit{remote command execution}, \textit{file searching} and \textit{uploading}, \textit{keylogging}, \textit{clipboard monitoring}, and the ability to retrieve and execute \textbf{\textit{bow}} script. The latter provides additional resilience in case \textbf{\textit{sys\_info.py}} fails to correctly download it.

This \textit{threat} also leverages various Python modules and \textit{Windows API} functions to interact with the system and maintain persistence by establishing a connection with the attacker's server. The presence of unused code suggests that the malware may have additional capabilities that are not currently active but maybe intended for future use.

\subsubsection{Code Analysis - bow}
\textbf{\textit{bow}}  was previously employed as a \textit{Browser credentials' dumper}. However, by deobfuscating this script, beside the aforementioned well-known malicious functionality, designed to steal browser's credentials, there was found, embedded and obfuscated, an additional malicious payload with the aim of delivery the \textit{Tsunami} toolset.

\begin{figure} [H]
    \centering
    \includegraphics[width=1\linewidth,frame]{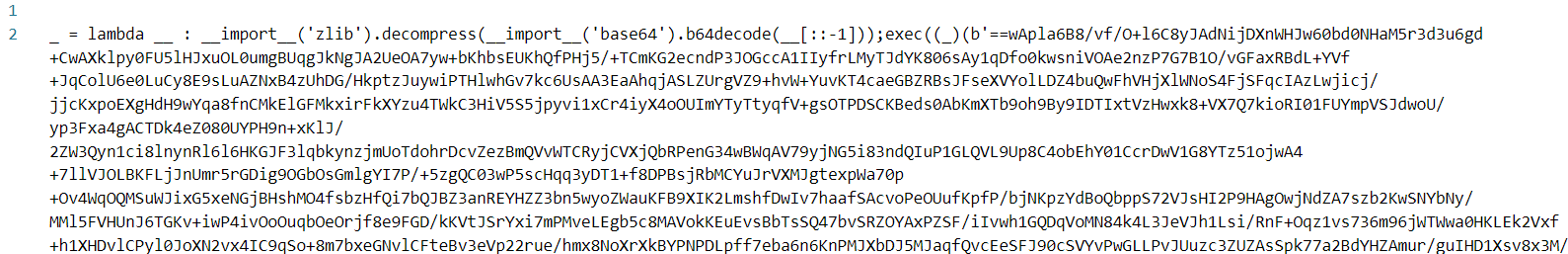}
    \caption{Snippet of the additional \textbf{\textit{Tsunami}} suite embedded in \textbf{\textit{Bow}} script.}
    \label{fig:58}
\end{figure}

As first, the credential stealing capabilities will be discussed, later also the newly identified functionalities will be analyzed as well.\newpage
\textbf{\textit{Browser Credentials Stealer}}\\
\textbf{\textit{bow}}  is a malicious program designed to extract sensitive information such as \textit{saved passwords} and \textit{credit card details} from various web browsers installed on a user's system. It targets multiple browsers, including \textit{Chrome}, \textit{Brave}, \textit{Opera}, \textit{Yandex}, and \textit{Microsoft Edge}, across different operating systems like \textit{Windows}, \textit{Linux}, and \textit{macOS}. The script decrypts the stored credentials and exfiltrates them to a remote server controlled by the attacker.

Starting from the main execution point, the script begins by importing necessary modules and setting up the environment. It attempts to import critical libraries required for its operation, and if they are not present, it installs them using \textit{pip} to ensure all dependencies are met. This includes libraries for \textit{HTTP requests}, cryptographic functions, and OS-specific modules for accessing system resources.

\begin{figure} [H]
    \centering
    \includegraphics[width=1\linewidth,frame]{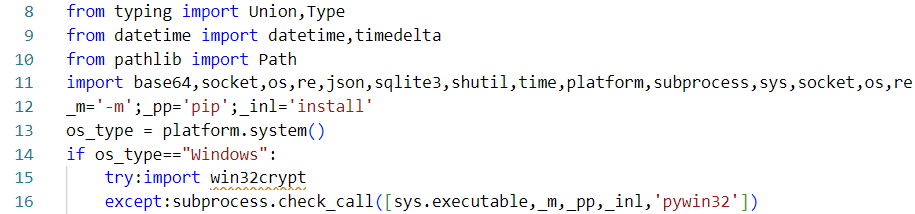}
    \caption{\textit{pip} imports and management of missing libraries.}
    \label{fig:59}
\end{figure}

The script sets up several global variables, including \textit{sType}, \textit{gType}, \textit{host1} and \textit{home}, which are used throughout the code for exfiltration and path resolution. It also determines the \textit{hostname} of the machine and constructs URLs for communication with the attacker's server.

\begin{figure} [H]
    \centering
    \includegraphics[width=0.55\linewidth,frame]{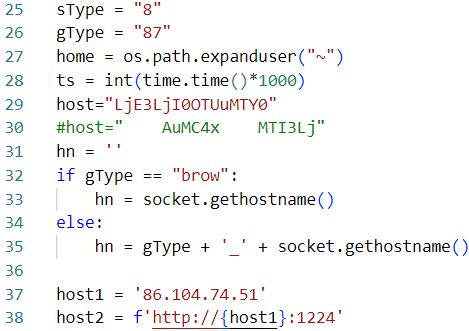}
    \caption{Global variables definition and \textit{C2 server} remote URL construction.}
    \label{fig:60}
\end{figure}

The script defines classes representing different browser versions it aims to target. Each class inherits from a base class \textit{BrowserVersion} and specifies the browser's base name along with version identifiers for \textit{Windows}, \textit{Linux}, and \textit{macOS}. An array \textit{available\_browsers} holds all the browser classes the script will attempt to extract data from.

\begin{figure} [H]
    \centering
    \includegraphics[width=1\linewidth,frame]{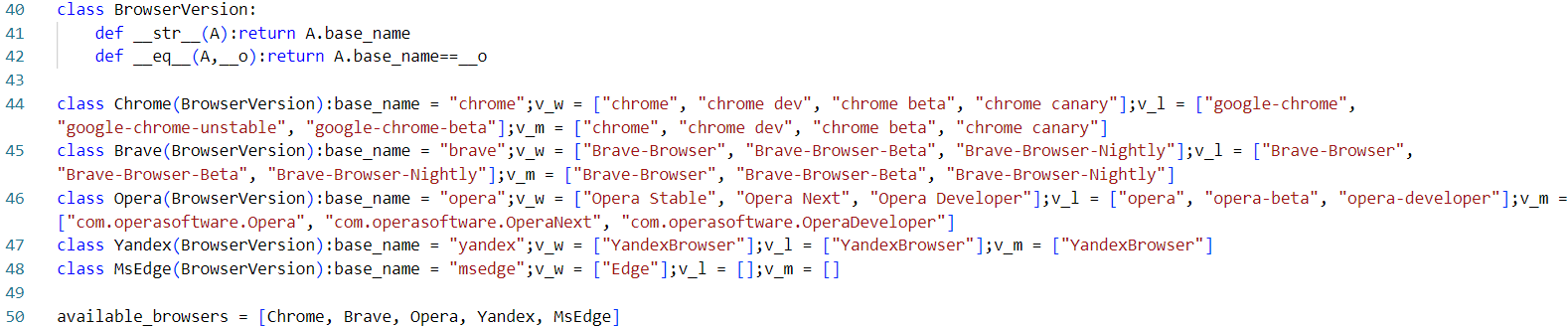}
    \caption{Classes defining all the targeted victims' browsers.}
    \label{fig:61}
\end{figure}

The core functionality resides within the \textit{ChromeBase} class and its subclasses for each operating system. This provides methods for decrypting stored credentials and retrieving data from browser databases.

In the \textit{ChromeBase} class, the \textit{get decorator} is used to dynamically update paths to the browser's data directories based on the operating system and browser versions.

\begin{figure} [H]
    \centering
    \includegraphics[width=1\linewidth,frame]{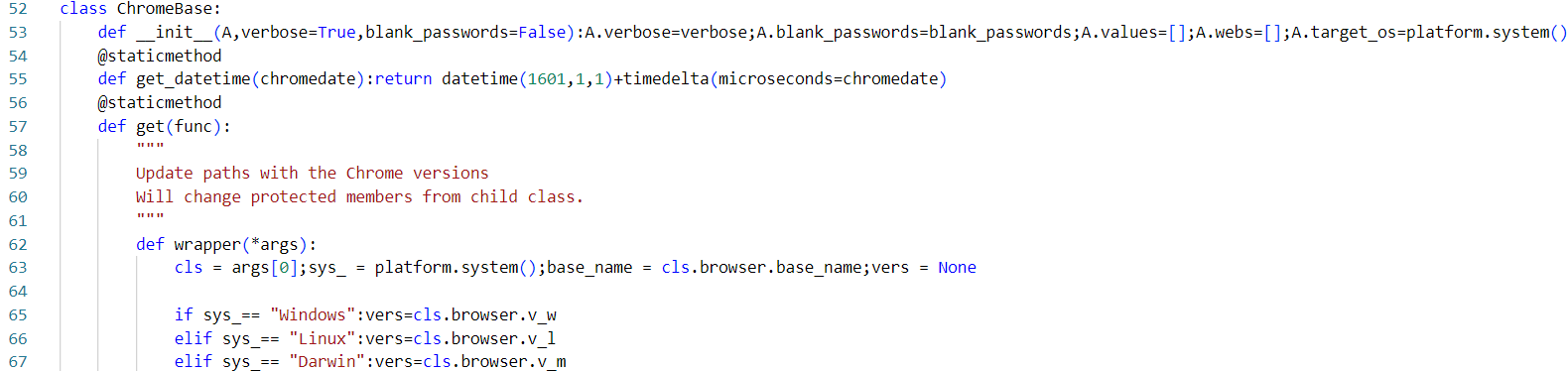}
    \caption{Snippet of the \textit{ChromeBase} class}
    \label{fig:62}
\end{figure}

The \textbf{\textit{retrieve\_database()}} method in \textit{ChromeBase} is responsible for copying the \textit{browser's login data database}, \textit{decrypting stored passwords}, and collecting them for \textit{exfiltration}.

\begin{figure} [H]
    \centering
    \includegraphics[width=1\linewidth,frame]{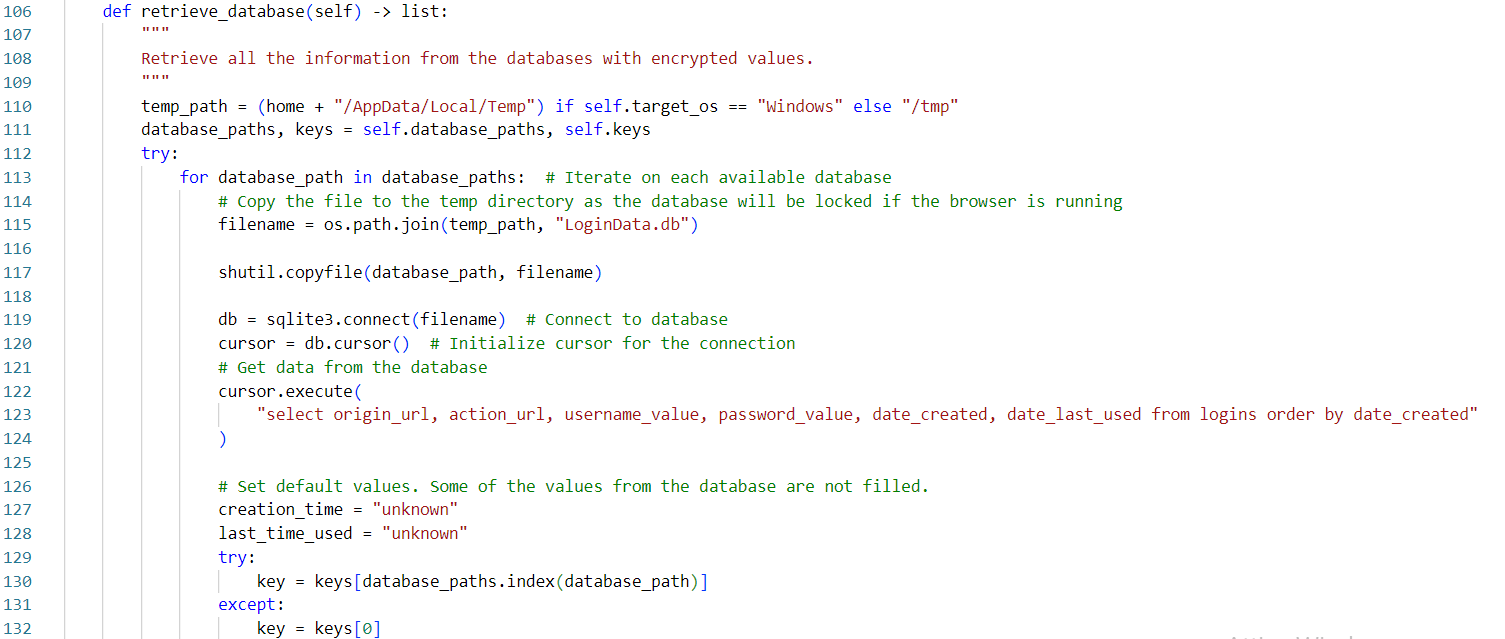}
    \caption{\textbf{\textit{retrieve\_database()}} targets \textit{Chrome} locally stored credentials.}
    \label{fig:63}
\end{figure}

Similarly, the \textbf{\textit{retrieve\_web()}} method extracts \textit{credit card information} stored by the browser.

\begin{figure} [H]
    \centering
    \includegraphics[width=1\linewidth,frame]{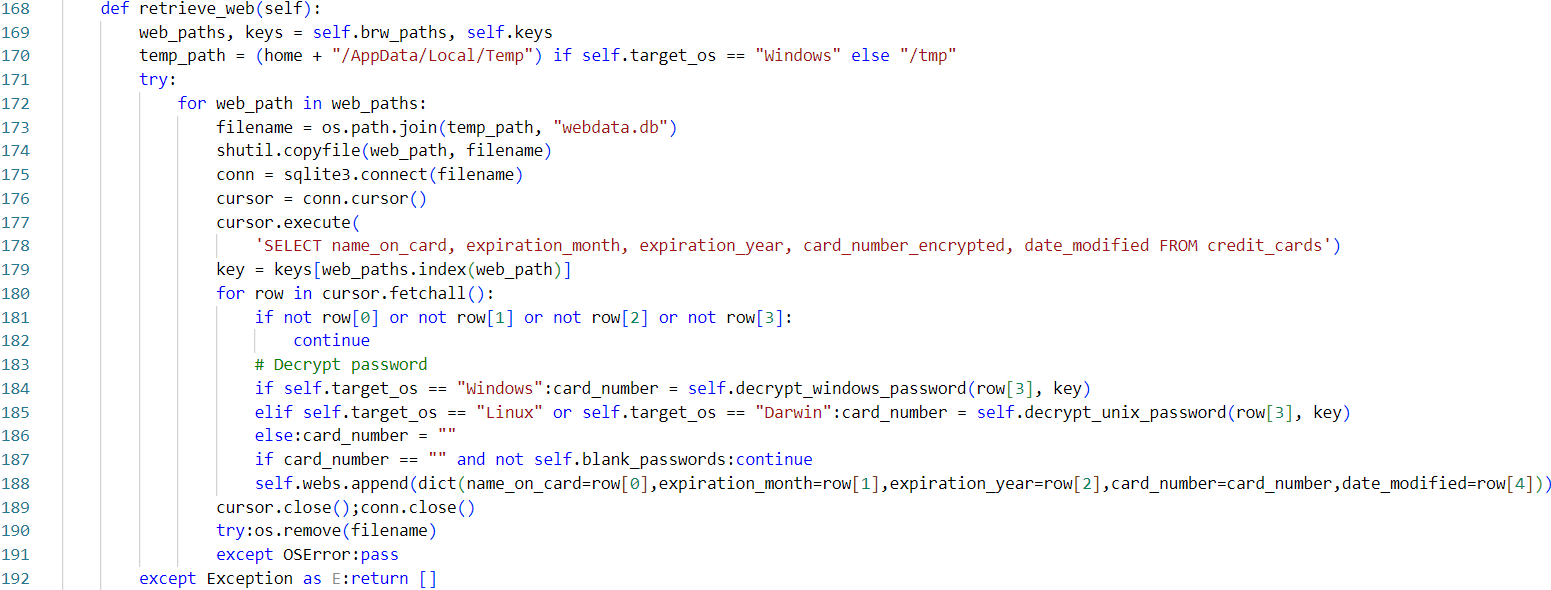}
    \caption{\textbf{\textit{retrieve\_web()}} targets credit cards information.}
    \label{fig:64}
\end{figure}

For Windows systems, the Windows class inherits from \textit{ChromeBase} and implements Windows-specific methods for \textit{decrypting passwords}. It uses the \textit{win32crypt} module to interact with \textit{Windows Data Protection API} (\textit{DPAPI}) for decryption.

\begin{figure}[H]
    \centering
    \includegraphics[width=1\linewidth,frame]{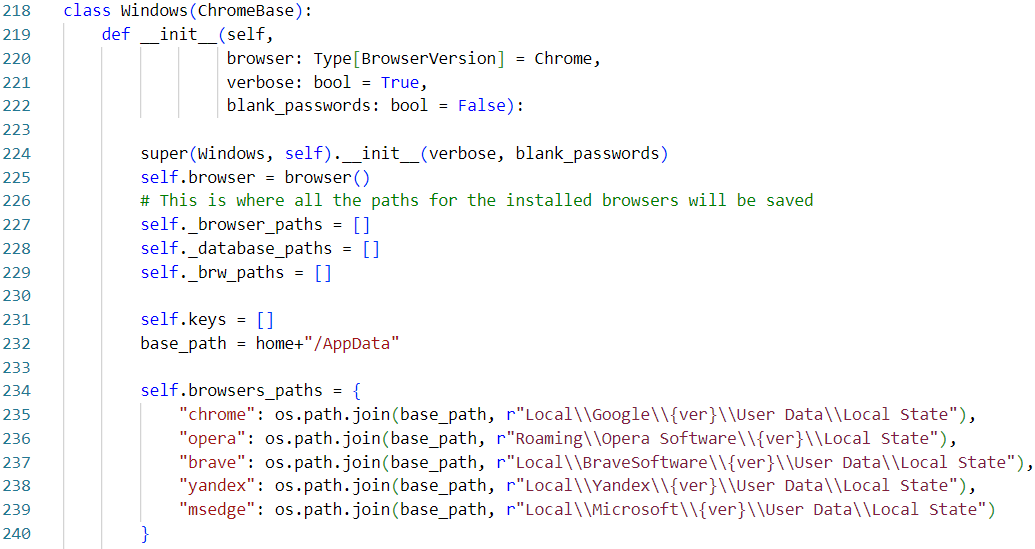}
    \caption{\textit{Windows} class initialization and browsers' paths.}
    \label{fig:65}
\end{figure}

For Linux systems, the \textit{Linux} class implements methods to retrieve the encryption key from the \textit{GNOME Keyring} using the \textit{secretstorage} module.

\begin{figure}[H]
    \centering
    \includegraphics[width=0.8\linewidth,frame]{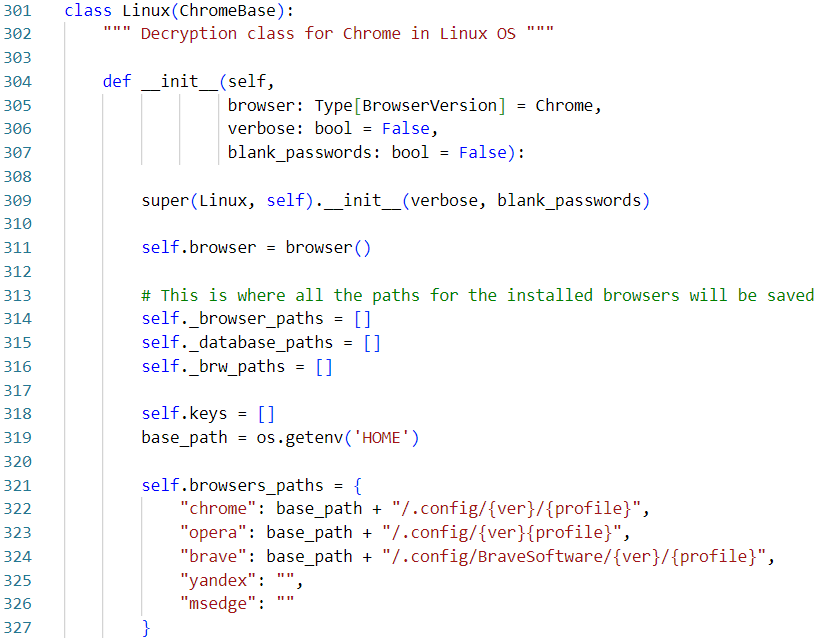}
    \caption{\textit{Linux} class initialization and browsers' paths.}
    \label{fig:66}
\end{figure}

For \textit{macOS} systems, the \textit{Mac} class retrieves the encryption key from the \textit{Keychain} using system commands.

\begin{figure}[H]
    \centering
    \includegraphics[width=0.8\linewidth,frame]{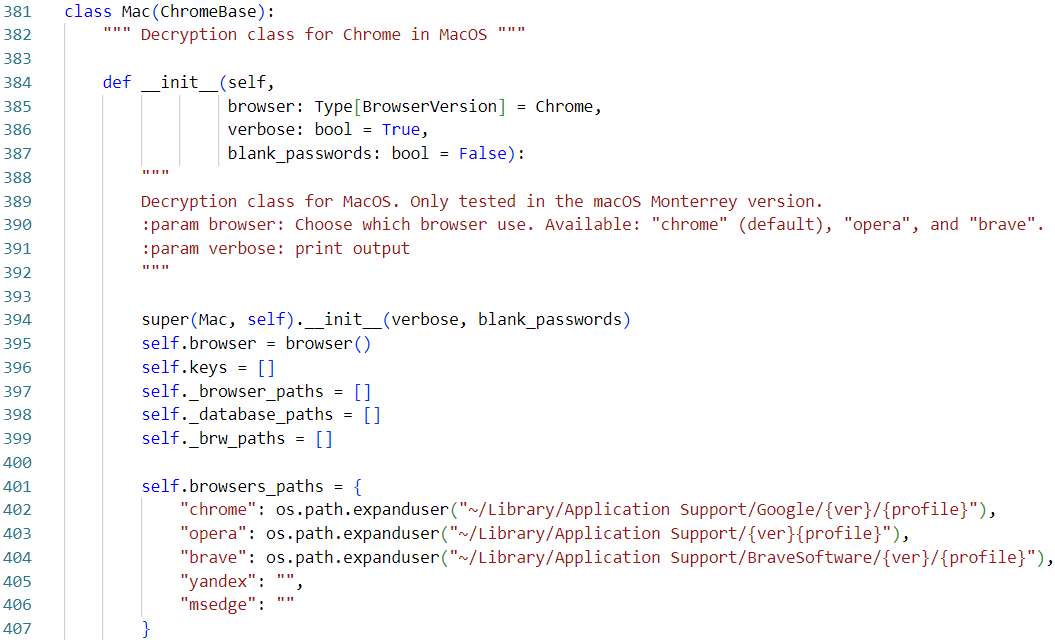}
    \caption{\textit{Mac} class initialization and browsers' paths.}
    \label{fig:67}
\end{figure}

At the end of the script, the main execution flow determines the operating system and initializes the appropriate class to perform the data extraction. It iterates over each available browser, retrieves stored credentials, and sends them to the attacker's server.

\begin{figure}[H]
    \centering
    \includegraphics[width=0.8\linewidth,frame]{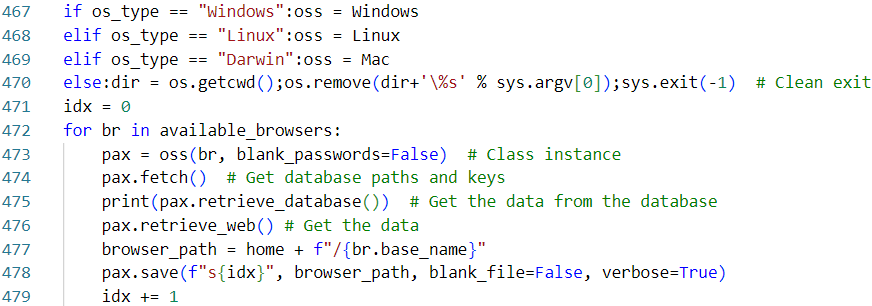}
    \caption{Main routine adapting its behavior within the identified OS.}
    \label{fig:68}
\end{figure}

The \textbf{\textit{save()}} method in \textit{ChromeBase} is responsible for exfiltrating the collected data by sending an \textit{HTTP POST request} to the attacker's server.

In this method, \textbf{\textit{self.pretty\_print()}} formats the extracted data into a readable string, which is then sent to the server specified by \textit{host2}. The data includes timestamps, host identifiers, and the collected credentials.

\begin{figure}[H]
    \centering
    \includegraphics[width=1\linewidth,frame]{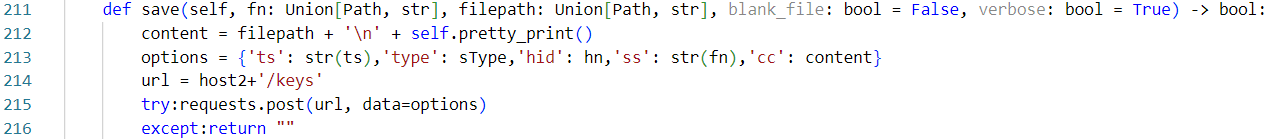}
    \caption{\textbf{\textit{save()}} function setups the exfiltration process.}
    \label{fig:69}
\end{figure}

Unused code in the script is minimal, with some commented out sections at the end that may have been used for debugging or cleanup purposes.

\begin{figure}[H]
    \centering
    \includegraphics[width=0.4\linewidth,frame]{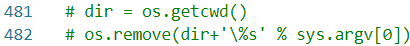}
    \caption{Commented clean-up last lines.}
    \label{fig:70}
\end{figure}

This most probably indicates an intention to remove the script after execution, possibly to cover its tracks, but it is commented out, so it doesn't execute.

In conclusion, this \textbf{\textit{bow}} script component operates by methodically \textit{accessing browser storage files}, \textit{decrypting sensitive information}, and sending it to a remote server without the user's consent. It uses platform-specific methods to \textit{handle encryption} and file paths, making it adaptable to various operating systems and browsers. The code is well-structured, leveraging object-oriented programming to encapsulate functionality for each operating system and browser type, which enhances its effectiveness as a malicious tool. Additionally, with some further \textit{OSINT} investigations, it has been possible to find another \textit{IoC} related to the same \textit{Threat Actor}, hosting this same script on another server in the past.

\begin{figure}[H]
        \centering
    \includegraphics[width=0.55\linewidth,frame]{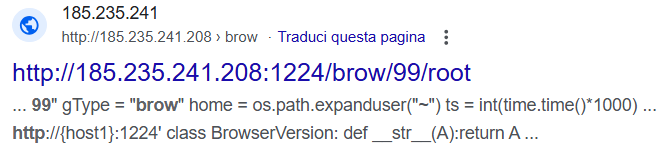}
    \caption{\textbf{\textit{Bow}} was hosted, in the past, on this server.}
    \label{fig:71}
\end{figure}

\textbf{\textit{Tsunami}}\\
With reference to \figurename~\ref{fig:58}, the first lines of the identified \textbf{\textit{bow}} script were embedding an additional malicious obfuscated payload. By applying the same 50-iterations deobfuscation process, as done for all the previously mentioned Python scripts, it was possible to gather its content.

The latter is a piece of malware designed to ensure that Python is installed on a Windows system and to persistently execute a secondary malicious script, referred to as the \textbf{\textit{TSUNAMI INJECTOR}}, by placing it in the system's \textit{startup folder}. The script \textit{employs obfuscation techniques} to conceal the secondary payload and attempts to gain \textit{elevated privileges} to install Python if it is not already present.

Starting from the \textit{main} execution point, the script begins by importing several modules necessary for its operation. These imports include standard libraries for system interaction, such as \textit{subprocess}, \textit{platform}, \textit{tempfile}, \textit{winreg}, \textit{ctypes}, \textit{random}, \textit{base64}, \textit{zlib}, \textit{time}, \textit{sys}, and \textit{os}. The script also attempts to suppress warnings to avoid drawing attention during execution. This suppression ensures that any warnings generated by the script are ignored, which is typical in malicious software to prevent the user from noticing unexpected behavior.

\begin{figure} [H]
    \centering
    \includegraphics[width=0.4\linewidth,frame]{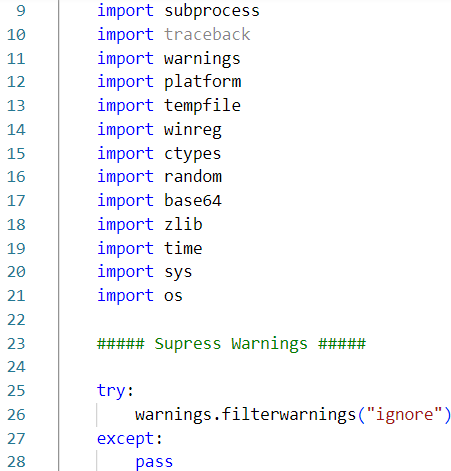}
    \caption{Script's imports}
    \label{fig:72}
\end{figure}

The script begins by defining several global variables that are essential to its operation. The \textit{DEBUG\_MODE} flag is initialized as \textit{False}, ensuring that the script suppresses debug output during execution unless explicitly enabled. This configuration emphasizes the malware's intent to operate covertly, minimizing any indicators of its presence.

Among the critical variables is the URL for downloading a Python installer, which points to an official Python repository. This mechanism enables the script to ensure that a Python interpreter is installed on the target system, a prerequisite for executing its subsequent stages. The inclusion of this step highlights the malware's adaptability and its capability to dynamically establish its required runtime environment.

The script also determines the path to the \textit{AppData Roaming} directory, a commonly utilized location in Windows for storing user-specific application data. This directory is leveraged to construct the storage path for the \textbf{\textit{TSUNAMI INJECTOR}}, the secondary malicious payload. The variables specify the name, folder, and full path where this payload will reside. Additionally, the \textbf{\textit{TSUNAMI\_INJECTOR\_SCRIPT}} variable is allocated to contain the actual code of this secondary stage, which serves a critical role in advancing the malware's objectives. A detailed examination of this payload and its functionality will be discussed in Sec. ~\ref{sec:tsu}.

\begin{figure}[H]
    \centering
    \includegraphics[width=0.9\linewidth,frame]{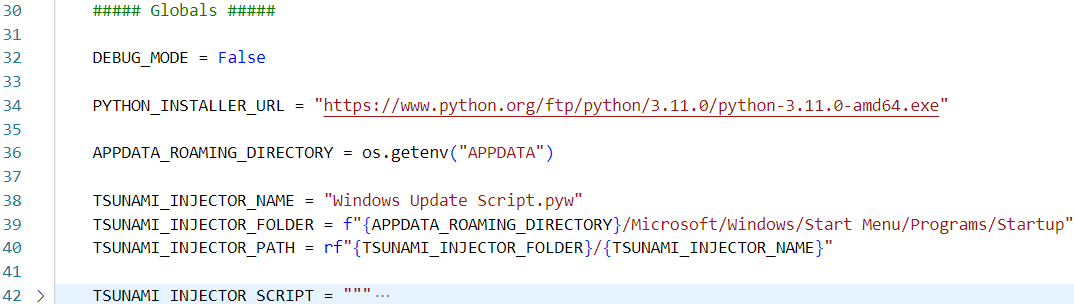}
    \caption{Script's global variables}
    \label{fig:73}
\end{figure}

The \textbf{\textit{obfuscate\_script()}} function takes the script data and a loop count to determine the level of obfuscation. It replaces a placeholder variable \textit{RandVar} with a random integer to ensure that the obfuscated script differs on each execution (avoiding an easy fingerprinting through hashing). In this function, the script repeatedly compresses and encodes the data, then reverses the encoded string. The obfuscation loop runs for the specified \textit{loop\_count}, which is set to 50 in the \textit{main block}, making the resulting script highly obfuscated and difficult to analyze. This technique is the same one used until now for all the identified Python scripts, here we can have a direct look on how the attacker implemented this by itself.

\begin{figure}[H]
    \centering
    \includegraphics[width=0.9\linewidth,frame]{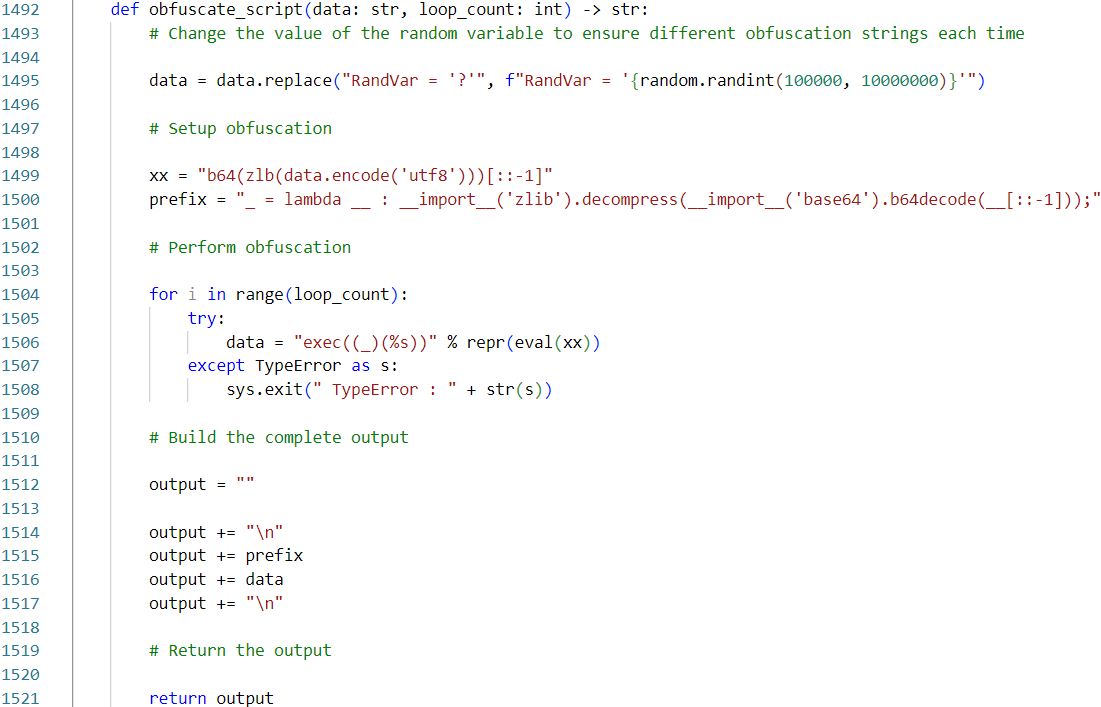}
    \caption{50-iterations obfuscation technique implementation.}
    \label{fig:74}
\end{figure}

Utility functions are defined to assist with the script's operations. The output function conditionally prints debug messages if \textit{DEBUG\_MODE} is enabled.

\begin{figure} [H]
    \centering
    \includegraphics[width=0.4\linewidth,frame]{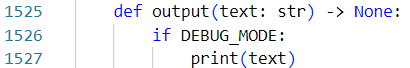}
    \caption{Debugging mode}
    \label{fig:75}
\end{figure}

The \textbf{\textit{download\_file()}} function uses \textit{PowerShell} to download a file from a given URL to a specified file path.

\begin{figure}[H]
    \centering
    \includegraphics[width=0.8\linewidth,frame]{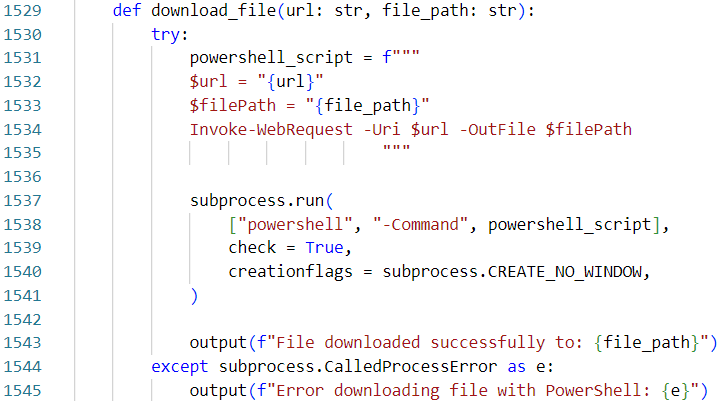}
    \caption{Function designed to download remote utilities.}
    \label{fig:76}
\end{figure}

By utilizing \textit{PowerShell}'s \textit{Invoke-WebRequest cmdlet}, the script avoids raising network-related flags that might occur with other methods.

The script proceeds to define functions under the \textit{Tsunami Infecter} section, which handle the installation of Python if it is not already present. The \textbf{\textit{is\_Python\_installed()}} function checks the Windows registry to determine if Python is installed on the system.

\begin{figure} [H]
    \centering
    \includegraphics[width=1\linewidth,frame]{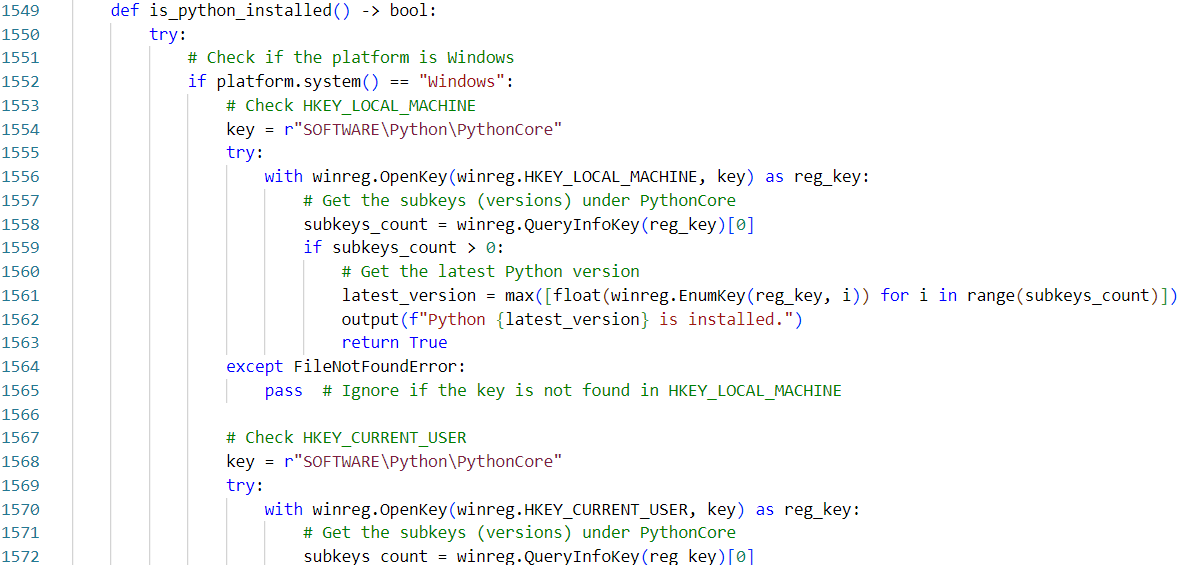}
    \caption{Function designed to check whether a Python interpreter is available on target machine.}
    \label{fig:77}
\end{figure}

This function attempts to open the \textit{PythonCore} registry key under both \textit{HKEY\_LOCAL\\\_MACHINE} and \textit{HKEY\_CURRENT\_USER} to check for installed Python versions. If no versions are found, it concludes that Python is not installed.

The \textbf{\textit{execute\_Python\_with\_uac()}} function tries to run the Python installer with administrative privileges using the Windows \textit{ShellExecute API}:

\begin{figure} [H]
    \centering
    \includegraphics[width=0.8\linewidth,frame]{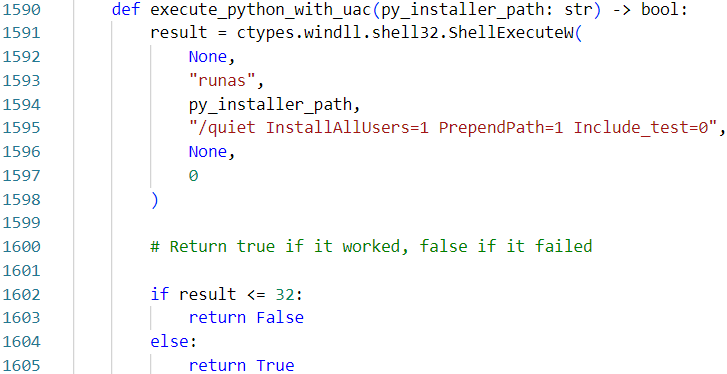}
    \caption{Function designed to \textit{runas} to install a Python interpreter.}
    \label{fig:78}
\end{figure}

By specifying the \textit{runas} verb, the script prompts the \textit{User Account Control} (\textit{UAC}) dialog to request elevated privileges. The installer is executed with silent installation parameters to avoid user interaction.

The \textbf{\textit{install\_Python()}} function orchestrates the download and installation of Python inside a newly created temporary file path, and attempts to execute it with elevated privileges. If the user denies the \textit{UAC} prompt, the script waits for a random interval between 10 and 30 seconds before retrying, persistently attempting to install Python.

\begin{figure} [H]
    \centering
    \includegraphics[width=0.8\linewidth,frame]{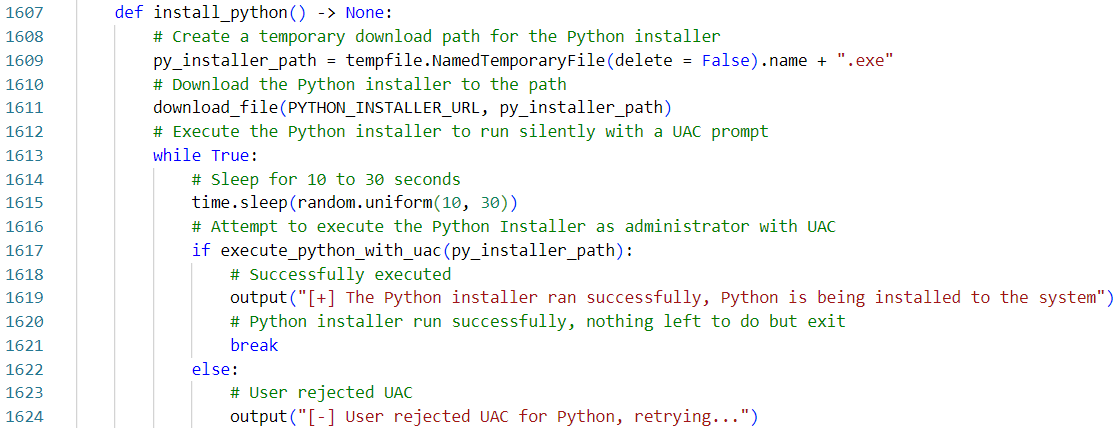}
    \caption{Function designed to run Python installer and prompting for user administrative permissions via \textit{UAC}.}
    \label{fig:79}
\end{figure}

In the main section of the script, the execution flow begins by checking if Python is installed. If Python is not installed, it proceeds to download and install it using the methods previously described. Once Python is confirmed to be installed, the script writes the obfuscated \textbf{\textit{TSUNAMI INJECTOR}} to the \textit{Windows Startup folder} to ensure persistence. The \textbf{\textit{obfuscate\_script()}} function is called with a \textit{loop\_count} of 50, resulting in a heavily obfuscated script that is difficult to analyze or detect by security software. The script is saved with a \textit{.pyw} extension, which allows Python scripts to run without opening a console window, further hiding its execution. The script includes a check for \textit{DEBUG\_MODE}, and if enabled, it waits for user input to keep the window open. The entire script is also wrapped in a try-except block that silently passes any exceptions.

\begin{figure} [H]
    \centering
    \includegraphics[width=0.7\linewidth,frame]{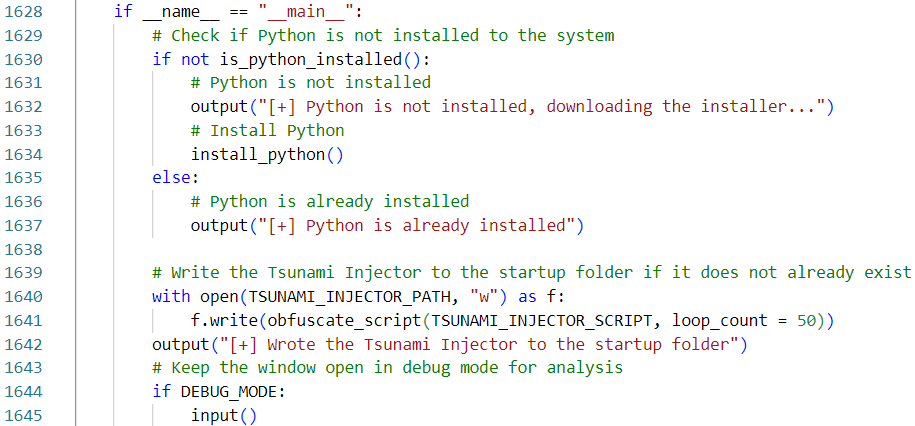}
    \caption{Script's \textit{main} routine}
    \label{fig:80}
\end{figure}

In conclusion, the script functions as a dropper that ensures Python is installed on the target Windows system, leveraging administrative privileges if necessary. It then installs a persistent, obfuscated secondary payload in the startup folder to achieve persistence and execute additional malicious activities each time the system boots. The use of obfuscation and silent error handling indicates an attempt to evade detection and analysis, which is characteristic of malicious software designed to compromise system security without the user's knowledge.

Moreover, as it is possible to see from the following image, the attacker posed, in the first lines of the \textbf{\textit{Windows Update Script.pyw}} script a peculiar citation.

\begin{figure}[H]
    \centering
    \includegraphics[width=1\linewidth,frame]{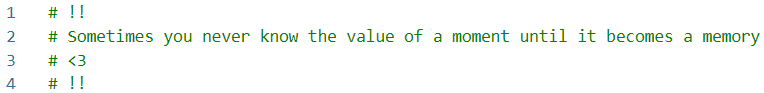}
    \caption{Interesting citation available inside \textbf{\textit{Windows Update Script.pyw}}.}
    \label{fig:CIT}
\end{figure}

The quote, \textit{Sometimes you never know the value of a moment until it becomes a memory}, is often attributed to Dr. Seuss, although its precise origins are uncertain. The phrase captures a universal truth about human experience: we often fail to recognize the significance of events as they happen and only appreciate them in hindsight. However, no additional insights about the usage of this were identified, either as associated to the \textit{threat} or the \textit{Threat Actor} itself.

\subsection{Fourth Stage}
\begin{figure} [H]
    \centering
    \includegraphics[width=0.9\linewidth,frame]{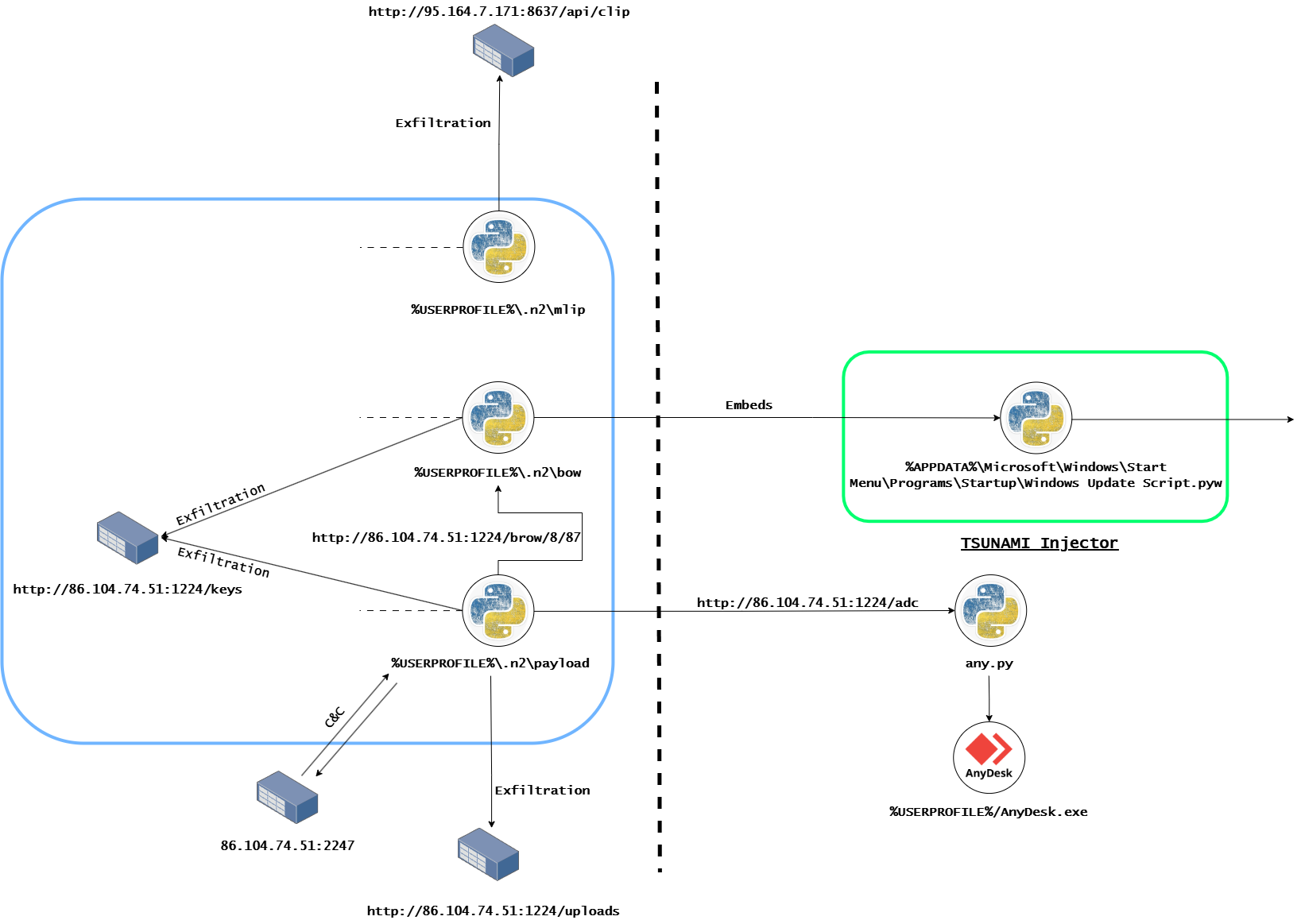}
    \caption{Moving from Third to Fourth Stage.}
    \label{fig:F0S}
\end{figure}
\subsubsection{Code Obfuscation}
In this stage, as yet reported in the previous section, \textbf{\textit{Windows Update Script.pyw}} was obfuscated with the well-known 50-iterations process. On the other hand, \textbf{\textit{any.py}} is not ubfuscated at all.

\subsubsection{Code Analysis - any.py}\label{Sec:any}
 \textbf{\textit{any.py}} is a malicious program designed to manipulate the configuration of \textit{AnyDesk}, a popular remote desktop application, on a target system. The script aims to \textit{modify AnyDesk's configuration files} to inject predetermined credentials, potentially allowing \textit{unauthorized remote access} to the system. It also attempts to download and execute \textit{AnyDesk} if it is not already present, and ensures that \textit{AnyDesk} is running with the manipulated configuration. Finally, the script cleans up by \textit{deleting itself from the system}. These imports include modules for system interaction (\textit{os}, \textit{platform}, \textit{subprocess}, \textit{sys}), networking (\textit{socket}, \textit{requests}), and data encoding/decoding (\textit{base64}, \textit{time}).

The script then determines the operating system type and retrieves environment variables essential for its execution. Starting from the main execution point, the script begins by importing necessary modules that facilitate its operation. The \textit{os\_type} variable holds the name of the operating system, which is crucial for setting file paths and executing OS-specific commands. The \textit{appdata} variable retrieves the path to the local application data directory on Windows systems. Next, the script defines variables that are used to construct the URL of a remote server controlled by the attacker. Here, host is a \textit{base64}-encoded string that, when decoded, provides the \textit{IP address} of the attacker's server. The \textit{hn} variable stores the \textit{hostname} of the victim's machine, and \textit{sType} is likely used to categorize the type of data being sent to the server. The script then decodes the \textit{host} string to obtain the actual server address. In the following snippet, this string is manipulated by rearranging its parts before decoding. The slicing \textit{host[8:] + host[:8]} swaps the first eight characters with the rest as a rudimentary obfuscation technique. After decoding, \textit{host1} contains the server address (95.164.17[.]24), and \textit{host2} constructs the full URL with a specific port (\textit{1224}).

\begin{figure} [H]
    \centering
    \includegraphics[width=0.7\linewidth,frame]{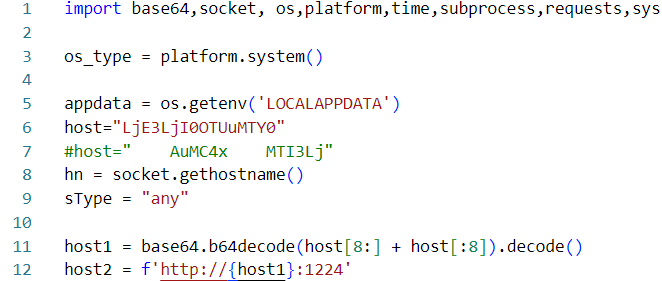}
    \caption{\textbf{\textit{any.py}} imports and global variables}
    \label{fig:81}
\end{figure}

The script then defines a function \textbf{\textit{save\_conf()}} that reads the contents of a given file and sends it to the attacker's server. This function checks if the file \textit{fn} exists. If it does, it reads the file's contents into \textit{buf}. If the latter is not empty, it constructs a data payload options containing the file content and sends it to the attacker's server via an \textit{HTTP POST request} to the \textit{/keys} endpoint. The script then sets up paths and variables necessary for interacting with \textit{AnyDesk}'s configuration. It defines the home directory and initializes an empty list files. The variable \textit{any\_path} specifies the default installation path of \textit{AnyDesk} on Windows systems.

\begin{figure}[H]
    \centering
    \includegraphics[width=0.8\linewidth,frame]{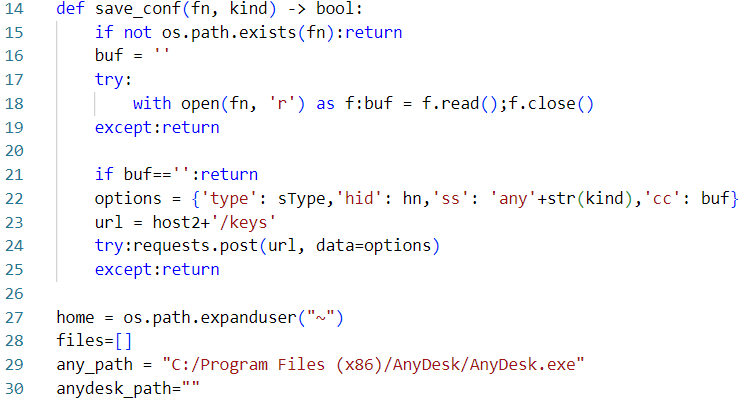}
    \caption{Defining \textit{AnyDesk} path and configuring \textit{C2} connection to share its settings.}
    \label{fig:82}
\end{figure}

A function \textbf{\textit{get\_anydesk\_path()}} is defined to locate or download \textit{AnyDesk} if it is not already installed.

\begin{figure}[H]
    \centering
    \includegraphics[width=1\linewidth,frame]{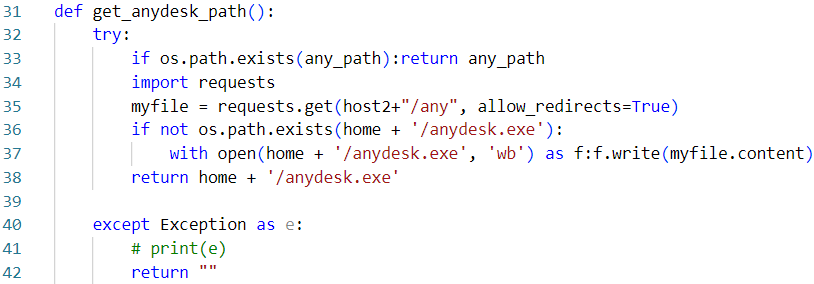}
    \caption{Funtion designed to establish \textit{AnyDesk} presence on target system.}
    \label{fig:83}
\end{figure}

This function first checks if \textit{AnyDesk} exists at the default path. If not, it attempts to download \textit{AnyDesk} from the attacker's server (\textit{host2 + /any}). The downloaded executable is saved in the user's home directory as \textit{anydesk.exe}. The function then returns the path to the \textit{AnyDesk} executable. The script proceeds to determine the paths to \textit{AnyDesk}'s configuration files based on the operating system. For Windows systems, it sets \textit{conf\_path1} and \textit{conf\_path2} to the possible locations of \textit{AnyDesk}'s \textit{service.conf} file. For non-Windows systems, it sets the paths accordingly. If neither configuration file exists on a Windows system, the script attempts to run \textit{AnyDesk}. This step ensures that \textit{AnyDesk} is running, potentially causing it to create the \textit{service.conf} file, which the script intends to modify.

\begin{figure} [H]
    \centering
    \includegraphics[width=1\linewidth,frame]{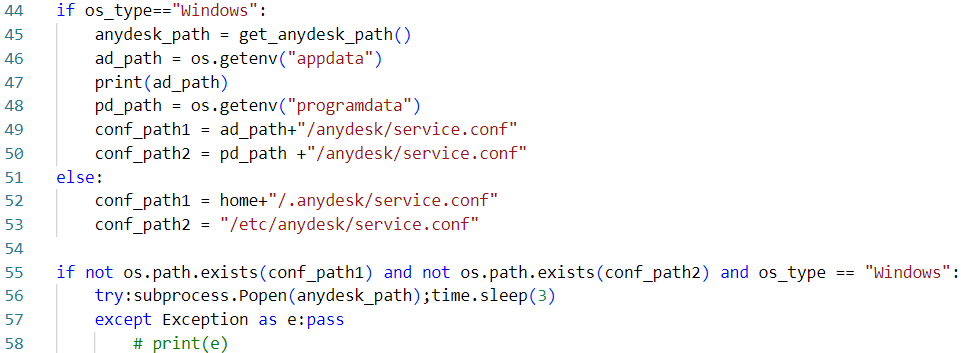}
    \caption{Script maps \textit{AnyDesk}'s configurations related paths.}
    \label{fig:84}
\end{figure}

It then defines a \textit{PowerShell} script as a multi-line string \textit{anydesk\_ps1}.

\begin{figure} [H]
    \centering
    \includegraphics[width=0.8\linewidth,frame]{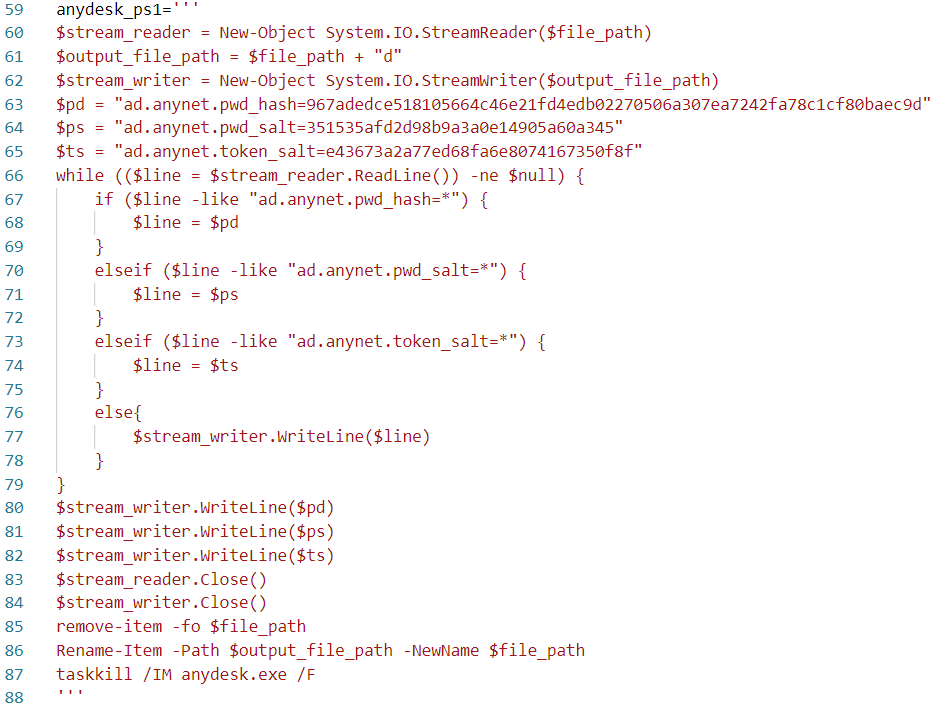}
    \caption{\textit{anydesk\_ps1} variable content}
    \label{fig:85}
\end{figure}

This script reads the \textit{AnyDesk} configuration file, replaces certain lines with predefined values (specifically \textit{pwd\_hash}, \textit{pwd\_salt}, and \textit{token\_salt}), and writes the changes back to the file. It then forcefully terminates \textit{AnyDesk}.

The core function that performs the configuration file modification is \textit{update\_conf}.

\begin{figure} [H]
    \centering
    \includegraphics[width=1\linewidth,frame]{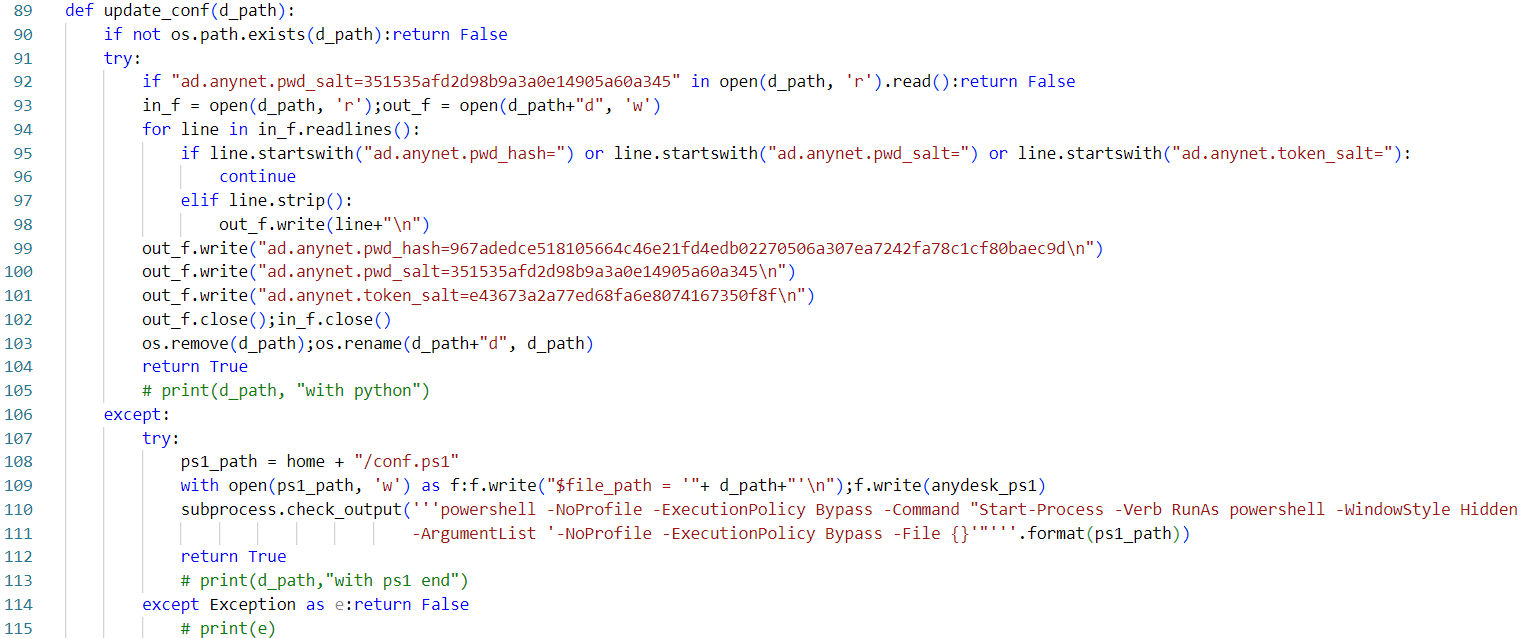}
    \caption{Function designed to update \textit{AnyDesk} configurations.}
    \label{fig:86}
\end{figure}

This function first checks if the configuration file at \textit{d\_path} exists. It then reads the file to see if it already contains the attacker's \textit{pwd\_salt}. If not, it proceeds to modify the file. It opens the existing configuration file for reading and a new file (\textit{d\_path + d}) for writing. It copies all lines except those starting with \textit{ad.anynet.pwd\_hash=}, \textit{ad.anynet.pwd\_salt=}, or \textit{ad.anynet.token\_salt=}. It then writes the attacker's predefined values for these settings to the new file.

If direct file modification fails (possibly due to permissions), the function attempts to execute the previously defined \textit{PowerShell} script with elevated privileges. It writes the \textit{PowerShell} script to a file (\textit{conf.ps1}) and executes it using a \textit{subprocess} call with \textit{Start-Process -Verb RunAs}, which prompts for administrative rights.

The script then calls \textbf{\textit{update\_conf()}} on both configuration file paths. After attempting to update the configuration files, the script defines a function \textit{restart\_anydesk} to restart the \textit{AnyDesk} application.

\begin{figure}[H]
    \centering
    \includegraphics[width=0.85\linewidth,frame]{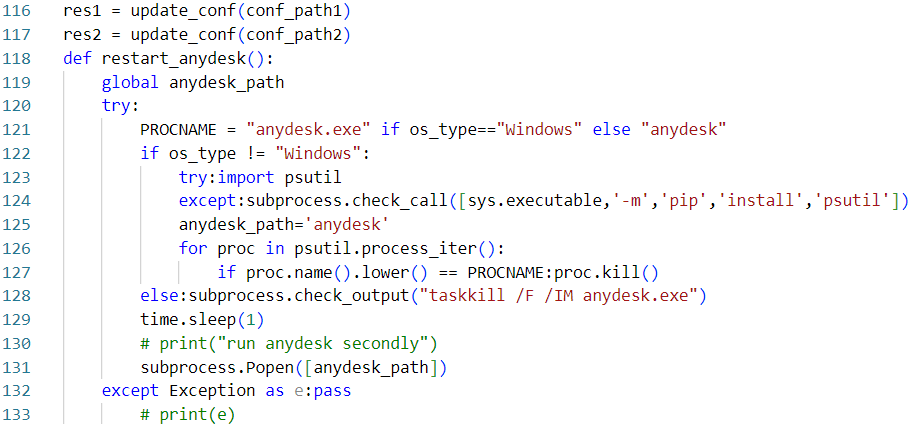}
    \caption{Configurations update and \textit{AnyDesk} restart.}
    \label{fig:87}
\end{figure}

This function kills any running \textit{AnyDesk} processes and restarts them subsequently. On non-Windows systems, it uses the \textit{psutil} library to iterate over running processes and terminate them. On Windows, it uses the \textit{taskkill} command. After killing the process, it waits for one second and restarts \textit{AnyDesk} using the \textit{anydesk\_path} determined earlier.

The script then saves the (possibly modified) configuration files to the attacker's server. By calling \textbf{\textit{save\_conf()}}, the script reads the contents of \textit{conf\_path1} and \textit{conf\_path2} and sends them to the server, allowing the attacker to retrieve the configuration files. Finally, the script restarts \textit{AnyDesk} and deletes itself.

\begin{figure} [H]
    \centering
    \includegraphics[width=0.7\linewidth,frame]{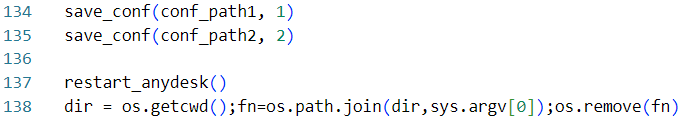}
    \caption{Manipulation of the \textit{AnyDesk} configuration and settings.}
    \label{fig:88}
\end{figure}

Deleting itself is a common tactic in malware to reduce forensic evidence and avoid detection.

Regarding unused code, the script includes commented-out print statements and exception handling that does not report errors. These comments suggest that during development, the script output errors for debugging purposes, but these were suppressed in the final version to avoid revealing its activities.

In conclusion, the script is a malicious tool designed to manipulate \textit{AnyDesk}'s configuration to insert known credentials, potentially granting the attacker unauthorized remote access to the victim's system. It ensures \textit{AnyDesk} is installed and running, modifies configuration files with predetermined values, restarts \textit{AnyDesk} to apply changes, and exfiltrates the configuration files to the attacker's server. The script takes measures to avoid detection by deleting itself after execution and suppressing error messages.

\subsubsection{Code Analysis - Windows Update Script.pyw} \label{sec:tsu}

This specific Python script is designed to establish persistence on a Windows system by creating \textit{scheduled tasks}, \textit{downloading} and \textit{executing additional malicious payloads}, and \textit{bypassing security measures} such as \textit{Windows Defender}. The script employs various obfuscation techniques to conceal its activities and evade detection. It attempts to \textit{escalate privileges} by prompting the \textit{User Account Control} (\textit{UAC}) dialog to gain administrative rights for executing its payloads.

Starting from the main execution point, the script begins by importing several modules necessary for its operation. These imports provide functionalities for \textit{network communication}, \textit{file handling}, \textit{system interaction}, \textit{encryption}, and \textit{obfuscation}. The script defines also a global variable \textit{RandVar}, which is assigned a random integer value. This variable is used within the obfuscation process to ensure that each deobfuscated script instance is unique. Next, the script sets up several global variables that determine paths and names used throughout its execution.

\begin{figure} [H]
    \centering
    \includegraphics[width=0.25\linewidth,frame]{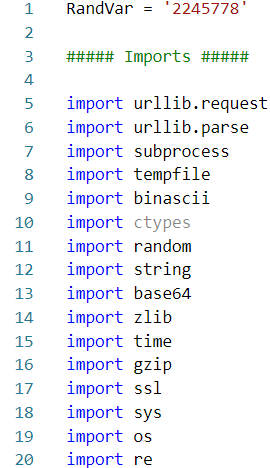}
    \caption{Script's imports and anti-fingerprinting variable \textit{RandVar}.}
    \label{fig:89}
\end{figure}

The script introduces several critical global variables that govern its behavior and facilitate the deployment of its malicious components. The \textit{DEBUG\_MODE} flag is used to toggle debug output, remaining disabled in its default state to minimize any detectable artifacts during execution.

Paths to the \textit{AppData} directories, both \textit{Roaming} and \textit{Local}, are retrieved using the variables \textit{ROAMING\_APPDATA\_PATH} and \textit{LOCAL\_APPDATA\_PATH}. These directories are commonly exploited by malware due to their accessibility and legitimate usage in Windows environments.

For the malicious payload, \textit{TSUNAMI\_PAYLOAD\_NAME} dynamically generates a random 16-character string to obfuscate the filename and evade static detection. The variables \textit{TSUNAMI\_PAYLOAD\_FOLDER} and \textit{TSUNAMI\_PAYLOAD\_PATH} are used to specify the temporary directory and complete file path for the payload's storage, reinforcing the attack’s stealth. Similarly, the names and paths for the malicious installer are defined using \textit{TSUNAMI\_INSTALLER\_NAME}, \textit{TSUNAMI\_INSTALLER\_FOLDER}, and \textit{TSUNAMI\_INSTALLER\_PATH}. These variables ensure precise control over the placement and execution of the installer within the compromised system.

Lastly, the script embeds a multi-line string containing the payload’s code, assigned to \textit{TSUNAMI\_PAYLOAD\_SCRIPT}. This design ensures that the payload is readily available for execution without requiring an immediate download, thus increasing the resilience and effectiveness of the attack.

\begin{figure} [H]
    \centering
    \includegraphics[width=0.85\linewidth,frame]{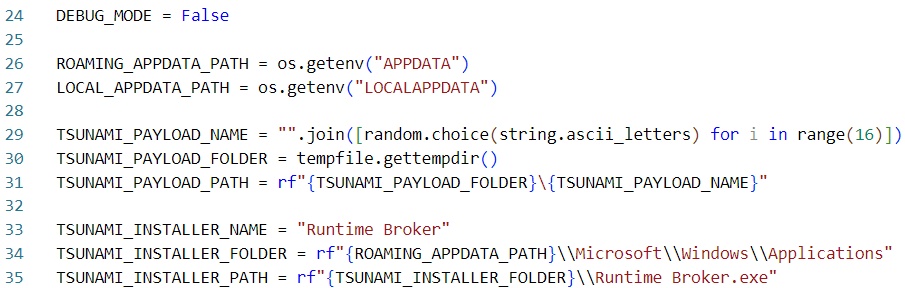}
    \caption{Global variables embedding additional payloads information and paths.}
    \label{fig:90}
\end{figure}

The script contains an embedded payload script as a multi-line string assigned to \textit{TSUNAMI\_PAYLOAD\_SCRIPT}, designed to be obfuscated and executed later.

\begin{figure} [H]
    \centering
    \includegraphics[width=0.85\linewidth,frame]{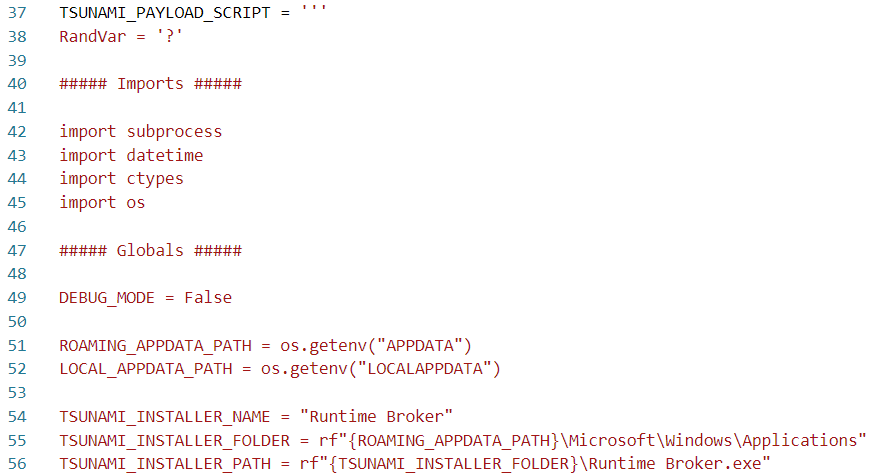}
    \caption{Code snippet of the embedded \textit{TSUNAMI\_PAYLOAD\_SCRIPT}.}
    \label{fig:91}
\end{figure}

Within this embedded script, the \textbf{\textit{add\_windows\_defender\_exception()}} function attempts to add specific file paths to the \textit{Windows Defender Exclusion List} by executing \textit{PowerShell} commands. The \textbf{\textit{create\_task()}} function creates a scheduled task named \textit{Runtime Broker} that executes the malicious installer at user logon with administrative privileges.

The \textbf{\textit{obfuscate\_script()}} function is responsible for obfuscating the payload script (identical to the one shown in \figurename~\ref{fig:74}). \textbf{\textit{Windows Update Script.pyw}} as first deploys and run this Python script to make arrangements for the next deploy of the \textbf{\textit{TSUNAMI INSTALLER}}. Indeed, it will apply \textit{AV} exclusions for the executable path and will also create a \textit{scheduled task} to allow its run at each user's login. At this point, the script will exploit the \textbf{\textit{is\_task\_scheduled()}} to check if this scheduled task exists with a \textit{PowerShell} query.

\begin{figure} [H]
    \centering
    \includegraphics[width=0.75\linewidth,frame]{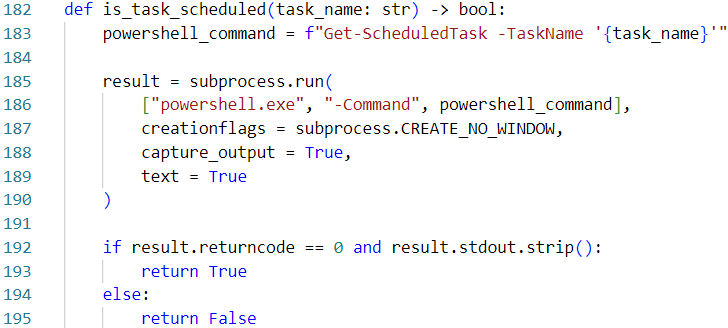}
    \caption{Function designed to check whether \textbf{\textit{Runtime Broker.exe}} is in a \textit{scheduled task}.}
    \label{fig:92}
\end{figure}

Then, the script defines functions to decrypt and decode an obfuscated URL from which it downloads an additional malicious payload. These functions perform \textit{xor} encryption/decryption (key: \textit{!!!HappyPenguin1950!!!}) and \textit{base64} decoding to retrieve the actual URL. These are encrypted and store in the \textit{URLS} array, which has a size of 1000 strings. Each one of these is composed of a \textit{Profile Name}, a '\_' and a \textit{File Name} (e.g. \textit{GlassesMagenta6644\_MassageRecorded9001}).

\begin{figure} [H]
    \centering
    \includegraphics[width=0.75\linewidth,frame]{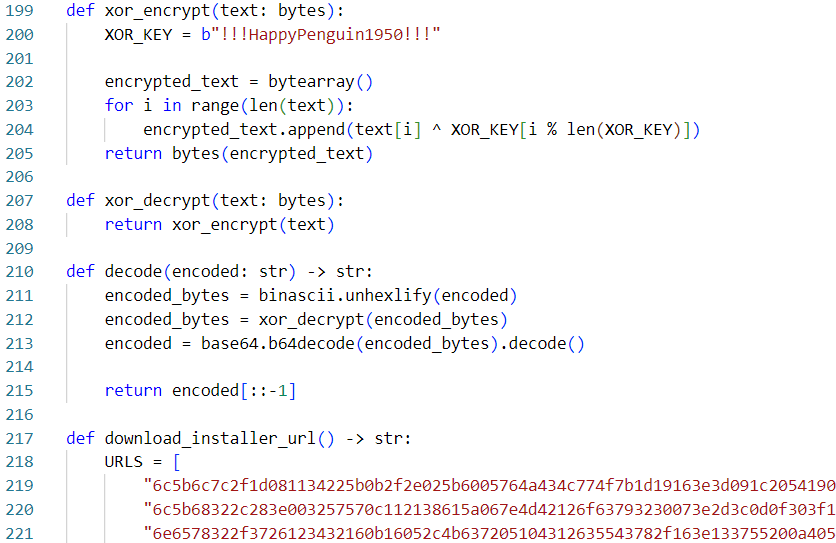}
    \caption{Functions designed to decrypt the strings embedded in the \textit{URLS} array.}
    \label{fig:95}
\end{figure}

\textbf{\textit{download\_installer\_url()}} shuffles the \textit{URLS} list and then begin looking for existing profiles and blacklisting non-existing ones. It also disables \textit{SSL} and employs as \textit{User-Agent} the string \textit{Mozilla/5.0}. In details, it retrieves from each single encrypted string the \textit{Profile Name}. Thus, looks for a document, named as the \textit{File Name} value, which will contain the path for the additional payload download, on \textit{Pastebin}.

\begin{figure}[H]
    \centering
    \includegraphics[width=0.75\linewidth,frame]{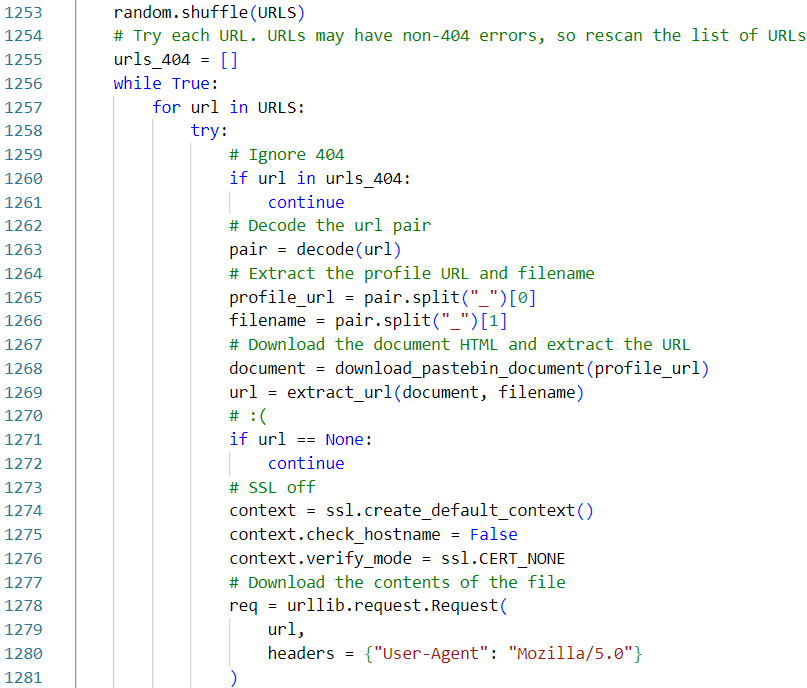}
    \caption{\textbf{\textit{download\_installer\_url()}} queries \textit{Pastebin} profiles and find existing ones.}
    \label{fig:96}
\end{figure}

\begin{figure} [H]
    \centering
    \includegraphics[width=0.9\linewidth,frame]{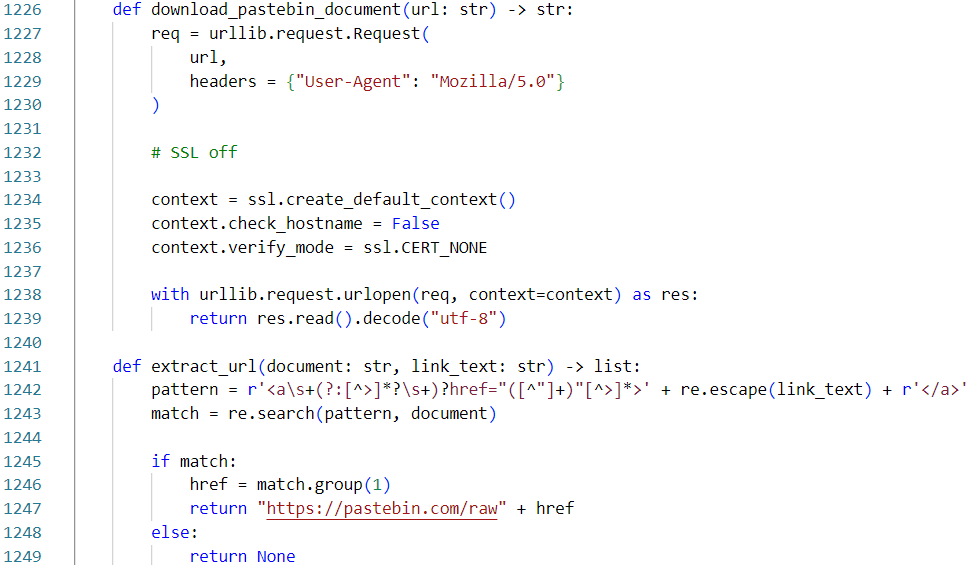}
    \caption{Function designed to download and decode data from \textit{Pastebin}.}
    \label{fig:94}
\end{figure}

During the \textit{dynamic analysis} of this sample, a hit was found among the 1000 possible profiles when attempting to connect to \textit{hxxps[:]//Pastebin[.]com/u/TwelveThrows2886}. As expected, \textit{TwelveThrows2886\_InductionInteriors4401} was the corresponding encrypted string and thus the only available file in this profile was named exactly \textit{InductionInteriors4401}. This file (\textit{hxxps[:]//pastebin.com/raw/suEqUQBY}) hosts an encoded string.

\begin{figure}[H]
    \centering
    \includegraphics[width=0.9\linewidth,frame]{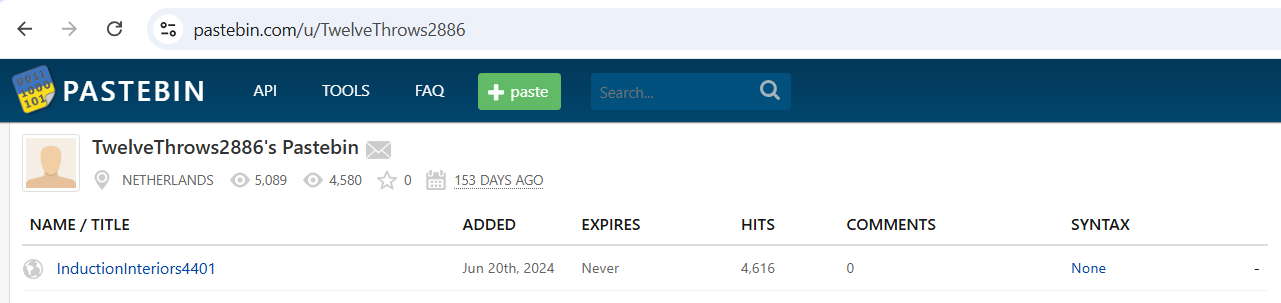}
    \caption{\textit{Pastebin} profile contacted to retrieve the additional payload.}
    \label{fig:102}
\end{figure}

\begin{figure}[H]
    \centering
    \includegraphics[width=0.9\linewidth,frame]{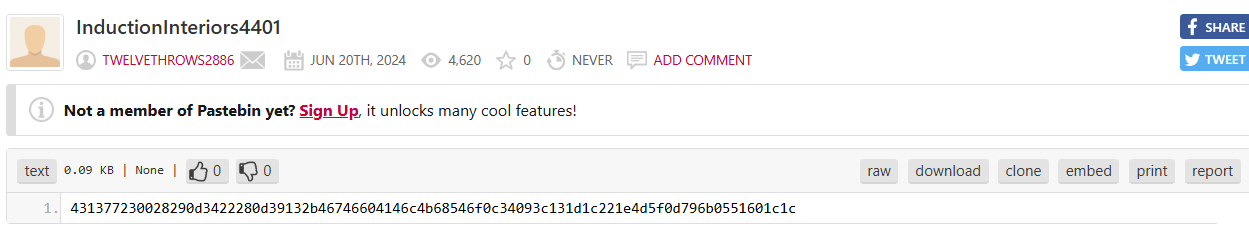}
    \caption{\textit{Pastebin} file containing the encoded URL for the additional payload location.}
    \label{fig:103}
\end{figure}

The decoded string translates to \textit{hxxp[:]//23.254.229.101/cat-video} and delivers a file named \textbf{\textit{cat video.mp4}}. This is instead a reversed \textit{gzip} archive which contains \textit{Runtime Broker.exe} and gets stored inside the following path: \textit{\%APPDATA\%\textbackslash Microsoft\textbackslash Windows\textbackslash\\ Applications\textbackslash Runtime Broker.exe}.

The script then defines functions to download the \textbf{\textit{TSUNAMI INSTALLER}} and execute the \textbf{\textit{TSUNAMI PAYLOAD}} with elevated privileges. \textbf{\textit{download\_installer()}} downloads the malicious installer, decodes it, and saves it to the specified path. \textbf{\textit{extract\_payload()}} writes the obfuscated payload script to a temporary file. \textbf{\textit{execute\_paylo\\ad\_with\_uac()}} attempts to execute the payload with administrative privileges by invoking \textit{ShellExecuteW} with the \textit{runas} verb.

\begin{figure}[H]
    \centering
    \includegraphics[width=0.8\linewidth,frame]{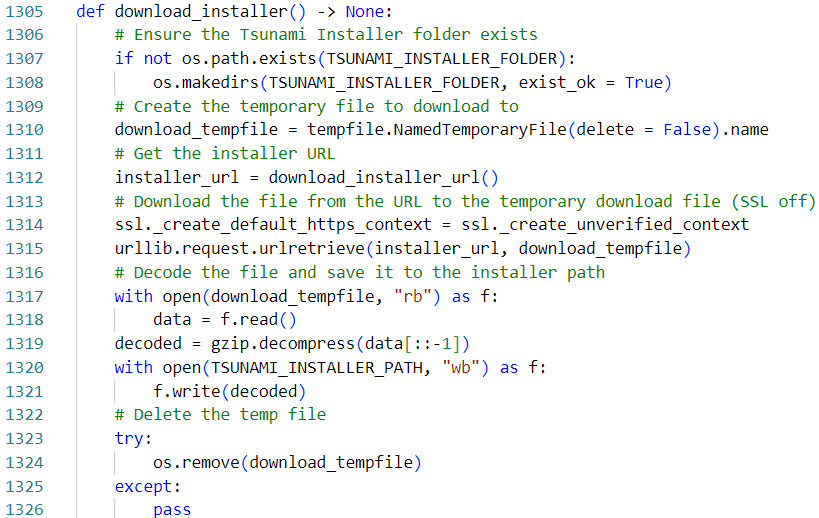}
    \caption{\textbf{\textit{download\_installer()}} code snippet}
    \label{fig:97}
\end{figure}

\begin{figure}[H]
    \centering
    \includegraphics[width=0.6\linewidth,frame]{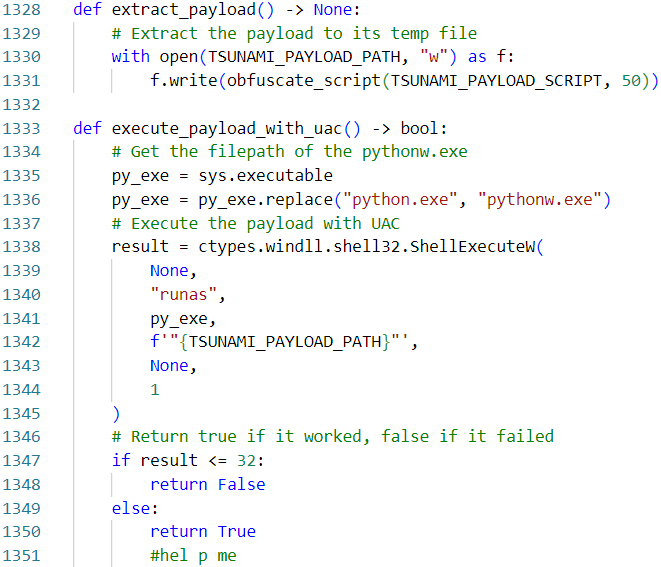}
    \caption{Function designed to employ \textit{runas} to install Python as admin.}
    \label{fig:98}
\end{figure}

In the \textit{main} section of the script, the execution flow is as follows.

\begin{figure}[H]
    \centering
    \includegraphics[width=0.9\linewidth,frame]{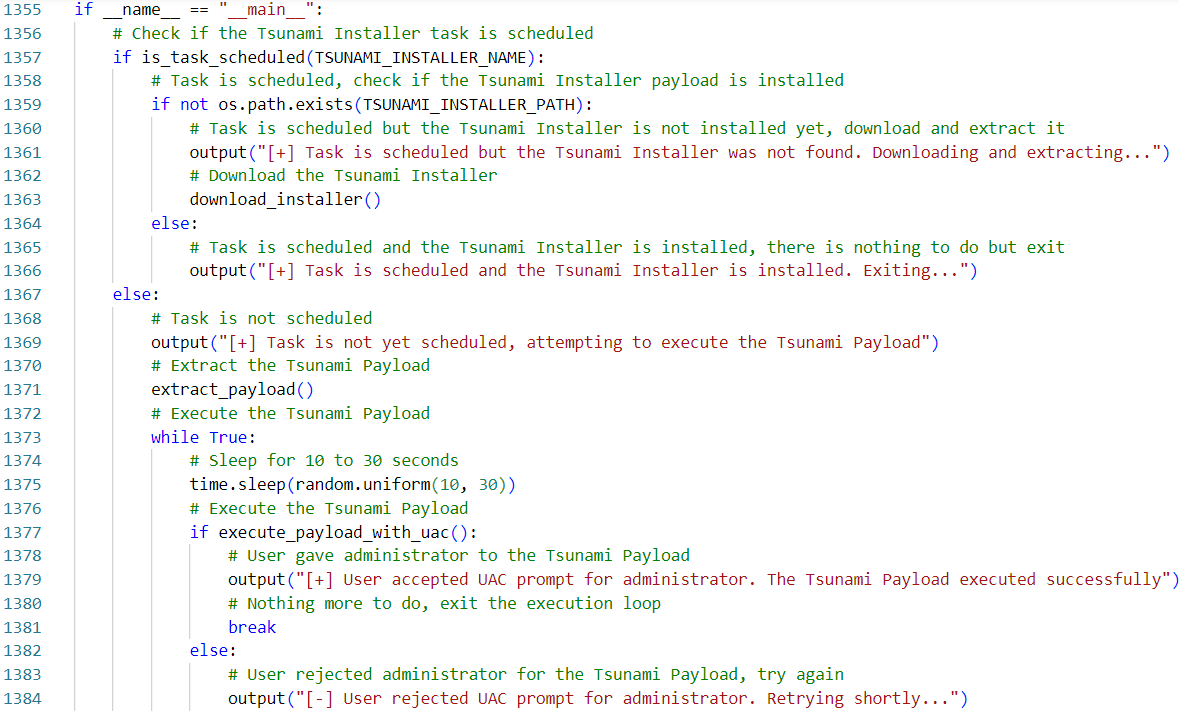}
    \caption{Script's main routine}
    \label{fig:99}
\end{figure}

The script checks if the \textit{scheduled task} \textit{Runtime Broker} exists.
If it does and the \textbf{\textit{TSUNAMI INSTALLER}} is not present, it downloads and installs this malicious executable. Otherwise, if it is present, it exits. Then, If a task for the \textbf{\textit{TSUNAMI INSTALLER}} is not scheduled, it attempts to execute the \textbf{\textit{TSUNAMI PAYLOAD}}, with elevated privileges, to schedule it. Thus, this script repeatedly prompts the \textit{UAC} dialog until the user grants administrative rights. Once the \textbf{\textit{TSUNAMI PAYLOAD}} executes successfully, it exits the loop.

While investigating the comments written inside this script, it is possible to find a reference about an extensive explanation of how the decryption URL schema works, hosted on the attacker's \textit{Youtube Channel}. However, this is just a joke since it redirects to the \textit{Never Gonna Give You Up} video (basically \textit{RickRolling} analysts).

\begin{figure}[H]
    \centering
    \includegraphics[width=1\linewidth,frame]{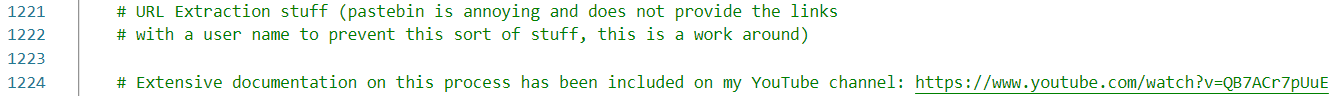}
    \caption{Developers \textit{RickRolling} analysts.}
    \label{fig:101}
\end{figure}

In conclusion, the script is a sophisticated piece of malware that aims to compromise a Windows system by \textit{installing malicious payloads}, \textit{achieving persistence}, and \textit{evading security measures}. It uses \textit{multiple layers of obfuscation} and \textit{encryption} to conceal its actions and relies on \textit{social engineering} (prompting \textit{UAC} dialogs) to gain \textit{elevated privileges}. The script's modular structure allows it to perform various malicious activities while making analysis and detection challenging.

\subsection{Fifth Stage}

\begin{figure} [H]
    \centering
    \includegraphics[width=0.6\linewidth,frame]{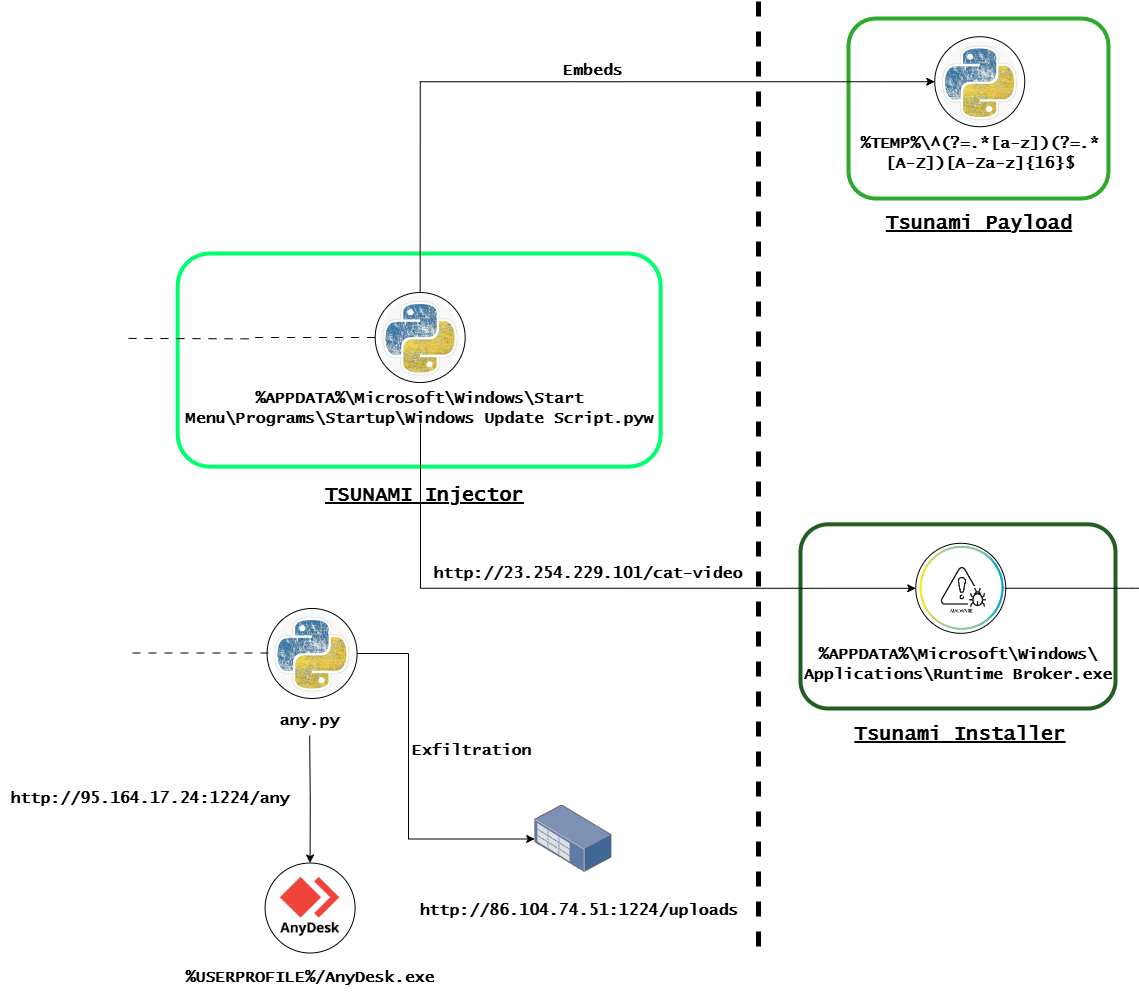}
    \caption{Moving from Fourth to Fifth Stage}
    \label{fig:FiS}
\end{figure}

\subsubsection{Code Obfuscation}
As discussed in the previous section, the \textbf{\textit{TSUNAMI CLIENT}} script is written to disk with the well-known 50-iterations obfuscation schema. On the other hand, \textbf{\textit{TSUNAMI INSTALLER}} executable is not a packed executable.

\subsubsection{Code Analysis - TSUNAMI PAYLOAD}
\textbf{\textit{TSUNAMI PAYLOAD}}, as mentioned above, is a malicious program designed to establish persistence on a Windows system by creating \textit{scheduled tasks} and modifying \textit{Windows Defender} settings to \textit{exclude certain files from scanning}. The script attempts to run with administrative privileges to \textit{modify system settings}, adds specific file paths to the \textit{Windows Defender Exclusion List}, and creates a \textit{scheduled task} that executes a malicious payload named \textit{Runtime Broker.exe} at user logon. This behavior allows the malware to \textit{evade detection} and \textit{maintain persistence across system reboots}.

Starting from the main execution point, the script begins by importing necessary modules that facilitate interaction with the operating system and system-level functions.

These imports enable the script to execute \textit{subprocesses} (such as \textit{PowerShell} commands), interact with \textit{Windows API} functions for privilege escalation checks, and manipulate file paths.

\begin{figure}[H]
    \centering
    \includegraphics[width=0.25\linewidth,frame]{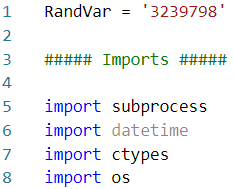}
    \caption{Script's imports}
    \label{fig:190}
\end{figure}

The script defines global variables that are crucial for its operation. \textit{DEBUG\_MODE} flag is set to False, indicating that debug output is suppressed during normal execution. The script retrieves the paths to the roaming and \textit{Local AppData} directories using environment variables. These paths are used to construct locations where the malicious payload and related files will be stored. The script specifies the name and paths for the \textbf{\textit{TSUNAMI INSTALLER}}, which is actually a disguised malicious executable. As a first analysis it is possible to have a look at this executable name, which is all but random, since it tries to mimic known Windows one \textit{RuntimeBroker.exe}. The latter is indeed a legitimate system process designed to manage permissions for modern \textit{Universal Windows Platform} (\textit{UWP}) applications. Its primary role is to act as a broker between these applications and the operating system, ensuring that apps operate within their defined permission boundaries. For instance, it monitors access to sensitive resources like location, microphone, and file systems, prompting the user when permissions are requested. The legitimate \textit{RuntimeBroker.exe} process is typically spawned by its parent process, \textit{svchost.exe}, which is responsible for hosting various system services and its path is located in the Windows system directory, specifically at \textit{C:\textbackslash Windows\textbackslash System32\textbackslash RuntimeBroker.exe}. This location is a key indicator of authenticity, as any instance of \textit{RuntimeBroker.exe} found outside the \textit{System32} directory is likely malicious or suspicious, just like in this specific case.

\begin{figure}[H]
    \centering
    \includegraphics[width=0.8\linewidth,frame]{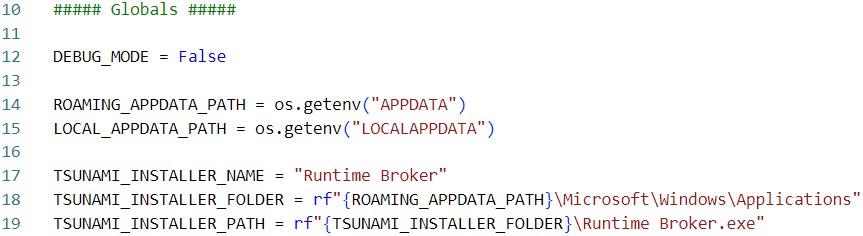}
    \caption{Global variables declarations}
    \label{fig:193}
\end{figure}

Through continued code analysis, it becomes evident that the \textbf{\textit{is\_admin()}} function is implemented to verify whether the script is executing with administrative privileges. This is achieved by invoking the \textbf{\textit{IsUserAnAdmin()}} function from the \textit{shell32} library. This function provides a straightforward mechanism to determine if the current user context has the necessary elevated permissions to perform privileged operations. If administrative privileges are not present, the script may encounter limitations in executing tasks that require such permissions, potentially resulting in failed operations or the bypassing of restricted functionality. This check ensures that the script can conditionally adapt its behavior based on the level of access available.

The script then defines functions that perform the core malicious activities. The \textbf{\textit{add\_windows\_defender\_exception()}} method adds specified file paths to the \textit{Windows Defender Exclusion List}.

\begin{figure}[H]
    \centering
    \includegraphics[width=0.7\linewidth,frame]{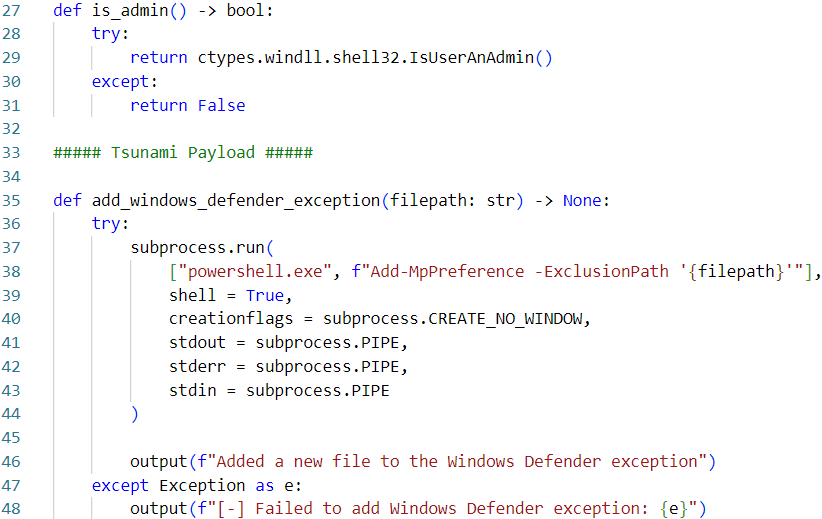}
    \caption{Functions designed to check user's permissions and apply \textit{AV} exclusions.}
    \label{fig:104}
\end{figure}

This function constructs a \textit{PowerShell} command that invokes \textit{Add-MpPreference} to exclude the specified \textit{filepath} from \textit{Windows Defender scans}. By doing so, the malware attempts to prevent its executable from being detected or removed by the antivirus software.

Instead, \textbf{\textit{create\_task()}} function creates a scheduled task that ensures the malicious payload runs at every user logon. In this function, a multi-line \textit{PowerShell} script is constructed to define a new \textit{scheduled task}. The task is configured with the following parameters:

\begin{itemize}
    \item \textit{Action}: Executes the malicious payload located at \textit{TSUNAMI\_INSTALLER\_PATH}.
    \item \textit{Trigger}: Set to trigger at user logon (\textit{-AtLogOn}).
    \item \textit{Principal}: Runs under the current user's context with interactive logon type and elevated privileges (\textit{RunLevel = 1}).
    \item \textit{Settings}: Configured to allow the task to start even if the system is on battery power and to not stop the task if the system switches power states.
\end{itemize}
The task is registered using \textit{Register-ScheduledTask}, ensuring that the malicious payload will persist and execute whenever the user logs in. 

\begin{figure}[H]
    \centering
    \includegraphics[width=1\linewidth,frame]{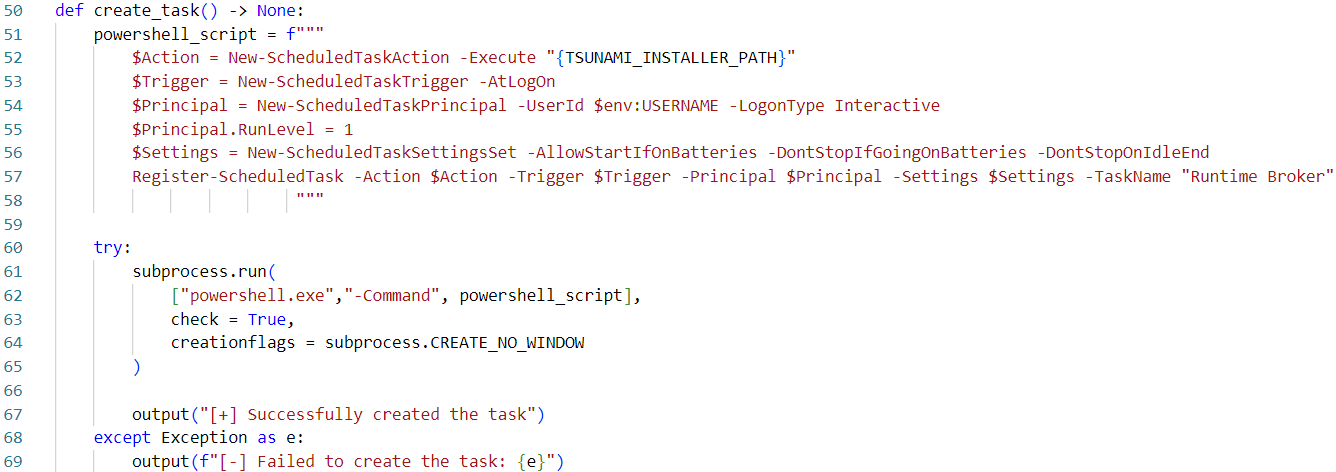}
    \caption{Function designed to add \textbf{\textit{Runtime Broker.exe}} as a \textit{scheduled task}.}
    \label{fig:105}
\end{figure}

The script first checks for administrative privileges by calling \textbf{\textit{is\_admin()}}. If the script is not running as an administrator, it outputs a warning message (if \textit{DEBUG\_MODE} is enabled). However, it proceeds with execution regardless of the privilege level, which may result in certain functions failing silently due to insufficient permissions. These paths include:

\begin{itemize}
    \item The main malicious payload (\textbf{\textit{Runtime Broker.exe}}) stored in the \textit{AppData Roaming} directory;
    \item A secondary payload or client component also named \textbf{\textit{Runtime Broker.exe}} in the \textit{AppData Local} directory;
    \item \textbf{\textit{msedge.exe}} which should host the \textit{XMRig} cryptocurrency miner.
\end{itemize}

By adding these paths to the exclusion list, the malware attempts to prevent \textit{Windows Defender} from scanning or quarantining these files, allowing malicious activities to proceed unhindered. The script iterates over the \textit{EXCEPTION\_PATHS} and calls \textbf{\textit{add\_windows\_defender\_exception()}} for each. After modifying the \textit{Windows Defender} settings, the script proceeds to create the \textit{scheduled task} by calling \textbf{\textit{create\_task()}}. This ensures that the malicious payload is executed at every user logon, establishing persistence on the system. Finally, if \textit{DEBUG\_MODE} is enabled, the script waits for user input before exiting, which is useful for testing or analysis purposes.

\begin{figure}[H]
    \centering
    \includegraphics[width=0.8\linewidth,frame]{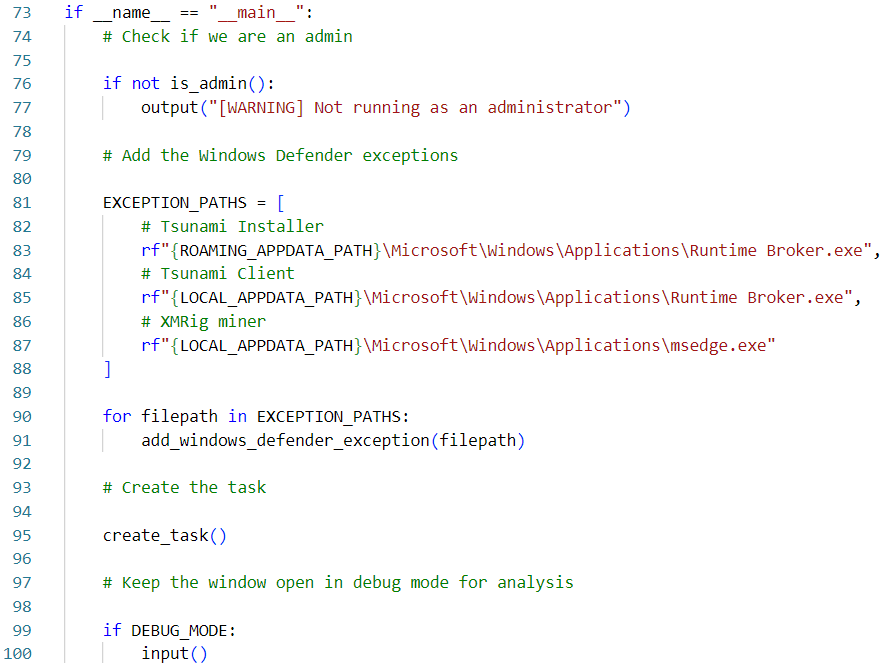}
    \caption{\textit{Main} routine}
    \label{fig:106}
\end{figure}

In conclusion, the script functions as a \textit{persistence mechanism} for a malicious payload on a Windows system. It attempts to \textit{elevate privileges}, modifies \textit{Windows Defender settings} to exclude its files from scanning, and creates a \textit{scheduled task} that executes the payload at user logon. The use of familiar names like \textbf{\textit{Runtime Broker.exe}} and placement within system-like directories aims to \textit{disguise the malware} and avoid raising suspicion. The script's ability to run without administrative privileges may limit its effectiveness, as certain operations require elevated permissions. The presence of unused code suggests that the malware may have additional capabilities that are not active in this version or that code has been removed or altered during obfuscation.

\subsubsection{Executable Analysis - TSUNAMI INSTALLER}
\textbf{\textit{Runtime Broker.exe}} acts as a central orchestrator of malicious operations. This process engages in a broad spectrum of activities that exploit native system utilities and functions, \textit{establishing a foothold in the system}, \textit{evading detection}, \textit{enabling persistence} and deploying a \textit{C2 TOR channel}.

\textbf{\textit{Static Analysis}}\\
This analysis reveals several advanced \textit{anti-analysis} techniques implemented within subjected executable. For instance, there are multiple matches indicating access to the \textit{Process Environment Block} (\textit{PEB}) to detect the presence of a \textit{debugger}, as logged in matches for \textit{PEB} access. This behavior aligns with previously observed \textit{anti-debugging} and \textit{anti-analysis} methods, emphasizing the malware’s intent to evade dynamic sandbox environments and indicates reliance on low-level system structures for evasion, likely preceding more overt malicious actions to ensure execution only in non-analytical environments (e.g. exploiting \textit{isDebuggerPresent} function). Also, it is possible to find execution delays trough \textit{Sleep}, \textit{Software Breakpoints} checks, \textit{Debug Break},  \textit{GetTickCount} and \textit{QueryPerformanceCounter} invokes.

\begin{figure} [H]
    \centering
    \includegraphics[width=0.9\linewidth,frame]{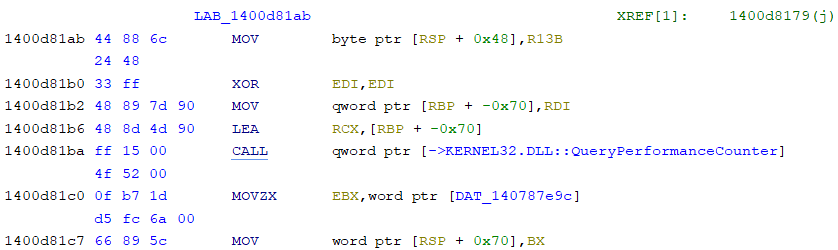}
    \caption{\textit{QueryPerfomanceCounter} invoke}
    \label{fig:107}
\end{figure}

\begin{figure}[H]
    \centering
    \includegraphics[width=0.9\linewidth,frame]{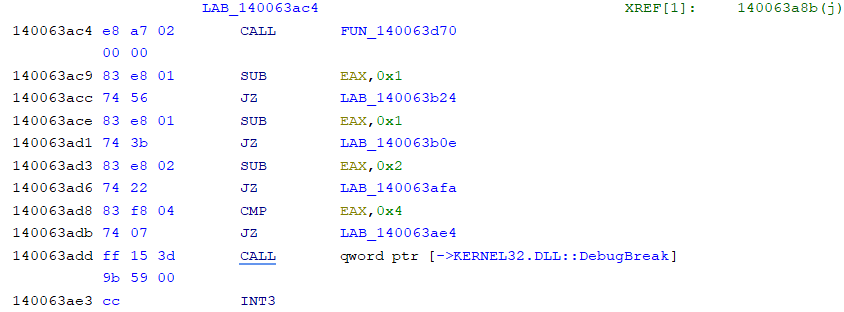}
    \caption{\textit{DebugBreak} invoke}
    \label{fig:108}
\end{figure}

Another significant discovery is the use of \textit{API} calls such as \textit{VirtualAlloc} and \textit{VirtualProtect} to allocate and modify memory permissions dynamically. These suggest the malware includes functionality for memory-based payload staging and execution, potentially leveraging reflective injection techniques. This capability allows the malware to inject code into other processes or execute \textit{shellcode} directly from allocated memory, increasing its stealth.

\begin{figure}[H]
    \centering
    \includegraphics[width=0.9\linewidth,frame]{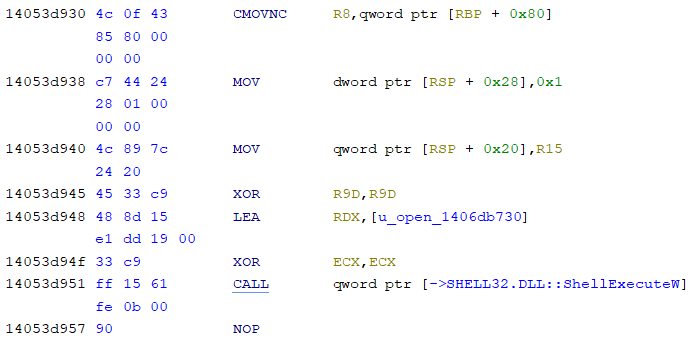}
    \caption{\textit{ShellExecuteW} invoke}
    \label{fig:109}
\end{figure}

The static analysis also identifies logic for delaying execution using \textit{API}s like \textit{SleepEx}, with the intention of bypassing automated sandboxes or security tools that rely on timeouts to detect malicious behavior. These deliberate delays enable the malware to outlast dynamic analysis environments that may prematurely conclude monitoring, ensuring its functionality is triggered only in live systems.

\begin{figure}[H]
    \centering
    \includegraphics[width=0.9\linewidth,frame]{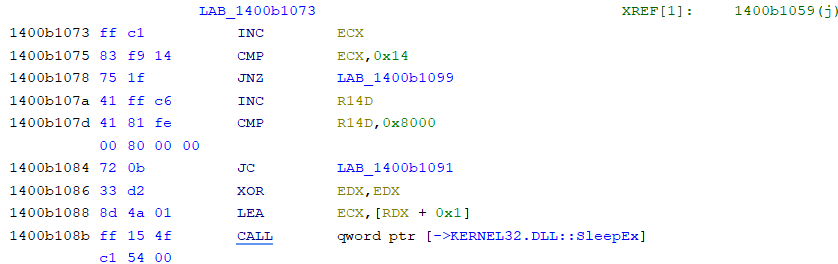}
    \caption{\textit{SleepEx} invoke}
    \label{fig:110}
\end{figure}

Furthermore, the file exhibits the capability to compress and decompress data using \textit{Zlib} (compress data via \textit{Zlib} inflate or deflate) and encode/encrypt data using \textit{base64} and \textit{xor}. These functionalities strongly correlate with obfuscation techniques observed during behavioral analysis, where repeated file and payload manipulation were recorded. For example, \textit{Zlib} compression is used in the malware’s payload delivery mechanism to reduce file size and disguise its contents.

\begin{figure}[H]
    \centering
    \includegraphics[width=0.9\linewidth,frame]{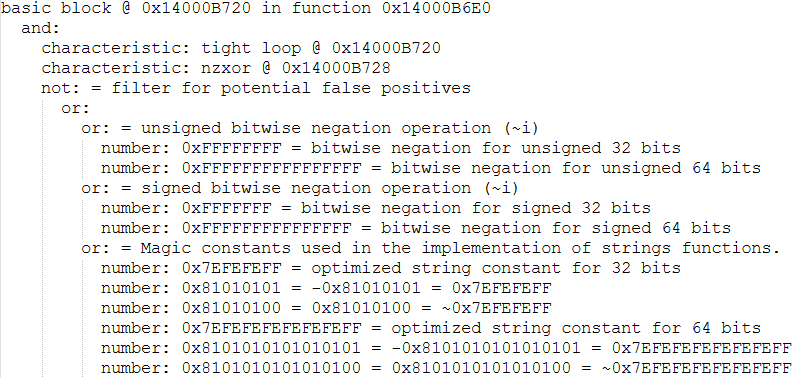}
    \caption{Set of values possibly associated with \textit{xor} activities.}
    \label{fig:111}
\end{figure}

New insights from the static analysis also highlight capabilities for obtaining \textit{system locale} and \textit{geographical} information, as seen in the following image. This discovery introduces the possibility that the malware is region-specific or dynamically adapts its behavior based on the host’s location.

\begin{figure}[H]
    \centering
    \includegraphics[width=0.9\linewidth,frame]{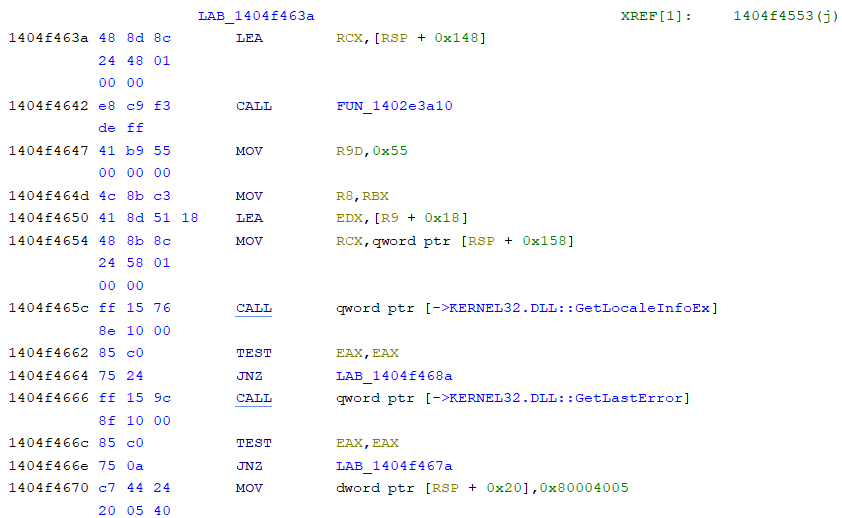}
    \caption{\textit{GetLocaleInfoEx} invoke}
    \label{fig:112}
\end{figure}

Further investigation of the malware’s embedded strings has uncovered the presence of debugging information, left behind by the developers. These artifacts provide valuable insights into the attacker's behavior and offer a deeper understanding of the development process behind this malicious tool. By analyzing these remnants, analysts can better fingerprint the attacker's techniques and gain additional intelligence about their testing environments, coding practices, and potential oversights. This evidence underscores the often iterative and sometimes rushed nature of malware development.

\begin{figure}[H]
    \centering
    \includegraphics[width=0.9\linewidth,frame]{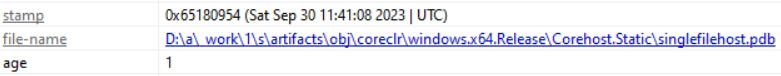}
    \caption{Debugging strings left behind by malware developers.}
    \label{fig:127}
\end{figure}

\textbf{\textit{Static Analysis - Runtime Broker.dll}}\\
The unusually large size of \textbf{\textit{RuntimeBroker.exe}} prompted an examination of its raw \textit{hex} code to uncover potential embedded components. This analysis revealed the presence of eighty-seven distinct executables embedded within the binary, including a substantial collection of statically linked known \textit{.NET DLLs}.

Among these embedded files, certain suspicious strings stood out, hinting at the presence of an unusual and potentially malicious \textit{library}. A deeper examination for content related to \textit{Tsunami} indeed revealed a subset of strings associated not only with the malware itself but also with Windows components being exploited to collect additional system information. This discovery underscores the likelihood that the binary conceals malicious payloads or extra functionalities, leveraging its considerable size and complexity to evade detection and analysis. These findings suggest that the identified suspicious\textit{DLL} may serve as a critical component of the malware, facilitating data gathering or other malicious operations. Further investigation into this \textit{library} is imperative to fully understand its purpose and its role within the broader malicious framework.

\begin{figure}[H]
    \centering
    \includegraphics[width=0.8\linewidth,frame]{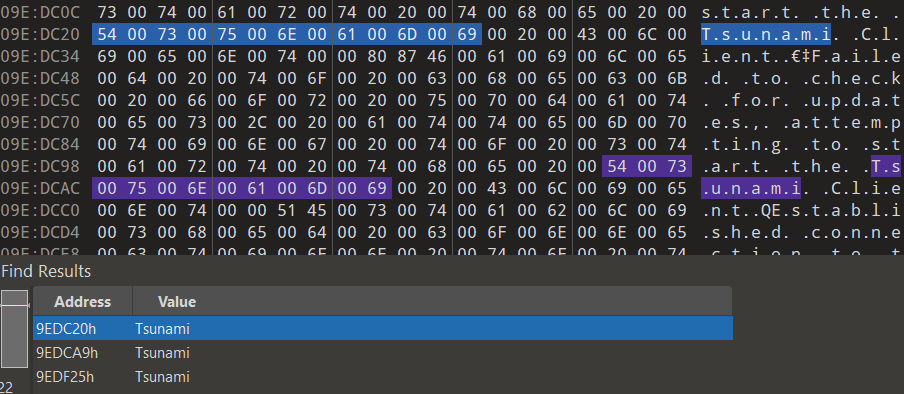}
    \caption{\textit{Tsunami} strings embedded in \textbf{\textit{Runtime Broker.exe}.}}
    \label{fig:142}
\end{figure}

By determining the address range associated with the most noteworthy strings and locating the specific executable segment containing this memory region, it became possible to isolate and extract the embedded component for standalone analysis. This meticulous extraction process revealed the core module of \textbf{\textit{RuntimeBroker.exe}}, identified as \textbf{\textit{RuntimeBroker.dll}}.

Analyzing \textbf{\textit{RuntimeBroker.dll}} independently provided a clearer view of its role within the larger binary. This module appeared to function as the central orchestrator, potentially handling key tasks such as \textit{Command-and-Control} communication, \textit{process injection}, and the \textit{execution of additional embedded payloads}. The identification and extraction of this core component were critical steps in unraveling the underlying structure and functionality of the malware, shedding light on its operational complexity and modular design.

\begin{figure}[H]
    \centering
    \includegraphics[width=0.8\linewidth,frame]{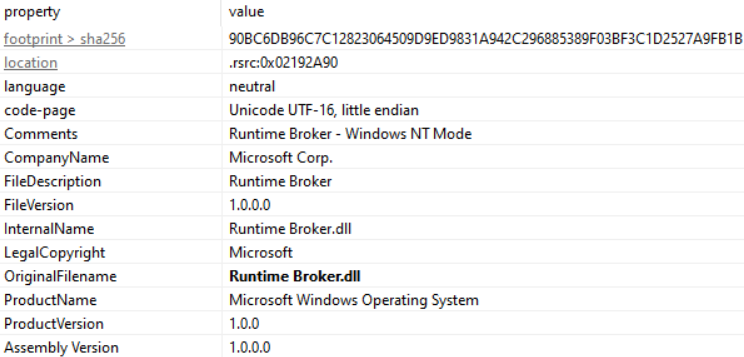}
    \caption{\textbf{\textit{Runtime Broker.dll}} overview}
    \label{fig:143}
\end{figure}

Since the library was written in \textit{.NET}, it was possible to load it into \textit{dnSpy} and examine its source code directly. Remarkably, debug information was still intact, and the code appeared completely unobfuscated, with human-readable functions, variables. This stark contrast highlights an inconsistency in the attacker’s efforts to conceal their operations. While the \textbf{\textit{error.js}} file, part of the initial stage, was heavily obfuscated, requiring significant effort for static analysis, the library hosting the core functionality of the first malicious executable dropped on the target system lacked any obfuscation or stripping.

This divergence suggests that, although the \textit{Threat Actor} has invested substantial resources in constructing a resilient, distributed, and flexible malicious architecture, their efforts to obscure their operations diminished in later stages of the infection chain. This could indicate either a rushed development cycle or a deliberate decision to prioritize obfuscation in earlier stages, leaving subsequent stages exposed. Unfortunately, these clues alone are insufficient to definitively determine whether this lapse was due to oversight or a calculated choice. Nonetheless, it underscores a critical aspect of the operation, revealing potential weaknesses in their approach to maintaining stealth and obfuscation consistency throughout the chain.

\begin{figure}[H]
    \centering
    \includegraphics[width=0.8\linewidth,frame]{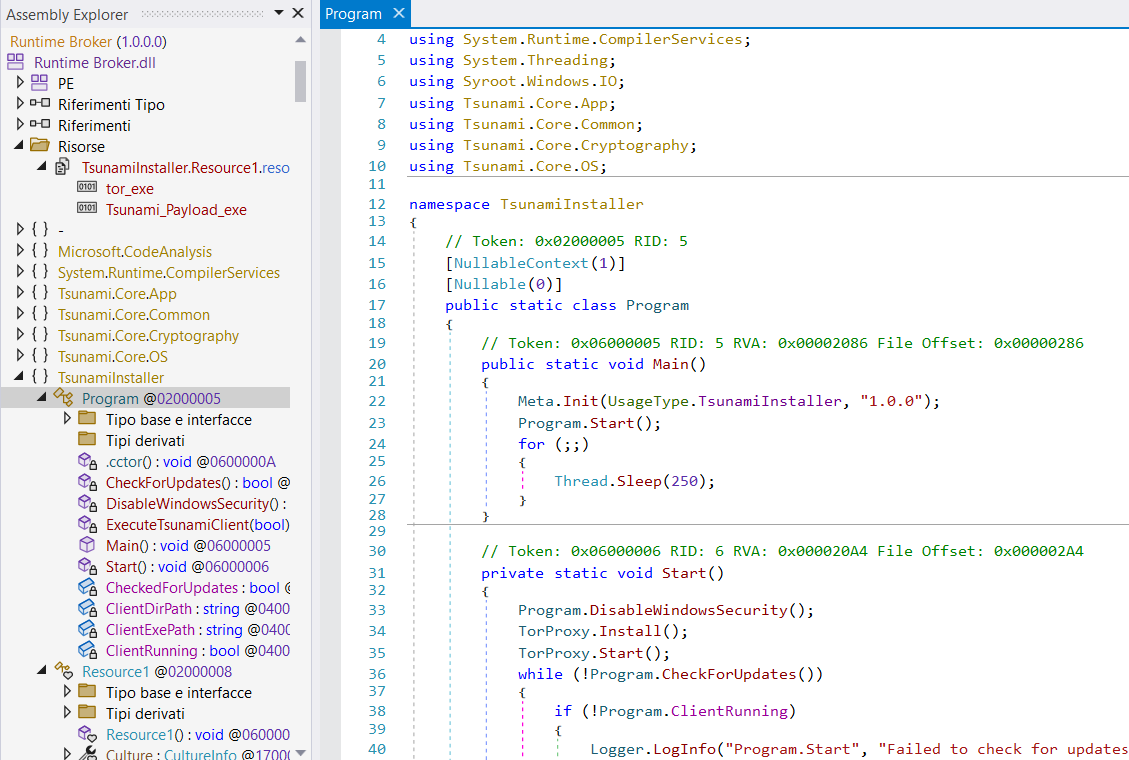}
    \caption{\textbf{\textit{Runtime Broker.dll}} reversed content}
    \label{fig:144}
\end{figure}

The \textit{Main} method initializes the program by invoking \textit{Meta.Init}, setting its usage type to \textit{TsunamiInstaller} with a specified version, i.e. \textit{1.0.0}, before invoking the \textbf{\textit{Start()}} method. The inclusion of an infinite loop at the end ensures that the program remains active, executing indefinitely and ready to retry failed operations as needed.

The workflow begins with the \textbf{\textit{Start()}} method, which initiates its operations by \textit{disabling key Windows security features} through a call to \textbf{\textit{Program.DisableWindowsSecu\\rity()}}. This step is likely aimed at neutralizing \textit{Windows Defender} and \textit{Firewall protections}, creating an environment where the malware can operate without interference from built-in security mechanisms. Following this, the program installs and starts the \textit{Tor proxy} using \textbf{\textit{TorProxy.Install()}} and \textbf{\textit{TorProxy.Start()}}, setting up an anonymized communication channel that obfuscates its connections to the \textit{Command-and-Control} server.

The program places a high priority on ensuring that its malicious payload is current and operational. It accomplishes this by repeatedly checking for updates with \textbf{\textit{Program.CheckForUpdates()}}. If updates are not available or the check fails, the program attempts to execute the \textbf{\textit{TSUNAMI CLIENT}} using \textbf{\textit{Program.ExecuteTsunamiCli\\ent()}}. This mechanism ensures that the payload remains functional and capable of adapting to the latest malicious features or patches. In the event that the client is not already running, the program logs its attempts to execute it.

Establishing a connection with the \textit{Command-and-Control} server is another critical aspect of the workflow. The program uses the \textit{Tor proxy} for this purpose, retrying every ten minutes if initial attempts fail. This persistence underscores the malware's resilience in maintaining communication with its operators. Once connected, it attempts to transmit telemetry data using \textbf{\textit{TelemetryUploader.SendApplicationLogs()}}, which likely includes \textit{runtime logs} and \textit{system information}. This data is valuable for profiling the compromised environment, assessing the malware’s deployment, or monitoring its operational state.

The program also incorporates a \textit{controlled shutdown mechanism}. After completing its tasks, such as verifying updates and transmitting telemetry, it logs a message indicating readiness to terminate and exits using \textbf{\textit{Environment.Exit(0)}}. This behavior suggests a level of sophistication in managing its lifecycle, ensuring it avoids unnecessary detection or conflicts with subsequent stages of its operation. The structured flow of actions, from disabling security to transmitting telemetry, demonstrates a calculated approach designed to maximize the malware's impact while maintaining stealth.

\begin{figure}[H]
    \centering
    \includegraphics[width=0.8\linewidth,frame]{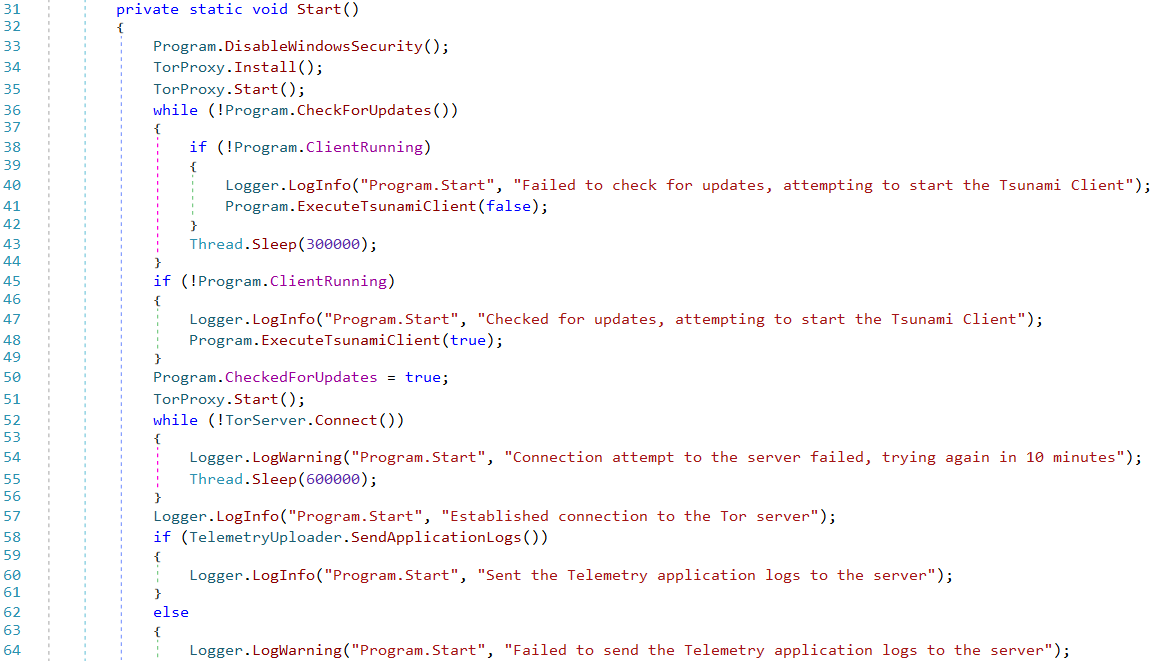}
    \caption{\textbf{\textit{Runtime Broker.dll}} \textit{Main} method.}
    \label{fig:145}
\end{figure}

At this point, each implemented class and its respective functionalities will be thoroughly examined, following a cascading order from the first to the last as they appear in the execution flow of the \textit{Main} method. This approach ensures a structured analysis, beginning with the foundational initialization and setup processes, and progressing through the subsequent operations, dependencies, and interactions. By dissecting the classes in the order they are invoked, it becomes possible to trace the logic, dependencies, and intent of the program, providing a comprehensive understanding of its architecture and behavior.

The \textit{Meta} class is a static utility designed to manage metadata for the application, providing essential details such as the application's \textit{usage type}, \textit{version}, \textit{session ID}, and \textit{server URL}. The \textbf{\textit{Init()}} method initializes these values, setting up the necessary environment for the application to operate. It assigns the \textit{UsageType} and \textit{AppVersion} based on the parameters passed during initialization. The \textit{AppSessionID} is dynamically generated as a unique identifier for each session using the \textbf{\textit{Guid.NewGuid()}} method, ensuring distinct identification for every instance. Additionally, the server URL is hardcoded to point to a \textit{.onion} address, which indicates the use of the \textit{Tor network} for communication, reinforcing the application's emphasis on anonymized operations (\textit{hxxp[:]//n34kr3z26f3jzp4ckmwuv5ipqyatumdxhgjgsmucc65jac56khdy5zqd[.]onion}).

\textit{Accessors} such as \textbf{\textit{GetUsageType()}}, \textbf{\textit{GetAppVersion()}}, \textbf{\textit{GetAppSessionID()}}, and \textbf{\textit{GetServerURL()}} provide controlled retrieval of these initialized values. These methods enable other components of the application to query the metadata without directly modifying it, ensuring data consistency and encapsulation. The class uses \textit{private static fields} to store these values, maintaining a centralized configuration structure that supports the application's runtime needs.

The design of the \textit{Meta} class reflects its critical role in orchestrating the application's configuration. By combining dynamic elements like the \textit{session ID} with predefined settings such as the \textit{server URL}, the class facilitates flexible yet consistent behavior across different stages of the application. The inclusion of a \textit{.onion URL} further aligns the class with the application's broader strategy of leveraging Tor for secure and anonymized communication.

\begin{figure}[H]
    \centering
    \includegraphics[width=0.8\linewidth,frame]{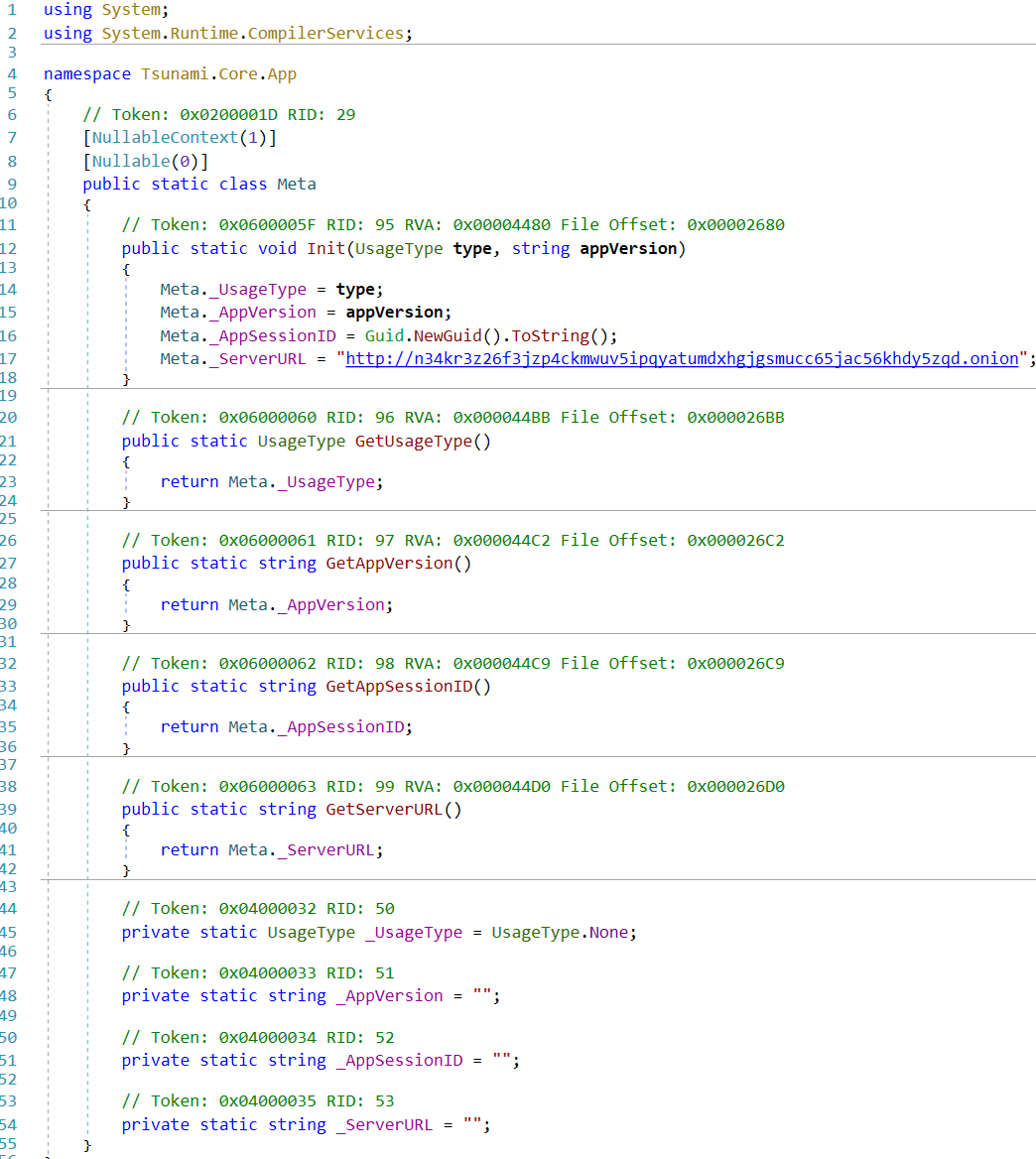}
    \caption{Overview of \textit{Meta} class}
    \label{fig:155}
\end{figure}

The \textbf{\textit{DisableWindowsSecurity()}} method is designed to \textit{neutralize Windows security features} by \textit{disabling both Windows Defender and Windows Firewall} through calls to the \textit{AntiDefender} class. The method begins by checking for the existence of an \textit{Anti Malware flag} using the \textbf{\textit{AntiDefender.FlagExists()}} method. This \textit{flag} acts as an indicator that the disabling operations have already been executed in a previous instance, allowing the program to adjust its behavior accordingly.

If the \textit{flag} exists, the program logs the detection and pauses execution for one minute, indicating a shorter delay when security features are presumed to have already been addressed. If the \textit{flag} does not exist, the program proceeds to \textit{disable Windows Defender} and \textit{Windows Firewall}, as implemented in the respective methods of the \textit{AntiDefender} class. Following this, it logs the absence of the \textit{flag} and introduces a longer delay of five minutes before continuing execution.

The use of conditional delays based on the \textit{flag}’s presence serves to reduce unnecessary re-execution of \textit{security-disabling routines} while providing a \textit{persistent mechanism} to \textit{disrupt or evade host protections}. By incorporating these actions early in the workflow, the program ensures that \textit{security defenses are neutralized}, enabling subsequent malicious operations to proceed unimpeded. The method’s detailed logging further demonstrates an emphasis on tracking the program’s progression, which aids in monitoring and debugging within the malware framework.

\begin{figure}[H]
    \centering
    \includegraphics[width=1\linewidth,frame]{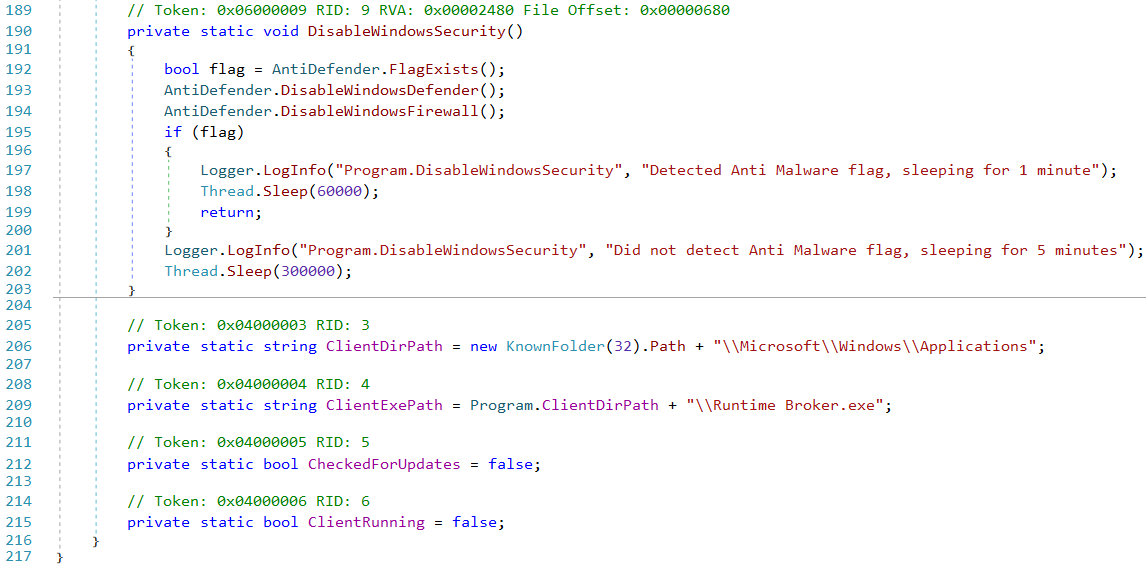}
    \caption{Overview of the \textbf{\textit{DisableWindowsSecurity()}} method}
    \label{fig:146}
\end{figure}

The \textit{AntiDefender} class represents a set of functions aimed at \textit{disabling key Windows security features}, specifically \textit{Windows Defender} and \textit{Windows Firewall}. The methods operate by adding exceptions to these defenses for specific applications, enabling the malware or potentially unwanted software to \textit{bypass detection and restriction mechanisms}.

The \textbf{\textit{DisableWindowsDefender()}} method is designed to add exclusions to \textit{Windows Defender} for a predefined list of applications, ensuring that these files are ignored by the antivirus. It retrieves the paths of these applications through the \textbf{\textit{GetApplicationList()}} method and iterates over them, invoking the \textbf{\textit{Shell.AddWindowsDefenderEx\\ception()}} function for each entry. This action allows the specified files to evade \textit{real-time scanning}, reducing the likelihood of detection. Logging is incorporated to document the process, recording successful additions of exceptions.

The \textbf{\textit{DisableWindowsFirewall()}} method performs a similar task but targets the \textit{Windows Firewall}. It first checks whether a \textit{flag} exists, indicating that the operation has already been performed. If the \textit{flag} is absent, it iterates over the same application list, invoking \textbf{\textit{Shell.AddWindowsFirewallException()}} for each entry. By adding \textit{firewall exceptions}, the method ensures that these applications can communicate over the network without restrictions. Once the exceptions are added, it creates the \textit{flag} file to avoid re-executing the process in subsequent runs.

The \textbf{\textit{CreateFlag()}} method generates a file named \textit{TsuAmFlag.txt} in the system's temporary directory. This file serves as an indicator that the \textit{firewall exception} process has already been completed. The method incorporates exception handling to ensure stability and logs the success or failure of the operation. The \textbf{\textit{FlagExists()}} method checks for the presence of this \textit{flag} file, returning a boolean value that determines whether the \textbf{\textit{DisableWindowsFirewall()}} method should proceed.

The \textbf{\textit{GetApplicationList()}} method defines a hardcoded list of paths to applications that require exceptions in both \textit{Windows Defender} and the \textit{Firewall}. These paths include various directories, such as \textit{temporary locations}, \textit{application folders}, and \textit{known Windows directories}, where components like \textbf{\textit{Runtime Broker.exe}}, \textbf{\textit{System Runtime Monitor.exe}}, and \textbf{\textit{msedge.exe}} are stored. By using the \textit{KnownFolder} class to retrieve specific system paths, the method adapts to the target system's environment dynamically.

\begin{figure}[H]
    \centering
    \includegraphics[width=0.9\linewidth,frame]{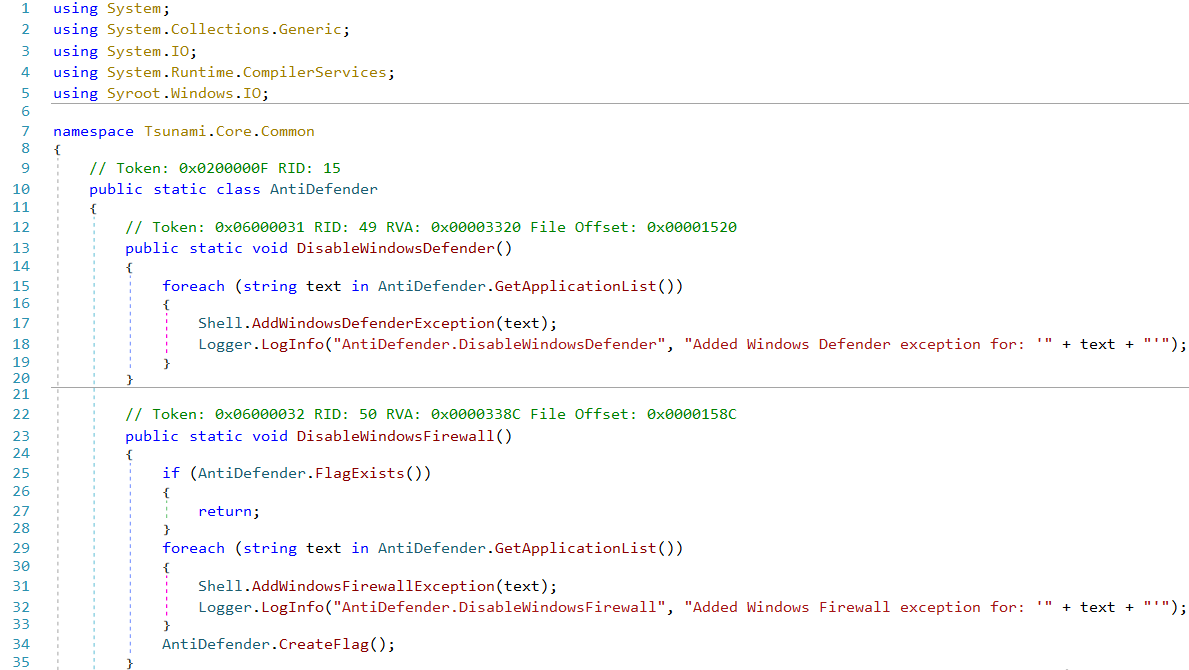}
    \caption{Overview of the \textit{AntiDefender} class}
    \label{fig:147}
\end{figure}

Upon further analysis of the \textit{AntiDefender} class, it becomes evident that it contains a hardcoded list of file paths that are subjected to the \textit{whitelisting} process. The paths in question include critical system directories and filenames that mimic legitimate applications, such as \textbf{\textit{Runtime Broker.exe}}, \textbf{\textit{System Runtime Monitor.exe}}, and other executables placed in standard or temporary directories.

This deliberate selection of paths indicates an effort to blend malicious components with legitimate system files, reducing the likelihood of detection. By targeting common system directories such as \textit{AppData}, \textit{WindowsApps}, and the temporary folder, the malware leverages locations that are often overlooked or trusted by security mechanisms. This whitelisting tactic ensures that key malware components can persist and execute their payloads without triggering alarms, further emphasizing the \textit{Threat Actor}’s focus on stealth and persistence.

A more detailed and comprehensive list of the paths corresponding to these folder identifiers will be presented in the subsequent dissection during the \textit{dynamic analysis} phase. This approach will enable the retrieval of runtime-resolved paths by observing the malware's behavior in a controlled environment, ensuring a thorough understanding of how these identifiers are translated into actual system directories.

\begin{figure}[H]
    \centering
    \includegraphics[width=0.9\linewidth,frame]{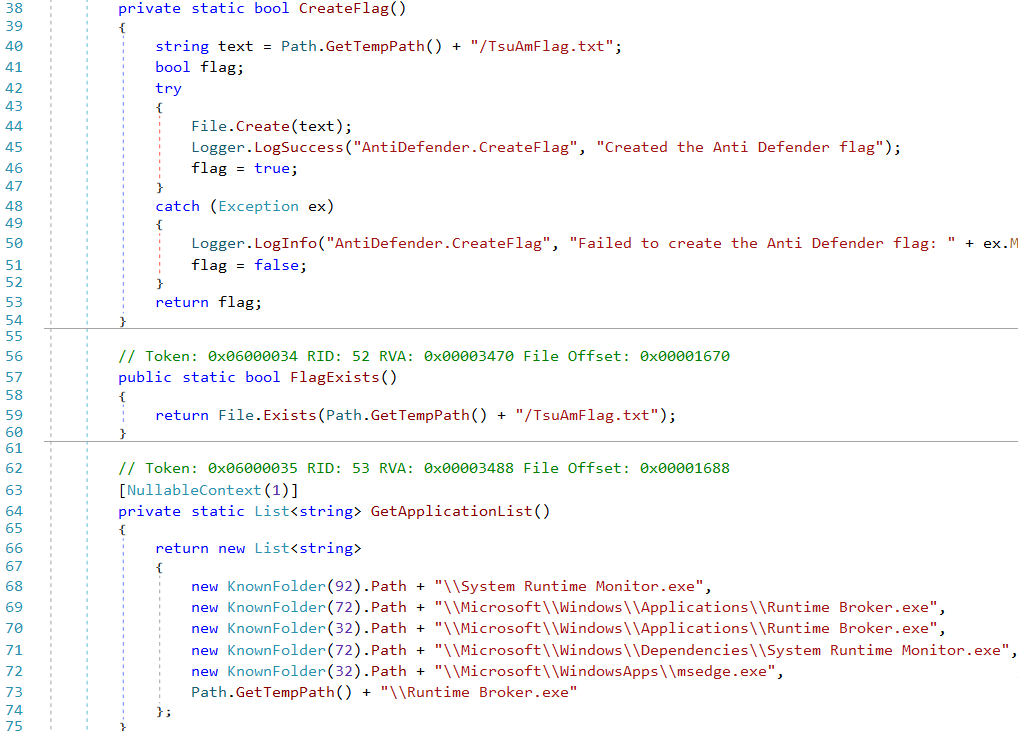}
    \caption{Hardcoded paths of additional payloads undergoing \textit{whitelisting} process.}
    \label{fig:148}
\end{figure}

The \textit{Shell} class provides utility functions to interact with the Windows system through \textit{PowerShell commands}. It includes methods to execute arbitrary commands, add exceptions to \textit{Windows Defender}, and \textit{create firewall rules}, primarily aiming to configure the system in favor of the malware's operations.

The \textbf{\textit{ExecutePowerShellCommand()}} method serves as a generic utility to execute \textit{PowerShell commands}. It creates a new \textit{Process} instance with \textit{powershell.exe} as the executable and the specified command as its argument. The process is configured to run without displaying a window (\textit{CreateNoWindow = true}), enabling it to execute silently. This generic command execution capability underpins the other methods in the class.

The \textbf{\textit{AddWindowsDefenderException()}} method uses a \textit{PowerShell command} to add a specified path to \textit{Windows Defender Exclusion List}, preventing the \textit{AV} from scanning or monitoring files in that location. The command is executed using \textit{powershell.exe} with elevated privileges (\textit{Verb = "runas"}), ensuring that administrative access is granted for modifying \textit{Defender} settings. This functionality is critical for the malware to \textit{bypass detection} and \textit{ensure the persistence} of its components.

Similarly, the \textbf{\textit{AddWindowsFirewallException()}} method constructs a \textit{PowerShell command} to create a \textit{firewall rule allowing inbound traffic} for a specified program. The rule is labeled with a generic name, such as \textit{Microsoft Edge WebEngine}, to avoid suspicion. Like the \textit{Defender exclusion method}, this command also runs with elevated privileges and suppresses any visible command window. The use of \textit{netsh} commands within \textit{PowerShell} highlights an effective approach to manipulate firewall rules programmatically.

This class demonstrates a deliberate focus on leveraging \textit{PowerShell} for system modifications, a common tactic in malware to evade detection and achieve operational goals. By embedding commands directly into the malware, the attackers reduce the reliance on external scripts, ensuring stealth and flexibility. The silent execution and elevation of privileges further underline the emphasis on maintaining a low profile while performing critical system changes.

\begin{figure}[H]
    \centering
    \includegraphics[width=0.85\linewidth,frame]{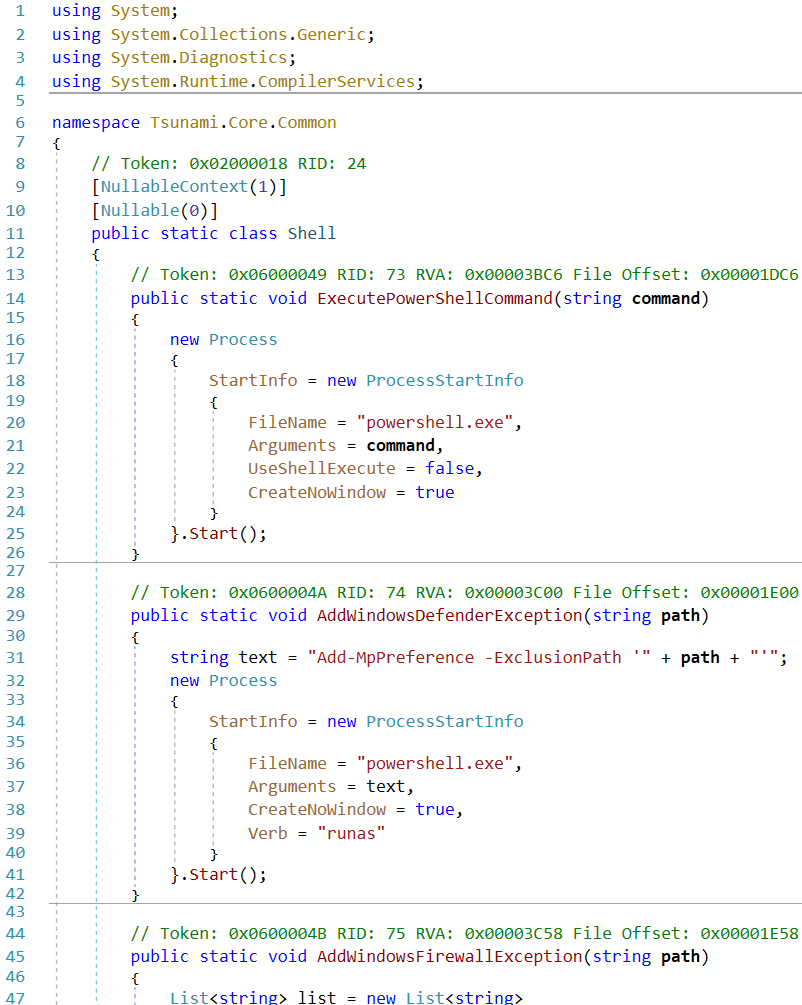}
    \caption{Overview of the \textit{Shell} class}
    \label{fig:156}
\end{figure}

The \textit{TorProxy} class provides a comprehensive implementation for managing a \textit{Tor proxy}, encompassing its installation, execution, and usage for network operations such as \textit{HTTP requests} and file downloads. The \textit{ExecutablePath} property specifies the location of the \textit{Tor proxy} executable as \textbf{\textit{Runtime Broker.exe}} within the system's \textit{temporary directory}. This choice of name and location raises suspicions of an attempt to masquerade as a legitimate Windows process, potentially aiding in evasion from detection mechanisms.

The \textbf{\textit{Install()}} method is responsible for deploying the \textit{Tor proxy} executable. It first checks if the proxy is already running, avoiding redundant installations. If the executable is absent, it retrieves the \textit{Tor binary} data from a resource loader and writes it to the specified location. The method is equipped with detailed logging to capture success or failure, reflecting the developer's attention to error handling and debugging capabilities. The \textbf{\textit{Start()}} method initiates the proxy process, configured to use a standard \textit{SOCKS} port (9050) and a temporary directory for its data storage. If an instance of the proxy is already active, the method attempts to terminate it before restarting, ensuring no conflicts arise from multiple running instances. Again, logging is extensively used to provide insights into process management.

The \textbf{\textit{Shutdown()}} method complements this functionality by stopping the \textit{Tor proxy}. It performs a check to confirm the process is running and, if so, attempts to terminate it. Detailed logs document whether the shutdown succeeds or fails, providing transparency and aiding troubleshooting.

Network communication is facilitated through the \textbf{\textit{SendRequest()}} method, which allows \textit{HTTP requests} to be routed through the \textit{Tor proxy}. This asynchronous function supports both \textit{GET} and \textit{POST} requests, with headers and payloads designed for \textit{JSON}-based data exchanges. By incorporating a custom \textit{SOCKS} port handler, the method ensures all traffic is anonymized. Comprehensive error handling and logging provide a detailed account of the request outcomes, including response status codes and content sizes. Similarly, the \textbf{\textit{DownloadFile()}} method enables file retrieval via the proxy. Using asynchronous streaming, it efficiently downloads files from specified URLs to designated file paths. Its reliance on the \textit{Tor network} for anonymizing traffic and the inclusion of robust error handling underscore its capability for secure and reliable file transfers.

The overall design of the \textit{TorProxy} class reflects a technically proficient implementation, leveraging asynchronous programming to ensure efficient and non-blocking operations. However, the choice to disguise the executable as \textbf{\textit{Runtime Broker.exe}} and deploy it in a temporary directory suggests potential misuse for malicious purposes. These characteristics, combined with the use of \textit{Tor} for anonymizing traffic, align with tactics commonly seen in malware aimed at concealing \textit{Command-and-Control} communications, \textit{data exfiltration}, or \textit{secondary payload delivery}.

\begin{figure}[H]
    \centering
    \includegraphics[width=0.9\linewidth,frame]{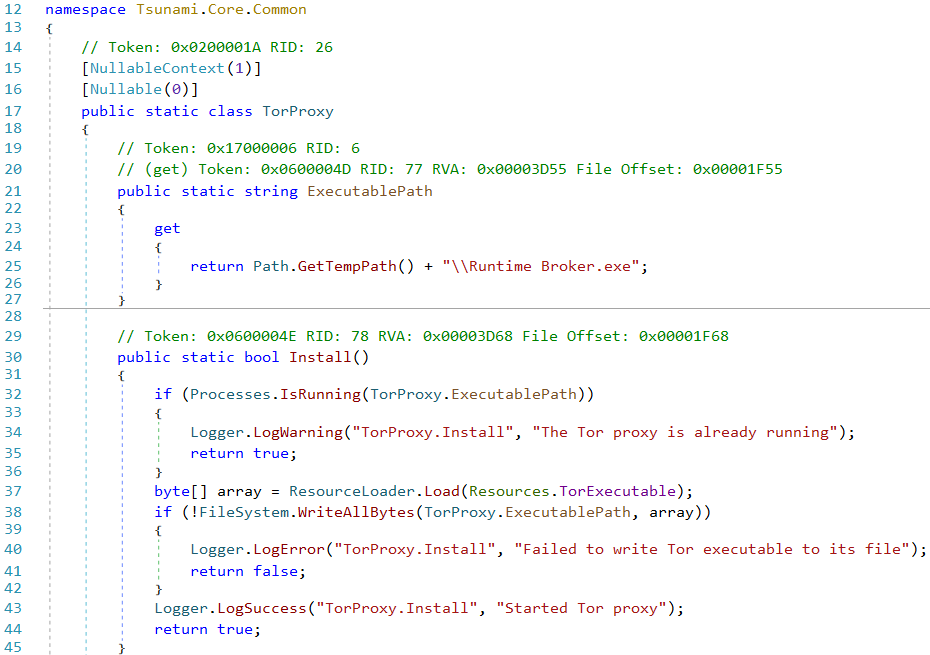}
    \caption{Snippet of \textit{TorProxy} class}
    \label{fig:149}
\end{figure}

The \textbf{\textit{CheckForUpdates()}} method is a robust implementation designed to manage updates for the \textit{Tsunami Client} application. It combines multiple functionalities to ensure the client executable is current, secure, and operational. The process begins by verifying the existence of the designated directory for the \textit{Tsunami Client}. If the directory is missing, it is created, and the operation is logged, ensuring the required environment is properly configured.

The method then requests the hash of the latest client version from the server via an \textit{HTTP GET request}, routed through a \textit{Tor proxy} for anonymized communication. The response from the server contains the \textit{hash} and a success status. If the request is successful, the received \textit{hash} is compared against the one of the currently installed client executable, computed using the \textit{SHA-256} algorithm. This step ensures the integrity of the existing file and determines whether an update is required. If the executable is missing or the hashes do not match, the method identifies the need for an update.

Before proceeding, the method checks whether the current version of the client is running. If it is, the method attempts to terminate the process to ensure a clean update environment. If termination fails, an error is logged, and the update process is aborted. Once the update is confirmed, the method downloads the latest compressed version of the client executable from the server using the \textit{Tor proxy}. The file is temporarily stored in the system’s temporary directory, and its contents are read, reversed, and decompressed using a \textit{GZIP} library. The decompressed data is then written to the client executable’s path, replacing the old version with the updated one. Finally, the temporary file is deleted, with any failure to delete it logged as a warning.

Throughout the process, the method incorporates comprehensive error handling and logging. Each step, whether successful or failed, is documented to ensure transparency and facilitate debugging. For instance, it logs successes for tasks such as fetching the hash and downloading the compressed file, and records warnings or errors for issues like hash mismatches, decompression failures, or file system errors. The use of \textit{SHA-256} hashing underscores the method’s focus on verifying update integrity, preventing corrupted or malicious files from being applied.

The reliance on the \textit{Tor proxy} for communication adds a layer of obfuscation, making it difficult to trace server interactions. The ability to dynamically download and apply updates allows for the deployment of new payloads or modifications, enhancing the adaptability and persistence of the system. The integration of \textit{GZIP} compression minimizes the size of update payloads, optimizing bandwidth usage while maintaining functionality through proper decompression. The \textbf{\textit{CheckForUpdates()}} method exemplifies careful and efficient design, incorporating advanced techniques for process management, error handling, and file integrity verification.

\begin{figure}[H]
    \centering
    \includegraphics[width=1\linewidth,frame]{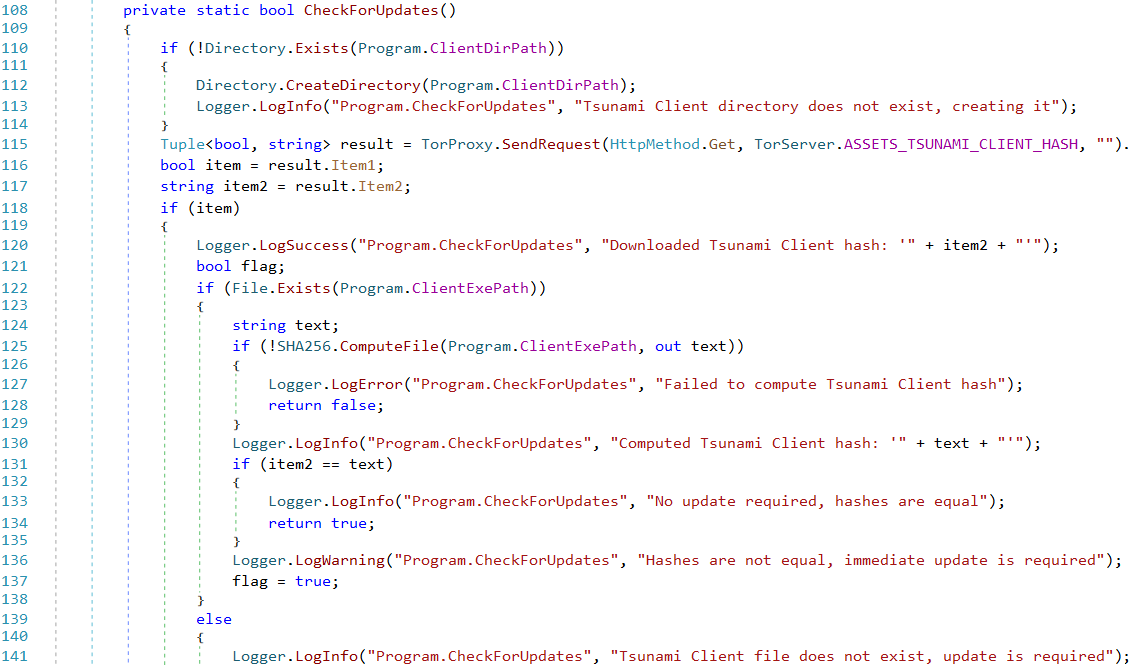}
    \caption{Overview of \textbf{\textit{CheckForUpdate()}} method}
    \label{fig:150}
\end{figure}

The susscessive analysis of the \textit{ResourceManager} component reveals the presence of two notable embedded resources: a tor.exe file and a \textbf{\textit{tsunami\_payload.exe}}. While the first file, \textbf{\textit{tor.exe}}, is actively extracted and utilized by the malware during execution, the latter appears to be embedded without any direct reference to its extraction or deployment within the program's logic. This discrepancy raises questions about the attacker's intent and the role of the unused \textbf{\textit{tsunami\_payload.exe}}.

The active usage of \textit{tor.exe} aligns with the malware's reliance on the \textit{Tor network} for anonymized communication. Conversely, the embedded \textbf{\textit{tsunami\_payload.exe}} stands out as an anomaly. Despite being included within the resource bundle, no references to its extraction or execution were identified in the program's workflow. This omission is particularly intriguing given the malware's reliance on hash-based comparison for deploying the most recent version of the \textit{Tsunami Client}. This update mechanism ensures that only the latest and potentially most secure version of the tool is deployed during the attack. The presence of this forgotten executable, a seemingly outdated or redundant payload, raises questions about its intended purpose.

One plausible explanation is that \textbf{\textit{tsunami\_payload.exe}} could have been a placeholder or backup resource intended for testing or as a contingency in case of a failure in the update process. Alternatively, its inclusion may have been unintentional, resulting from oversight or rushed development during the malware's construction. The lack of references to its deployment leaves its intended role ambiguous and opens the possibility that it was meant to serve in a future iteration of the malware but was left dormant in this version.

Nevertheless, its presence allows for standalone analysis. This dormant payload provides an additional opportunity to uncover details about the attacker's broader toolkit or objectives. Its embedded status, while curious, does not detract from the malware's operational efficiency but instead offers valuable insights into the development practices and potential missteps of the threat actor.

\begin{figure}[H]
    \centering
    \includegraphics[width=0.6\linewidth,frame]{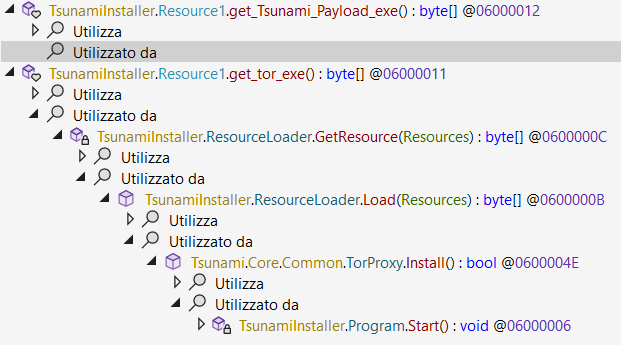}
    \caption{\textbf{\textit{tsunami\_payload.exe}} availability with no reference to its deployment.}
    \label{fig:160}
\end{figure}

The \textit{TorServer} class provides functionality for establishing and managing communication with a remote server over the Tor network. It facilitates tasks such as \textit{session initialization}, \textit{environment information submission}, and \textit{data transmission}. The implementation exhibits a deliberate focus on maintaining persistent and anonymized communication, leveraging the \textit{Tor proxy} for network routing.

The \textbf{\textit{Connect()}} method serves as the entry point for establishing communication with the remote server. It sequentially calls the \textbf{\textit{SendInit()}} and \textbf{\textit{SendEnvironmentInfo()}} methods to initialize the session and transmit the host system's environment details. The method ensures that both steps are successful, logging any failures and terminating the connection attempt if errors occur. Upon successful completion, a \textit{session key} is obtained, which is critical for subsequent interactions.

The \textbf{\textit{SendInit()}} method initializes the connection by sending an \textit{empty payload} (\{\}) to the server's \textit{API} initialization endpoint. The server responds with a \textit{session key}, which is parsed and stored for later use. This acts as an \textit{authentication token}, binding subsequent requests to a specific session. The method logs the outcome of the initialization, ensuring transparency in the connection process.

The \textbf{\textit{SendEnvironmentInfo()}} method collects detailed system information, including application version, \textit{system specifications} (e.g., processor, RAM, display size, operating system), and \textit{geographic location} (e.g., city and country). This information is compiled into a dictionary and transmitted to the server via the \textbf{\textit{SendData()}} method. The latter ensures that critical system attributes are accurately collected and sent, potentially aiding in profiling the victim's environment for tailored malicious activities.

The \textbf{\textit{SendData()}} method is a generalized function for transmitting data to the server. It serializes the data into a \textit{JSON} object, incorporating the \textit{session key} for authentication. The payload is then sent via the \textbf{\textit{TorProxy.SendRequest()}} method, which routes the request through the \textit{Tor network}. Analyzed method provides detailed logging for successful transmissions, including the size of the data sent and the response received.

This class also defines several constant URLs for various \textit{API} endpoints, including those for \textit{telemetry}, \textit{browser passwords}, \textit{session data}, and other assets. These endpoints reflect a comprehensive framework for data exfiltration and telemetry reporting, likely intended for managing stolen information and maintaining control over the infected system.

A noteworthy aspect of this class is its use of Tor for anonymizing communication. By routing all requests through the Tor network, it obscures the server's location and the nature of the communication, complicating detection and attribution efforts. The implementation of detailed logging and error handling ensures that failures are documented, facilitating debugging and operational resilience.

The \textit{TorServer} class demonstrates a well-structured approach to managing communication within a malicious framework. Its integration of session management, environment profiling, and anonymized data transmission reflects a high degree of sophistication. This class is likely a critical component of a broader malware architecture designed for \textit{data exfiltration}, \textit{telemetry}, and \textit{maintaining remote control} over compromised systems.

\begin{figure}[H]
    \centering
    \includegraphics[width=0.75\linewidth,frame]{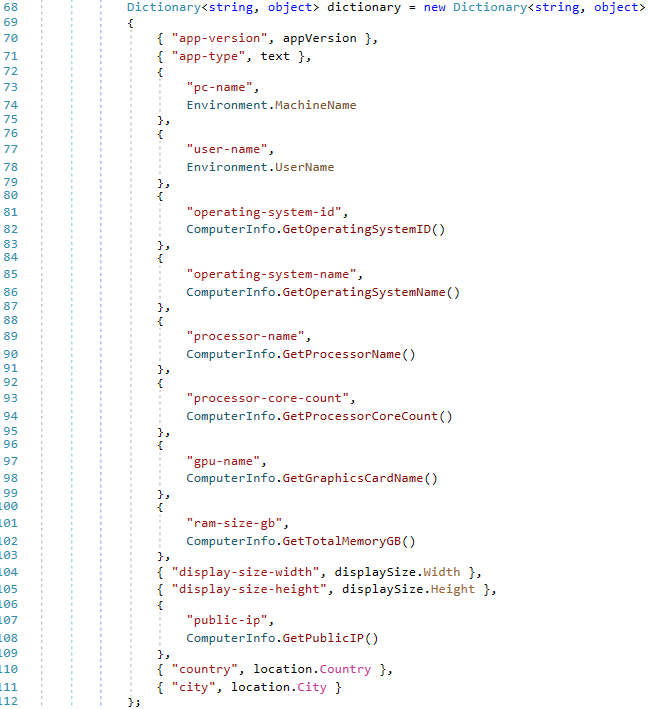}
    \caption{\textit{JSON}-based template with acquired information to exfiltrate.}
    \label{fig:151}
\end{figure}

\begin{figure}[H]
    \centering
    \includegraphics[width=1\linewidth,frame]{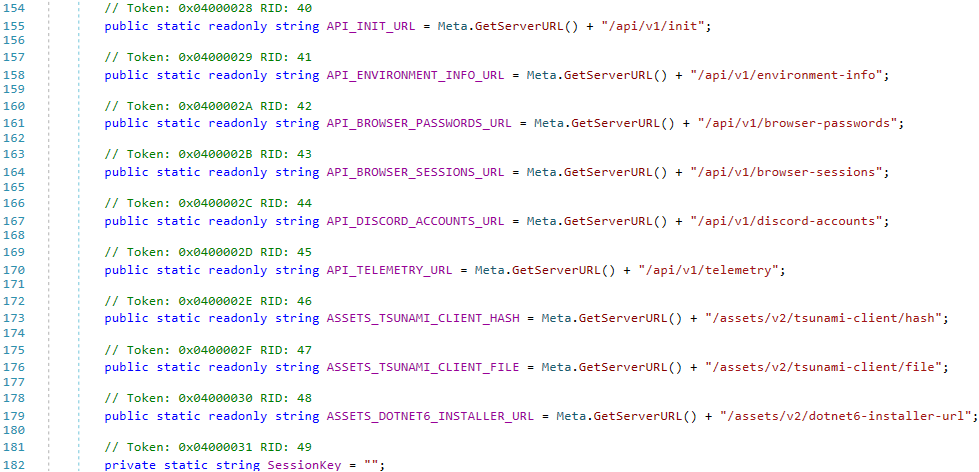}
    \caption{\textit{API} endpoint paths for each single activity the malware takes care of.}
    \label{fig:152}
\end{figure}

\begin{itemize}
    \item \textit{hxxp[:]//n34kr3z26f3jzp4ckmwuv5ipqyatumdxhgjgsmucc65jac56khdy5zqd[.]onion/\\assets/v2/dotnet6-installer-ur}
    \item \textit{hxxp[:]//n34kr3z26f3jzp4ckmwuv5ipqyatumdxhgjgsmucc65jac56khdy5zqd[.]onion/\\api/v1/discord-accounts}
    \item \textit{hxxp[:]//n34kr3z26f3jzp4ckmwuv5ipqyatumdxhgjgsmucc65jac56khdy5zqd[.]onion/\\api/v1/browser-passwords}
    \item \textit{hxxp[:]//n34kr3z26f3jzp4ckmwuv5ipqyatumdxhgjgsmucc65jac56khdy5zqd[.]onion/\\api/v1/init}
    \item \textit{hxxp[:]//n34kr3z26f3jzp4ckmwuv5ipqyatumdxhgjgsmucc65jac56khdy5zqd[.]onion/\\assets/v2/tsunami-client/file}
    \item \textit{hxxp[:]//n34kr3z26f3jzp4ckmwuv5ipqyatumdxhgjgsmucc65jac56khdy5zqd[.]onion/\\api/v1/browser-sessions}
    \item \textit{hxxp[:]//n34kr3z26f3jzp4ckmwuv5ipqyatumdxhgjgsmucc65jac56khdy5zqd[.]onion/\\api/v1/telemetry}
    \item \textit{hxxp[:]//n34kr3z26f3jzp4ckmwuv5ipqyatumdxhgjgsmucc65jac56khdy5zqd[.]onion/\\assets/v2/tsunami-client/hash}
    \item \textit{hxxp[:]//n34kr3z26f3jzp4ckmwuv5ipqyatumdxhgjgsmucc65jac56khdy5zqd[.]onion/\\api/v1/environment-info}
\end{itemize}

As observed in previous instances, nearly all components within the identified list, except for the \textit{Discord} and \textit{Browser} related paths, are actively utilized in at least one of the malicious functions implemented in the analyzed \textit{DLL}. This notable exception raises similar questions to those posed earlier, as it may represent a remnant of a prior iteration of the module, initially developed for a different purpose and subsequently repurposed or adapted to fit its current scope. It might be a leftover artifact from an earlier stage of development, where the module was designed with broader or alternative functionalities. This could imply that the malware’s architecture has evolved, discarding certain features while adapting others to serve the campaign’s objectives. Alternatively, it might offer a glimpse into future intentions, signaling the attacker’s plans to incorporate \textit{Discord} and \textit{Browser} focused features into subsequent versions of the module.

Such patterns reflect the iterative nature of the \textit{Threat Actor}’s development process, where modularity and flexibility play key roles. The inclusion of potentially deprecated or yet-to-be-deployed components demonstrates the evolving scope of their malicious toolkit. While it is possible that the \textit{Discord} and \textit{Browser} related paths was left in unintentionally due to rushed development, it cannot be dismissed as a mere oversight. Instead, it provides valuable insight into the attacker’s design philosophy and the lifecycle of their malicious tools.

This dormant paths, much like other unexplored functionalities or components, highlights the importance of monitoring the malware's development over time. Analyzing such artifacts can reveal potential shifts in the attacker’s focus, providing early warning of new techniques or targets that may emerge in future campaigns.

\begin{figure}[H]
    \centering
    \includegraphics[width=0.8\linewidth,frame]{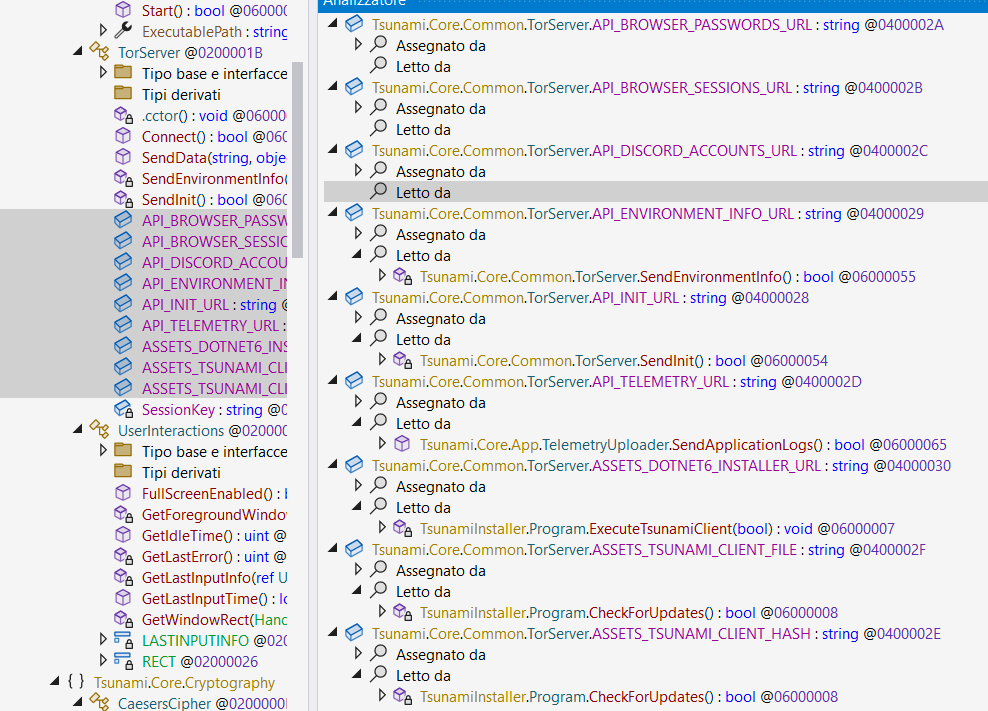}
    \caption{\textit{Discord} and \textit{Browser} paths are not read by any function.}
    \label{fig:161}
\end{figure}

By proceeding with the code analysis, it is possible to focus on the \textit{ComputerInfo} class. The latter is designed to gather detailed system information, leveraging both managed \textit{.NET} functionality and native Windows \textit{API}s. It provides methods to extract data about \textit{hardware}, \textit{operating system}, and \textit{display settings}, as well as \textit{geolocation} and \textit{public IP address} information. The methods combine \textit{command-line utilities}, \textit{registry queries}, and \textit{API calls} to compile a comprehensive profile of the host system.

The class includes methods such as \textbf{\textit{GetProcessorName()}}, \textbf{\textit{GetProcessorCoreCount()}}, and \textbf{\textit{GetGraphicsCardName()}} to retrieve details about the system's CPU and GPU. These methods execute Windows Management Instrumentation \textit{Command-line} (\textit{WMIC}) queries through the command prompt and parse the output. For instance, \textbf{\textit{GetProcessorName()}} retrieves the \textit{CPU} name by running the \textit{WMIC} command for processor details, extracting and formatting the output string. Similarly, \textbf{\textit{GetProcessorCoreCount()}} uses \textit{WMIC} to determine the number of \textit{CPU cores}, and \textbf{\textit{GetGraphicsCardName()}} queries the \textit{GPU} name.

To determine if a dedicated \textit{GPU} exists, the \textbf{\textit{DedicatedGraphicsCardExists()}} method uses \textit{WMIC} to fetch video controller descriptions and searches for keywords like \textit{Nvidia} or \textit{Radeon} in the output. This method provides insight into the graphical capabilities of the system, which can be useful for tailoring payloads or assessing the target's computational power.

The class includes \textbf{\textit{GetTotalMemoryGB()}}, which retrieves the system's \textit{physical memory} using the \textbf{\textit{GetPhysicallyInstalledSystemMemory()}} function from \textit{kernel32.dll}. This \textit{API} call ensures accurate memory reporting in \textit{GB}, independent of the system's \textit{OS} or configuration. Display size is obtained through the \textbf{\textit{EnumDisplaySettings()}} function from \textit{user32.dll}, which retrieves the screen resolution for the primary monitor.

Operating system details are retrieved via methods such as \textbf{\textit{GetOperatingSystemName()}} and \textbf{\textit{GetOperatingSystemID()}}. The former uses \textit{WMIC} to fetch the \textit{OS} caption and formats it as a user-friendly string. The latter queries the \textit{Windows registry} for the \textit{product ID} using predefined paths, demonstrating its ability to gather licensing information or unique identifiers tied to the operating system.

The \textit{geolocation} capabilities of the class are implemented in \textbf{\textit{GetLocation()}}, which combines \textit{public IP} retrieval with location services such as \textit{ipinfo.io}. The method sends HTTP requests to these \textit{API}s, fetching data about the system's \textit{public IP}, \textit{country}, and \textit{city}. The \textbf{\textit{GetPublicIP()}} method offers similar functionality, querying multiple online services for the \textit{public IP address}.

Internally, the class uses helper methods to parse and extract relevant information from the outputs of \textit{WMIC commands}, \textit{registry queries}, and \textit{API responses}. The use of both \textit{.NET} libraries and \textit{unmanaged} code illustrates a hybrid approach, enabling the class to access a wide range of system information.

This class serves as a robust tool for profiling the host system, with applications ranging from hardware and software inventory to geolocation and network assessment. While such capabilities can be legitimate in administrative or diagnostic contexts, in this context they are used to fingerprint the targeted machine and possibly to tailor an evental deploy of the previously referenced \textbf{\textit{XMRig Miner}}.

\begin{figure}[H]
    \centering
    \includegraphics[width=1\linewidth,frame]{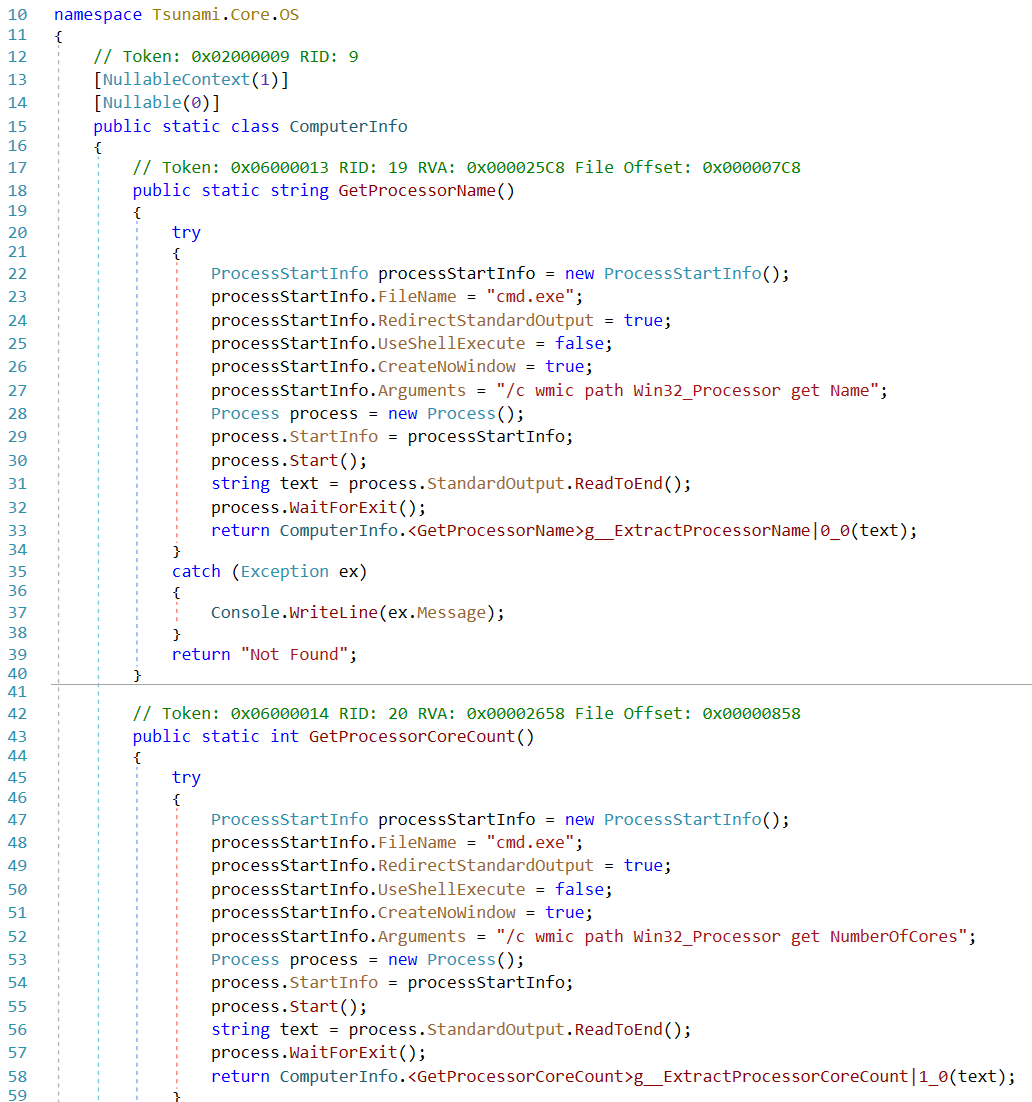}
    \caption{Snippet of the \textit{ComputerInfo} class}
    \label{fig:158}
\end{figure}

The \textbf{\textit{ExecuteTsunamiClient()}} method manages the execution of the \textit{Tsunami Client}, with a focus on ensuring the necessary runtime environment, such as \textit{.NET 6}, is installed and operational. It begins by verifying if the \textit{.NET 6} framework is present on the system. If not, it attempts to retrieve the installer URL from the server via a Tor proxy, provided the server is online. This step underscores its reliance on dynamic dependencies, highlighting its adaptability but also its dependency on external infrastructure.

If the server is offline or the installer URL cannot be retrieved, the method logs an error and aborts the process, reflecting the criticality of \textit{.NET 6} to the client's functionality. Once the URL is obtained, the method invokes the \textbf{\textit{DotNet6.Install()}} function to download and install the framework. Any failure during this installation process is logged, emphasizing robust error reporting.

After ensuring the runtime environment is ready, the method attempts to launch the \textit{Tsunami Client} executable. If successful, it logs the initiation of the client and sets the \textit{ClientRunning} flag to \textit{true}, indicating operational status. Conversely, a failure to start the client is logged as an error, ensuring transparency in operation status.

This method demonstrates a structured approach to dependency management and execution control. The integration of dynamic installation for \textit{.NET 6} enables the malware to adapt to a variety of environments, ensuring compatibility regardless of the target system's initial configuration. Its reliance on the \textit{Tor proxy} for obtaining dependencies highlights an emphasis on obfuscating communication, aligning with tactics commonly employed by malicious software.

The presence of robust error handling and detailed logging provides insights into its operational logic but also reveals its potential misuse. By ensuring dependencies are dynamically resolved and operational status is closely monitored, the method reflects a design aimed at maintaining resilience and adaptability, potentially in support of a larger malicious framework.

\begin{figure}[H]
    \centering
    \includegraphics[width=1\linewidth,frame]{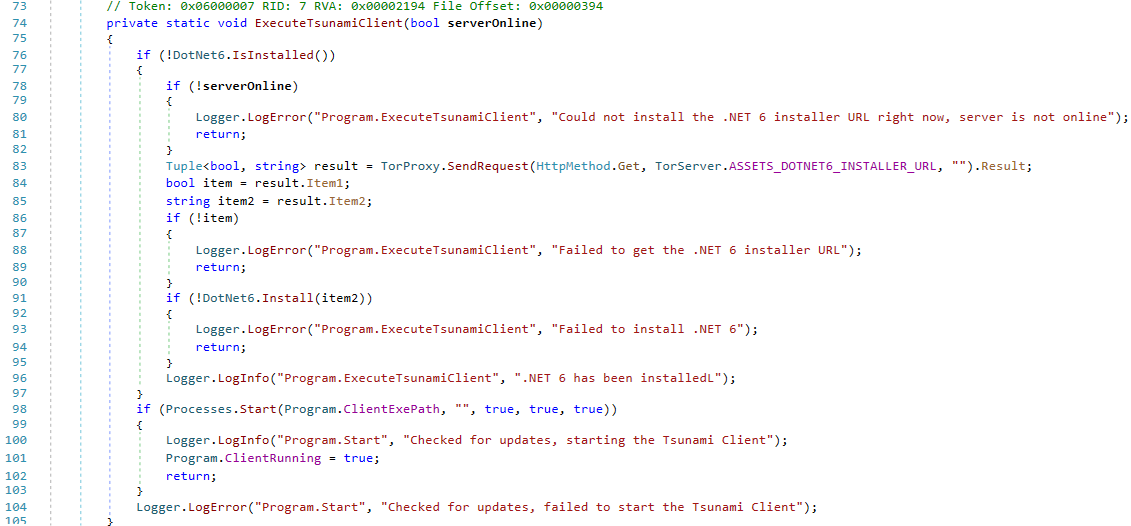}
    \caption{Overview of the \textbf{\textit{ExecuteTsunamiClient()}}}
    \label{fig:154}
\end{figure}

The \textit{TelemetryUploader} class appears to be designed for aggregating and transmitting application logs to a remote server under the guise of legitimate telemetry functionality. The \textbf{\textit{SendApplicationLogs()}} method processes runtime logs by categorizing them into Success, Info, Warning, and Error types, creating both a summary and a detailed report of the application’s activity. These logs are dynamically categorized based on the application’s role (e.g., \textit{ClientAppLogs} or \textit{InstallerAppLogs}) to ensure contextual relevance, further suggesting a tailored approach to data collection.

A telemetry object encapsulates the \textit{session ID}, \textit{log categories}, and detailed \textit{runtime data}, which is transmitted to a remote server via the \textbf{\textit{TorServer.SendData()}} method. 

The robust design, detailed logging, and anonymized communication suggest that its likely intent is to \textit{gather intelligence} from compromised hosts, either for \textit{system profiling}, \textit{operational oversight}, or \textit{further exploitation}. The sophistication of this class underlines the need for thorough investigation and monitoring to mitigate its potential impacts.

\begin{figure}[H]
    \centering
    \includegraphics[width=0.9\linewidth,frame]{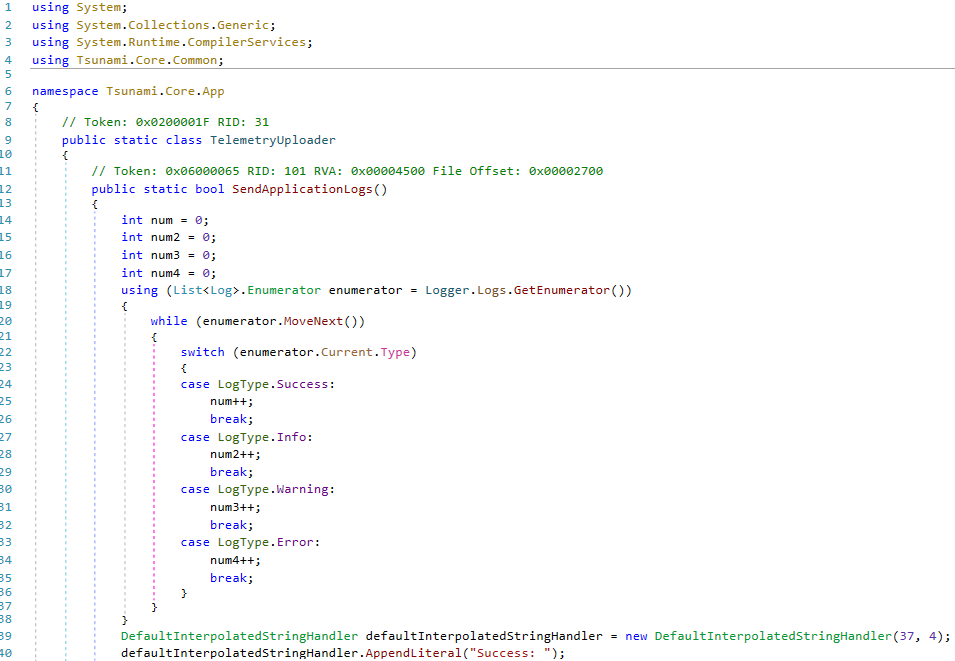}
    \caption{Overview of the \textit{TelemetryUploader} class}
    \label{fig:153}
\end{figure}

The \textit{UserInteractions} class is a utility designed to monitor and analyze user activity and system interaction states. It relies on Windows \textit{API} calls to retrieve \textit{idle time}, detect \textit{fullscreen applications}, and assess the \textit{user's last input}. Despite being implemented in the source code, this class \textit{remains unused} within the provided execution flow, raising questions about its intended purpose and whether it was meant for testing, debugging, or future expansion.

The class includes methods such as \textbf{\textit{GetIdleTime()}} and \textbf{\textit{GetLastInputTime()}}, which determine the duration since the last user interaction. These methods leverage the \textbf{\textit{GetLastInputInfo()}} function from \textit{User32.dll} to fetch the timestamp of the most recent input. \textbf{\textit{GetIdleTime()}} calculates the elapsed time in milliseconds, while \textbf{\textit{GetLastInputTime()}} provides this information in seconds, incorporating error handling to manage API call failures.

The \textbf{\textit{FullScreenEnabled()}} method evaluates whether the currently active application is running in \textit{fullscreen} mode. It retrieves the dimensions of the primary display using the \textbf{\textit{ComputerInfo.GetDisplaySize()}} method and compares them with the dimensions of the foreground window, obtained via \textbf{\textit{GetWindowRect()}} and \textbf{\textit{GetForegroundWindow()}} from \textit{User32.dll}. By constructing and comparing rectangles, this method determines if the foreground window occupies the entire screen.

The class relies on two internal structs, \textit{RECT} and \textit{LASTINPUTINFO}, which act as data containers for \textit{API} calls. \textit{RECT} stores the dimensions of a window, while \textit{LASTINPUTINFO} holds details about the last user input. These structures facilitate seamless integration with the Windows \textit{API}, enabling the class's functionality.

Despite its sophisticated design, the absence of this class from the operational codebase suggests it was either deprecated, unfinished, or reserved for future use. The presence of such a class indicates an interest in user activity profiling, potentially to tailor malicious actions based on the victim's behavior. For example, detecting \textit{fullscreen} mode might signal a gaming or media application, potentially delaying certain malware activities to avoid detection.

The unused state of the \textit{UserInteractions} class could also hint at incomplete development or a deliberate exclusion from the main code to reduce detection risk. Its capabilities align with broader reconnaissance and behavioral monitoring goals, but without active invocation, it remains an artifact that offers insights into the malware's potential design objectives and development process.

\begin{figure}[H]
    \centering
    \includegraphics[width=0.8\linewidth,frame]{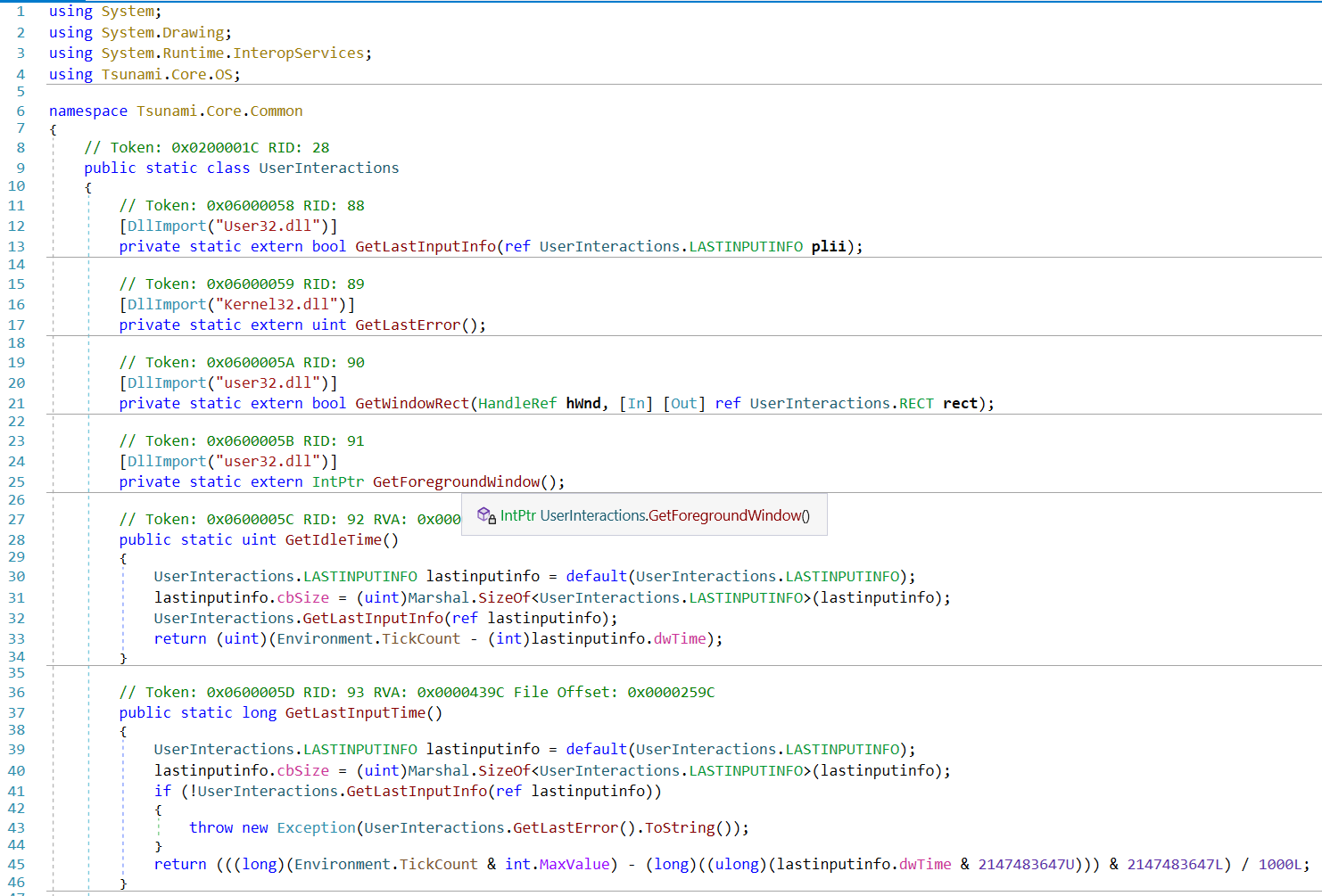}
    \caption{Overview of the unused \textit{UsersInteractions} class}
    \label{fig:157}
\end{figure}

The \textit{CaesersCipher} class implements a classical \textit{Caesar cipher encryption} and \textit{decryption} algorithm, providing basic functionality for shifting letters in a string by a specified number of steps. Despite its simplicity and potential utility, this class remains unused within the provided codebase, suggesting it may have been intended for testing, debugging, or as part of a feature that was ultimately removed or deferred.

The \textbf{\textit{Encrypt()}} method transforms a given string by shifting each alphabetical character forward in the alphabet by the specified number of steps (\textit{step}). It preserves the case of the letters, ensuring uppercase and lowercase characters are shifted within their respective ranges, and leaves non-alphabetic characters unchanged. For example, the letter 'A' shifted by one \textit{step} would become 'B', while 'z' shifted by one \textit{step} would wrap around to 'a'.

Similarly, the \textbf{\textit{Decrypt()}} method reverses the transformation by shifting characters backward by the specified number of steps, also preserving case and ignoring non-alphabetic characters. The implementation uses modular arithmetic to handle the wrapping of letters at the boundaries of the alphabet.

The unused state of this class raises questions about its intended role within the malware. Its implementation suggests it might have been designed for lightweight obfuscation of strings or data, such as encoding configuration settings, URLs, or commands to evade simple detection mechanisms. However, the simplicity of the Caesar cipher makes it unsuitable for robust cryptographic purposes, as it is easily broken through frequency analysis or brute force due to the limited \textit{keyspace}.

The inclusion of the \textit{CaesersCipher} class, despite its non-use, provides insight into the potential development process of the malware. It could indicate that the developers experimented with or considered alternative encryption mechanisms before settling on more complex or secure methods elsewhere in the code. Alternatively, it might reflect a placeholder or backup implementation, highlighting the iterative nature of the malware's development lifecycle.

\begin{figure}[H]
    \centering
    \includegraphics[width=0.8\linewidth,frame]{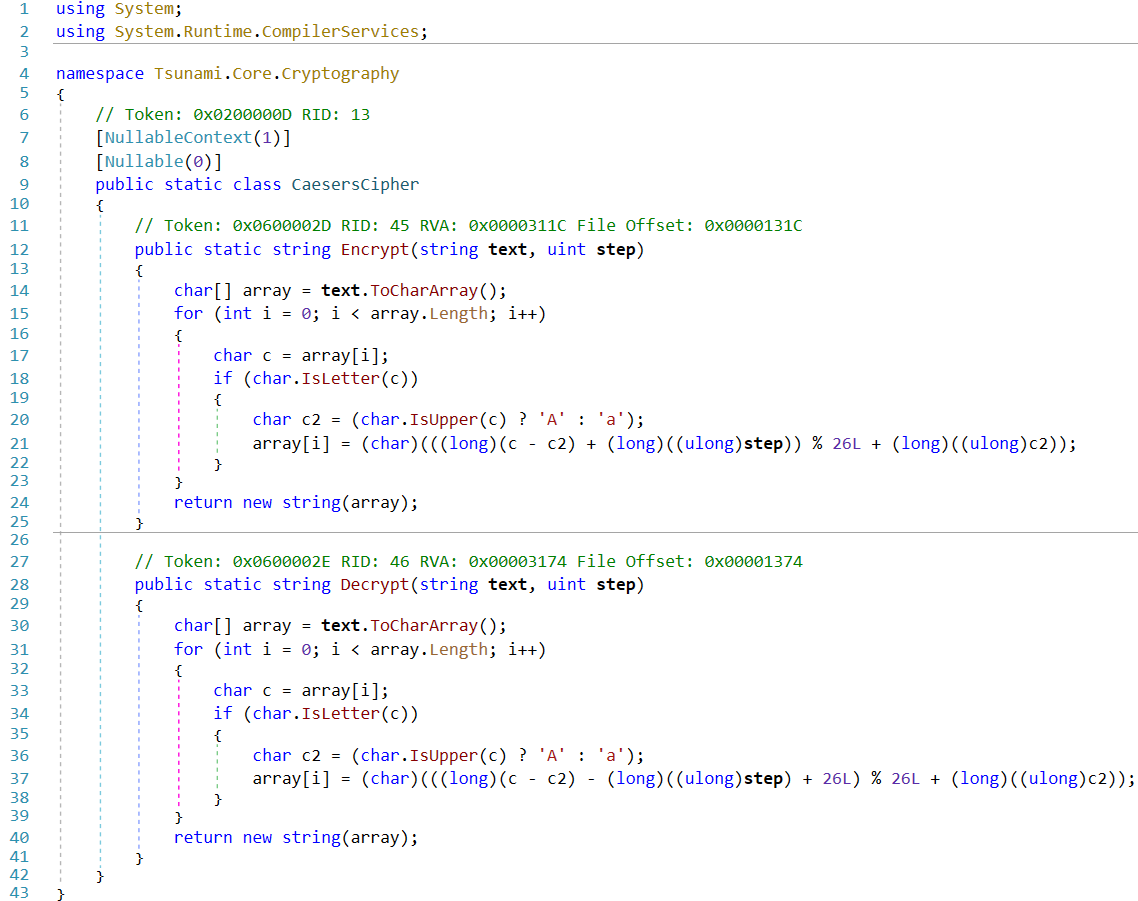}
    \caption{Overview of the unused \textit{CaesarsCipher} class}
    \label{fig:159}
\end{figure}

\textbf{\textit{Dynamic Analysis}}\\
The execution of Runtime Broker.exe shows, as first, the executable being accessed from the \textit{\%APPDATA\%\textbackslash Roaming\textbackslash Microsoft\\\textbackslash Windows} directory. This unconventional execution path immediately raises suspicions, as it deviates from standard system directory conventions, as previously mentioned. Subsequent interactions with system libraries like \textit{KernelBase.dll} and \textit{kernel32.dll} suggest that the process is preparing its runtime environment, loading functions critical for system-level interactions. These methods likely include capabilities for memory manipulation, process injection, or thread management, which are common in malicious processes aiming to extend their reach within the system.

\begin{figure}[H]
    \centering
    \includegraphics[width=0.9\linewidth,frame]{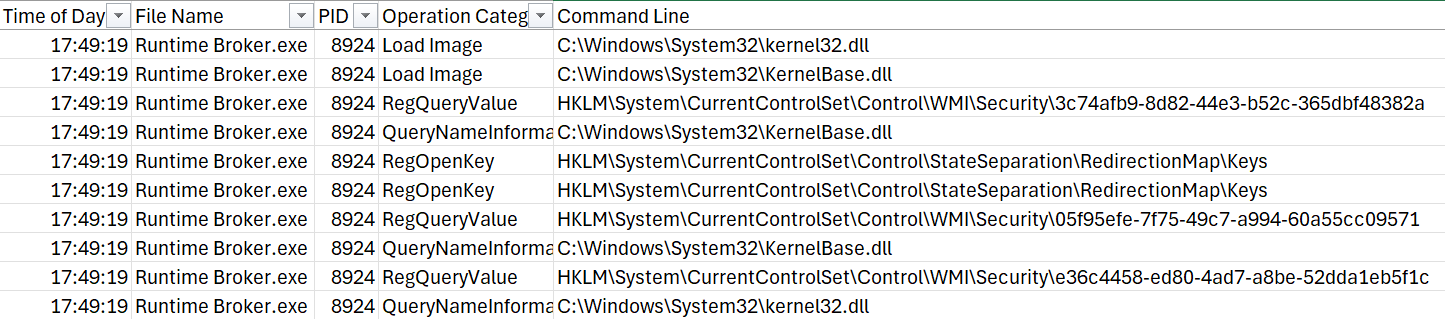}
    \caption{\textbf{\textit{Runtime Broker.exe}} loading system libraries.}
    \label{fig:113}
\end{figure}

There are also Registry operations appearing particularly noteworthy. Analyzing registry operations reveals access to
\textit{HKLM\textbackslash System\textbackslash CurrentControlSet\textbackslash Services\textbackslash bam\textbackslash State \textbackslash UserSettings}, a registry key that tracks user-level application activity. This query suggests reconnaissance activities aimed at gathering information about system usage patterns or identifying running applications for potential injection or exploitation. What has been recorded and shown below indicates that the process accessed \textit{HKLM\textbackslash System \textbackslash CurrentControlSet\textbackslash Control\textbackslash Session Manager}, a key integral to managing system boot configurations. By querying this key, the malware likely intends to evaluate or modify startup behaviors, ensuring that it executes automatically upon system reboot.

\begin{figure}[H]
    \centering
    \includegraphics[width=0.9\linewidth,frame]{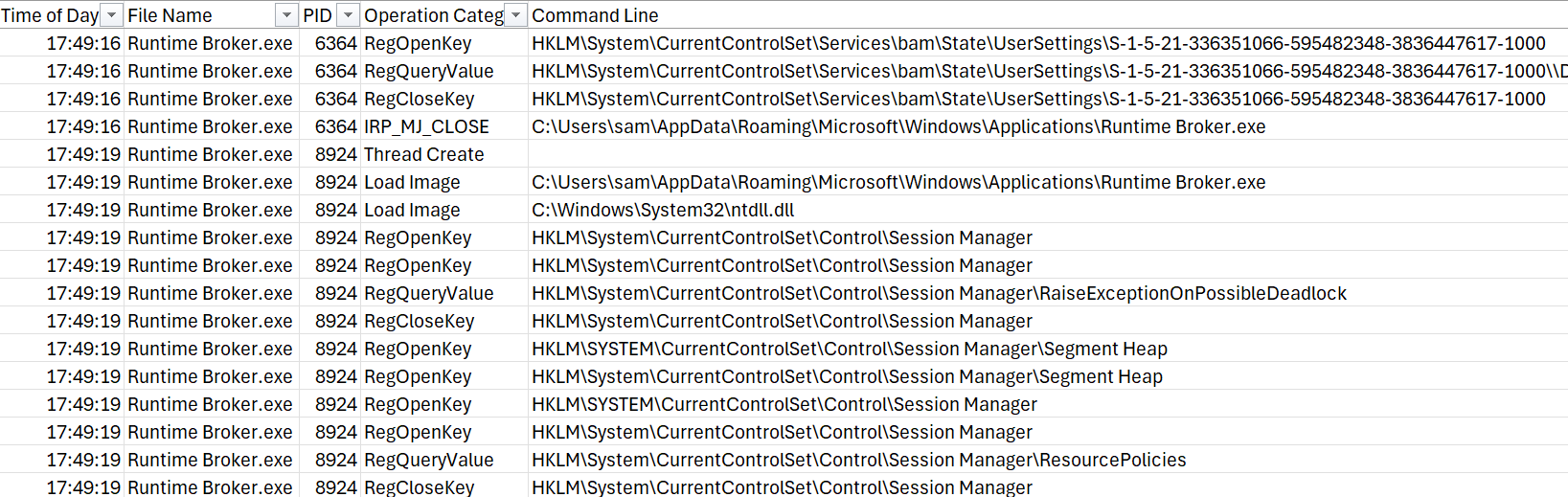}
    \caption{\textbf{\textit{Runtime Broker.exe}} querying interesting registry keys.}
    \label{fig:114}
\end{figure}

Additional file operations involve interactions with \textit{apphelp.dll}, a library often associated with compatibility and application support in Windows. This may indicate attempts to exploit or modify application compatibility settings as part of its malicious strategy.

\begin{figure}[H]
    \centering
    \includegraphics[width=0.8\linewidth,frame]{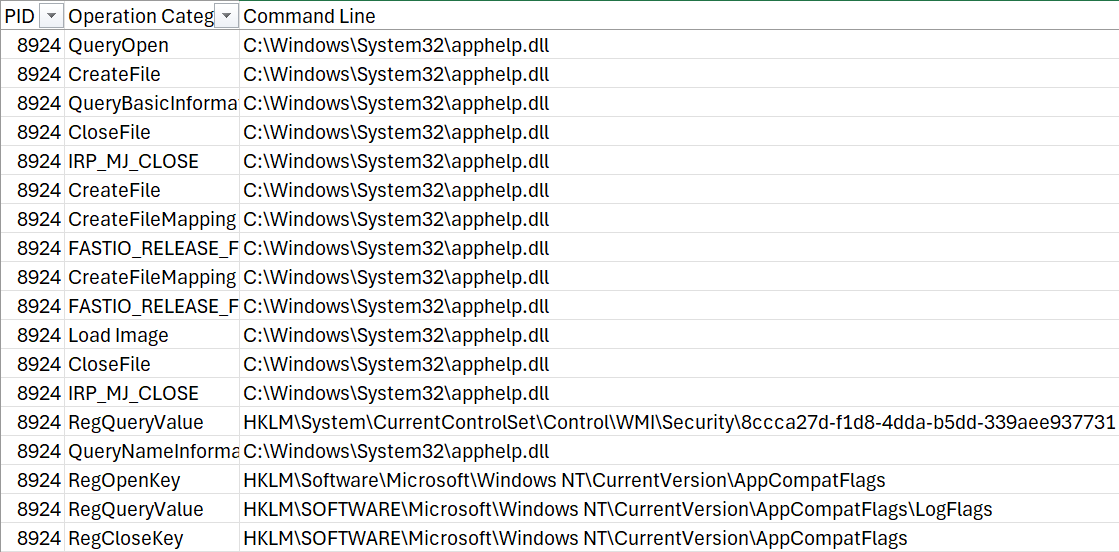}
    \caption{\textbf{\textit{Runtime Broker.exe}} interacting with \textit{apphelp.dll}.}
    \label{fig:115}
\end{figure}

Then, there are several attempts to access specific registry keys under \textit{HKLM\textbackslash Software \textbackslash Policies\textbackslash Microsoft\textbackslash Windows\textbackslash Display} and \textit{HKLM\textbackslash SOFTWARE\textbackslash Microsoft\textbackslash Windows NT \textbackslash CurrentVersion}. These actions frequently result in a \textit{NAME NOT FOUND} detail, indicating the queried registry entries do not exist. The desired access permissions are predominantly read-related, with some operations querying values and enumerating \textit{subkeys}. This phase suggests that the process is performing system reconnaissance, as previously identified in the analysis of \textbf{\textit{Runtime Broker.dll}}.

\begin{figure}[H]
    \centering
    \includegraphics[width=1\linewidth,frame]{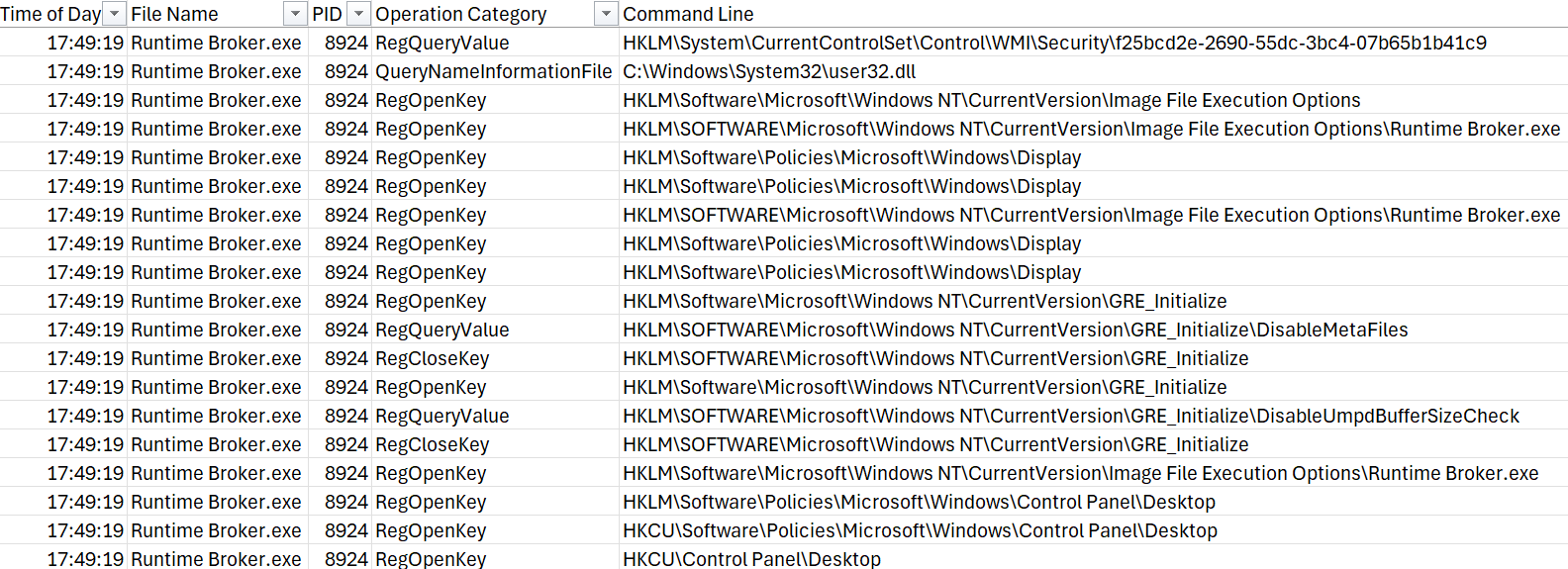}
    \caption{Executable querying extensively the \textit{HKLM} hive.}
    \label{fig:117}
\end{figure}

Later, activities shift toward file handling and memory management. Operations like \textit{CreateFileMapping} and \textit{FASTIO\_RELEASE\_FOR\_SECTION\_SYNCHRONIZATION} appear, signaling interaction with memory-mapped files. These are common in processes attempting to share memory between applications or manage large datasets efficiently. Additionally, thread creation events (\textit{Thread Create}) indicate that new execution threads are being initialized, hinting at multitasking or concurrency within the process. The interaction with system libraries, such as \textit{rpcss.dll}, and the presence of \textit{FAST IO DISALLOWED} suggest potential privilege or capability constraints imposed on the process.

\begin{figure}[H]
    \centering
    \includegraphics[width=1\linewidth,frame]{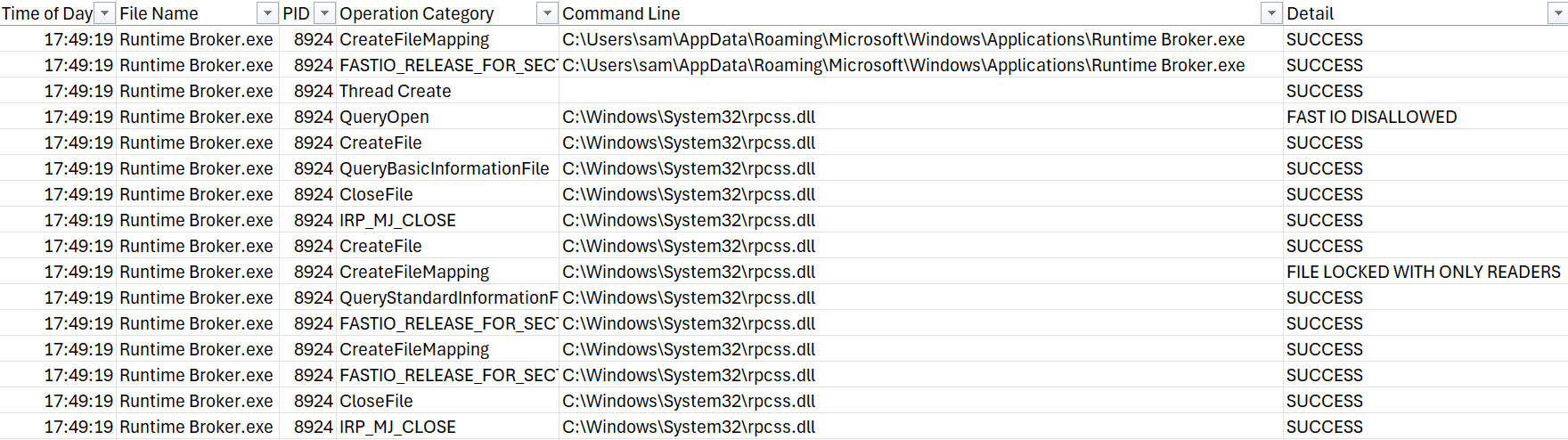}
    \caption{Executable interacting with \textit{rpcss.dll}.}
    \label{fig:118}
\end{figure}

The process queries and opens multiple registry keys under paths such as \textit{HKLM \textbackslash Software\textbackslash Microsoft\textbackslash Windows\textbackslash CurrentVersion} and \textit{HKLM\textbackslash System\textbackslash CurrentControlSet}. The successful results for these actions indicate that the queried keys exist, and the desired access permissions, predominantly read permissions, are granted. These operations likely aim to retrieve system or application configurations, such as file paths, environment settings, or user preferences.

\begin{figure}[H]
    \centering
    \includegraphics[width=1\linewidth,frame]{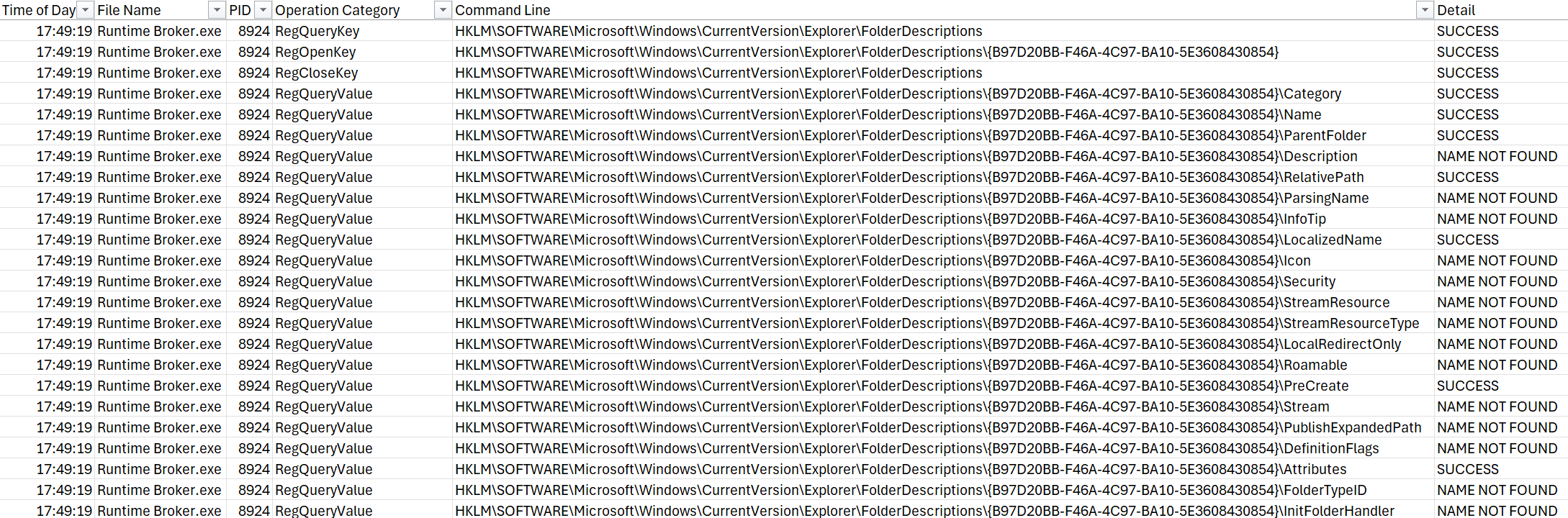}
    \caption{Executable continues to map the \textit{HKLM} hive looking for keys of interest.}
    \label{fig:119}
\end{figure}

Registry-related events dominate this range, with key activities including \textit{RegQueryKey}, \textit{RegOpenKey}, and \textit{RegCloseKey}. The keys being accessed, such as those under \textit{Control\textbackslash Hvsi} and \textit{Nls\textbackslash Sort}, suggest the process is targeting configurations related to hardware-assisted virtualization and system sorting behaviors, respectively. These entries might be leveraged for compatibility checks, feature detection, or runtime behavior adjustments.

\begin{figure}[H]
    \centering
    \includegraphics[width=1\linewidth,frame]{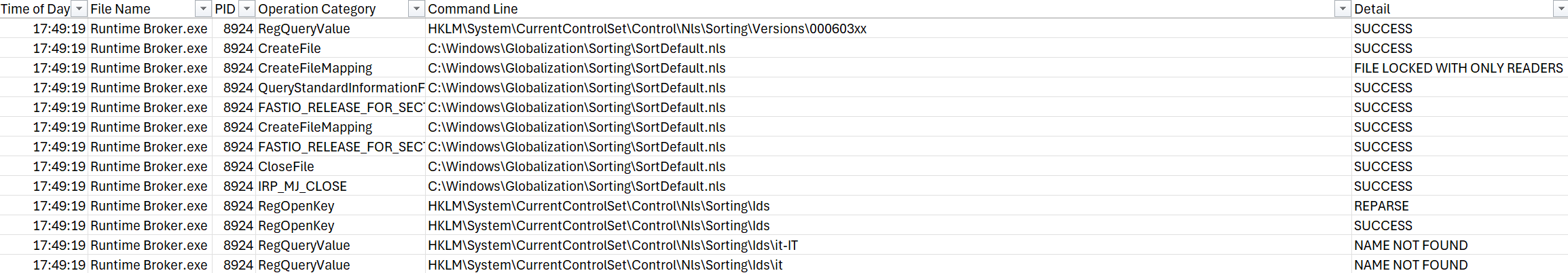}
    \caption{Executable interactions with \textit{Control\textbackslash Hvsi} and \textit{Nls\textbackslash Sort}.}
    \label{fig:120}
\end{figure}

The occasional \textit{NAME NOT FOUND} details for specific queries, such as in the \textit{RegQueryValue} operation under \textit{Control\textbackslash Hvsi\textbackslash IsHvsiContainer}, indicate that some queried values are absent, perhaps revealing conditional checks within the process's logic. Indeed, this registry key is associated with \textit{Hypervisor-based Security Isolation} (\textit{HVSI}) and is typically used to indicate whether a system or process is running inside an \textit{HVSI} container. \textit{Hypervisor-based Security Isolation} (\textit{HVSI}) is a feature enabled by \textit{virtualization-based security} (\textit{VBS}) and \textit{Hyper-V} on Windows systems. It isolates critical system components and certain processes within containers that are protected by the hypervisor. This enhances security by preventing unauthorized access and code execution, even in the event of a \textit{kernel compromise}. This allows the subjected executable to both adapt its behavior, basing on the security measures available on the system, and acquire system's security configuration to later exfiltrate to the remote \textit{Threat Actor}.

\begin{figure}[H]
    \centering
    \includegraphics[width=1\linewidth,frame]{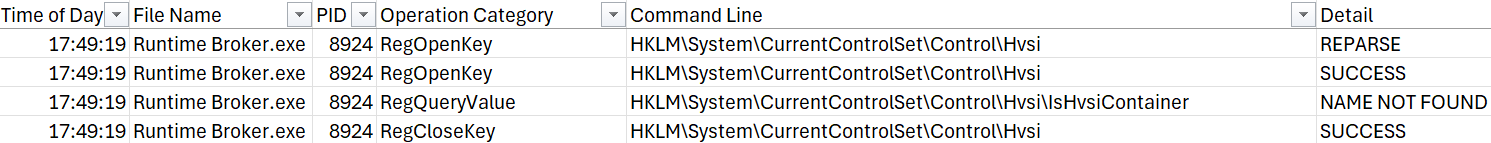}
    \caption{\textbf{\textit{Runtime Broker.exe}} tries to identify the presence of \textit{HVSI} container.}
    \label{fig:121}
\end{figure}

There is also evidence of deeper system exploration, such as the retrieval of data related to \textit{kernel32.dll}. This could imply attempts to verify core system library availability or extract runtime parameters that depend on the system's localization and sorting configuration.

The process, at this point, attempts to open or query specific files, related to \textit{PowerShell} instances. Each of them posed in a different folder, and related to different application (e.g. \textit{Chocolatey}). Additional details are provided in the following image.

\begin{figure}[H]
    \centering
    \includegraphics[width=1\linewidth,frame]{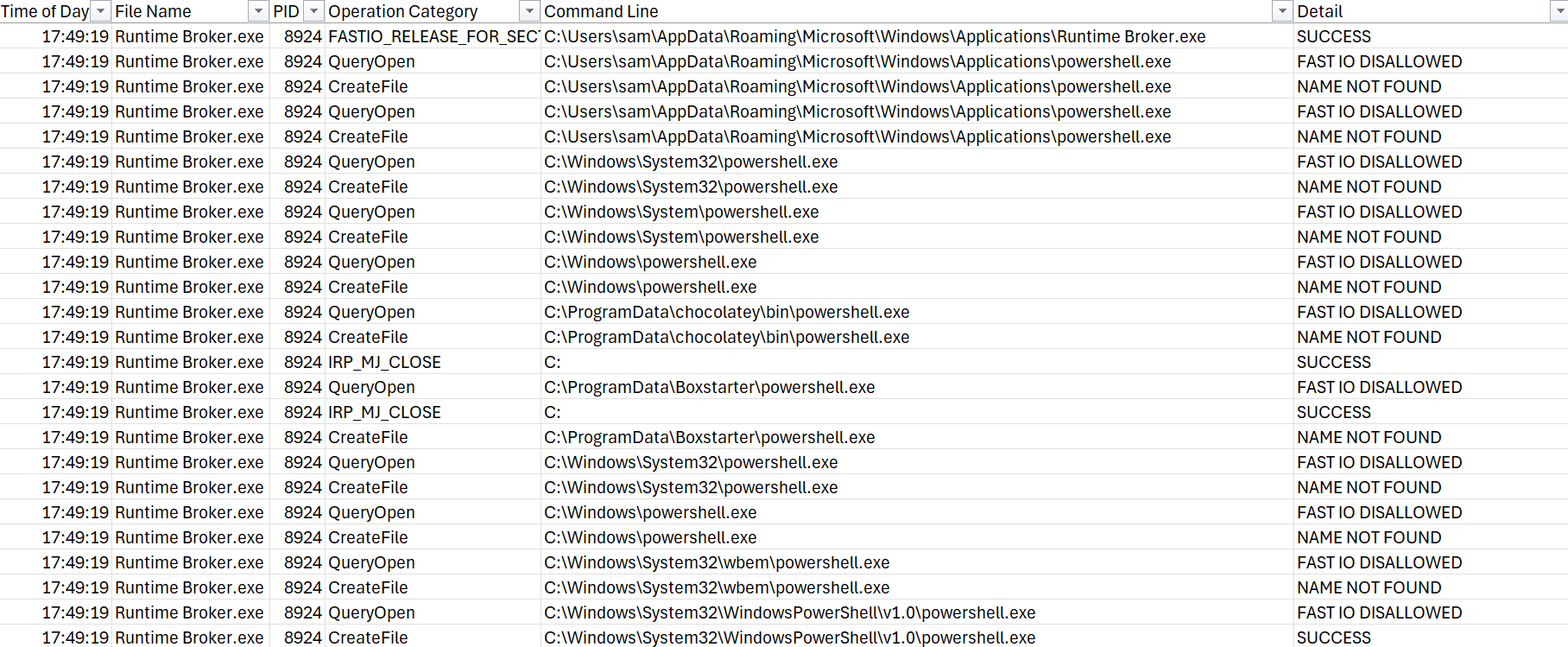}
    \caption{\textbf{\textit{Runtime Broker.exe}} tries to map \textit{PowerShell.exe} instances.}
    \label{fig:122}
\end{figure}

There are also interactions with files related to system patching and \textit{PowerShell}, such as \textit{sysmain.sdb} in the \textit{C:\textbackslash Windows\textbackslash apppatch} directory and \textit{powershell.exe} in the \textit{C:\textbackslash Windows\textbackslash System32\textbackslash WindowsPowerShell\textbackslash v1.0} path. These successful interactions, like \textit{FASTIO\_RELEASE\_FOR\_SECTION\_SYNCHRONIZATION} and \textit{QueryStandardInformationFile}, suggest that the process is inspecting system utilities and environment details, possibly for compatibility checks or preparatory tasks.

\begin{figure}[H]
    \centering
    \includegraphics[width=1\linewidth,frame]{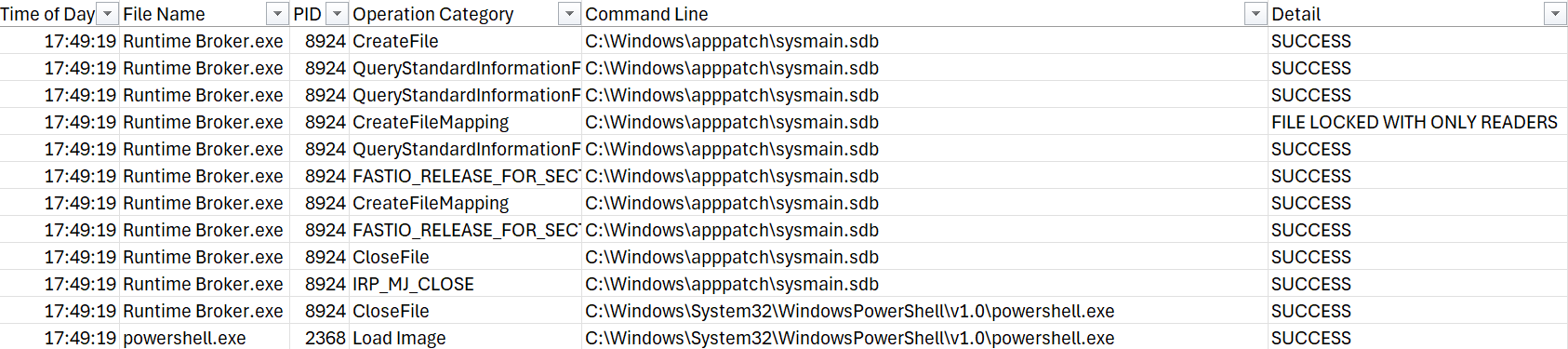}
    \caption{Additional system queries made by \textbf{\textit{Runtime Broker.exe}}.}
    \label{fig:123}
\end{figure}

As depicted in the accompanying image, the establishment of these exclusions occurs in two distinct phases, executed by separate components. Initially, upon the execution of \textbf{\textit{Runtime Broker.exe}}, all six new \textit{firewall rules} are applied (paths correspond to the one identified in \figurename~\ref{fig:148}). Subsequently, after a delay of approximately 15 seconds, these same exclusions are reapplied by a child \textit{PowerShell} process, spawned by \textbf{\textit{Runtime Broker.exe}}. This evidence underscores the heightened level of resilience and redundancy embedded by the developers across their toolset.
\begin{itemize}
    \item \textit{\%APPDATA\%\textbackslash Microsoft\textbackslash Windows\textbackslash Start Menu\textbackslash Programs\textbackslash Startup\textbackslash System Runtime Monitor.exe}
    \item \textit{\%APPDATA\%\textbackslash Microsoft\textbackslash Windows\textbackslash Applications \textbackslash Runtime Broker.exe}
    \item \textit{\%LOCALAPPDATA\%\textbackslash Microsoft\textbackslash Windows\textbackslash Applications\textbackslash Runtime Broker.exe}
    \item \textit{\%APPDATA\%\textbackslash Microsoft\textbackslash Windows\textbackslash Dependencies\textbackslash System Runtime Monitor.exe}
    \item \textit{\%LOCALAPPDATA\%\textbackslash Microsoft\textbackslash Windows\textbackslash WindowsApps\textbackslash msedge.exe}
    \item \textit{\%TEMP\%\textbackslash Runtime Broker.exe}
\end{itemize}

At the onset of the malware’s activity, some of the most noteworthy behaviors pertain to the manipulation of \textit{Firewall policies} and \textit{Antivirus exclusions}. These actions provide analysts with critical insights into the additional payloads that the \textit{Threat Actor} intends to deploy within the target systems. One of the initial observations involves six \textit{inbound allow rules} introduced by the \textbf{\textit{Runtime Broker.exe}} executable within the \textit{Windows Firewall}. These rules are deceptively labeled as \textit{Microsoft Edge WebEngine} as previously identified in the analysis of the \textbf{\textit{Runtime Broker.dll}}.

\begin{figure}[H]
    \centering
    \includegraphics[width=1\linewidth,frame]{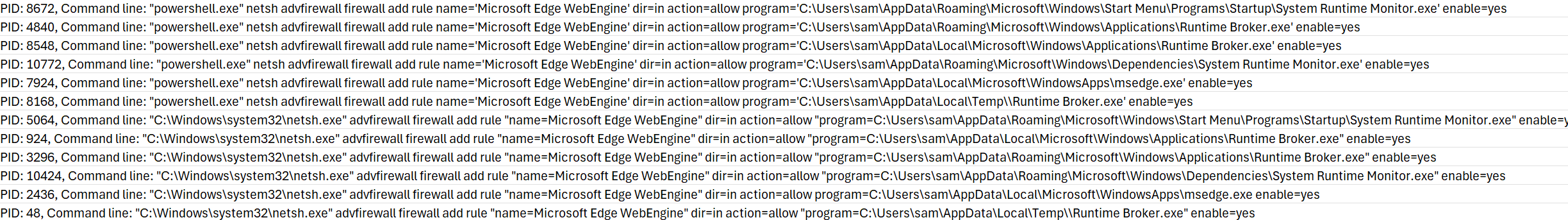}
    \caption{Windows Firewall exclusions}
    \label{fig:126}
\end{figure}

A similar fail-safe rationale is evident in the implementation of \textit{AV exclusions}. Prior to the active execution of \textit{Runtime Broker.exe}, the 	\textit{TSUNAMI PAYLOAD} script was responsible for modifying \textit{Defender’s} policies and registering \textbf{\textit{Runtime Broker.exe}} as a \textit{scheduled task} (\figurename~\ref{fig:115}). As illustrated in the subsequent image, both the three file paths managed by the \textit{TSUNAMI PAYLOAD} and an additional four paths introduced later are excluded from \textit{Defender}'s scans. This layered approach ensures that, even in scenarios where the Python script might fail to execute its intended tasks, the executable can independently enforce the exclusions. Such robust and redundant design highlights the meticulous planning and sophistication employed by the malware’s developers.

\begin{figure}[H]
    \centering
    \includegraphics[width=1\linewidth,frame]{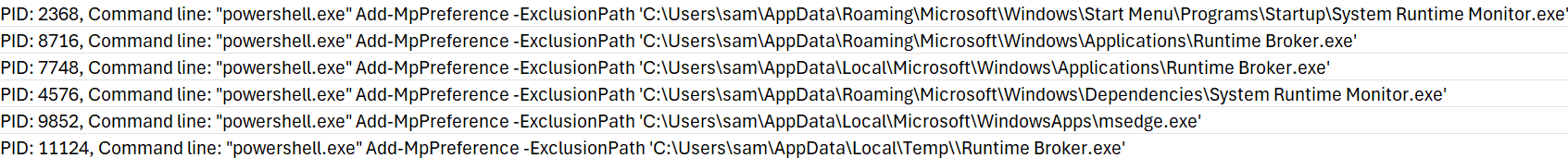}
    \caption{Defender's exclusions}
    \label{fig:125}
\end{figure}

Furthermore, it is also interesting how the \textbf{\textit{TSUNAMI CLIENT}} refers to the \textit{XMRig Miner} path as \textit{\%LOCALAPPDATA\%\textbackslash Microsoft\textbackslash Windows\textbackslash Applications\textbackslash msedge.exe}, at the same time, this path is not embedded inside the \textbf{\textit{Runtime Broker.exe}} code, which instead whitelists \textit{\%LOCALAPPDATA\%\textbackslash Microsoft\textbackslash WindowsApps\textbackslash msedge.exe}. It is not possible, as per the achieved analysis, to distinguish between the existence of two different payloads or a change in the attacker's behavior which was not consistent between these two applications.

After around 34 seconds of execution, identified \textit{threat} went silent for around 4 minutes. This behavior is consistent within the expected malware capabilities and what identified during the static analysis. \textbf{\textit{Runtime Broker.exe}} slows down its execution to avoid being detected within \textit{Sandbox analyses}, which usually employ shorter analysis time frames.

\begin{figure}[H]
    \centering
    \includegraphics[width=1\linewidth,frame]{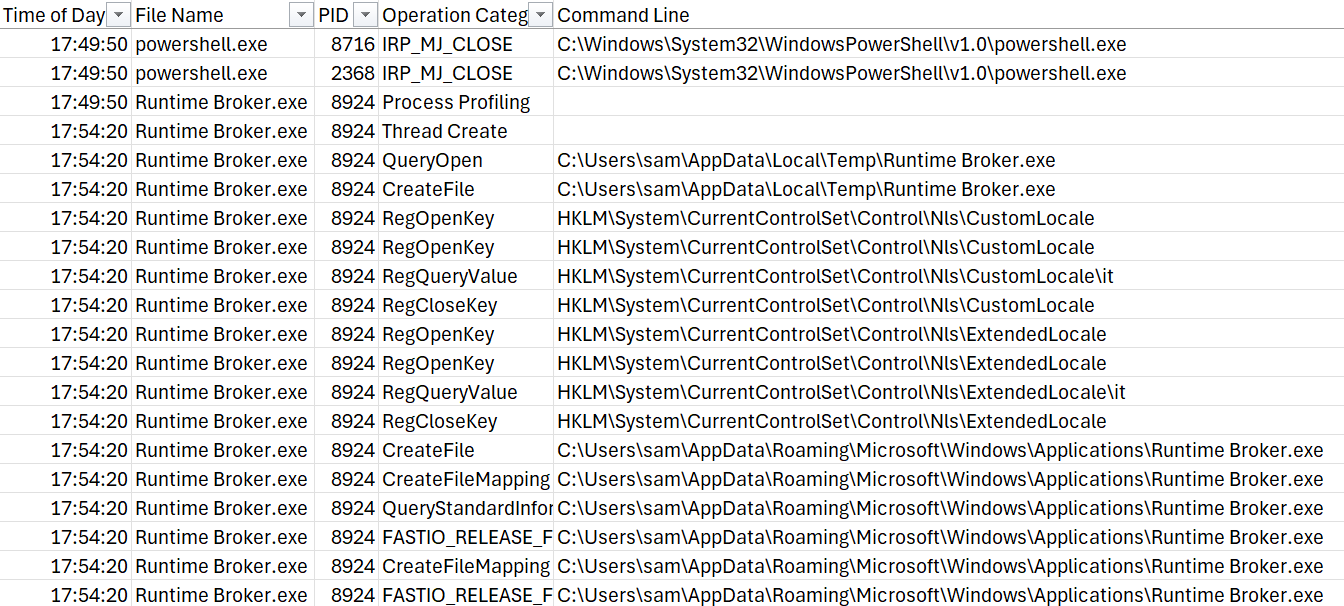}
    \caption{Malware execution stops around 17:49:50 to the restart at 17:54:20.}
    \label{fig:116}
\end{figure}

Once the malware got unfrozen, one of the first activities it carries out on the system is to drop \textbf{\textit{tor.exe}} inside path \textit{\%TEMP\%\textbackslash Runtime Broker.exe}. This executable was indeed previously whitelisted from \textit{Defender}'s scan engine and allowed to receive inbound connections from \textit{Windows Firewall}. 

\begin{figure}[H]
    \centering
    \includegraphics[width=1\linewidth,frame]{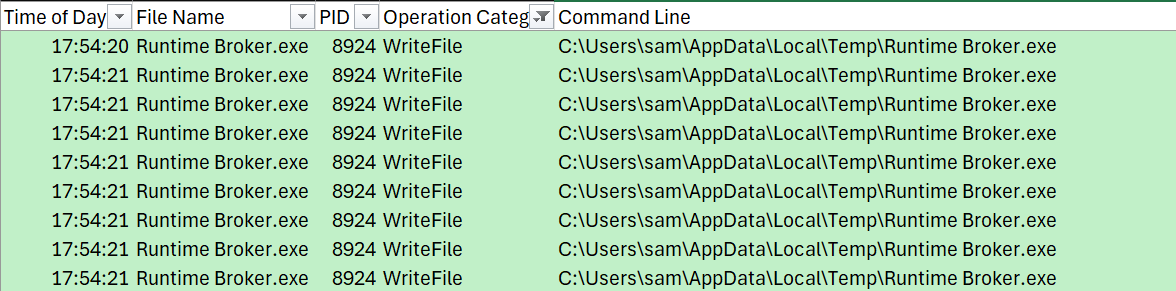}
    \caption{\textbf{\textit{Runtime Broker.exe}} drops an embedded malicious executable in \textbf{\textit{\%TEMP\%\textbackslash Runtime Broker.exe}}.}
    \label{fig:128}
\end{figure}

Once deployed, this additional payload is also executed to achieve a \textit{TOR} connections towards remote networks.

\begin{figure}[H]
    \centering
    \includegraphics[width=1\linewidth,frame]{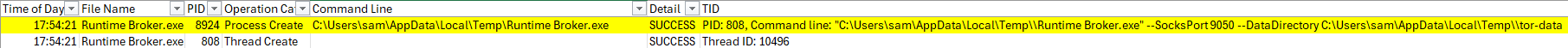}
    \caption{\textbf{\textit{\%TEMP\%\textbackslash Runtime Broker.exe} is executed}}
    \label{fig:129}
\end{figure}

From the initiation of the execution until it was terminated, spanning a total duration of eight minutes and resulting in the logging of over 241,000 events, the initial \textbf{\textit{Runtime Broker.exe}} process actively transmitted data from the host's port 63300 to port 9050, designated as the \textit{TOR SocksPort}. This activity, as depicted in \figurename~\ref{fig:129}, confirms that port 9050 was specifically utilized by the \textbf{\textit{\%TEMP\%\textbackslash Runtime Broker.exe}} as a \textit{Inter Process Communication} (\textit{IPC}) alternative, compared to standard ones (i.e. \textit{named pipes}).

\begin{figure}[H]
    \centering
    \includegraphics[width=1\linewidth,frame]{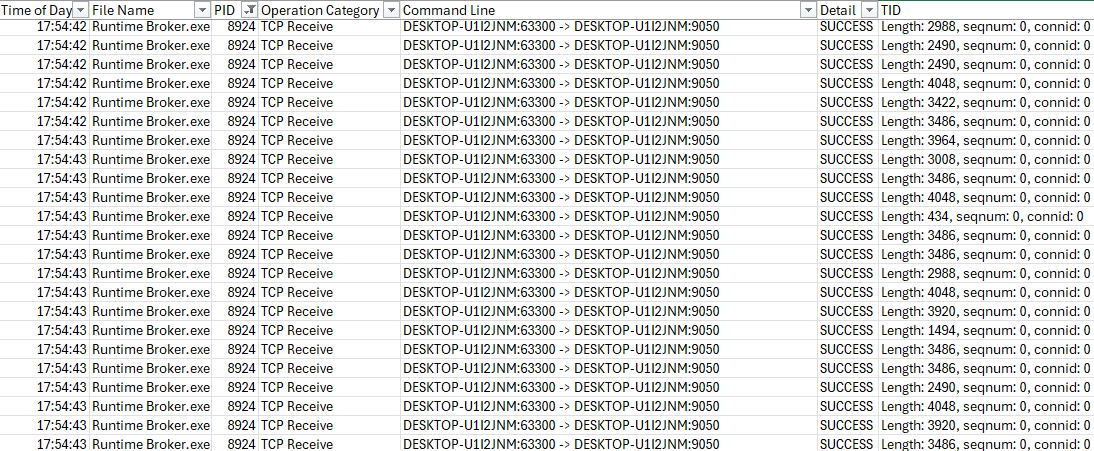}
    \caption{\textbf{\textit{Runtime Broker.exe}} sends acquired data to the local \textit{TOR SocksPort}.}
    \label{fig:130}
\end{figure}

The behavior involving the parent process sending data through a child process that runs \textit{Tor} represents an interesting and deliberate design choice to use \textit{Tor} as a local proxy to exchange data between processes. This setup offers various technical gains as well as drawbacks when compared to traditional \textit{Inter-Process Communication} (\textit{IPC}) mechanisms, such as \textit{named pipes}, \textit{shared memory}, or \textit{sockets}.

The use of \textit{Tor} as a means to handle local \textit{Inter-Process Communication} (\textit{IPC}) presents significant advantages. The primary gain lies in the inherent obfuscation and anonymization that the \textit{Tor} network provides. By routing data between processes over \textit{Tor}, the malware developer ensures that even local communication appears as part of a legitimate \textit{Tor} network flow. This not only obfuscates the purpose of the communication but also effectively anonymizes its endpoints, making network-based detection difficult. This is particularly effective because network analysis often focuses on identifying unusual connections to external addresses, while \textit{Tor} is widely recognized for privacy purposes, which may lead security tools to treat it with less scrutiny. Furthermore, by communicating over a local \textit{SOCKS proxy} on port 9050, the malware can easily convert internal messages into externally routable data, offering a seamless transition between local activity and external control or exfiltration.

This separation between the parent process (responsible for payload execution or information gathering) and the child process running \textit{Tor} as a proxy also creates a modular approach. In software design, modularity provides flexibility and scalability, which allows each component to be independently modified or updated without affecting the overall functionality. In this scenario, the \textit{Tor proxy} module handles network anonymity, while the parent process focuses on the core malicious operations. This architecture also decouples the anonymization and routing logic from the malicious payload itself, allowing for greater flexibility and code reuse. The \textit{Tor} process can be used by multiple malicious modules, potentially even in parallel, to handle diverse communication needs, which increases the versatility of the malware.

Another important advantage is the simplicity of implementation for cross-platform compatibility. \textit{Tor}-based local communication relies on \textit{network sockets}, which are inherently cross-platform. This means the malware developer can easily adapt the code to work on different operating systems (e.g., Windows, Linux, macOS) with minimal changes. This contrasts sharply with named pipes, which are Windows-specific and require entirely different implementations if the malware is to function on a non-Windows environment. By using \textit{Tor} and \textit{network sockets}, the malware becomes highly adaptable, reducing development overhead for maintaining multiple versions of the same malware for different operating systems.

However, despite these advantages, using \textit{Tor} as a \textit{local proxy} for \textit{IPC} also comes with some drawbacks that must be considered. One of the fundamental drawbacks is the inherent overhead associated with using the \textit{Tor network}. \textit{Tor}’s routing mechanism is designed to provide anonymity by encrypting and routing traffic through multiple nodes, which introduces latency and computational overhead. Even though the \textit{Tor} proxy in this scenario is operating locally, it still retains the characteristics of the network’s design, which may result in slower communication between processes compared to the direct nature of \textit{named pipes} or \textit{shared memory}. Standard \textit{IPC} mechanisms, like \textit{named pipes} or \textit{shared memory}, are optimized for \textit{low-latency}, \textit{high-throughput} data exchange between processes on the same machine. \textit{Tor}, on the other hand, is optimized for privacy, which means performance is not a priority.

Additionally, using \textit{Tor} introduces complexity, both in deployment and maintenance. The \textit{Tor client} requires certain configurations, such as creating and managing the data directory, handling key files, and maintaining network state. This setup may increase the chance of detection by endpoint monitoring tools that look for non-standard directory structures or unauthorized executables, especially when these executables exhibit behavior associated with network anonymization. In a scenario where security policies are configured to monitor for unauthorized use of \textit{Tor} or similar software, such behavior may raise an alarm, leading to further investigation.

From a technical standpoint, using \textit{Tor} also poses risks of failure related to network components. For example, if the local \textit{Tor} process crashes or is terminated by endpoint security software, the entire communication channel would be disrupted, effectively disabling any data flow between the parent and child processes. In contrast, \textit{IPC} mechanisms like named pipes or shared memory are more tightly integrated into the operating system, and thus less prone to being disrupted by network-related issues. This dependence on the local \textit{Tor} process introduces an additional point of failure that may make the malware less resilient in certain environments.

Using \textit{Tor} as a local means of inter-process data exchange also complicates the task of maintaining persistence. \textit{Persistence mechanisms} like \textit{registry modifications} or \textit{scheduled tasks} must be crafted to not only ensure that the malware payload is reinstated after a reboot, but also that the \textit{Tor} component remains operational. If the \textit{Tor client} is blocked, disabled, or deleted, the entire communication strategy collapses. This makes the malware inherently more brittle compared to implementations relying on more native \textit{IPC} approaches, where persistence and functionality could be maintained more seamlessly within the operating system's standard features.

Furthermore, the use of \textit{Tor} introduces a visibility challenge for the malware itself. Network security analysts and advanced detection tools often flag \textit{Tor}-related processes or network activity for closer examination, given \textit{Tor}’s common use by malware for command-and-control communication. In an environment where network monitoring is performed actively, the presence of a \textit{Tor client}, even if it is just used locally, can serve as an \textit{Indicator of Compromise} (\textit{IoC}) and might invite forensic analysis of the host system. Traditional \textit{IPC} methods, such as \textit{named pipes}, tend to blend in with other operating system activity, making them inherently more covert from an analyst’s perspective.

In conclusion, the decision to use \textit{Tor} as a \textit{local proxy} for \textit{Inter-Process Communication} involves a trade-off between the desire for anonymity and modularity versus the efficiency and resilience provided by standard \textit{IPC} mechanisms. The advantages of using \textit{Tor} include enhanced anonymity, modular separation of network responsibilities, and cross-platform adaptability. However, these benefits come at the cost of increased complexity, reduced communication efficiency, and the risk of raising suspicions due to the inherently recognizable and often monitored presence of Tor components. This approach is effective in highly targeted attacks where the benefits of obfuscation and anonymity outweigh the drawbacks, but it may be counterproductive in environments with strong network monitoring and endpoint protections, where the presence of \textit{Tor} can itself trigger alerts.

At the same time, the \textit{Tor Client} performs remote connections towards \textit{TOR} nodes and employing \textit{DGA} domains to hide its real destination.

\begin{figure}[H]
    \centering
    \includegraphics[width=1\linewidth,frame]{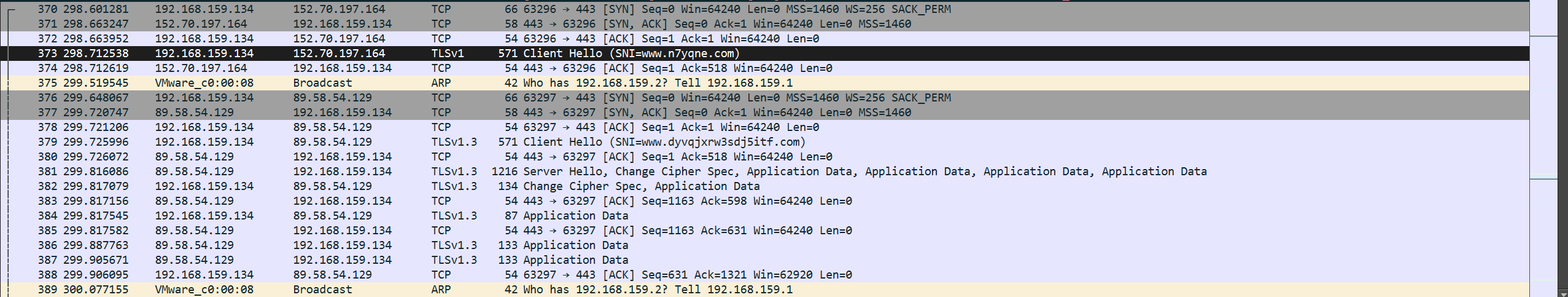}
    \caption{\textit{TOR Client} connecting towards \textit{TOR Network}.}
    \label{fig:162}
\end{figure}

By trying to load the executable inside \textit{ILSpy}, it is also possible to gather the presence of the \textit{DotNetTor DLL} (v.2.3.3.0) as an additional reference to the discussion provided above.

\begin{figure}[H]
    \centering
    \includegraphics[width=0.8\linewidth,frame]{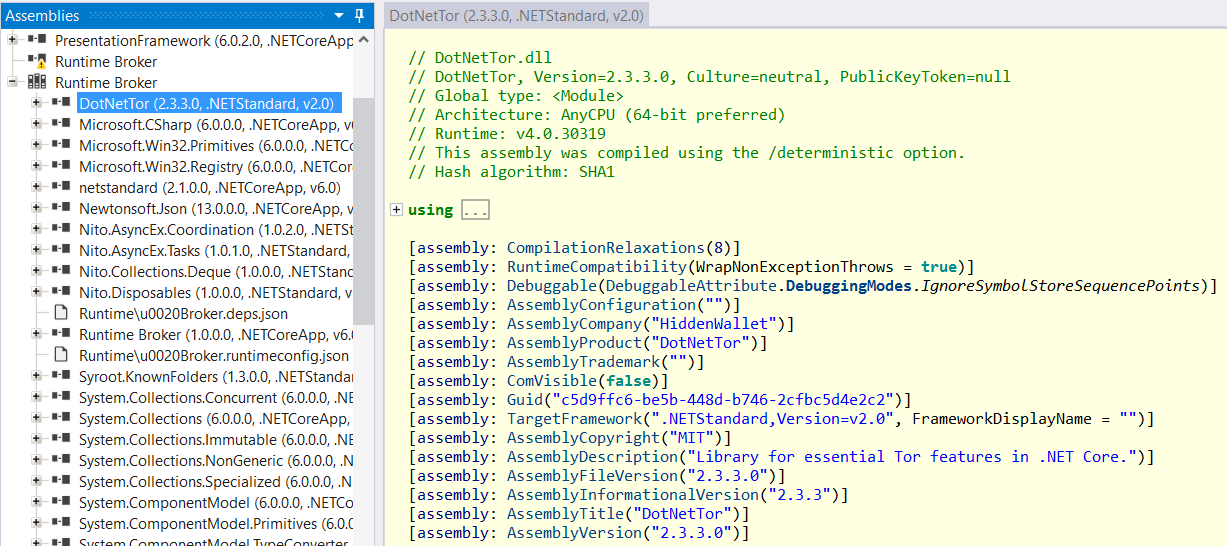}
    \caption{\textbf{\textit{Runtime Broker.exe}} implements \textit{DotNetTor} library.}
    \label{fig:140}
\end{figure}

In conclusion, the observed execution demonstrates the malware's primary objective: to comprehensively \textit{map the victim's system asset}, \textit{exfiltrate valuable information}, and \textit{deploy additional payloads}. However, it is evident that the malware's capabilities extend beyond those exhibited during this analysis. This observation suggests that either prolonged analysis durations are required or that certain features, such as \textit{Process Injection} or \textit{Shellcode Execution}, necessitate activation via attacker-issued commands.

\subsection{Sixth Stage}
\begin{figure}[H]
    \centering
    \includegraphics[width=1\linewidth,frame]{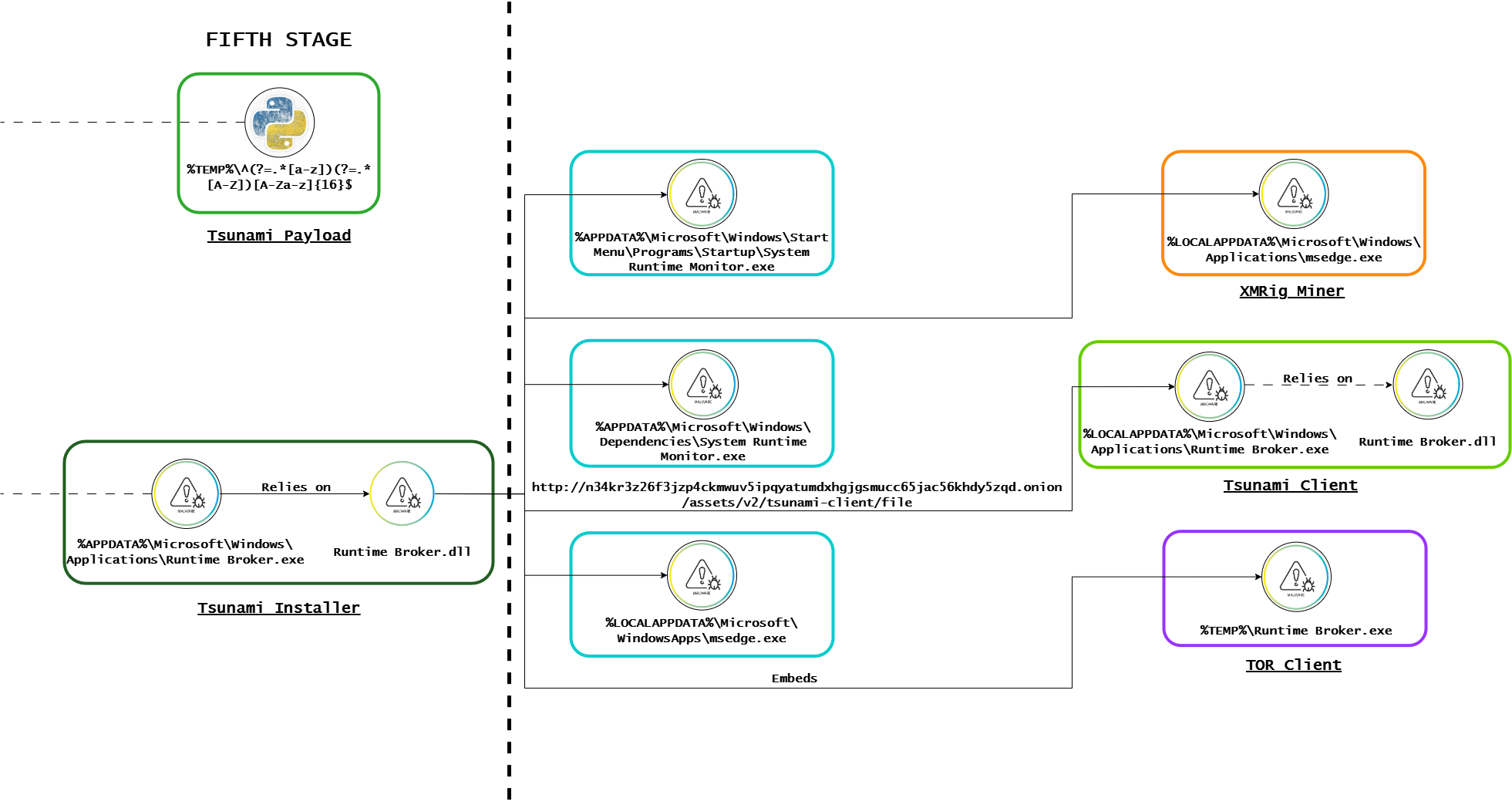}
    \caption{Moving from Fifth-Stage to Sixth-Stage.}
    \label{fig:Sis}
\end{figure}

\subsubsection{Code Obfuscation}
With respect to different six executables identified as possible additional \textit{threats}, only one of them was actively deployed on the analyzed system, \textbf{\textit{\%TEMP\%\textbackslash RuntimeBroker.exe}}, \textbf{\textit{tor.exe}} and is not a packed executable. On teh other hand, it is of interest to analyze the embedded and not used \textbf{\textit{tsunami\_payload.dll}}

\subsubsection{Code Analysis - tsunami\_payload.exe}

As with the previously identified executable, this additional payload similarly embeds a \textit{.NET DLL} within its code. This practice reflects a recurring design choice by the threat actors, indicating a preference for incorporating modular components directly into their executables. By embedding such a library, the attackers can encapsulate specific functionalities, likely to maintain modularity and ensure that critical operations remain within the same binary, reducing dependencies on external files.

The inclusion of a \textit{.NET DLL} suggests that the payload is leveraging the capabilities of the .NET framework to implement complex functionalities, which may include system-level operations, network communication, or further stages of malicious behavior. This approach enables the attackers to streamline their deployment process, as the embedded library eliminates the need for downloading or unpacking additional resources during runtime, which could otherwise expose the malware to detection.

However, the embedded nature of the \textit{.NET DLL} also presents opportunities for static analysis. Analysts can isolate and extract the library for closer examination, potentially uncovering the specific functionalities it provides or its interactions with the larger payload. Such insights could offer valuable intelligence into the attacker’s objectives, methodologies, or even allow for the creation of signatures to detect the malware.

The reuse of this technique in multiple payloads underscores the attackers’ methodical approach to constructing their malware, emphasizing modularity and reusability across their toolset. It also raises questions about the specific role and necessity of embedding such a library in this particular case, suggesting either a deliberate redundancy to ensure functionality or a potential oversight during the payload’s development.

\begin{figure}[H]
    \centering
    \includegraphics[width=0.9\linewidth,frame]{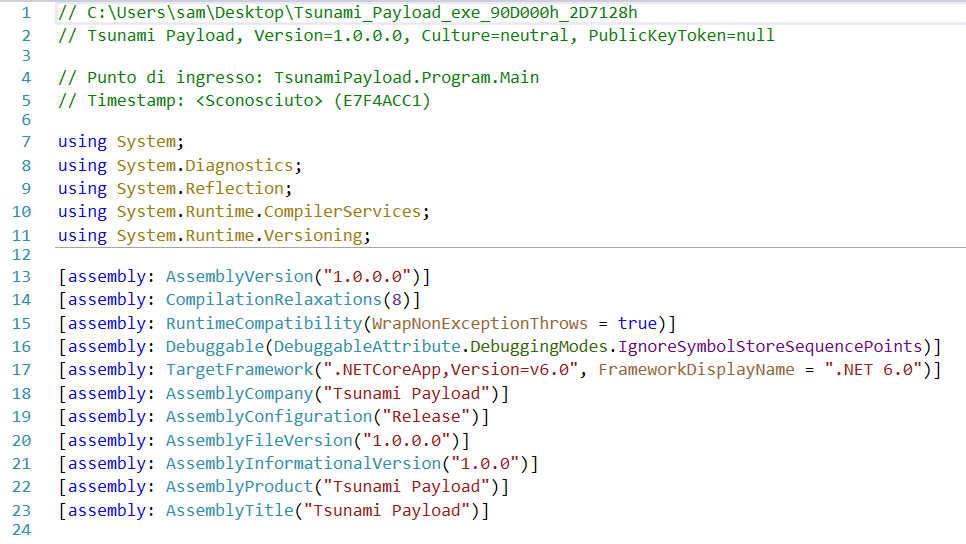}
    \caption{Overview of the \textbf{\textit{TSUNAMI PAYLOAD}} embedded \textit{.NET DLL}}
    \label{fig:163}
\end{figure}

The code demonstrates clear intentions to \textit{disable system security features}, \textit{establish persistence through a scheduled task}, initiate \textit{Tor-based communication}, and \textit{send telemetry data} to a remote server.

The \textit{Main} method initializes the program by calling the \textit{Meta.Init} function with the usage type set to \textit{TsunamiPayload}, signaling its role within the malware’s architecture. It then invokes the \textbf{\textit{Start()}} method, which orchestrates the core functionality of the payload. It begins by disabling\textit{ Windows Defender} and \textit{Firewall} through the \textbf{\textit{DisableWindowsSecurity()}} function, which leverages the \textit{AntiMalware} class. This ensures that critical security mechanisms are neutralized, allowing the malware to operate with minimal resistance and performs it in the same way as it was achieved previously by the \textbf{\textit{Runtime Broker.dll}}.

\textit{Persistence} is established by creating a \textit{scheduled task} named \textit{Runtime Broker}. Using the \textit{TaskService} library, the malware registers this task to execute the previous stage \textbf{\textit{Runtime Broker.exe}}, \textbf{\textit{TSUNAMI INSTALLER}}, located in \textit{AppData Roaming}. This ensures the payload is executed at every user logon, effectively embedding itself into the system's startup process. The configuration of the task, such as enabling it to run with \textit{administrative privileges} (\textit{RunLevel = 1}) and allowing multiple instances, highlights the attacker’s efforts to ensure resilience and continuous operation.

After establishing persistence, the \textit{TorProxy} component is installed and started, while \textbf{\textit{TelemetryUploader.SendApplicationLogs()}} is used to share telemetry data within the \textit{C2 server}. All of these actions perfectly mimic what was previously achieved with \textbf{\textit{Runtime Broker.dll}}.

Error handling within the Start method ensures the program remains functional even if certain operations, such as creating the scheduled task, fail. However, the logging of success messages for failed operations (\textbf{\textit{Logger.LogSuccess()}} in the catch block) appears to be a misleading or incorrectly implemented feature, possibly intended to confuse or mislead analysts.

\begin{figure}[H]
    \centering
    \includegraphics[width=0.9\linewidth,frame]{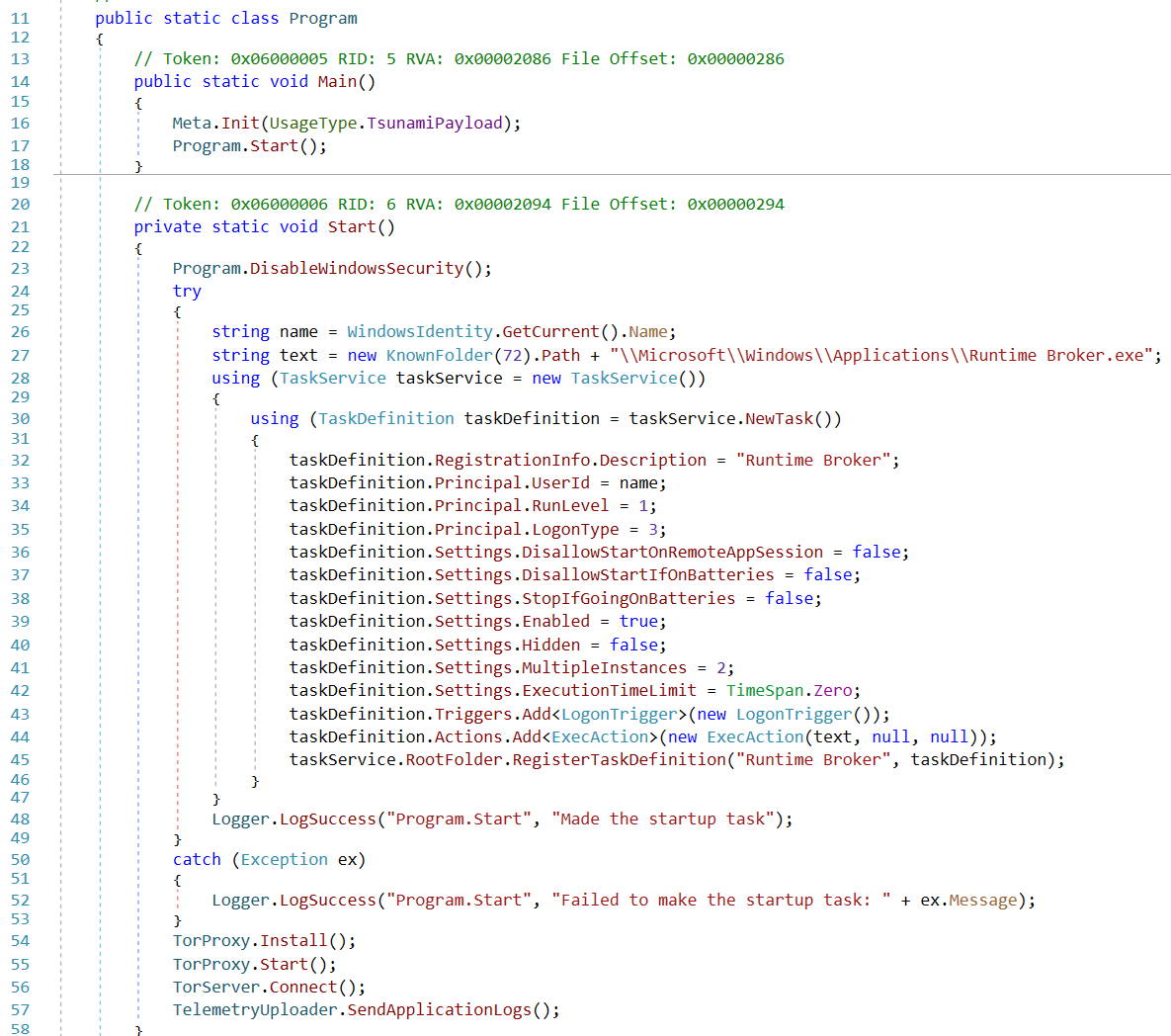}
    \caption{\textbf{\textit{Tsunami\_payload.dll}} \textit{Main} method}
    \label{fig:164}
\end{figure}

In summary, the \textbf{\textit{tsunami\_payload.dll}} performs a narrowed subset of the actions seen in its preceding stage, while embedding a significant portion of the same source code. Despite this overlap, a few critical differences are notable. One of the most significant changes is the method of \textit{persistence}, which is now achieved through the creation of a \textit{scheduled task} specifically targeting the \textbf{\textit{TSUNAMI INSTALLER}}. This mechanism ensures that the installer is executed at every user logon, embedding the payload firmly into the system’s startup sequence.

Another key distinction lies in the selective \textit{whitelisting} of executables. Unlike previous stages, where broader security exceptions were made, this stage restricts the whitelist to a more curated set of executables. This modification could reflect an attempt to minimize detection or streamline the malware’s operations by focusing only on components deemed essential for its functionality.

These changes highlight a potential evolution in the attacker’s methodology, aiming for efficiency and stealth while maintaining the core capabilities of the malware. The persistence mechanisms, combined with the adjusted scope of whitelisting, indicate a refined approach to ensuring the payload’s longevity and operational success on compromised systems.

\begin{itemize}
    \item \textit{\%APPDATA\%\textbackslash Microsoft\textbackslash Windows\textbackslash Start Menu\textbackslash Programs\textbackslash Startup\textbackslash System Runtime Monitor.exe}
    \item \textit{\%APPDATA\%\textbackslash Microsoft\textbackslash Windows\textbackslash Applications \textbackslash Runtime Broker.exe}
    \item \textit{\%LOCALAPPDATA\%\textbackslash Microsoft\textbackslash Windows\textbackslash Applications\textbackslash Runtime Broker.exe}
    \item \textit{\%LOCALAPPDATA\%\textbackslash Microsoft\textbackslash Windows\textbackslash WindowsApps\textbackslash msedge.exe}
\end{itemize}

\newpage

\section{Additional Analysis of Attacker's Infrastructure}
By moving around attacker's Webserver hosted at 86.104.74[.]51 it has been possible to gather additional information on tits setup, by looking at the \textit{PHPInfo} page. This provides a detailed overview of the attacker's server environment, exposing vulnerabilities and potential exploitation points that are critical for tracking their infrastructure. By correlating this information with the activity and characteristics of the identified \textit{IP}s, a coherent picture of the attacker's tactics, techniques, and infrastructure management emerges.

The server hosting the \textit{PHPInfo} page operates on \textit{Windows Server 2016} and employs a lightweight \textit{XAMPP} stack, consisting of \textit{Apache 2.4.58} and \textit{PHP 8.0.30}. This configuration points to a possible development or staging environment, as indicated by the paths (\textit{C:/xampp/php, C:/xampp/apache}) and default settings, such as \textit{postmaster@localhost} for the server administrator. The exposure of the \textit{PHPInfo} page itself demonstrates poor operational security, which either reflects an oversight or deliberate disregard for stealth, potentially indicating a rushed or less sophisticated deployment.

\begin{figure}[H]
    \centering
    \includegraphics[width=0.9\linewidth,frame]{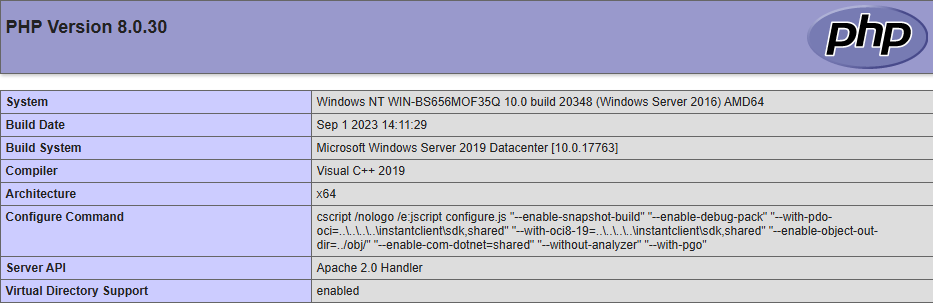}
    \caption{Overview of the \textbf{\textit{PHPInfo()}} available on attacker's main Webserver.}
    \label{fig:131}
\end{figure}

Further examination reveals that key configurations, such as the enabled \textit{allow\_url\_fopen} directive and permissive upload and execution parameters (\textit{upload\_max\_filesize=1024M}), could facilitate malicious activities like remote file inclusion or large payload execution. The combination of high resource allowances, enabled error reporting, and lack of critical function restrictions suggests that the server is configured to handle resource-intensive or long-running scripts, such as those used for data exfiltration or payload unpacking. The presence of multiple enabled PHP extensions, including \textit{cURL}, \textit{zlib}, and \textit{bz2}, further demonstrates capabilities for advanced data handling and compressed payload manipulation, which are hallmarks of modern malicious operations.

The \textit{PHPInfo} file also provides insight into the network environment, exposing registered streams and protocols that include \textit{HTTP2}, \textit{SSL/TLS}, and other transports. These details suggest that the server is equipped for complex and secure network communication, a requirement for modern \textit{Command-and-Control} (\textit{C2}) frameworks. Such configurations enhance the attacker's ability to execute multi-layered campaigns, though they also offer indicators that can be leveraged for detection and tracking.

\begin{figure}[H]
    \centering
    \includegraphics[width=0.9\linewidth,frame]{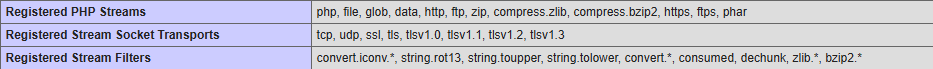}
    \caption{Some additional parameters}
    \label{fig:132}
\end{figure}

Furthermore, this same configuration file allows to gather very interesting additional insights also on the Windows system running behind this Webserver. The asset itself is a Windows-based server operating with \textit{administrative privileges}, named \textit{WIN-BS656MOF35Q}, and configured to allow \textit{Remote Desktop Protocol} (\textit{RDP}) access. The presence of \textit{SESSIONNAME} set to \textit{RDP-Tcp\#0} indicates that the attacker is actively managing the server using \textit{RDP}, originating from a client machine named \textit{DESKTOP-V0U7LU6}. The use of RDP for connecting to the server implies that the attacker requires manual control, allowing them to directly execute commands, manage files, and make real-time adjustments to the malicious infrastructure.

The client machine name, \textit{DESKTOP-V0U7LU6}, appears to follow a default Windows naming convention, suggesting that this client system is either newly configured or intentionally generic. This default configuration could indicate a throwaway device being used for malicious purposes while minimizing any personalized trace that might link back to the attacker's identity or reveal additional information. This is a common tactic used to maintain operational security (\textit{OPSEC}), as a non-descriptive system name helps avoid drawing attention during investigations or when interacting with compromised systems.

\begin{figure}[H]
    \centering
    \includegraphics[width=0.9\linewidth,frame]{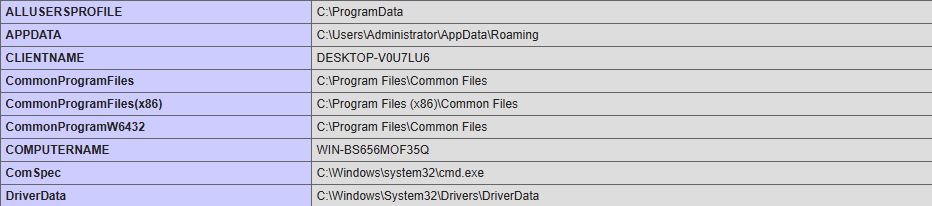}
    \caption{Information about the underlying Windows Server system.}
    \label{fig:133}
\end{figure}

The server hardware itself is a powerful Windows machine, with the \textit{PROCESSOR\_AR\\CHITECTURE} set to \textit{AMD64} and \textit{NUMBER\_OF\_PROCESSORS} set to 32. The processor is identified as \textit{Intel64 Family 6 Model 79 Stepping 1, GenuineIntel}, highlighting that this asset has substantial computational resources, possibly indicating a server-grade machine or a high-end workstation. This level of computing power suggests that the system is capable of supporting demanding operations, such as \textit{encryption}, \textit{network relays}, or \textit{multi-threaded control} of a large number of compromised clients.

The attacker has configured the server using \textit{XAMPP}, a popular development environment that includes \textit{Apache}, \textit{PHP}, and \textit{MySQL}. This configuration is evident from paths like \textit{DOCUMENT\_ROOT} set to \textit{C:/xampp/htdocs} and the use of \textit{PHP} version \textit{8.0.30}, \textit{Apache} \textit{2.4.58}, and \textit{OpenSSL 3.1.3}. The use of \textit{XAMPP} is particularly significant as it points to a development or testing server configuration that may not be appropriately secured for a production environment. \textit{XAMPP} is designed for ease of use, and default configurations often lack the security features necessary to protect the system in a live deployment. This provides a window of opportunity for defenders, as these configurations may expose vulnerabilities or lead to misconfigurations that could be exploited to regain control of the system or disrupt the attacker's operations.

\begin{figure}[H]
    \centering
    \includegraphics[width=0.9\linewidth,frame]{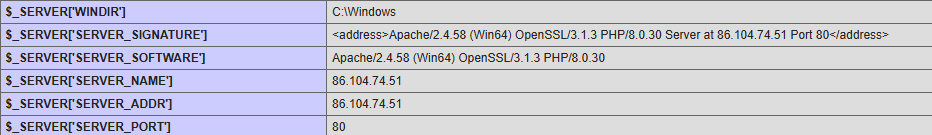}
    \caption{Server's Software lists}
    \label{fig:134}
\end{figure}

The server \textit{IP address} is confirmed by the \textit{SERVER\_NAME}, \textit{SERVER\_ADDR}, and \textit{HTTP\_HOST} variables. The \textit{SERVER\_SIGNATURE} reveals the software stack being used, which includes \textit{Apache} and \textit{PHP}, while running on a Windows (\textit{Win64}) environment. This stack's details are critical for identifying potential vulnerabilities that may be exploited by defenders. Additionally, the use of \textit{HTTP} on port 80 (\textit{SERVER\_PORT} set to 80) implies that the server may not enforce secure (\textit{HTTPS}) communications, leaving it potentially vulnerable to \textit{Man-in-the-Middle} (\textit{MITM}) attacks.

The presence of a web-based dashboard (HTTP\_REFERER set to \textit{hxxp[:]//86.104.74[.]\\51/dashboard/}) implies that the server is being used to host a control panel, which may be central to managing the infrastructure or interacting with compromised clients. Such dashboards are often used in \textit{Command-and-Control} (\textit{C2}) operations, providing an interface for the attacker to manage their campaigns, send commands, and exfiltrate data. The fact that this dashboard is accessible over \textit{HTTP} further suggests lax security and could provide an opportunity for defenders to exploit weaknesses in the interface or intercept unencrypted data.

In addition to \textit{XAMPP}, the presence of \textit{Node.js} and \textit{NVM} (Node Version Manager) installed on the server, with directories like \textit{C:\textbackslash Program Files\textbackslash nodejs} and \textit{C:\textbackslash Users\textbackslash Adminis\\trator\textbackslash AppData\textbackslash Roaming\textbackslash nvm} included in the system Path, suggests that the attacker is using \textit{JavaScript}-based tools or services. \textit{Node.js} is often used for executing lightweight scripts, hosting web services, or automating various aspects of a campaign. The inclusion of both \textit{XAMPP} and \textit{Node.js} illustrates the versatility of the attacker's infrastructure, which is configured to support multiple scripting environments, potentially allowing for rapid adaptation to different tasks and objectives. This highlights the server’s capability to execute multiple types of workloads, from traditional web hosting to script-based operations.

The environment variables also reveal that the attacker is operating with administrative privileges, as indicated by the USERNAME being set to \textit{Administrator} and \textit{USERPROFILE} pointing to \textit{C:\textbackslash Users\textbackslash Administrator}. Administrative privileges give the attacker a high degree of control over the server, allowing them to install additional tools, make system modifications, and persist within the system. Such privileges also suggest that the attacker might have used privilege escalation techniques to gain control over the server, possibly leveraging existing vulnerabilities or weak configurations. The presence of PowerShell modules in the \textit{PSModulePath} (\textit{C:\textbackslash Program Files\textbackslash WindowsPowerShell\textbackslash Modules}) implies that PowerShell scripts are available, which are frequently used by attackers to automate various post-exploitation tasks, including enumeration, data exfiltration, and lateral movement within the network.

\begin{figure}[H]
    \centering
    \includegraphics[width=0.9\linewidth,frame]{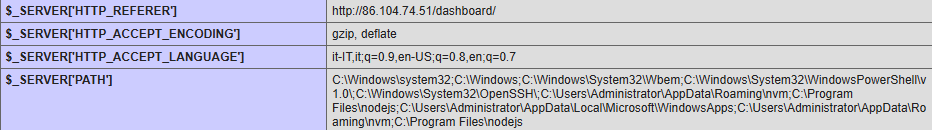}
    \caption{Server's Environmental variables}
    \label{fig:135}
\end{figure}

The \textit{CLIENTNAME}, \textit{SESSIONNAME}, and \textit{LOGONSERVER} values collectively confirm that the attacker has direct, manual access to the server, which might indicate an interest in maintaining control of the asset beyond automated scripts. This manual intervention could involve more sophisticated or targeted operations that require real-time decision-making or adjustment based on network conditions, responses from defenders, or the progress of their activities. The presence of \textit{RDP} (\textit{RDP-Tcp\#0}) further emphasizes the attacker's active presence on the system, managing and operating the infrastructure through a graphical user interface.

The network details, including the \textit{REMOTE\_ADDR} value of \textit{85.190.233[.]54}, suggest active client connections to the server, which could represent either compromised victims or an intermediate attacker device interacting with the hosted infrastructure. This interaction indicates ongoing activity, potentially involving monitoring or controlling compromised clients through the web-based dashboard or other means.

Temporary paths such as \textit{TEMP} and \textit{TMP} set to \textit{C:\textbackslash Users\textbackslash ADMINI\~1\textbackslash AppData\textbackslash Local\\\textbackslash Temp\textbackslash 2} are indicative of locations that may be used by the attacker for staging payloads or storing intermediary files before exfiltration. These directories are commonly used due to their writable nature and are easily accessible by all processes, making them ideal for temporarily holding malicious payloads without raising suspicion.

\begin{figure}[H]
    \centering
    \includegraphics[width=0.9\linewidth,frame]{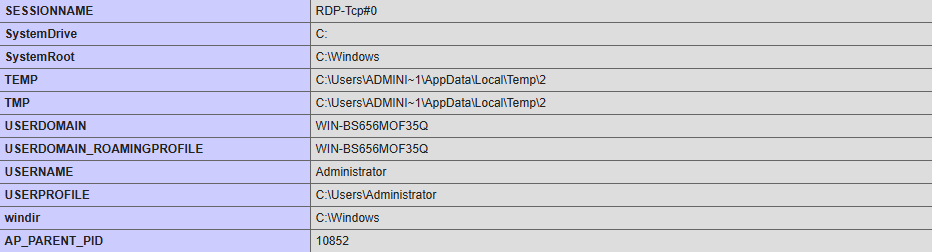}
    \caption{Server's remote \textit{RDP} connection details and \textit{Temp} folders paths.}
    \label{fig:136}
\end{figure}

In conclusion, the asset under analysis is a Windows server, powerful and versatile, configured for both web hosting and script execution, with active \textit{RDP}-based control by an attacker using administrative privileges. The server leverages \textit{XAMPP} for web services, \textit{Node.js} for scripting, and has direct, potentially insecure web interfaces that expose management capabilities through an HTTP-based dashboard. The asset is accessible via RDP from a generic client machine, indicating an effort by the attacker to maintain an active, low-profile presence. While the setup provides the attacker with significant flexibility and capability, it also exposes several security weaknesses. The use of \textit{XAMPP} with default configurations, a publicly accessible \textit{HTTP} dashboard, and reliance on \textit{RDP} all present potential points of vulnerability that could be exploited by defenders to disrupt the attacker’s control over the infrastructure, gather further intelligence, or mitigate the ongoing malicious activities.

\section{Mitigation Strategies}
To mitigate the risks posed by threats of this nature, organizations must adopt a comprehensive and proactive approach to cybersecurity. Enhancing employee awareness through regular training can significantly reduce the effectiveness of social engineering tactics, as educated staff are less likely to fall prey to deceptive schemes like fictitious job offers. Implementing advanced security solutions capable of detecting and responding to obfuscated and multi-stage malware is essential. Regular system updates and the application of security patches can close vulnerabilities that attackers might exploit.

Strengthening authentication processes by adopting multi-factor authentication can add an additional layer of security, for sensitive accounts, making unauthorized access more difficult especially for attackers exfiltrating administrative credentials from low-privilege systems. Monitoring network activity for anomalies and establishing robust incident response plans can further enhance an organization's ability to detect and respond to intrusions promptly. Collaborating with cybersecurity professionals and participating in information-sharing initiatives can help organizations stay informed about emerging threats and adapt their defenses accordingly.

By fostering a security-conscious culture and investing in advanced protective measures, organizations can better safeguard themselves against sophisticated cyber adversaries like the Lazarus Group. Remaining vigilant and adaptive is crucial in the ever-evolving landscape of cyber threats, ensuring that defenses evolve in tandem with the tactics employed by attackers.

Additionally, by taking into account identified \textit{IoCs} and \textit{TTPs}, reported inside the \textit{Appendix} section (App. \ref{App:IoC}), both a proactive approach and a Threat Intelligence based one can be implemented. These allows to track possible already established compromise and block malicious files which could be exploited by the \textit{Threat Actor} to have a foothold inside the victim's network.
\newpage
\section{Conclusion}
This report highlights a sophisticated and meticulously constructed\textit{ multi-stage threat campaign}, demonstrating technical expertise and a focused intent on long-term system compromise and financial data theft. The campaign unfolds through a series of infection stages, each building upon the last with enhanced functionality and advanced obfuscation techniques. This layered approach reflects the attackers' careful planning and understanding of security mechanisms, ensuring that each stage remains both functional and resistant to detection.

\textit{Obfuscation} emerges as a cornerstone of this campaign, with techniques such as \textit{multi-layered encoding} and \textit{control flow manipulation} employed to hinder reverse engineering and evade standard detection methods. These methods not only complicate analysis but also underscore the attackers’ efforts to protect their malware from scrutiny and countermeasures. The modular design of the malware further enhances its adaptability, allowing it to dynamically incorporate additional components, update its functionalities, and tailor its operations to specific environments. This flexibility demonstrates a level of sophistication that is characteristic of advanced threat actors.

The malware’s focus on targeting \textit{sensitive data} is particularly notable. It employs a range of techniques, including \textit{credential harvesting}, \textit{clipboard monitoring}, and \textit{direct file extraction}, to exfiltrate information such as \textit{browser-stored credentials}, \textit{cryptocurrency wallet details}, and \textit{system configurations}. This breadth of capability reflects a deliberate intent to maximize the value of compromised systems. \textit{Persistence mechanisms}, such as the creation of \textit{scheduled tasks} and the use of \textit{startup folder scripts}, further reinforce this intent, ensuring the malware remains operational even after system restarts.

An intriguing aspect of the analysis is the integration of \textit{open-source components} and legitimate tools, such as \textit{Python} and \textit{AnyDesk}, into the malware’s architecture. By embedding publicly available utilities, the attackers not only extend the malware’s capabilities but also exploit the trust associated with these legitimate tools to evade detection. However, the presence of \textit{unused code}, \textit{debugging information}, and \textit{redundant artifacts} within the malware suggests a degree of oversight or a rushed deployment. These remnants offer valuable insight into the attackers' development processes and potential areas of improvement.

The \textit{Tactics, Techniques, and Procedures} observed in this campaign strongly align with those associated with the \textbf{\textit{Lazarus Group}}, a \textit{North Korean state-sponsored} \textit{Threat Actor} known for targeting financial institutions and engaging in cyber-espionage. The campaign’s focus on \textit{financial} and \textit{cryptocurrency-related} data, combined with its advanced design and execution, aligns with the group’s established objectives and operational patterns.

The analysis underscores the growing sophistication of modern cyber threats and the necessity for enhanced defensive measures. It highlights the importance of proactive \textit{threat hunting}, robust monitoring for \textit{Indicators of Compromise}, and comprehensive user education to mitigate risks. This campaign exemplifies the evolving nature of advanced persistent threats, revealing a highly adaptive adversary capable of leveraging both technical innovation and strategic planning to achieve its objectives.
\newpage
\appendix
\section{Appendix}
\subsection{IoCs, TTPs \& Yara Rules} \label{App:IoC}
The entire set of \textit{IoCs}, \textit{TTPs} and few \textit{Yara} Rules, gathered through-out this entire analysis, are available inside the following \textit{AlienVault OTX} \href{https://otx.alienvault.com/pulse/673a203ecde8542897c3eea0}{pulse}.

\begin{figure}[H]
    \centering
    \includegraphics[width=1\linewidth,frame]{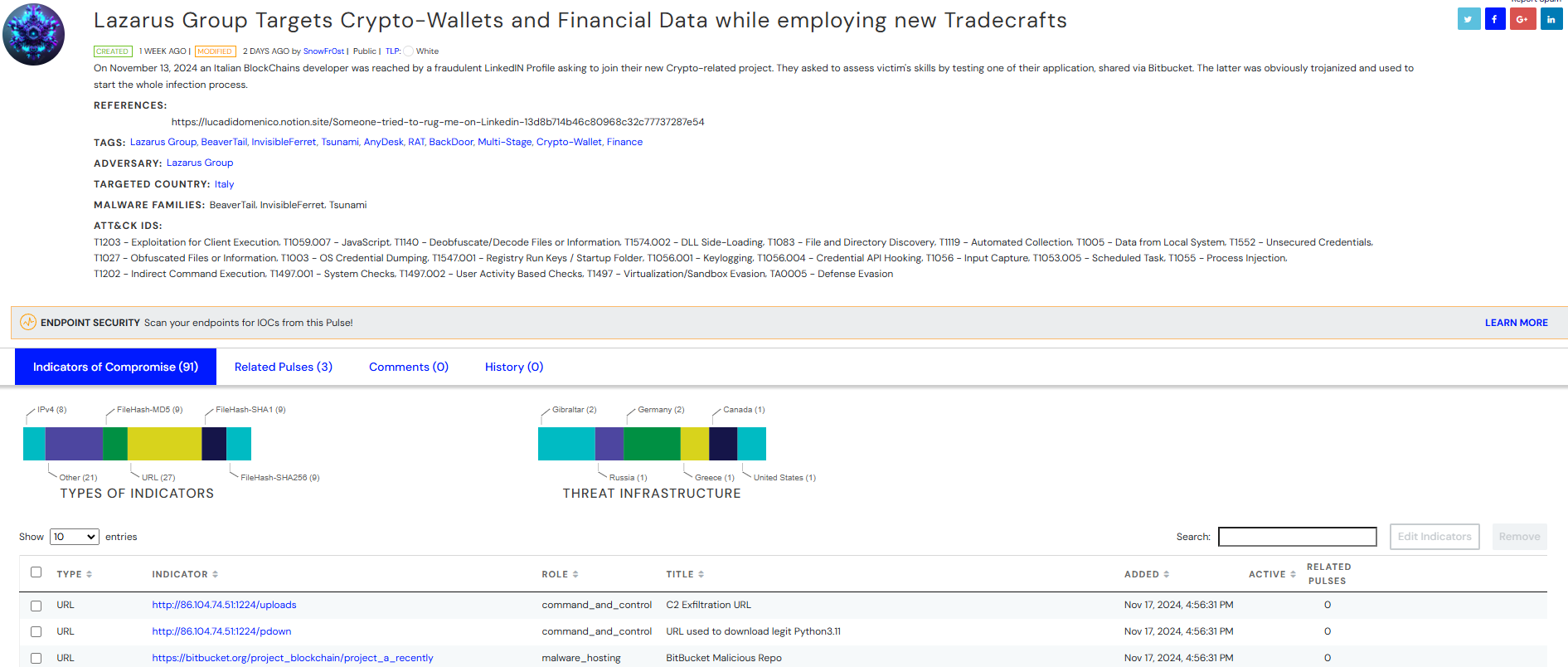}
    \caption{Overview of the \textit{AlienVault OTX pulse}}
    \label{fig:165}
\end{figure}
\newpage

\subsection{Sigma Rules}
\vspace*{\fill} 
\begin{center}
\begin{lstlisting}[language=yaml]
title: Detection of Suspicious AnyDesk File Modification and Termination via PowerShell
id: 1234abcd-5678-efgh-ijkl-9012mnopqrst
description: Detects suspicious PowerShell activity involving AnyDesk file modification and process termination when specific command patterns are observed.
status: experimental
author: Alessio Di Santo
date: 2024-11-26
logsource:
  category: process_creation
  product: windows
detection:
  selection:
    Image: '*\powershell.exe'
    CommandLine|all:
      - 'ad.anynet.pwd_hash='
      - 'ad.anynet.pwd_salt='
      - 'ad.anynet.token_salt='
      - 'taskkill /IM anydesk.exe /F'
  condition: selection
fields:
  - CommandLine
  - ParentCommandLine
  - ParentImage
  - Image
  - User
level: high
tags:
  - attack.persistence
  - attack.t1562.001
  - attack.t1098
falsepositives:
  - Legitimate administrative maintenance involving AnyDesk
mitre:
  - T1562.001
  - T1098
\end{lstlisting}
\end{center}
\vspace*{\fill} 
\newpage
\vspace*{\fill} 
\begin{center}
\begin{lstlisting} [language=yaml]
title: Detection of Suspicious Scheduled Task for Runtime Broker.exe
id: abcd1234-efgh-5678-ijkl-9012mnopqrst
description: Detects the creation of a scheduled task targeting Runtime Broker.exe located in %APPDATA%\Microsoft\Windows\Applications for persistence.
status: experimental
author: Alessio Di Santo
date: 2024-11-26
logsource:
  category: process_creation
  product: windows
detection:
  selection:
    Image: '*\powershell.exe'
    CommandLine|all:
      - 'New-ScheduledTaskAction -Execute'
      - 'Register-ScheduledTask'
      - 'TaskName "Runtime Broker"'
      - 'LogonType Interactive'
      - '*\Microsoft\Windows\Applications\Runtime Broker.exe'
  condition: selection
fields:
  - CommandLine
  - ParentCommandLine
  - ParentImage
  - Image
  - User
  - FileName
level: high
tags:
  - attack.persistence
  - attack.t1053.005
falsepositives:
  - Legitimate scheduled task creation by administrators targeting similar paths
mitre:
  - T1053.005
\end{lstlisting}
\end{center}
\vspace*{\fill} 
\newpage
\vspace*{\fill} 
\begin{center}
\begin{lstlisting}[language=yaml]
title: Detect Specific Windows Firewall Rule Exclusions
id: 5678abcd-ef01-2345-ghij-klmnopqrstuv
status: experimental
description: Detects suspicious Windows Firewall rule additions that include specific paths for exclusion, such as `Runtime Broker.exe`, `msedge.exe`, and `System Runtime Monitor.exe`.
author: Alessio Di Santo
date: 2023-11-26
logsource:
  product: windows
  service: sysmon
detection:
  selection:
    EventID: 1
    CommandLine|contains|all:
      - 'netsh advfirewall firewall add rule'
      - 'action=allow'
    CommandLine|contains:
      - '\System Runtime Monitor.exe'
      - '\Microsoft\Windows\Applications\Runtime Broker.exe'
      - '\Microsoft\Windows\Applications\msedge.exe'
      - 'C:\Users\*\AppData\Local\Temp\Runtime Broker.exe'
  condition: selection
fields:
  - CommandLine
  - Image
  - ParentCommandLine
  - User
  - HostName
falsepositives:
  - Legitimate configuration of Windows Firewall rules for trusted applications.
  - Administrative scripts for deploying or updating legitimate software.
level: high
tags:
  - attack.defense-evasion
  - attack.t1562.004
  - windows-firewall
  - netsh
  - known-folder-paths
modifications:
  - Tailored rule to focus on known suspicious paths being excluded via firewall rules.
  - Excludes benign patterns based on environment-specific baselines.
\end{lstlisting}
\end{center}
\vspace*{\fill} 
\newpage
\vspace*{\fill} 
\begin{center}
\begin{lstlisting}[language=yaml]
title: Detection of Malicious Windows Defender Exclusion Paths
id: 5678efgh-1234-abcd-ijkl-9012mnopqrst
description: Detects suspicious usage of the Add-MpPreference PowerShell command to add specific paths to Windows Defender exclusion list.
status: experimental
author: Alessio Di Santo
date: 2024-11-26
logsource:
  category: process_creation
  product: windows
detection:
  selection:
    CommandLine|contains:
      - "Add-MpPreference -ExclusionPath"
  paths:
    CommandLine|contains:
      - "\System Runtime Monitor.exe"
      - "\Microsoft\Windows\Applications\Runtime Broker.exe"
      - "\Microsoft\Windows\Applications\msedge.exe"
  condition: selection and paths
fields:
  - CommandLine
  - ParentCommandLine
  - ParentImage
  - Image
  - User
level: high
tags:
  - attack.persistence
  - attack.t1562.001
  - attack.defense_evasion
falsepositives:
  - Legitimate administrative usage
mitre:
  - T1562.001
  - T1070.006
  - T1098
\end{lstlisting}
\end{center}
\vspace*{\fill} 
\newpage
\vspace*{\fill} 
\begin{center}
\begin{lstlisting}[language=yaml]
title: Malicious System Information Collection via WMIC and Registry Queries
id: e3b8c5f4-1d2e-43d9-8748-82b8cbe3c28a
description: Detects suspicious WMIC and registry queries used for system reconnaissance or enumeration. Intended for use with SIEM aggregation to identify all activities over time.
status: experimental
author: Alessio Di Santo
date: 2024-11-26
logsource:
  category: process_creation
  product: windows
detection:
  selection_wmic_processor_name:
    CommandLine|contains: 'wmic path Win32_Processor get Name'
  selection_wmic_processor_cores:
    CommandLine|contains: 'wmic path Win32_Processor get NumberOfCores'
  selection_wmic_videocontroller:
    CommandLine|contains: 'wmic path Win32_VideoController get Name'
  selection_wmic_os:
    CommandLine|contains: 'wmic os get Caption'
  selection_reg_query_productid_32bit:
    CommandLine|contains: 'reg query "HKEY_LOCAL_MACHINE\SOFTWARE\Microsoft\Windows NT\CurrentVersion" /v ProductID'
  selection_reg_query_productid_64bit:
    CommandLine|contains: 'reg query "HKEY_LOCAL_MACHINE\SOFTWARE\Wow6432Node\Microsoft\Windows NT\CurrentVersion" /v ProductID'
  condition: selection_wmic_* or selection_reg_*
fields:
  - CommandLine
  - ParentCommandLine
  - ParentImage
  - Image
  - User
level: high
tags:
  - attack.discovery
  - attack.t1082
falsepositives:
  - Legitimate administrative tools or scripts
mitre:
  - T1082
\end{lstlisting}
\end{center}
\vspace*{\fill} 
\newpage
\subsection{Infection Chain}
\begin{figure}[H]
    \centering
    \includegraphics[width=0.95\textheight, angle=270,frame]{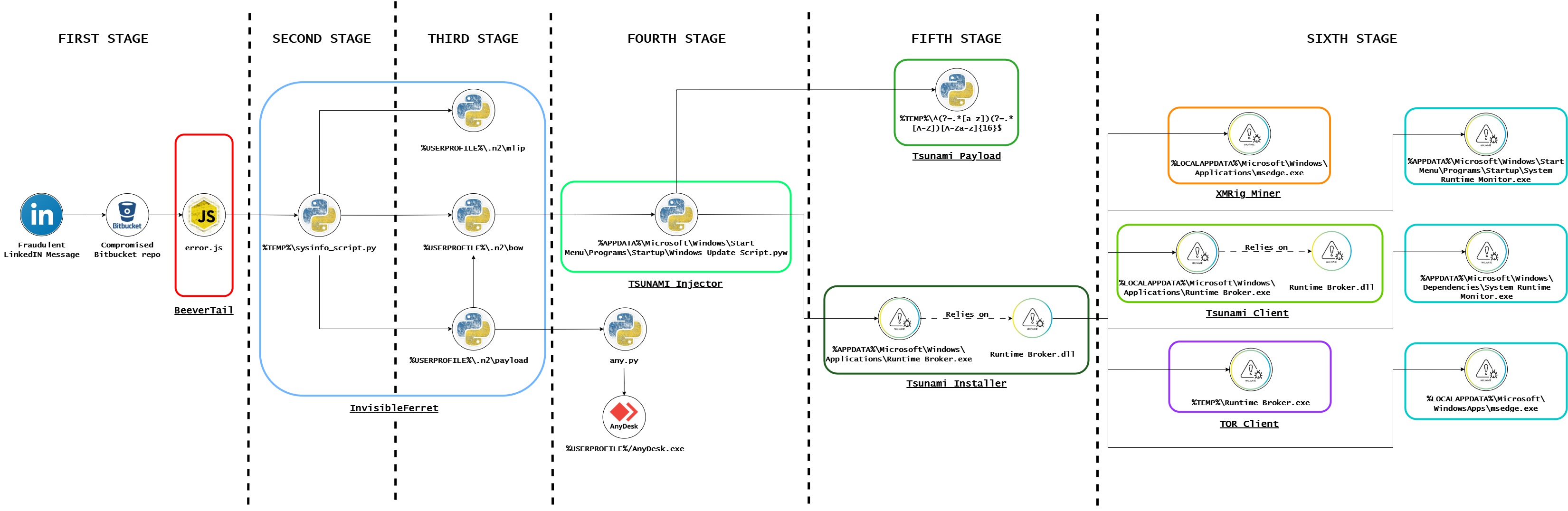} 
\end{figure}

\subsection{Diamond Model}
\begin{figure}[H]
    \centering
    \includegraphics[width=0.95\textheight, angle=270,frame]{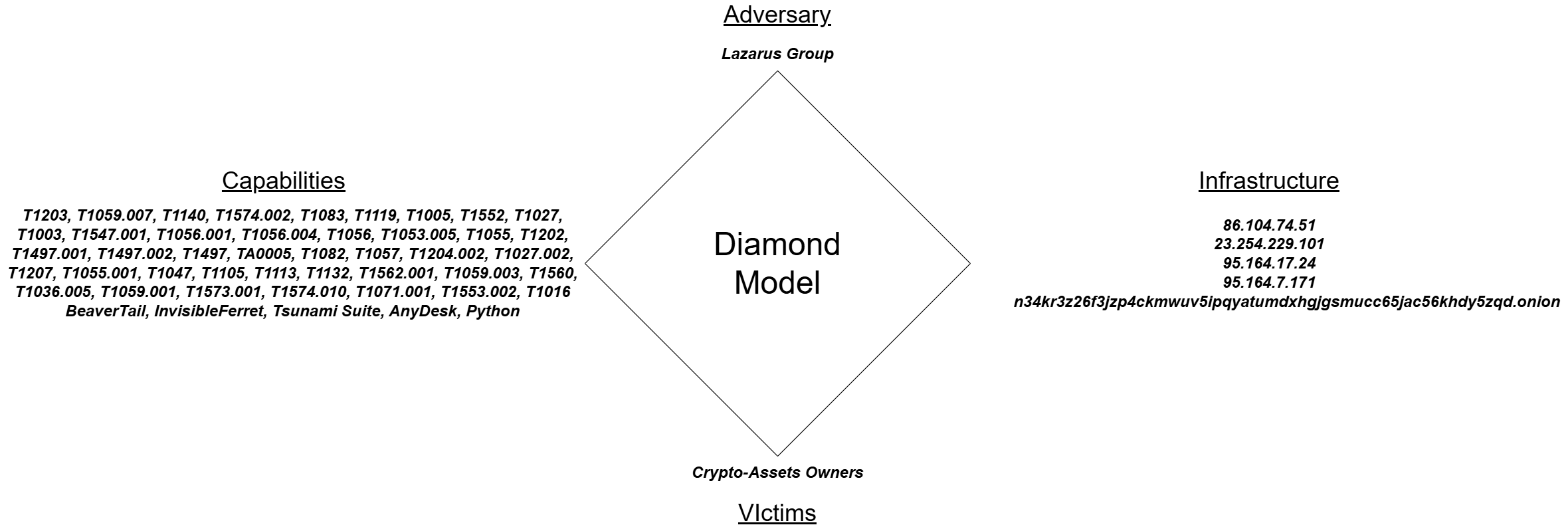}
    \label{fig:DM}
\end{figure}

\end{document}